# Unified Theory of Bivacuum, Particles Duality, Fields & Time.
## New Fundamental Bivacuum Mediated Interaction

Alex Kaivarainen

University of Turku, Department of physics
Vesilinnantie 5, FIN-20014, Turku, Finland
H2o@karelia.ru
http://web.petrsu.ru/~alexk/new_articles/index.html

## CONTENTS

# EXTENDED SUMMARY

The original Bivacuum concept developed in this work, like Dirac theory of vacuum, admit the equal probability of positive and negative energy. The Unified theory (UT) represents efforts of this author to create the Hierarchical picture of the World, starting from specific Bivacuum superfluid matrix, providing the elementary particles origination and fields, excited by particles **Corpuscle** ⇌ **Wave** pulsation.

Bivacuum is introduced, as a dynamic matrix of the Universe, composed from non mixing subquantum particles and antiparticles. The subquantum particles and antiparticles are considered, as the minimum stable vortical structures of Bivacuum with dimensions about or less than $10^{-19} m$ of opposite direction of rotation (clockwise and anticlockwise) of zero mass and charge. Their spontaneous collective paired vortical excitations represent Bivacuum dipoles in form of strongly correlated pairs: torus($\mathbf{V}^+$) + antitorus($\mathbf{V}^-$), separated by energetic gap. Three kinds of Bivacuum dipoles are named Bivacuum fermions, antifermions and Bivacuum bosons. Their torus and antitorus in primordial Bivacuum are characterized by the opposite mass and charge, compensating each other and making Bivacuum neutral with zero energy density. The radiuses of torus and antitorus of dipoles in symmetric primordial Bivacuum are equal to each other and determined by Compton radiuses of three generation of $e, mu, tau$ electrons. The infinitive number of Bivacuum fermions and antifermions: $\mathbf{BVF}^\uparrow \equiv [\mathbf{V}^+ \uparrow\uparrow \mathbf{V}^-]^i$ and $\mathbf{BVF}^\downarrow \equiv [\mathbf{V}^+ \downarrow\downarrow \mathbf{V}^-]^i$ and Bivacuum bosons: $\mathbf{BVB}^\pm = [\mathbf{V}^+ \uparrow\downarrow \mathbf{V}^-]^i$, as intermediate state between $\mathbf{BVF}^\uparrow$ and $\mathbf{BVF}^\downarrow$ form superfluid matrix of Bivacuum ($i = e, \mu, \tau$). The correlated torus $\mathbf{V}^+$ and antitorus $\mathbf{V}^-$ of these triple dipoles has the opposite energy, mass, charge and magnetic moments.

The symmetric primordial Bivacuum can be considered as the *Universal Reference Frame* (**URF**), i.e. *Ether*, in contrast to *Relative Reference Frame* (**RRF**), used in special relativity (SR) theory. The elements of *Ether* - correspond to our Bivacuum dipoles. It will be shown in our work, that the result of Michelson - Morley experiment is a consequence of *ether drug* by the Earth or Virtual Replica of the Earth in terms of our theory.

The *1st stage* of elementary particles origination is a formation of *sub-elementary* fermions or antifermions. This is a result of Bivacuum fermions and antifermions ($\mathbf{BVF}^\updownarrow$) symmetry shift towards the positive or negative energy, correspondingly, as a result their pairs rotation around common axis. Due to relativistic dependence of Bivacuum dipoles on tangential velocity of such rotation (**v**), their symmetry shift is accompanied by uncompensated *mass and charge origination*.

The *2nd stage* of elementary particles formation is a fusion of triplets $< [\mathbf{F}_\uparrow^+ \bowtie \mathbf{F}_\downarrow^-] + \mathbf{F}_\updownarrow^\pm >^i$ from sub-elementary fermions and antifermions of corresponding lepton generation ($i = e, \mu, \tau$), representing the electrons, muons and protons/neutrons. The triplets are stabilized by three factors: a) the resonance exchange interaction of Bivacuum virtual pressure waves ($\mathbf{VPW}_{q=1}^\pm)^i$ with pulsing sub-elementary fermions of Compton angular frequency: $\omega_{q=1}^i = \mathbf{m}_{q=1}^i \mathbf{c}^2/\hbar$; b) the Coulomb attraction between sub-elementary fermions of the opposite charges; c) the gluons (pairs of cumulative virtual clouds in terms of our theory) exchange between sub-elementary fermions (quarks in the case of protons and neutrons).

Both of stages of triplets formation - symmetry shift and fusion occur at Golden mean (GM) conditions: $(\mathbf{v}/\mathbf{c})^2 = 0.618$.

The fusion of elementary fermions from sub-elementary ones can be accompanied by energy release, determined by the value of mass defect. A stable triplets of sub-elementary fermions have some similarity with three Borromean rings, interlocing with each other - a symbol, popular in Medieval Italy.

The boson like *photon* in out theory $\langle 2[\mathbf{F}_\uparrow^- \bowtie \mathbf{F}_\downarrow^+]_{S=0} + (\mathbf{F}_\updownarrow^+ + \mathbf{F}_\updownarrow^-)_{S=\pm 1} \rangle$ is a result of fusion/annihilation of two triplets: [**electron** + **positron**], turning two asymmetric fermions to quasi-symmetric boson. More common way of photons origination is due to acceleration of elementary charges - triplets, following by sufficient symmetry shift in Cooper pairs: $3[\mathbf{BVF}^\uparrow \bowtie \mathbf{BVF}^\downarrow]$, representing *secondary anchor sites* for [W] phase of these triplets. The latter mechanism works, for example, in the process of atoms and molecules transitions from their excited to the ground state. The electromagnetic field, is a result of Corpuscle - Wave pulsation of photons and their fast rotation with angular frequency ($\omega_{rot}$) in [C] phase, equal in symmetric Bivacuum to photons $[\mathbf{C} \leftrightarrows \mathbf{W}]$ pulsation frequency. The pair of sub-elementary fermions of photon with similar spins $\left(\mathbf{F}_\uparrow^+ + \mathbf{F}_\uparrow^-\right)_{S=\pm 1}$ determines its integer value of spin. The clockwise or anticlockwise direction of photon rotation, as respect to direction of its propagation, corresponds to spin sign: $S = \pm \hbar$.

It is shown, that the [corpuscle (C) $\leftrightarrows$ wave (W)] duality of fermions is a result of modulation of quantum beats between the asymmetric 'actual' (torus) and 'complementary' (antitorus) states of sub-elementary fermions and antifermions by de Broglie wave (wave B) frequency of these particles. The frequency of wave B is equal to frequency of $[\mathbf{C} \leftrightarrows \mathbf{W}]$ pulsations of the primary *'anchor'* Bivacuum fermion $(\mathbf{BVF}_{anc}^\updownarrow)^i$ of unpaired $\mathbf{F}_\uparrow^{\pm} >^i$ directly related to translational kinetic energy and momentum of triplets. The [C] phase of each sub-elementary fermions of triplets $< \left[\mathbf{F}_\uparrow^+ \bowtie \mathbf{F}_\downarrow^-\right] + \mathbf{F}_\uparrow^{\pm} >^i$ exists as a mass, electric and magnetic asymmetric dipoles. The total energy, charge and spin of particle, moving in space with velocity ($\mathbf{v}$) is determined by the unpaired sub-elementary fermion $\left(\mathbf{F}_\uparrow^{\pm}\right)_z$, as far the paired ones in $[\mathbf{F}_\uparrow^+ \bowtie \mathbf{F}_\downarrow^-]_{x,y}$ of triplets compensate each other. In the case of bosons, like *photons*, propagating in space with light velocity, the contribution of the rest mass is zero or very close to zero.

The $[\mathbf{C} \to \mathbf{W}]$ transition of fermions is a result of two stages superposition. *The 1st stage* is a reversible dissociation of [C] phase to Cumulative virtual cloud $(\mathbf{CVC}^{\pm})_{\mathbf{F}_\uparrow^{\pm}}$ of subquantum particles and the '*anchor*' Bivacuum fermion $(\mathbf{BVF}_{anc}^\updownarrow)$:

$$(\mathbf{I}): \ \left[\left(\mathbf{F}_\uparrow^{\pm}\right)_{\mathbf{C}} \overset{\mathbf{Recoil/Antirecoil}}{\underset{\mathbf{E,H,G-fields}}{\Longleftrightarrow}} \mathbf{BVF}_{anc}^\updownarrow + (\mathbf{CVC}^{\pm})_{\mathbf{F}_\uparrow^{\pm}} \right]_{\mathbf{W}}^i$$

*The 2nd stage* of $[\mathbf{C} \to \mathbf{W}]$ transition is a reversible dissociation of the anchor Bivacuum fermion $(\mathbf{BVF}_{anc}^\updownarrow)^i = [\mathbf{V}^+ \updownarrow \mathbf{V}^-]_{anc}^i$ to symmetric $(\mathbf{BVF}^\updownarrow)^i$ and the anchor cumulative virtual cloud $(\mathbf{CVC}^{\pm})_{\mathbf{BVF}_{anc}^\updownarrow}$, with linear dimension and frequency, equal to of de Broglie wave length and frequency of particle, correspondingly:

$$(\mathbf{II}): \ \left[\mathbf{BVF}_{anc}^\updownarrow\right]_{\mathbf{C}}^i \overset{\mathbf{Recoil/Antirecoil}}{\underset{\mathbf{E,H,G-fields}}{\Longleftrightarrow}} \left[\mathbf{BVF}^\updownarrow + (\mathbf{CVC}^{\pm})_{\mathbf{BVF}_{anc}^\updownarrow}\right]_W^i$$

This second stage of reaction of transition of [C] phase to [W] phase determines the empirical parameters of wave B of elementary particle. The relativistic effects are provided by the increasing of symmetry shift of the primary 'anchor' $\mathbf{BVF}_{anc}^\updownarrow$ with external translational velocity of particle. The effects, accompanied emission $\leftrightarrows$ absorption of cumulative virtual clouds $(\mathbf{CVC}^{\pm})_{\mathbf{F}_\uparrow^{\pm}}^i$ and $(\mathbf{CVC}^{\pm})_{\mathbf{BVF}_{anc}^\updownarrow}^i$ on the 1st and 2nd stages of $[\mathbf{C} \leftrightarrows \mathbf{W}]$ pulsation and rotation of triplets stand for origination of electric, magnetic and gravitational fields.

The 1st stage of particle duality is a consequence of the rest mass influence on dynamics of fermions. In the case of bosons, like photons, propagating in space with light velocity, the contribution of the rest mass and 1st stage to process is negligible. *The mechanism of photon duality* is determined by the 2nd stage only. In general case the process of $[\mathbf{C} \leftrightarrows \mathbf{W}]$ pulsation is accompanied by reversible conversion of rotational energy of elementary particles in [C] phase to their translational energy in [W] phase.

It is shown, that Principle of least action is a consequence of forced combinational resonance of elementary particles and quantized virtual pressure waves $(\mathbf{VPW}_{q=1,2,3}^{\pm})^i$ of Bivacuum. The latter provides propagation of wave packet of particle in [W] phase between activated *secondary anchor sites,* where the [C] phase is realized.

The mechanism of microscopic and macroscopic quantum entanglement between remote coherent particles via bundles of Virtual Guides $[\mathbf{N(t,r)} \times \sum \mathbf{VirG}_{SME}(\mathbf{S} <==> \mathbf{R})]_{x,y,z}^i$ of spin, momentum and energy is proposed also. The $\mathbf{VirG}_{SME}^i$ represent quasi one-dimensional Bose condensate, assembled form Cooper pairs of Bivacuum fermions $[\mathbf{BVF}^\uparrow \bowtie \mathbf{BVF}^\downarrow]^i$ or Bivacuum bosons $(\mathbf{BVB}^{\pm})^i$. The tuning of $[\mathbf{C} \leftrightarrows \mathbf{W}]$ pulsation of particles, necessary for entanglement is realized under $\left(\mathbf{VPW}_{q=1,2,3}^{\pm}\right)^i$ action. The bundles of Virtual Guides in superfluid Bivacuum have some similarity with vortical filaments in superfluid liquid helium and can be activated by rotating elementary particles.

It is demonstrated, that the charge and spin equilibrium oscillation in Bivacuum matrix in form of spherical elastic waves, provide the electric and magnetic fields origination. These excitations are the consequence of reversible $\left[\textit{diverging} \rightleftarrows \textit{converging}\right]$ of Cumulative Virtual Clouds $(\mathbf{CVC}^{\pm})$, involving the *recoil$\rightleftarrows$ antirecoil* effects, accompanied [*Corpuscle* ⇋ *Wave*] pulsation of sub-elementary fermions/antifermions of triplets and their fast rotation. The particle *recoil*⇌ *antirecoil* oscillation of elementary particles, responsible for electromagnetism and gravitation, are induced by their [**C** ⇌ **W**] pulsation. The most probable velocity of these oscillation for the rest mass or zero-point conditions where calculated.

The tendency of Bivacuum fermions and antifermions of *opposite* spins and charges to formation of *Cooper pairs* [**BVF**$^\uparrow_\downarrow$⋈ **BVF**$^\downarrow_\uparrow$]$^i_{as}$, decreasing the resulting Bivacuum dipoles asymmetry with *decreasing* the separation between particles of opposite charges, is responsible for Coulomb attraction between particles. The Coulomb repulsion between particles of similar sign of charge is also a result of Bivacuum to decrease its resulting asymmetry in space between charges by *increasing* the separation.

The mechanism of Pauli repulsion between triplets of *similar* spins is shown to be a consequence of the effect of excluded volume, tending to be occupied by two **CVC**$^\pm$ at the same time emitted by unpaired sub-elementary fermions of the same phase of [**C** ⇌ **W**] pulsation. The energy of Pauli repulsion is about $1/\alpha \simeq 137$ times stronger, that Coulomb interaction. The Pauli repulsion is most effective on the distances between fermions equal or less than de Broglie wave length of these particles: $\lambda_B = h/\mathbf{p}$.

The magnetic field and **N** or **S** poles origination is a result of shift of equilibrium [**BVF**$^\uparrow$⇌ **BVB**$^\pm$⇌ **BVF**$^\downarrow$] to the left or right, correspondingly, depending on clockwise or anticlockwise rotation of triplets. The direction of fermions rotation is correlated with direction of their propagation and sign of charge. The magnetic poles attraction or repulsion, like in the case of Coulomb interaction is also dependent on possibility of *Cooper pairs* of Bivacuum dipoles in space between them to assembly or disassembly. However, this process can be independent on the internal symmetry shifts between torus and antitorus of **BVF**$^\uparrow$ or **BVF**$^\downarrow$, responsible for electric field.

The gravitational waves and G-field are the result of positive and negative energy virtual pressure waves excitation (**VPW**$^+_q$ and **VPW**$^-_q$)$^i$ by the in-phase [**C** ⇌ **W**] pulsation of unpaired sub-elementary fermion **F**$^\pm_\uparrow$⟩, counterphase with pulsation of paired ones [**F**$^-_\uparrow$ ⋈ **F**$^+_\uparrow$]$^i$ in elementary particles. These virtual pressure waves provide the *attraction* or *repulsion/antigravitation* between pulsing remote particles, depending on the phase shift of their pulsation. Our gravitation theory has a common with hydrodynamic *Bjerknes* attraction or repulsion force between pulsing spheres. The *antigravitation* generated by counterphase [**C** ⇌ **W**] pulsation of unpaired sub-elementary fermion **F**$^\pm_\uparrow$⟩ in very remote objects can be responsible for mysterious *negative pressure energy or dark energy*. For the other hand, the potential positive/attraction gravitational energy of huge number of symmetric Bivacuum dipoles exists even in the absence of matter in the *empty space*. This energy can be provided by positive and negative virtual pressure waves, excited as a result of symmetric transitions of tori and antitori of **BVF**$^\updownarrow$. These transitions, compensating the energy of each other, can be considered as zero-point oscillation of Bivacuum dipoles, in contrast to zero-point oscillation of elementary particles at **T** = **0**, induced by their [**C** ⇌ **W**] pulsation. This attraction effect of *'dark matter'*, provided by these symmetric oscillation of Bivacuum dipoles, is determined by sum of the absolute values of energies of excited torus and antitorus of **BVF**$^\updownarrow_q$ = [**V**$^+$ ⇕ **V**$^-$]$_q$:

$$E^0_G = \sum^{N\to\infty} \beta^i(\mathbf{m}^+_V + \mathbf{m}^-_V)^i\mathbf{c}^2 = \sum^{N\to\infty} \beta^i\mathbf{m}^i_0\mathbf{c}^2(2n+1)$$

This gravitational energy of empty Bivacuum may be responsible for Casimir effect and *dark matter effect*. As far the energies of tori **V**$^+_{j-k}$ and antitori **V**$^-_{j-k}$ pulsation are in-phase, symmetric and opposite by sign, they compensate each other and do not violate the energy conservation law.

It follows from our UT, that the pace of time for any closed system is determined by pace of kinetic energy change of this system particles. The new approach to time problem, based on Bivacuum, as the Universal Reference Frame, is more advanced than that, following from relativistic theory, based on Relative Reference Frames. The time of action in our formula is dependent not only on velocity of particle/object, but also on its acceleration. It works not only for inertial systems, but also for *inertialess conservative systems*, which are much more common in Nature, than inertial. Our theory of time, as a part of Unified theory, explains the same experiments, which where used for confirmation of special and general relativity, otherwise.

The validity of Unified Theory is confirmed by logical coherence of many of its consequences and ability to explain a lot of important conventional and unconventional phenomena. Among the first scope are two-slit experiment, radiation of photons by accelerated charges, Michelson - Morley, Hefele-Keating and Pound-Rebeka experiments, etc. The so-called 'paranormal' phenomena (incompatible with conventional paradigm), like Kozyrev, Shnoll and Tiller data, remote genetic transmutations and psi phenomena, involving remote vision, remote healing, telepathy, telekinesis, etc. turns to 'normal' in the framework of UT.

 The *specific character* of telepathic signal transmission from [S] to [R] may be provided by modulation of $\mathbf{VR}_{MT}^S$ of microtubules by $\mathbf{VR}_{DNA}^S$ of sender's chromosomes and vice versa in neuron ensembles, responsible for subconsciousness, imagination and consciousness. It looks, that in cells, including neurons, the system:

$$[\textbf{pair of orthogonal Centrioles + Chromosomes}]$$

stands for *sending* and *receiving* of specific genetic and neurons state active information via bundles of $[\mathbf{N(t, r)} \times \sum \mathbf{VirG}_{SME}(\mathbf{S} <==> \mathbf{R})]_{x,y,z}^i$. It is a crucial stage in proposed in our work mechanism of Induced Remote Genetic Transmutation (RT), Induced Remote Morphogenesis (RM) and Remote Healing (RH), discovered experimentally by Dzang (1981) and Gariaev (2001). The resonance - most effective remote informational/energy exchange between two living organisms or psychics is dependent on corresponding 'tuning' of their [Centrioles + Chromosomes] systems and corresponding neuron ensembles. In accordance to our theory of elementary act of consciousness, the modulation of dynamics of [assembly $\rightleftharpoons$ disassembly] of microtubules by influence on probability of cavitational fluctuations and corresponding [*gel* $\rightleftharpoons$ *sol*] transitions in the 'tuned' nerve cells ensembles in [Receiver] by directed mental activity of [Sender] can provide *telepathic contact and remote viewing* between [Sender] and [Receiver]. The mechanism of *remote healing* could be the same, but the local targets in the body of patient [R] should not be necessarily the [MTs + DNA] systems of nerve cells, but those in cells of the ill organs: heart, liver, etc.
The *telekinesis*, as example of mind-matter interaction, should be accompanied by strong collective nonequilibrium process (excitation) in the nerve system of Sender. Corresponding momentum and kinetic energy are transmitted to 'Receiver' or 'Target' via multiple bundles of Virtual Guides: $[\mathbf{N(t, r)} \times \sum \mathbf{VirG}_{SME}(\mathbf{S} <==> \mathbf{R})]_{x,y,z}^i$, connecting $[\text{MTs+DNA}]_{S,R}$ of [S] and [R], which can be termed a *Psi- channels*.
 We may conclude, that our UT is able to explain a lot of unconventional experimental data, like Kozyrev, Shnoll and Tiller ones, remote genetic transmutation, remote vision, mind-matter interaction, etc. without contradictions with fundamental laws of Nature. For details see: http://arxiv.org/abs/physics/0103031.

**Keywords**: vacuum, Bivacuum, torus, antitorus, virtual Bose condensation, Bivacuum-mediated interaction (BMI), universal reference frame, nonlocality, virtual fermions and bosons, sub-elementary fermions, symmetry shift, golden mean, mass, charge, fusion of elementary particles triplets, corpuscle - wave duality, de Broglie wave, electromagnetism, gravitation, entanglement, principle of least action, tuning energy, time, virtual spin waves, virtual pressure waves, virtual guides, Pauli principle, virtual replica, quantum Psi, telepathy, telekinesis, remote genetic transmutations, remote healing, remote vision.

# Abbreviations and Definitions, Introduced in Unified theory*

- $(\mathbf{V}^+)$ and $(\mathbf{V}^-)$ are correlated actual torus and complementary antitorus (pair of 'donuts') of Bivacuum of the opposite energy, charge and magnetic moment, formed by collective excitations of non mixing subquantum particles and antiparticles of opposite angular momentums;
- $(\mathbf{BVF}^\uparrow = \mathbf{V}^+ \uparrow\uparrow \mathbf{V}^-)^i$ and $(\mathbf{BVF}^\downarrow = \mathbf{V}^+ \downarrow\downarrow \mathbf{V}^-)^i$ are virtual dipoles of three opposite poles: actual (inertial) and complementary (inertialess) mass, positive and negative charge, positive and negative magnetic moments, separated by energetic gap, named Bivacuum fermions and Bivacuum antifermions. The opposite half integer spin $S = \pm\frac{1}{2}\hbar$ of $(\mathbf{BVF}^\updownarrow)^i$, notated as ($\uparrow$ **and** $\downarrow$), depends on direction of clockwise or anticlockwise in-phase rotation of pairs of [torus $(\mathbf{V}^+)$ + antitorus $(\mathbf{V}^-)$], forming them. The index: $i = e, \mu, \tau$ define the energy and Compton radiuses of $(\mathbf{BVF}^{\updownarrow})^i$ of three electron generations;

- $(\mathbf{BVB}^{\pm} = \mathbf{V}^+ \Updownarrow \mathbf{V}^-)^i$ are Bivacuum bosons, representing the intermediate transition state between Bivacuum fermions of opposite spins: $\mathbf{BVF}^{\uparrow} \rightleftharpoons \mathbf{BVB}^{\pm} \rightleftharpoons \mathbf{BVF}^{\downarrow}$;

- $|\mathbf{m}_V^+|\mathbf{c}^2$ and $|\mathbf{m}_V^-|\mathbf{c}^2$ are the energies of torus and antitorus of Bivacuum dipoles: $\left[\mathbf{BVF}^{\updownarrow}\right]_{j,k}^i$ and $\left[\mathbf{BVB}^{\pm}\right]_{j,k}^i$;

- $(\mathbf{VC}_{j,k}^+ \sim \mathbf{V}_j^+ - \mathbf{V}_k^+)^i$ and $(\mathbf{VC}_{j,k}^- \sim \mathbf{V}_j^- - \mathbf{V}_k^-)^i$ are virtual clouds and anticlouds, composed from subquantum particles and antiparticles, correspondingly. Virtual clouds and anticlouds emission/absorption accompany the correlated transitions between different excitation energy states $(j$ and $k)$ of torus $(\mathbf{V}_{j,k}^+)^i$ and antitorus $(\mathbf{V}_{j,k}^-)^i$ of Bivacuum dipoles: $\left[\mathbf{BVF}^{\updownarrow}\right]_{j,k}^i$ and $\left[\mathbf{BVB}^{\pm}\right]_{j,k}^i$;

- **VirP**$^{\pm}$ is *virtual pressure,* resulted from the process of subquantum particles density oscillation, accompanied the virtual clouds $(\mathbf{VC}_{j,k}^{\pm})$ emission and absorption in the process of torus and antitorus transitions between their $j$ and $k$ states;

- $\mathbf{\Delta VirP}_{j,k}^{\pm} = |\mathbf{VirP}^+ - \mathbf{VirP}^-|_{j,k} \sim \||\mathbf{m}_V^+| - |\mathbf{m}_V^-|\|\mathbf{c}^2 \geq 0$ means the excessive virtual pressure, being the consequence of Bivacuum dipoles asymmetry. It determines the *kinetic energy* of Bivacuum, which can be positive or zero;

- $\sum \mathbf{VirP}_{j,k}^{\pm} = |\mathbf{VirP}^+ + \mathbf{VirP}^-|_{j,k} \sim \||\mathbf{m}_V^+| + |\mathbf{m}_V^-|\|\mathbf{c}^2 > 0$ is the total virtual pressure. It determines the potential energy of Bivacuum and always is positive;

- $\mathbf{VPW}_{q=1,2..}^+$ and $\mathbf{VPW}_{q=1,2..}^-$ are the *positive and negative virtual pressure waves,* related with oscillations of $\mathbf{VirP}_{j,k}^{\pm}$. In symmetric primordial Bivacuum the energy of these oscillations compensate each other;

- $\mathbf{F}_{\updownarrow}^+$ and $\mathbf{F}_{\updownarrow}^-$ are sub-elementary *fermions and antifermions* of the opposite charge (+/-) and energy. They emerge due to stable symmetry shift of the *mass and charge* between the *actual* $(\mathbf{V}^+)$ and *complementary* $(\mathbf{V}^-)$ torus of $\mathbf{BVF}^{\updownarrow}$ dipoles, providing the rest mass and charge origination: $[\mathbf{m}_V^+ - \mathbf{m}_V^-]^{\phi} = \pm \mathbf{m}_0$ and $[\mathbf{e}_V^+ - \mathbf{e}_V^-]^{\phi} = \pm \mathbf{e}_0$ to the left or right, correspondingly. Their stabilization and fusion to triplets, represented by electrons and protons, is accompanied by big energy release, determined by mass defect, occur when the velocity of rotation of Cooper pairs $[\mathbf{BVF}^{\uparrow} \bowtie \mathbf{BVF}^{\downarrow}]$ around the common axis corresponds to Golden mean: $(\mathbf{v}/\mathbf{c})^2 = 0.618$;

- *Hidden Harmony* condition means the equality of the internal and external group and phase velocities of Bivacuum fermions and Bivacuum bosons: $\mathbf{v}_{gr}^{in} = \mathbf{v}_{gr}^{ext}$; $\mathbf{v}_{ph}^{in} = \mathbf{v}_{ph}^{ext} = \mathbf{v}$. It is proved that this condition is a natural background of Golden mean realization in physical systems: $\mathbf{\phi} = (\mathbf{v}^2/\mathbf{c}^2)^{ext,in} = 0.6180339887$;

- $\langle[\mathbf{F}_{\updownarrow}^+ \bowtie \mathbf{F}_{\updownarrow}^-] + \mathbf{F}_{\updownarrow}^{\pm}\rangle^{e^-,p^+}$ are the coherent triplets of fused sub-elementary fermions and antifermions of $\mu$ and $\tau$ generations, representing the electron/positron or proton/antiproton. In the latter case a sub-elementary fermions and antifermions corresponds to $u$ and $d$ quarks;

- $(\mathbf{CVC}^+$ and $\mathbf{CVC}^-)$ are the *cumulative virtual clouds* of subquantum particles and antiparticles, standing for [W] phase of sub-elementary fermions and antifermions, correspondingly. The reversible quantum beats $[\mathbf{C} \rightleftharpoons \mathbf{W}]$ between asymmetric torus and antitorus of sub-elementary fermions are accompanied by [emission $\rightleftharpoons$ absorption] of $\mathbf{CVC}^{\pm}$. The stability of triplets of leptons and partons is determined by the resonant interaction of sub-elementary fermions and antifermions by $\mathbf{CVC}^{\pm}$ exchange in the process of [**Corpuscle** $\rightleftharpoons$ **Wave**] pulsations. The virtual pairs $[\mathbf{CVC}^+ \bowtie \mathbf{CVC}^-]_{e,p,n}$ display the gluons (bosons) properties, stabilizing the electrons, protons and neutrons;

- **VirBC** means *virtual Bose condensation* of Cooper - like pairs $[\mathbf{BVF}^{\uparrow} \bowtie \mathbf{BVF}^{\downarrow}]$ and/or $[\mathbf{BVB}^{\pm}]$ with external translational momentum close to zero: $\mathbf{p} \simeq \mathbf{0}$ and corresponding de Broglie wave length close to infinity: $\mathbf{\lambda}_B = (\mathbf{h}/\mathbf{p}) \simeq \infty$, providing the nonlocal properties of huge Bivacuum domains;

- **TE** and **TF** are *Tuning Energy and Tuning Force* of Bivacuum, realized by means of forced resonance of basic Bivacuum pressure waves $(\mathbf{VPW}_{q=1}^{\pm})$ with $[\mathbf{C} \rightleftharpoons \mathbf{W}]$ pulsation of elementary particles, driving the matter to Golden Mean conditions and slowing down (cooling) the thermal dynamics of particles, driving their mass to the rest mass value. Such Bivacuum - Matter interaction is responsible for realization of principle of Least action, 2nd and 3d laws of thermodynamics;

- **VirSW**$^{\pm 1/2}$ are the *Virtual spin waves*, excited as a consequence of angular momentums of cumulative virtual clouds $(\mathbf{CVC}^{\pm})$ of sub-elementary particles in triplets $\langle[\mathbf{F}_{\updownarrow}^- \bowtie \mathbf{F}_{\updownarrow}^+] + \mathbf{F}_{\updownarrow}^{\pm}\rangle$ due to angular momentum conservation law. The **VirSW**$^{\pm 1/2}$ are highly anisotropic, depending on orientation of triplets in space and their rotational/librational dynamics, being the physical background of torsion field;

- **VirG$_{SME}^S$** is the nonlocal virtual spin-momentum-energy guide (quasi-1D virtual microtubule), formed primarily by standing **VirSW**$_S^{S=+1/2}$ $\overset{BVB^\pm}{\underset{BVF^\uparrow \bowtie BVF^\downarrow}{<=\Rightarrow>}}$ **VirSW**$_R^{S=-1/2}$ of opposite spins and induced self-assembly of Bivacuum bosons $(BVB^\pm)^i$ or Cooper pairs of $[BVF^\uparrow \bowtie BVF^\downarrow]^i$, representing quasi one-dimensional Bose condensate. The bundles of virtual guides $[N(t, r) \times \sum VirG_{SME} (S <==> R)]_{x,y,z}^i$ connect the remote coherent triplets $\langle [F_\uparrow^- \bowtie F_\downarrow^+] + F_\updownarrow^\pm \rangle^{e,p}$, representing elementary particles, like protons and electrons in free state or in composition of atoms or their coherent groups, providing remote nonlocal interaction - microscopic and macroscopic ones;

- (**mBC**) means *mesoscopic molecular Bose condensate* in the volume of condensed matter with dimensions, determined by the length of 3D standing de Broglie waves of molecules, related to their librations and translations;

- **VR** means three-dimensional (3D) *Virtual Replica* of elementary, particles, atoms, molecules and macroscopic objects, including living organisms. The primary **VR** of macroscopic object is a consequence of complex system of excitations of Bivacuum dipoles. It represents a superposition of Bivacuum virtual standing waves **VPW**$_m^\pm$ and **VirSW**$_m^{\pm 1/2}$, modulated by $[C \rightleftharpoons W]$ pulsation of elementary particles and translational and librational de Broglie waves of molecules of macroscopic object;

- **VRM$^i$(r, t)** means the *primary* **VR** multiplication/iteration in space and time. The infinitive multiplication of primary **VR$^i$** in space in form of 3D packets of virtual standing waves is a result of interference of all pervading external coherent basic *reference waves* - Bivacuum Virtual Pressure Waves (**VPW**$_{q=1}^\pm$) and Virtual Spin Waves (**VirSW**$_{q=1}^{\pm 1/2}$)$^i$ with similar kinds of modulated standing waves, like that, forming the primary **VR**. The latter has a properties of the *object waves* in terms of holography. Consequently, the **VRM** can be named **Holoiteration** by analogy with hologram (in Greece *'holo'* means the 'whole' or 'total'). The spatial **VRM(r)** may stand for *remote vision* of psychic. The ability of enough complex system of **VRM(t)** to self-organization in nonequilibrium conditions, make it possible multiplication of primary VR not only in space but as well, in time in both time direction - positive (evolution) and negative (devolution). The feedback reaction between most probable/stable **VRM(t,r)** and nerve system of psychic, including visual centers of brain, can be responsible for *clairvoyance;*

- **Psi − channels** are multiple correlated bundles of virtual guides $[N(t, r) \times \sum VirG_{SME} (S <==> R)]_{x,y,z}^i$, connecting coherent particles of nerve cells of [S]-*psychic* and [R] - *target* in superimposed **VRM(r, t)**$_S$ $\bowtie$ **VRM(r, t)**$_R$. This combination of Bivacuum mediated interactions (BMI), providing the transmission of not only information, but as well the momentum and energy, can be responsible for *telekinesis and remote healing*;

- **BMI** is a new fundamental *Bivacuum Mediated Interaction,* additional to electromagnetic, gravitational, weak and strong ones. It is a result of superposition of Virtual replicas of Sender [S] and Receiver [R] in nonequilibrium state, provided by **VRM(r,t)** and formation of bundles $[N(t, r) \times \sum VirG_{SME} (S <==> R)]_{x,y,z}^i$ between coherent atoms of [S] and [R]. Just **BMI** is responsible for remote ultraweak nonlocal interaction between entangled systems and so-called paranormal phenomena, which appears to be quite 'normal'.

*********************************************************************

*$^\star$The abbreviations are not in alphabetic, but in logical order to make this glossary more useful for perception of new notions, introduced in Unified theory.*

# Introduction

The Dirac's equation points to equal probability of positive and negative energy (Dirac, 1947). In asymmetric Dirac's vacuum its realm of negative energy is saturated with infinitive number of electrons. However, it was assumed that these electrons, following Pauli principle, have not any gravitational or viscosity effects. Positrons and electron in his model represent the 'holes', originated as a result of the electrons jumps in realm of positive energy over the energetic gap: $\Delta = \mathbf{2m}_0 c^2$. Currently it becomes clear, that the Dirac type model of vacuum is not general enough to explain all known experimental data, for example, the bosons emergency. The model of Bivacuum, presented in this paper and previous works of this author (Kaivarainen, 1995; 2000; 2004; 2005, 2006) is more advanced. However, it use the same starting point of equal probability of positive and negative energy, confined in each of Bivacuum elements, named Bivacuum dipoles.

Aspden (2003) introduced in his aether theory the basic unit, named Quon, as a pair of virtual muons of opposite charges, i.e. [muon + antimuon]. This idea has some common with our model of Bivacuum dipoles. Each dipole represents collective excitations of sub-quantum particles and antiparticles, composing vortical pair: *torus + antitorus* of opposite energy/mass, charge and magnetic moments with three Compton radiuses, corresponding to three lepton generation: electron, muon and tauon (Kaivarainen, 2004-2006).

Our notions of strongly correlated torus ($\mathbf{V}^+$) and antitorus ($\mathbf{V}^-$) of Bivacuum dipoles have also some similarity with 'phytons', introduced by Akimov and Shipov for explanation of torsion field action. After Akimov (1995): "In non polarized condition, physical vacuum contains in each of its elements a *phyton*, which is a kind of circle shape - two wave packets, which are rotating in *opposite* directions, corresponding to right and left spin. Primarily phytons are compensated, as far the sum of their angular momentums is zero. This is a reason, why the vacuum does not manifest nonzero angular momentum. But, if in the vacuum the spinning object appears, then the *phytons*, with axes of rotation, coinciding with that of the object, will keep the same rotation, and phytons which' rotational axes were originally in the opposite direction, will be inverted partly under the influence of the spinning object.

Two subclasses of Bivacuum dipoles where introduced: Bivacuum bosons ($\mathbf{BVB}^\pm)_{S=0}$ with torus and antitorus, rotating in opposite direction and virtual Cooper pairs of Bivacuum fermions and antifermions with torus + antitorus both rotating clockwise or anticlockwise, correspondingly $[\mathbf{BVF}^\uparrow \bowtie \mathbf{BVF}^\downarrow]_{S=0,\pm1...}$. The ability of Bivacuum dipoles to form virtual Bose condensate from the bundles of quasi one-dimensional virtual microtubules (single and doubled) is demonstrated in our theory. These bundles, like vortical structures in liquid $^4\mathbf{He}$ and $^3\mathbf{He}$ (superfluid turbulence), makes it possible consider Bivacuum as a two component liquid with superfluid and normal properties. The superfluid model of vacuum, composed from pairs of fermions of opposite spins and charge where discussed earlier by Sinha et. al., (1976; 1976a; 1978) and also by Boldyreva and Sotina (1999).

In accordance with Planck aether hypothesis of Winterberg (2002), the vacuum is a superfluid made up of positive and negative Planck mass particles. The Planck mass plasma model makes the following assumptions:

1. The ultimate building blocks are positive and negative Planck mass particles. The interaction obeys the laws of Newtonian mechanics, except for *lex tertia*, which under the assumed force law is violated during the collision between a positive and a negative Planck mass particle. The violation of the *lex tertia* means that during the mutually attractive collision between a positive and a negative Planck mass particle, the momentum, not the energy, fluctuates. This establishes Heisenberg uncertainty principle at the most fundamental level, explaining why quantum mechanics can be derived from the Planck mass plasma.

2. A Planck mass particles of the same sign repel and those of opposite sign attract each other, with the magnitude and range of the force equal to the Planck force $\mathbf{M}_P\mathbf{c}^2/\mathbf{R}_P = \mathbf{c}^4/\mathbf{G}$ and the Planck length $\mathbf{R}_P = \mathbf{h}/(\mathbf{M}_P\mathbf{c})$.

3. Space - vacuum is filled with an equal number of positive and negative Planck mass particles whereby each Planck length volume is in the average occupied by one Planck mass particle. The collision of positive and negative Plank mass particles is a source of *zitterbewegung* in Winterberg model of vacuum.

In its ground state the Planck aether is a two component positive-negative mass superfluid with a phonon - roton energy spectrum for each component. Assuming that the phonon - roton spectrum measured in superfluid helium is universal, this would mean that in the Planck aether this spectrum has the same shape.

Rotons can be viewed as small vortex rings with the ring radius of the same order as the vortex core radius. A fluid with cavitons is in a state of negative pressure, and the same is true for a fluid with vortex rings. In vortices the centrifugal force creates a vacuum in the vortex core, making a vortex ring to behave like a caviton.

In Winterberg model the positive and negative Plank masses are not considered as a unified mass dipoles with possibility of polarization and symmetry shift. The mechanism of origination of mass, charge, magnetic moment and spin of elementary particles, the background of three lepton generation where not analyzed and proposed.

Nonetheless of some common features with models of Aspden, Akimov - Shipov's ' and Winterberg, the concept of Bivacuum and it elements: Bivacuum bosons ($\mathbf{BVB}^\pm$) and fermions ($\mathbf{BVF}^\updownarrow$) is more advanced. It explains the origination of *mass and charge* of sub-elementary fermions, as a result of torus $\mathbf{V}^+$ and antitorus $\mathbf{V}^-$ of Bivacuum dipoles symmetry shift, the mechanism of *corpuscle ⇌ wave* pulsation and fusion of elementary particles from triplets of sub-elementary fermions and antifermions. The electric, magnetic and gravitational fields are shown to be a result of elastic *recoil ⇌ antirecoil* effects and *zitterbewegung,* induced by these pulsation in Bivacuum matrix. In the framework of our approach all fundamental physical phenomena are hierarchically interrelated and unified.

David Bohm made the first one, who made an attempt to explain wholeness of the Universe, without loosing the causality principle. Experimental discovery: "Aharonov-Bohm effect" (1950) pointing that electron is able to "feel" the presence of a magnetic field even in a regions where the probability of field existing is zero, was stimulating. For explanation of nonlocality Bohm introduced in 1952 the notion of *quantum potential*, which pervaded all of space. But unlike gravitational and electromagnetic fields, its influence did not decrease with distance. All the particles are interrelated by very sensitive to any perturbations quantum potential. This means that signal transmission between particles may occur instantaneously. The idea of *quantum potential or active information* is close to notion of *pilot wave*, proposed by de Broglie at the Solvay Congress in 1927. In fact, Bohm develops the de Broglie idea of pilot wave, applying it for many-body system.

In 1957 Bohm published a book: Causality and Chance in Modern Physics. Later he comes to conclusion, that Universe has a properties of giant, flowing hologram. Taking into account its dynamic nature, he prefer to use term: **holomovement**. In his book: Wholeness and the Implicate Order (1980) he develops an idea that our *explicated unfolded reality is a product of enfolded (implicated) or hidden order of existence. He consider the manifestation of all forms in the universe, as a result of enfolding and unfolding exchange between two orders, determined by super quantum potential.*

In book, written by D. Bohm and B. Hiley (1993): "THE UNDIVIDED UNIVERSE. An ontological interpretation of quantum theory" the electron is considered, as a particle with well- defined position and momentum which are, however, under influence of special wave (quantum potential). Elementary particle, in accordance with these authors, is a *sequence of incoming and outgoing waves*, which are very close to each other. However, particle itself does not have a wave nature. Interference pattern in double slit experiment after Bohm is a result of periodically "bunched" character of quantum potential.

After Bohm, the manifestation of corpuscle - wave duality of particle is dependent on the way, which observer interacts with a system. He claims that both of this properties are always enfolded in particle. *It is a basic difference with our model, assuming that the wave and corpuscle phase are realized alternatively with high frequency during two different semiperiods of sub-elementary particles, forming particles in the process of quantum beats between sublevels of positive (actual) and negative (complementary) energy. This frequency is amplitude and phase modulated by experimentally revealed de Broglie wave of particles.*

The important point of Bohmian philosophy, coinciding with our concept, is that everything in the Universe is a part of dynamic continuum. Neurophysiologist Karl Pribram does made the next step in the same direction as Bohm: *"The brain is a hologram enfolded in a holographic Universe"*.

The good popular description of Bohm and Pribram ideas are presented in books: "The Bell's theorem and the curious quest for quantum reality" (1990) by David Peat and "The Holographic Universe" (1992) by Michael Talbot. Such original concepts are interesting and stimulating, indeed, but should be considered as a first attempts to transform intuitive perception of duality and quantum wholeness into clear geometrical and mathematical models.

Some common features with our and Bohm-Hiley models has a Unitary Quantum Theory (UQT), proposed by Sapogin (1982). In the UQT any elementary particle is not a point and source of field like in the ordinary quantum mechanics, but represents a wave packet of a certain unified field (Sapogin and Boichenko, 1991). The dispersion equation of such a nonlinear field turned out to be such, that the wave packet (particle) during its movement periodically appears and disappears, and the envelope of this process coincides with the de Broglie wave. Numerous particles during their periodic disappearance (spreading in the Universe) and repeated appearance represent vacuum fluctuations. The corresponding transversal self-focusing of the wave packet is possible only in conditions if the refraction index of space/vacuum is dependent of particle velocity. The square of wave packet describes the oscillating charged particle mass and energy (Sapogin, et.al., 2002), following the conventional Newton equations. The essential in UQT is the absence of the energy and the momentum conservation laws for single particles.

In 1950 John Wheeler and Charles Misner published Geometrodynamics, a new description of space-time properties, based on topology. Topology is more general than Euclidean geometry and deeper than non-Euclidean, used by Einstein in his General theory of relativity. Topology does not deal with distances, angles and shapes. Drawn on a sheet of stretching rubber, a circle, triangle and square are indistinguishable. A ball, pyramid and a cube also can be transformed one to the other. However, objects with holes in them can never be transformed by stretching and deforming into objects without holes. For example black hole can be described in terms of topology. It means that massive rotating body behave as a space-time hole. Wheeler supposed that *elementary particles and antiparticles, their spins, positive and negative charges can be presented as interconnected black and white holes.* Positron and electron pair correspond to such model. The energy, directed to one of the hole, goes throw the connecting tube -"handle" and reappears at the other. The connecting tube exist in another space-time than holes itself. Such a tube is undetectable in normal space and the process of energy transmission looks as instantaneous. In conventional space-time two ends of tube, termed 'wormholes' can be a vast distant apart. It gives an explanation of quantum nonlocality.

The same is true for introduced in our theory nonlocal Virtual spin-momentum-energy guides (**VirG**$_{SME}$). The mono or paired **VirG**$_{SME}$, formed by Bivacuum bosons (**BVB**$^{\pm}$) or Cooper pairs of Bivacuum fermions, correspondingly, may connect not only particles and antiparticles, like positrons and electrons, but also the same kind of particles (electrons, protons, neutrons) with opposite spins and 'tuned' frequency of Corpuscle $\rightleftharpoons$ Wave pulsation.

Sidharth (1998, 1999) considered *elementary particle as a relativistic vortex of Compton radius, from which he recovered its mass and quantized spin* ($s = \frac{1}{2}\hbar$). He pictured a particle as a fluid vortex steadily circulating with light velocity along a 2D ring or spherical 3D shell with radius

$$L = \frac{\hbar}{2mc} \qquad\qquad 1$$

Inside such vortex the notions of negative energy, superluminal velocities and nonlocality are acceptable without contradiction with conventional theory.

Bohm's hydrodynamic formulation and substitution

$$\psi = \mathrm{R}e^{iS} \qquad\qquad 2$$

where $R$ and $S$ are real function of $\vec{r}$ and $t$, transforms the Schrödinger equation to

$$\frac{\partial\rho}{\partial t} + \vec{\nabla}(\rho\vec{\mathbf{v}}) = 0 \qquad\qquad 3$$

$$or: \hbar\frac{\partial S}{\partial t} + \frac{\hbar^2}{2m}(\vec{\nabla}S)^2 + V = \frac{\hbar^2}{2m}(\nabla^2 R/R) \equiv Q \qquad\qquad 4$$

where: $\rho = R^2$; $\vec{\mathbf{v}} = \frac{\hbar^2}{2m}\vec{\nabla}S$ *and* $Q = \frac{\hbar^2}{2m}(\nabla^2 R/R)$

Sidharth comes to conclusion that the energy of nonlocal quantum potential ($Q$) is determined by inertial mass ($m$) of particle:

$$Q \equiv -\frac{\hbar^2}{2m}(\nabla^2 R/R) = mc^2 \qquad\qquad 5$$

He treated also a charged Dirac fermions, as a Kerr-Newman black holes. Within the region of Compton vortex the superluminal velocity and negative energy are possible after Sidharth. If measurements are averaged over time $t \sim mc^2/\hbar$ and over space $L \sim \hbar/mc$, the imaginary part of particle's position disappears and we are back in usual Physics (Sidharth, 1998).

Barut and Bracken (1981) considered *zitterbewegung* - rapidly oscillating imaginary part of particle position, leading from Dirac theory (1947), as a harmonic oscillator in the Compton wavelength region of particle. The Einstein (1971, 1982) and Shrödinger (1930) also spoke about oscillation of the electron with frequency: $\mathbf{v} = \mathbf{m}_0\mathbf{c}^2/h$ and the amplitude: $\zeta_{max} = \hbar/(2\mathbf{mc})$. It was demonstrated by Shrödinger, that position of free electron can be presented as: $\mathbf{x} = \overline{\mathbf{x}} + \zeta$, where $\overline{\mathbf{x}}$ characterize the average position of the free electron, and $\zeta$ its instant position, related to its oscillations. Hestness (1990) proposed, that *zitterbewegung* arises from self-interaction, resulting from wave-particle duality.

This ideas are close to our explanation of elementary particles zero-point oscillations, as a recoil $\rightleftharpoons$ antirecoil vibrations, accompanied corpuscle $\leftrightarrow$ wave pulsations. Corresponding oscillations of each particle kinetic energy, in accordance to our theory of time (Kaivarainen, 2005), is related with oscillations of *instant* time for this closed system. We came here to concept of *space-time-energy discreet trinity*, generated by corpuscle − wave duality.

Serious attack on problem of quantum nonlocality was performed by Roger Penrose (1989) with his twister theory of space-time. After Penrose, quantum phenomena can generate space-time. The twisters, proposed by him, are lines of infinite extent, resembling twisting light rays. Interception or conjunction of twistors lead to origination of particles. In such a way the local and nonlocal properties and particle-wave duality are interrelated in twistors geometry. The analysis of main quantum paradoxes was presented by Asher Peres (1992) and Charles Bennett et. al., (1993).

In our Unified model the *local* properties of sub-elementary particles are resulted from their Bivacuum symmetry shift, accompanied by their uncompensated mass and charge origination. The *nonlocal* interaction of two or more particles of the same kinds (photons, electrons, protons, neutrons) in state of entanglement, are the consequence of *Bivacuum gap* oscillation between torus ($\mathbf{V}^+$) and antitorus ($\mathbf{V}^-$) of $\mathbf{BVF}^\ddagger$, $\mathbf{BVB}^\pm$ and corresponding pulsation of radiuses of $\mathbf{BVB}^\pm$ or Cooper pairs of Bivacuum fermions [$\mathbf{BVF}^\uparrow \bowtie \mathbf{BVF}^\downarrow$]. This kind of signals are mediated by quasi one-dimensional Bose condensation of Bivacuum dipoles, assembling virtual guides ($\mathbf{VirG}_{SME}$) of spin, momentum and energy, connecting these particles with close frequency and phase of [$\mathbf{C} \rightleftharpoons \mathbf{W}$] pulsation.

The quite different approach, using computational derivation of quantum relativistic systems with forward-backward space-time shifts, developed by Daniel Dubois (1999), led to some results, similar to ours (Kaivarainen, 1995, 2001, 2003, 2004). For example, the group and phase masses, introduced by Dubois, related to internal group and phase velocities, has analogy with actual and complementary masses, introduced in our Unified theory (UT). In both approaches, the product of these masses is equal to the particle's rest mass squared. The notion of discrete time interval, used in Dubois approach, may correspond to PERIOD of [$C \rightleftharpoons W$] pulsation of sub-elementary particles in UT. The positive internal time interval, in accordance to our model, corresponds to forward $C \rightarrow W$ transition and the negative one to the backward $W \rightarrow C$ transition.

Puthoff (2001) developed the idea of 'vacuum engineering', using hypothesis of polarizable vacuum (PV). The electric permittivity ($\varepsilon_0$) and magnetic permeability ($\mu_0$) is interrelated in 'primordial' symmetric vacuum, as: $\varepsilon_0 \mu_0 = 1/c^2$. It is shown that changing of vacuum refraction index: $\mathbf{n} = \mathbf{c}/\mathbf{v} = \varepsilon^{1/2}$, for example in gravitational or electric potentials, is accompanied by variation of lot of space-time parameters.

Fock (1964) and Puthoff (2001), explained the bending of light beam, induced by gravitation near massive bodies also by vacuum refraction change, i.e. in another way, than General theory of relativity. However, the mechanism of vacuum polarization and corresponding refraction index changes in electric and gravitational fields remains obscure. Our Unified theory propose such mechanism (see section 8.11).

The transformation of neutron to proton and electron, in accordance to Electro - Weak (EW) theory, developed by Glashov (1961), Weinberg (1967) and Salam (1968), is mediated by negative *massless* $W^-$ boson. The reverse reaction in EW theory: proton → neutron is mediated by positive *massless* $W^+$ boson. Scattering of the electron on neutrino, not accompanied by charge transferring, is mediated by third *massless* neutral boson $Z^0$.

In (EW) theory the Higgs field was introduced for explanation of spontaneous symmetry violation of intermediate vector bosons: charged $W^\pm$ and neutral $Z^0$ with spin 1, accompanied by origination of big mass of these particles. The EW theory needs also the quantum of Higgs field, named Higgs bosons with big mass, zero charge and integer spin. The fusion of Higgs bosons with $W^\pm$ and $Z^0$ particles is accompanied by increasing of their mass up to 90 mass of protons. The experimental discovery of heavy $W^{\pm}$ and $Z^0$ particles in 1983 after their separation, accompanied getting the system a big external energy, was considered as a conformation of EW theory.

The spontaneous symmetry violation of vacuum, in accordance to Goldstone theorem, turns two virtual particles with imaginary masses ($i\mathbf{m}$) to one real particle with mass: $\mathbf{M}_1 = \sqrt{2}\ \mathbf{m}$ and one real particle with zero mass: $\mathbf{M}_2 = \mathbf{0}$. However, the Higgs field and Higgs bosons are still not found. "We have eliminated most of hunting area", confirms Neil Calder from CERN recently. *This author propose another explanation of mass and charge origination.*

In conventional approach, described above, two parameters of $\mathbf{W}^{\pm}$ particles, like charge and mass are considered, as independent.

Thomson, Heaviside and Searl supposed that mass is an electrical phenomena. In theory of Haisch, Rueda and Puthoff (1994), Rueda and Haish (1998) it was proposed, that the inertia is a reaction force, originating in a course of dynamic interaction between the electromagnetic zero-point field (ZPF) of vacuum and charge of elementary particles. However, it's not clear in this approach, how the charge itself originates.

Our Unified theory is an attempt to unify mass and charge with magnetic moment, spin and symmetry shift of sub-elementary fermions, induced by external translational-rotational motion (see chapter 4). This theory unifies the origination of elementary particles, their rest mass and charge, electromagnetism and gravitation with particles corpuscle-wave duality, standing also for their zero-point oscillations. In accordance to formalism of our theory, the rest mass and charge of elementary fermions origination are both the result of Bivacuum fermions (BVF) symmetry shift, corresponding to Golden mean conditions, i.e. equality of the ratio of external velocity of BVF to light velocity squared to: $(\mathbf{v}/\mathbf{c})^2 = 0.618 = \phi$. At this condition the asymmetric Bivacuum dipole turns to sub-elementary fermion. The electric, magnetic and gravitational fields are the result of huge number of Bivacuum dipoles symmetry shift oscillation, excited by *recoil⇌ antirecoil* dynamics, accompanied the corpuscle ⇌ wave pulsation of sub-elementary particles, forming the elementary particles (chapter 8).

In our approach, the resistance of particle to acceleration (i.e. inertia force), proportional to its mass (second Newton's law) is a consequence of resistance of frequency of particle's $\mathbf{C} \rightleftharpoons \mathbf{W}$ pulsation to change, keeping the equilibrium (tuned state) with frequency of surrounding Bivacuum dipoles symmetry - energy oscillation. We named this resistance to equilibrium shift between dynamics of particles and dynamics of Bivacuum - *"The generalized principle of Le Chatelier's".*

In contrast to nonlocal Mach's principle, our theory of particle - Bivacuum interaction explains the existence of inertial mass of even single particle in empty Universe.

*The main goals of our work can be formulated as follows:*

1. Development of superfluid Bivacuum model, as the dynamic matrix of dipoles, formed by pairs of virtual torus and antitorus of the opposite energy/mass, charge and magnetic moments, compensating each other. The explanation of fusion of the electrons, positrons, muons, protons, neutrons and photons, as a triplets of asymmetric Bivacuum sub-elementary fermions of tree lepton generation $(e, \mu, \tau)$. The *external* properties of such elementary particles are still described by the existing formalism of quantum mechanics and Maxwell equations;

2. Development of the dynamic model of wave-corpuscle duality of sub-elementary particles/antiparticles, composing elementary particles and antiparticles. Explanation of the entanglement, based on new theory;

3. Generalization of the Einstein's and Dirac's formalism for free relativistic particles, considering the correlated pairs of *inertial - actual torus* and *inertialess - complementary* *antitorus* of sub-elementary fermions, forming elementary particles;

4. Finding analytical equations, unifying the internal and external parameters of sub-elementary particles. Elucidation the conditions of triplets (elementary fermions) fusion from sub-elementary fermions. Origination of the rest mass and elementary charge. Understanding the mechanisms of triplets stabilization;

5. Explanation of the absence of Dirac's monopole in Nature;

6. Understanding the nature of zero-point oscillations and recoil⇌antirecoil effects, accompanied the [**Corpuscle ⇌ Wave**] pulsation of fermions, responsible for electric, magnetic and gravitational fields origination;

7. Unification of the Principle of least action, the time, the 2nd and 3d laws of thermodynamics with Principle of least action and action of Bivacuum virtual pressure waves ($\mathbf{VPW}^{\pm}$), on the dynamics of elementary particles;

8. Elaboration a concept of Virtual Replica (VR) of any material object and its spatial multiplication in Bivacuum, as a consequence of superposition of the *reference* basic Bivacuum virtual pressure waves ($\mathbf{VPW}^{\pm}_{q=1}$) and virtual spin waves ($\mathbf{VirSW}^{\pm 1/2}_{q=1}$) with the *object* virtual waves ($\mathbf{VPW}^{\pm}_{\mathbf{m}}$) and ($\mathbf{VirSW}^{\pm 1/2}_{\mathbf{m}}$), modulated by de Broglie waves of particles (nucleons), forming this object;

9. Working out the new mechanism of Bivacuum mediated nonlocal remote interaction between the remote coherent microscopic and macroscopic systems via introduced Virtual guides of spin, momentum and energy ($\mathbf{VirG}_{S,M,E}$) and their coherent bundles;

10. Explanation of Kozyrev's, Shnoll and Tiller experiments and mechanisms of overunity devices action and other phenomena, incompatible with mainstream paradigm, which may be considered as *paranormal*, following from our Unified theory;

11. The validation of Unified Theory, based on logical coherence of many of its consequences and ability to explain a lot of fundamental not only the conventional, but as well the unconventional/paranormal experimental results, including getting the free energy from Bivacuum, cold fusion, etc.

## 1. New Hierarchical Model of Bivacuum, as a Superfluid Multi-Dipole Structure

### *1.1. Properties of Bivacuum dipoles - Bivacuum fermions and Bivacuum bosons*

The Bivacuum concept is a result of new interpretation and development of Dirac theory (Dirac, 1958), pointing to equal probability of positive and negative energy in Nature.

The Bivacuum is introduced, as a dynamic superfluid matrix of the Universe, composed from non-mixing *subquantum particles* of opposite polarization and three nonquantized spin values, separated by an energy gap. The hypothetical *microscopic* subquantum particles and antiparticles have a dimensions about or less than ($10^{-19}$ m), zero mass, spin and charge. They spontaneously self-organize in infinite number of *mesoscopic* paired vortices - Bivacuum dipoles of three generations with Compton radii, corresponding to electrons ($e$), muons ($\mu$) and tauons ($\tau$), corresponding to three different spin values. Only such *mesoscopic* collective excitations of subquantum particles in form of pairs of rotating fast *torus and antitorus* are quantized. In turn, these Bivacuum 'molecules' compose the *macroscopic* superfluid ideal liquid, representing the infinitive Bivacuum matrix.

Each of two strongly correlated 'donuts' of Bivacuum dipoles acquire the opposite mass charge and magnetic moments, compensating each other in the absence of symmetry shift between them. The latter condition is valid only for symmetric *primordial* Bivacuum, where the influence of matter and fields on Bivacuum is negligible.

The symmetric primordial Bivacuum can be considered as the *Universal Reference Frame* (**URF**), i.e. *Ether*, in contrast to *Relative Reference Frame* (**RRF**), used in special relativistic (SR) theory. The elements of *Ether - ethons* correspond to our Bivacuum dipoles. It will be shown in our work, that the result of Michelson - Morley experiment is a consequence of *ether drug* by the Earth or Virtual Replica of the Earth in terms of our theory.

The sub-elementary fermion and antifermion origination is a result of the Bivacuum dipole symmetry shift toward the torus or antitorus, correspondingly. The correlation between paired vortical structures in a liquid medium was theoretically proved by Kiehn (1998).

The infinite number of paired vortical structures: [torus ($\mathbf{V}^+$) + antitorus ($\mathbf{V}^-$)] with the in-phase clockwise or anticlockwise rotation are named Bivacuum fermions ($\mathbf{BVF}^\uparrow = \mathbf{V}^+ \upuparrows \mathbf{V}^-)^i$ and Bivacuum antifermions ($\mathbf{BVF}^\downarrow = \mathbf{V}^+ \downdownarrows \mathbf{V}^-)^i$, correspondingly. Their intermediate - transition states are named Bivacuum bosons of two possible polarizations: ($\mathbf{BVB}^+ = \mathbf{V}^+\uparrow\downarrow \mathbf{V}^-)^i$ and ($\mathbf{BVB}^- = \mathbf{V}^+\downarrow\uparrow \mathbf{V}^-)^i$ The *positive and negative energies of torus and antitorus* ($\pm \mathbf{E}_{\mathbf{V}^\pm}$) of three lepton generations ($i = e, \mu, \tau$), interrelated with their radiuses ($\mathbf{L}_{\mathbf{V}^\pm}^n$), are quantized as quantum harmonic oscillators of opposite energies:

$$[\mathbf{E}_{\mathbf{V}^\pm}^n = \pm \mathbf{m}_0 \mathbf{c}^2 (\frac{1}{2} + \mathbf{n}) = \pm \hbar \boldsymbol{\omega}_0 (\frac{1}{2} + \mathbf{n})]^i \qquad \mathbf{n} = 0, 1, 2, 3 \ldots \qquad 1.1$$

$$or : \left[ \; \mathbf{E}_{\mathbf{V}^\pm}^n = \frac{\pm \hbar \mathbf{c}}{\mathbf{L}_{\mathbf{V}^\pm}^n} \; \right]^i \qquad where : \qquad \left[ \; \mathbf{L}_{\mathbf{V}^\pm}^n = \frac{\pm \hbar}{\mathbf{m}_0 \mathbf{c}(\frac{1}{2} + \mathbf{n})} = \frac{\mathbf{L}_0}{\frac{1}{2} + \mathbf{n}} \; \right]^i \qquad 1.1a$$

where: $[\mathbf{L}_0 = \hbar/\mathbf{m}_0 \mathbf{c}]^{e, \mu, \tau}$ is a Compton radii of the electron of corresponding lepton generation ($i = e, \mu, \tau$) and $\mathbf{L}_0^e \gg \mathbf{L}_0^\mu > \mathbf{L}_0^\tau$. The Bivacuum fermions ($\mathbf{BVF}^\updownarrow)^{\mu, \tau}$ with smaller Compton radiuses can be located inside the bigger ones ($\mathbf{BVF}^\updownarrow)^e$.

The absolute values of increments of torus and antitorus energies ($\Delta \mathbf{E}_{\mathbf{V}^\pm}^i$), interrelated with increments of their radii ($\Delta \mathbf{L}_{\mathbf{V}^\pm}^i$) in primordial Bivacuum (i.e. in the absence of matter and field influence), resulting from in-phase symmetric fluctuations are equal:

$$\Delta \mathbf{E}_{\mathbf{V}^\pm}^i = -\frac{\hbar c}{\left( \mathbf{L}_{\mathbf{V}^\pm}^i \right)^2} \Delta \mathbf{L}_{\mathbf{V}^\pm}^i = -\mathbf{E}_{\mathbf{V}^\pm}^i \frac{\Delta \mathbf{L}_{\mathbf{V}^\pm}^i}{\mathbf{L}_{\mathbf{V}^\pm}^i} \qquad or : \qquad 1.2$$

$$-\Delta \mathbf{L}_{\mathbf{V}^\pm}^i = \frac{\pi \left( \mathbf{L}_{\mathbf{V}^\pm}^i \right)^2}{\pi \hbar c} \Delta \mathbf{E}_{\mathbf{V}^\pm}^i = \frac{\mathbf{S}_{\mathbf{BVF}^\pm}^i}{2 \hbar \mathbf{c}} \Delta \mathbf{E}_{\mathbf{V}^\pm}^i = \mathbf{L}_{\mathbf{V}^\pm}^i \frac{\Delta \mathbf{E}_{\mathbf{V}^\pm}^i}{\mathbf{E}_{\mathbf{V}^\pm}^i} \qquad 1.2a$$

where: $\mathbf{S}_{\mathbf{BVF}^\pm}^i = \pi \left( \mathbf{L}_{\mathbf{V}^\pm}^i \right)^2$ is a square of the cross-section of torus and antitorus, forming Bivacuum fermions ($\mathbf{BVF}^\updownarrow$) and Bivacuum bosons ($\mathbf{BVB}^\pm$).

The virtual *mass*, *charge* and *magnetic moments* of torus and antitorus of $\mathbf{BVF}^\updownarrow$ and $\mathbf{BVB}^\pm$ are opposite and in symmetric *primordial* Bivacuum compensate each other in their basic ($\mathbf{n} = 0$) and excited ($\mathbf{n} = 1, 2, 3 \ldots$) states.

The Bivacuum 'atoms': $\mathbf{BVF}^\updownarrow = [\mathbf{V}^+ \updownarrow \mathbf{V}^-]^i$ and $\mathbf{BVB}^\pm = [\mathbf{V}^+\uparrow\downarrow \mathbf{V}^-]^i$ represent dipoles of three different poles - the mass ($\mathbf{m}_V^+ = |\mathbf{m}_{V}^-| = \mathbf{m}_0)^i$, electric ($e_+$ and $e_-$) and magnetic ($\boldsymbol{\mu}_+$ and $\boldsymbol{\mu}_-$) dipoles.

The torus and antitorus ($\mathbf{V}^+ \updownarrow \mathbf{V}^-)^i$ of Bivacuum fermions of opposite spins $\mathbf{BVF}^\uparrow$ and $\mathbf{BVF}^\downarrow$ are both rotating in the same direction: clockwise or anticlockwise. This determines the positive and negative spins ($\mathbf{S} = \pm 1/2\hbar$) of Bivacuum fermions. Their opposite spins may compensate each other, forming virtual Cooper pairs: $[\mathbf{BVF}^\uparrow \bowtie \mathbf{BVF}^\downarrow]$ with neutral boson properties. The rotation of adjacent $\mathbf{BVF}^\uparrow$ and $\mathbf{BVF}^\downarrow$ in Cooper pairs is *side- by- side* in opposite directions, providing zero resulting spin of such pairs and ability to virtual Bose condensation. The torus and antitorus of Bivacuum bosons $\mathbf{BVB}^\pm = [\mathbf{V}^+\uparrow\downarrow \mathbf{V}^-]^i$ with resulting spin, equal to zero, are rotating in opposite directions.

The *energy gap* between the torus and antitorus of symmetric $(\mathbf{BVF}^{\updownarrow})^i$ or $(\mathbf{BVB}^{\pm})^i$ is:

$$[\mathbf{A}_{BVF} = \mathbf{E}_{\mathbf{V}^+} - (-\mathbf{E}_{\mathbf{V}^-}) = \hbar\boldsymbol{\omega}_0(1 + 2\mathbf{n})]^i = \mathbf{m}_0^i \mathbf{c}^2(1 + 2\mathbf{n}) = \frac{h\mathbf{c}}{[\mathbf{d}_{\mathbf{V}^+\updownarrow\mathbf{V}^-}]_n^i} \qquad 1.3$$

where the characteristic distance between torus $(\mathbf{V}^+)^i$ and antitorus $(\mathbf{V}^-)^i$ of Bivacuum dipoles *(gap dimension)* is a quantized parameter:

$$[\mathbf{d}_{\mathbf{V}^+\updownarrow\mathbf{V}^-}]_n^i = \frac{h}{\mathbf{m}_0^i \mathbf{c}(1 + 2\mathbf{n})} \qquad 1.4$$

From (1.2) and (1.2a) we can see, that at $\mathbf{n} \to \mathbf{0}$, the energy gap $\mathbf{A}_{BVF}^i$ is decreasing till $\hbar\boldsymbol{\omega}_0 = \mathbf{m}_0^i \mathbf{c}^2$ and the spatial gap dimension $[\mathbf{d}_{\mathbf{V}^+\updownarrow\mathbf{V}^-}]_n^i$ is increasing up to the Compton length $\boldsymbol{\lambda}_0^i = \mathbf{h}/\mathbf{m}_0^i \mathbf{c}$. On the contrary, the infinitive symmetric excitation of torus and antitorus is followed by tending the spatial gap between them to zero: $[\mathbf{d}_{\mathbf{V}^+\updownarrow\mathbf{V}^-}]_n^i \to 0$ at $\mathbf{n} \to \infty$. This means that the quantization of space and energy of Bivacuum elements are interrelated and discreet.

### 1.2 The basic (carrying)Virtual Pressure Waves ($VPW_q^{\pm}$) and Virtual spin waves ($VirSW_q^{\pm 1/2}$) of Bivacuum

The emission and absorption of Virtual clouds $(\mathbf{VC}_{j,k}^+)^i$ and anticlouds $(\mathbf{VC}_{j,k}^-)^i$ in primordial Bivacuum, i.e. in the absence of matter and fields or where their influence on symmetry of Bivacuum is negligible, are the result of correlated transitions between different excitation states $(j,k)$ of torus $(\mathbf{V}_{j,k}^+)^i$ and antitoruses $(\mathbf{V}_{j,k}^-)^i$, forming symmetric $[\mathbf{BVF}^{\updownarrow}]^i$ and $[\mathbf{BVB}^{\pm}]^i$, corresponding to three lepton generations ($i = e, \mu, \tau$) :

$$(\mathbf{VC}_q^+)^i \equiv [\mathbf{V}_j^+ - \mathbf{V}_k^+]^i \;-\; virtual\ cloud \qquad 1.5$$

$$(\mathbf{VC}_q^-)^i \equiv [\mathbf{V}_j^- - \mathbf{V}_k^-]^i \;-\; virtual\ anticloud \qquad 1.5a$$

where: $j > k$ are the integer quantum numbers of torus and antitorus excitation states; $q = j - k$.

The virtual clouds: $(\mathbf{VC}_q^+)^i$ and $(\mathbf{VC}_q^-)^i$ exist in form of collective excitation of *subquantum* particles and antiparticles of opposite energies, correspondingly. They can be considered as 'drops' of virtual Bose condensation of subquantum particles of positive and negative energy and similar in case of $[\mathbf{BVF}^{\updownarrow}]^i$ and opposite in case of $[\mathbf{BVB}^{\pm}]^i$ angular momentums.

The process of [*emission* $\rightleftharpoons$ *absorption*] of virtual clouds should be accompanied by oscillation of *virtual pressure (*$\mathbf{VirP}^{\pm}$*) and excitation of positive and negative virtual pressure waves:* $\mathbf{VPW}_q^+$ *and* $\mathbf{VPW}_q^-$. In primordial Bivacuum the energies of opposite virtual pressure waves totally compensate each other: $\mathbf{VPW}_q^+ + \mathbf{VPW}_q^- = 0$. However, in asymmetric secondary Bivacuum, in presence of matter and fields, the total compensation is absent and the resulting virtual pressure is nonzero (Kaivarainen, 2005): $(\Delta\mathbf{VirP}^{\pm} = |\mathbf{VirP}^+| - |\mathbf{VirP}^-|) > 0$.

In accordance with our model of Bivacuum, virtual particles and antiparticles represent the asymmetric Bivacuum dipoles $(\mathbf{BVF}^{\updownarrow})^{as}$ and $(\mathbf{BVB}^{\pm})^{as}$ of three electron generations ($i = e, \mu, \tau$) in unstable state, not corresponding to Golden mean conditions (see section 2.1). Virtual particles and antiparticles are the result of correlated and opposite Bivacuum dipole symmetry fluctuations. Virtual particles, like the real sub-elementary particles, may exist in Corpuscular and Wave phases (see section 5). The Corpuscular [C]- phase, represents strongly correlated pairs of asymmetric torus ($V^+$) and antitorus ($V^-$) of two different by absolute values energies. The Wave [W]- phase, results from quantum beats between these states, which are accompanied by emission or absorption of Cumulative Virtual Clouds ($CVC^+$ or $CVC^-$), formed by subquantum particles.

Virtual particles have a mass, charge, spin, etc., but they differs from real sub-elementary ones by their lower stability (short life-time) and inability for fusion to stable triplets (see section 3). They are a singlets or very unstable triplets or other clusters of Bivacuum dipoles ($\mathbf{BVF}^\updownarrow$)$^{as}$ in contrast to real fermions-triplets.

For Virtual Clouds ($\mathbf{VC}^\pm$) and virtual pressure waves ($\mathbf{VPW}^\pm$) excited by them, the relativistic mechanics is not valid. *Consequently, the causality principle also does not work in a system (interference pattern) of* $\mathbf{VPW}^\pm$.

The energies of positive and negative $\mathbf{VPW}_q^+$ and $\mathbf{VPW}_q^-$, emitted $\rightleftharpoons$ absorbed by Bivacuum dipoles, as a result of their torus ($V^+$) and antitorus ($V^-$) transitions between $\mathbf{j}$ and $\mathbf{k}$ quantum states can be presented as:

$$\mathbf{E}^i_{\mathbf{VPW}_q^+} = \hbar\omega_0^i(\mathbf{j-k})_{V^+} = \mathbf{m}_0^i\mathbf{c}^2(\mathbf{j-k}) \qquad 1.6$$

$$\mathbf{E}^i_{\mathbf{VPW}_q^-} = -\hbar\omega_0^i(\mathbf{j-k})_{V^-} = -\mathbf{m}_0^i\mathbf{c}^2(\mathbf{j-k}) \qquad 1.6a$$

The quantized fundamental Compton frequency of $\mathbf{VPW}_q^\pm$ is:

$$\mathbf{q}\,\omega_0^i = \mathbf{q}\,\mathbf{m}_0^i\mathbf{c}^2/\hbar \qquad 1.7$$

where: $\mathbf{q} = \mathbf{j-k} = \mathbf{1,2,3}..$ is the quantization number of $\mathbf{VPW}_{j,k}^\pm$ energy;

In symmetric primordial Bivacuum the total compensation of positive and negative Virtual Pressure Waves takes a place:

$$\mathbf{q}\mathbf{E}^i_{\mathbf{VPW}_{j,k}^+} = \left| -\mathbf{q}\mathbf{E}^i_{\mathbf{VPW}_{j,k}^-} \right| = \mathbf{q}\,\hbar\omega_0^i \qquad 1.8$$

This means that the coherent excitation of $\mathbf{VPW}_{j,k}^+$ and $\mathbf{VPW}_{j,k}^-$ do not violate the energy conservation law. This is important for explanation of Bivacuum properties, as a source of 'free' energy for overunity devices (see chapter 19).

The density oscillation of $\mathbf{VC}_{j,k}^+$ and $\mathbf{VC}_{j,k}^-$ and virtual particles and antiparticles represent *positive and negative virtual pressure waves* ($\mathbf{VPW}_{j,k}^+$ and $\mathbf{VPW}_{j,k}^-$). The symmetric excitation of positive and negative energies/masses of torus and antitorus means increasing of primordial Bivacuum potential energy, corresponding to increasing of energy gap between them (see eq. 1.3):

$$[\mathbf{A}_{BVF}(\mathbf{n}) = \mathbf{E}^n_{V^+} - (-\mathbf{E}^n_{V^-}) = \hbar\omega_0(1+2\mathbf{n})]^i = \mathbf{m}_0^i\mathbf{c}^2(1+2\mathbf{n}) \qquad 1.8a$$

where quantum number: $\mathbf{n} = \mathbf{0,1,2,3}\ldots$ is equal to both - the actual torus ($V_n^+$) and complementary antitorus ($V_{\bar{n}}^-$).

The symmetric transitions/beats between the excited and basic states of torus and antitorus are accompanied by virtual pressure waves excitation of corresponding frequency (1.6 and 1.6a).

The correlated *virtual Cooper pairs* of adjacent Bivacuum fermions ($\mathbf{BVF}_{S=\pm1/2}^\updownarrow$), rotating in opposite direction with resulting spin, equal to zero and Bosonic properties, can be presented as:

$$[\mathbf{BVF}_{S=1/2}^\uparrow \bowtie \mathbf{BVF}_{S=-1/2}^\downarrow]_{S=0} \equiv [(\mathbf{V}^+\uparrow\uparrow\ \mathbf{V}^-)\ \bowtie (\mathbf{V}^+\downarrow\downarrow\ \mathbf{V}^-)]_{S=0} \qquad 1.9$$

Such a pairs, as well as Bivacuum bosons ($\mathbf{BVB}^\pm$) in conditions of ideal equilibrium,

like the *Goldstone bosons,* have zero mass and spin: $S = 0$. The virtual clouds $(\mathbf{VC}_q^{\pm})$, emitted and absorbed in a course of correlated transitions of $[\mathbf{BVF}^{\uparrow} \bowtie \mathbf{BVF}^{\downarrow}]_{S=0}^{j,k}$ between (j) and (k) sublevels: $q = j - k$, excite the virtual pressure waves $\mathbf{VPW}_q^+$ and $\mathbf{VPW}_q^-$, carrying the opposite angular momentums. They compensate the energy and momentums of each other totally in primordial Bivacuum and partly in *secondary Bivacuum* - in presence of matter and fields.

Some similarity is existing between virtual Cooper pair and Falaco vertex pair. The Falaco vertex is a topological defect in a viscous fluid, but due to its coherence it can form a long-lived metastable state in which two opposite spins are paired together. These two dimensional topological surface defects are connected by a string - one dimensional topological defect and form stabilized stationary state. Such an object can be also as the topological equivalent of pair of sub-elementary fermion and sub-elementary antifermion $[\mathbf{F}^{\uparrow} \bowtie \mathbf{F}^{\downarrow}]_{S=0}^{j,k}$, as a basic element of elementary particles (see chapter 5).

*The nonlocal virtual spin waves* $(\mathbf{VirSW}_{j,k}^{\pm 1/2})$, with properties of massless collective Nambu-Goldstone modes, like a real spin waves, represent the oscillation of angular momentum equilibrium of Bivacuum fermions with opposite spins via "flip-flop" mechanism, accompanied by origination of intermediate states - Bivacuum bosons $(\mathbf{BVB}^{\pm})$:

$$\mathbf{VirSW}_{j,k}^{\pm 1/2} \sim \left[ \mathbf{BVF}^{\uparrow}(\mathbf{V}^+ \uparrow\uparrow \mathbf{V}^-) \rightleftharpoons \mathbf{BVB}^{\pm}(\mathbf{V}^+ \Updownarrow \mathbf{V}^-) \rightleftharpoons \mathbf{BVF}^{\downarrow}(\mathbf{V}^+ \downarrow\downarrow \mathbf{V}^-) \right] \qquad 1.10$$

The $\mathbf{VirSW}_{j,k}^{+1/2}$ and $\mathbf{VirSW}_{j,k}^{-1/2}$ are excited by $(\mathbf{VC}_q^{\pm})_{S=1/2}^{\cup}$ and $(\mathbf{VC}_q^{\pm})_{S=-1/2}^{\cup}$ of opposite angular momentums, $S_{\pm 1/2} = \pm \frac{1}{2}\hbar = \pm \frac{1}{2}\mathbf{L}_0 \mathbf{m}_0 \mathbf{c}$ and frequency, equal to $\mathbf{VPW}_q^{\pm}$ (1.7):

$$\mathbf{q}\boldsymbol{\omega}_{\mathbf{VirSW}^{\pm 1/2}}^i = \mathbf{q}\boldsymbol{\omega}_{\mathbf{VPW}^{\pm}}^i = \mathbf{q}\mathbf{m}_0^i \mathbf{c}^2/\hbar = \mathbf{q}\boldsymbol{\omega}_0^i \qquad 1.10a$$

The most probable basic virtual pressure waves $\mathbf{VPW}_{q=1}^{\pm}$ and virtual spin waves $\mathbf{VirSW}_{q=1}^{\pm 1/2}$ correspond to minimum quantum number $\mathbf{q} = (\mathbf{j} - \mathbf{k}) = \mathbf{1}$.

The $\mathbf{VirSW}_q^{\pm 1/2}$, like so-called torsion field, can serve as a carrier of the phase/spin (angular momentum) and information - *qubits,* but not the energy.

The Bivacuum bosons $(\mathbf{BVB}^{\pm})$, may have two polarizations $(\pm)$, determined by spin state of their actual torus $(\mathbf{V}^+)$:

$$\mathbf{BVB}^+ = (\mathbf{V}^+ \uparrow\downarrow \mathbf{V}^-), \quad when \; \mathbf{BVF}^{\uparrow} \rightarrow \mathbf{BVF}^{\downarrow} \qquad 1.11$$

$$\mathbf{BVB}^- = (\mathbf{V}^+ \downarrow\uparrow \mathbf{V}^-), \quad when \; \mathbf{BVF}^{\downarrow} \rightarrow \mathbf{BVF}^{\uparrow} \qquad 1.11a$$

The Bose-Einstein statistics of energy distribution, valid for system of weakly interacting bosons (ideal gas), do not work for Bivacuum due to strong coupling of pairs $[\mathbf{BVF}^{\uparrow} \bowtie \mathbf{BVF}^{\downarrow}]_{S=0}$ and $(\mathbf{BVB}^{\pm})$, forming virtual Bose condensate $(\mathbf{VirBC})$ with nonlocal properties. The Bivacuum nonlocal properties can be proved, using the Virial theorem (Kaivarainen, 2004, 2005).

### 1.3 Virtual Bose condensation (VirBC), as a base of Bivacuum superfluid properties and nonlocality

It follows from our model of Bivacuum, that the infinite number of Cooper pairs of Bivacuum fermions $[\mathbf{BVF}^{\uparrow} \bowtie \mathbf{BVF}^{\downarrow}]_{S=0}^i$ and their intermediate states - Bivacuum bosons $(\mathbf{BVB}^{\pm})^i$, as elements of Bivacuum, have zero or very small (in presence of fields and matter) translational momentum: $\mathbf{p}_{\mathbf{BVF}^{\uparrow} \bowtie \mathbf{BVF}^{\downarrow}}^i = \mathbf{p}_{\mathbf{BVB}}^i \rightarrow 0$ and corresponding de Broglie wave length tending to infinity: $\boldsymbol{\lambda}_{\mathbf{VirBC}}^i = \mathbf{h}/\mathbf{p}_{\mathbf{BVF}^{\uparrow} \bowtie \mathbf{BVF}^{\downarrow}, \mathbf{BVB}}^i \rightarrow \infty$. It leads to origination of 3D net of virtual adjacent pairs of double virtual microtubules from Cooper pairs $[\mathbf{BVF}^{\uparrow} \bowtie \mathbf{BVF}^{\downarrow}]_{S=0}$, and $(\mathbf{BVB}^{\pm})_{S=0}$, which may form single microtubules. The longitudinal momentum of Bivacuum dipoles forming such virtual microfilaments and their bundles/beams can be close to zero and corresponding de Broglie wave length exceeding the distance between neighboring dipoles lot of times. Consequently, the 3D system of these twin and single microtubules, termed Virtual Guides ($\mathbf{VirG^{BVF^\uparrow \bowtie BVF^\downarrow}}$ and $\mathbf{VirG^{BVB^\pm}}$), represent Bose condensate with superfluid properties. Consequently Bivacuum, like liquid helium, can be considered as a liquid, containing two components: the described superfluid and normal, representing fraction of Bivacuum dipoles not involved in virtual guides (VirG). The radiuses of VirG are determined by the Compton radiuses of the electrons, muons and tauons. Their length is limited by decoherence effects, related to Bivacuum symmetry shift. In highly symmetric Bivacuum the length of **VirG** with nonlocal properties, connecting remote coherent elementary particles, may have the order of stars and galactics separation. However, the virtual microfilaments of **VirG** may form also a closed - ring like rotating structures with perimeter, determined by resulting de Broglie wave length of this ring elements. The life-time of such closed structures can be very big, as far they represent standing and non dissipating virtual de Broglie waves of Bivacuum dipoles.

**Nonlocality**, as the independence of potential energy on the distance from energy source in 3D net filaments of virtual (and real) Bose condensate, follows from application of the Virial theorem to systems of Cooper pairs of Bivacuum fermions $[\mathbf{BVF^\uparrow \bowtie BVF^\downarrow}]_{S=0}$ and Bivacuum bosons ($\mathbf{BVB^\pm}$) (Kaivarainen, 1995; 2004-2006).

The Virial theorem in general form is correct not only for classical, but also for quantum systems. It relates the averaged kinetic $\overline{\mathbf{T}}_k(\vec{\mathbf{v}}) = \sum_i \overline{\mathbf{m}_i \mathbf{v}_i^2/2}$ and potential $\overline{\mathbf{V}}(\mathbf{r})$ energies of particles, composing these systems:

$$2\overline{\mathbf{T}}_k(\vec{\mathbf{v}}) = \sum_i \overline{\mathbf{m}_i \mathbf{v}_i^2} = \sum_i \vec{\mathbf{r}}_i \partial \overline{\mathbf{V}}/\partial \vec{\mathbf{r}}_i \qquad 1.12$$

If the potential energy $\overline{\mathbf{V}}(\mathbf{r})$ is a homogeneous $\mathbf{x} - order$ function like:

$$\overline{\mathbf{V}}(\mathbf{r}) \sim \mathbf{r^x}, \quad \text{then} \quad \mathbf{n} = \frac{2\overline{\mathbf{T}}_k}{\overline{\mathbf{V}}(\mathbf{r})} \qquad 1.12a$$

For example, for a harmonic oscillator, when $\overline{\mathbf{T}}_k = \overline{\mathbf{V}}$, we have $\mathbf{x} = \mathbf{2}$. For Coulomb interaction: $\mathbf{x} = -\mathbf{1}$ and $\tilde{\mathbf{T}} = -\tilde{\mathbf{V}}/\mathbf{2}$.

The important consequence of the Virial theorem is that, if the average kinetic energy and momentum ($\overline{\mathbf{p}}$) of particles in a certain volume of a Bose condensate (BC) tends to zero:

$$\overline{\mathbf{T}}_k = \overline{\mathbf{p}}^2/2\mathbf{m} \rightarrow \mathbf{0} \qquad 1.13$$

the interaction between particles in the volume of BC, characterized by the radius: $\mathbf{L}_{BC} = (\hbar/\overline{\mathbf{p}}) \rightarrow 0$, becomes nonlocal, as independent on distance between them:

$$\overline{\mathbf{V}}(\mathbf{r}) \sim \mathbf{r^x} = \mathbf{1} = \mathbf{const} \quad \text{at} \quad \mathbf{x} = 2\overline{\mathbf{T}}_k/\overline{\mathbf{V}}(\mathbf{r}) = \mathbf{0} \qquad 1.14$$

Consequently, it is shown, that nonlocality, as independence of potential on the distance from potential source, is the inherent property of macroscopic Bose condensate. The individual particles (real, virtual or subquantum) in a state of Bose condensation are spatially indistinguishable due to the uncertainty principle. The Bivacuum dipoles $[\mathbf{BVF^\uparrow \bowtie BVF^\downarrow}]_{S=0}$ and $(\mathbf{BVB^\pm})_{S=0}$ due to quasi one-dimensional Bose condensation are tending to self-assembly by 'head-to-tail' principle. They compose very long virtual microtubules - Virtual Guides with wormhole properties. In special cases they form a closed structures - rotating rings with radius, dependent on velocity of rotation. The 3D net of these two kind of Virtual Guides (double $\mathbf{VirG^{BVF^\dagger \bowtie BVF^\downarrow}}$ and mono $\mathbf{VirG^{BVB^\pm}}$) bundles represent the nonlocal and superfluid fraction of Bivacuum..

## 2. Virtual Particles and Antiparticles

Generally accepted difference of virtual particles from the actual ones, is that the former, in contrast to latter, does not follow the laws of relativistic mechanics:

$$\mathbf{m} = \frac{\mathbf{m}_0}{\left[1 - (\mathbf{v/c})^2\right]^{1/2}} \qquad 2.1$$

For actual free particle with rest mass ($\mathbf{m}_0$) and relativistic mass ($\mathbf{m}$), the formula, following from (2.1) is:

$$\mathbf{E}^2 - \vec{p}^2\mathbf{c}^2 = \mathbf{m}_0^2\mathbf{c}^4 \qquad 2.2$$

where $\mathbf{E}^2 = (\mathbf{mc}^2)^2$ is the total energy squared and $\vec{\mathbf{p}} = \mathbf{m}\,\vec{\mathbf{v}}$ is the momentum of particle.

*In accordance to our model of Bivacuum, virtual particles represent asymmetric Bivacuum dipoles (BVF)$^{as}$ and (BVB)$^{as}$ of three electron's generation ($i = e, \mu, \tau$) in unstable state far from Golden mean conditions. Virtual particles, like the real sub-elementary particles, may exist in two phase: Corpuscular [C]- phase, representing correlated pairs of asymmetric torus ($V^+$) and antitorus ($V^-$) of two different energy states and Wave [W]- phase, resulting from quantum beats between these states. Corresponding transitions are accompanied by emission $\rightleftharpoons$ absorption of Cumulative Virtual Cloud (CVC$^+$ or CVC$^-$), formed by subquantum particles and antiparticles. For virtual particles the equality (2.2) is invalid.*

Virtual particles differs from real sub-elementary ones by their lower stability (short and uncertain life-time) and inability for fusion to triplets, as far their symmetry shift, determined by their external velocity and corresponding relativistic effects are not big enough to follow the Golden Mean condition (see section 5).

For Cumulative Virtual Clouds ($\mathbf{CVC}^\pm$) and excited by them periodic subquantum particles and antiparticles density oscillation in Bivacuum - virtual pressure waves ($\mathbf{VPW}_q^\pm$), the relativistic mechanics and equality (2.2) are not valid. *Consequently, the causality principle also do not works in a system of* $\mathbf{VPW}_q^\pm$, representing oscillations of subquantum particles density.

The [electron - proton] interaction is generally considered, as a result of virtual photons exchange (the cumulative virtual clouds $\mathbf{CVC}^\pm$ exchange in terms of our theory- section 13.2), when the electron and proton total energies do not change. Only the directions of their momentums are changed. In this case the energy of virtual photon in the equation (2.2) $E = 0$ and:

$$\mathbf{E}^2 - \vec{p}^2\mathbf{c}^2 = -\vec{p}^2\mathbf{c}^2 < \mathbf{0} \qquad 2.3$$

The measure of virtuality ($\mathbf{Vir}$) is a measure of Dirac's relation validity:

$$(\mathbf{Vir}) \sim \mid \mathbf{m}_0^2\mathbf{c}^4 - (\mathbf{E}^2 - \vec{p}^2\mathbf{c}^2) \mid \geq 0 \qquad 2.4$$

In contrast to actual particles, the virtual ones have a more limited radius of action. The more is the virtuality ($\mathbf{Vir}$), the lesser is the action radius. Each of emitted virtual quantum (virtual cloud) must be absorbed by the same particle or another in a course of their

[**C** ⇌ **W**] pulsations.

A lot of process in quantum electrodynamics can be illustrated by Feynman diagrams (Feynman, 1985). In these diagrams, *actual* particles are described as infinitive rays (lines) and virtual particles as stretches connecting these lines (Fig. 1).

Each peak (or angle) in Feynman diagrams means emission or absorption of quanta or particles. The energy of each process (electromagnetic, weak, strong) is described using correspondent fine structure constants.

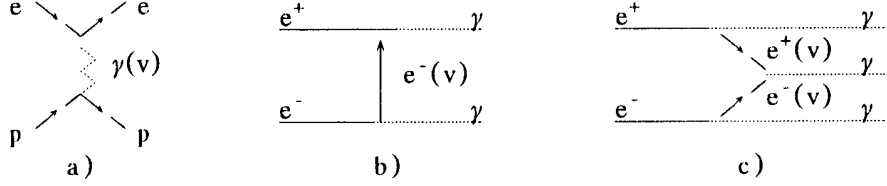

**Fig. 1.** Feynman diagrams describing electron-proton scattering (interaction), mediated by virtual photons: **a)** - annihilation of electron and positron by means of virtual electron $e^-(v)$ and virtual positron $e^+(v)$ with origination of *two* and *three* actual photons (**γ**) : diagrams **b)** and **c)** correspondingly.

### 3  Three conservation rules for asymmetric Bivacuum fermions $(\mathbf{BVF}^{\updownarrow})_{as}$ and Bivacuum bosons $(\mathbf{BVB}^{\pm})_{as}$

There are three basic postulates in our theory, interrelated with each other:

**I**.  The absolute values of internal rotational kinetic energies of torus and antitorus are equal to each other and to the half of the rest mass energy of the electrons of corresponding lepton generation, independently on the external group velocity (**v**), turning the symmetric Bivacuum fermions (**BVF**$^{\updownarrow}$) to asymmetric ones:

$$[\mathbf{I}] : \quad \left( \frac{1}{2}\mathbf{m}_V^+(\mathbf{v}_{gr}^{in})^2 = \frac{1}{2}|-\mathbf{m}_V^-|(\mathbf{v}_{ph}^{in})^2 = \frac{1}{2}\mathbf{m}_0\mathbf{c}^2 = \mathbf{const} \right)_{in}^{i} \qquad 3.1$$

where the positive $\mathbf{m}_V^+$ and negative $-\mathbf{m}_V^- = i^2\mathbf{m}_V$ are the 'actual' - inertial and 'complementary' (imaginary) - inertialess masses of torus (**V**$^+$) and antitorus (**V**$^-$); the $\mathbf{v}_{gr}^{in}$ and $\mathbf{v}_{ph}^{in}$ are the *internal* angular group and phase velocities of subquantum particles and antiparticles, forming torus and antitorus, correspondingly. In symmetric conditions of *primordial* Bivacuum and its virtual dipoles, when the influence of matter and fields is absent: $\mathbf{v}_{gr}^{in} = \mathbf{v}_{ph}^{in} = \mathbf{c}$ and $\mathbf{m}_V^+ = |-\mathbf{m}_V^-| = \mathbf{m}_0$.

It will be proved in section (7.1) of this paper, that the above condition means the infinitive life-time of torus and antitorus of **BVF**$^{\updownarrow}$ and **BVB**$^{\pm}$.

**II**.  The internal magnetic moments of torus (**V**$^+$) and antitorus (**V**$^-$) of asymmetric Bivacuum fermions $\mathbf{BVF}_{as}^{\uparrow} = [\mathbf{V}^+\uparrow\uparrow\ \mathbf{V}^-]$ and antifermions: $\mathbf{BVF}_{as}^{\downarrow} = [\mathbf{V}^+\downarrow\downarrow\ \mathbf{V}^-]$, when $\mathbf{v}_{gr}^{in} \neq \mathbf{v}_{ph}^{in}$, $\mathbf{m}_V^+ \neq |-\mathbf{m}_V^-|$ and $|\mathbf{e}_+| \neq |\mathbf{e}_-|$, are equal to each other and to that of Bohr magneton: $[\mathbf{\mu}_B = \mathbf{\mu}_0 = \frac{1}{2}|\mathbf{e}_0|\frac{\hbar}{\mathbf{m}_0\mathbf{c}}]$, independently on their internal $(\mathbf{v}_{gr,ph}^{in})_{rot}$ and external translational velocity (**v** > **0**) and mass and charge symmetry shifts.

In contrast to permanent magnetic moments of **V**$^+$ and **V**$^-$, their actual and complementary masses $\mathbf{m}_V^+$ and $|-\mathbf{m}_V^-|$, internal angular velocities ($\mathbf{v}_{gr}^{in}$ and $\mathbf{v}_{ph}^{in}$) and electric charges $|\mathbf{e}_+|$ and $|\mathbf{e}_-|$, are dependent on the external and internal velocities, however, in such a way, that they compensate each other variations:

$$[\textbf{II}] : \left( \begin{array}{c} |\pm\boldsymbol{\mu}_+| \equiv \frac{1}{2}\,|\mathbf{e}_+|\,\dfrac{|\pm h|}{|\mathbf{m}_V^+|(\mathbf{v}_{gr}^{in})_{rot}} = \ |\pm\boldsymbol{\mu}_-| \equiv \frac{1}{2}\,|-\mathbf{e}_-|\,\dfrac{|\pm h|}{|-\mathbf{m}_V^-|(\mathbf{v}_{ph}^{in})_{rot}} = \\[4mm] = \boldsymbol{\mu}_0 = \frac{1}{2}\,|\mathbf{e}_0|\,\dfrac{\hbar}{\mathbf{m}_0\mathbf{c}} = \textbf{ const} \end{array} \right)^{i} \qquad 3.2$$

This postulate reflects the condition of the invariance of magnetic moments $|\pm\boldsymbol{\mu}_\pm|$ and spin values ($\mathbf{S} = \pm\frac{1}{2}\hbar$) of torus and antitorus of Bivacuum dipoles with respect to their internal and external velocity, i.e. the absence of these parameters symmetry shifts;

**III**. The equality of Coulomb attraction force between torus and antitorus $\mathbf{V}^+ \Updownarrow \mathbf{V}^-$ of primordial Bivacuum dipoles of all three lepton generations $i = e,\ \mu,\ \tau$ (electrons, muons and tauons), providing uniform equilibrium electric energy distribution in Bivacuum:

$$[\textbf{III}] : \ \mathbf{F}_0^i = \left( \frac{\mathbf{e}_0^2}{[\mathbf{d}_{\mathbf{V}^+ \Updownarrow \mathbf{V}^-}^2]_n} \right)^e = \left( \frac{\mathbf{e}_0^2}{[\mathbf{d}_{\mathbf{V}^+ \Updownarrow \mathbf{V}^-}^2]_n} \right)^\mu = \left( \frac{\mathbf{e}_0^2}{[\mathbf{d}_{\mathbf{V}^+ \Updownarrow \mathbf{V}^-}^2]_n} \right)^\tau \qquad 3.2a$$

where: $[\mathbf{d}_{\mathbf{V}^+ \Updownarrow \mathbf{V}^-}]_n^i = \dfrac{h}{\mathbf{m}_0^i \mathbf{c}(1+2\mathbf{n})}$ is the separation between torus and antitorus of Bivacuum three pole dipoles (1.4) at the same state of excitation ($n$). A similar condition is valid as well for opposite magnetic poles interaction; $|\mathbf{e}_+|\,|\mathbf{e}_-| = \mathbf{e}_0^2$.

The important consequences of postulate **III** are the following equalities:

$$(\mathbf{e}_0\mathbf{m}_0)^e = (\mathbf{e}_0\mathbf{m}_0)^\mu = (\mathbf{e}_0\mathbf{m}_0)^\tau = \sqrt{|\mathbf{e}_+\|\mathbf{e}_-\|\mathbf{m}_V^+\mathbf{m}_V^-|} = const$$

$$\text{or:} \quad \mathbf{e}_0^\mu = \mathbf{e}_0^e(\mathbf{m}_0^e/\mathbf{m}_0^\mu); \quad \mathbf{e}_0^\tau = \mathbf{e}_0^e(\mathbf{m}_0^e/\mathbf{m}_0^\tau) \qquad 3.2b$$

This means that the toruses and antitoruses of symmetric Bivacuum dipoles of generations with bigger mass: $\mathbf{m}_0^\mu = 206,7\ \mathbf{m}_0^e;\ \ \mathbf{m}_0^\tau = 3487,28\ \mathbf{m}_0^e$ have correspondingly smaller charges.

As is shown in the next section, just these conditions provide *the same charge symmetry shift* of Bivacuum fermions of three generations ($i = e,\mu,\tau$) at the different mass symmetry shift between corresponding torus and antitorus, determined by Golden mean.

From (4.5) and (4.5a) we get, that relations, similar to 3.2b are true also for asymmetric Bivacuum dipoles of different generations if they have the same external velocities ($\mathbf{v}$):

$$\mathbf{e}_+^\mu = \mathbf{e}_+^e(\mathbf{m}_0^e/\mathbf{m}_0^\mu); \quad \mathbf{e}_+^\tau = \mathbf{e}_+^e(\mathbf{m}_0^e/\mathbf{m}_0^\tau) \qquad 3.2c$$

It follows from the second postulate (II), that the resulting magnetic moment of sub-elementary fermion or antifermion ($\boldsymbol{\mu}^\pm$), equal to the Bohr's magneton, is interrelated with the actual spin of Bivacuum fermion or antifermion as:

$$\boldsymbol{\mu}^\pm = (|\pm\boldsymbol{\mu}_+\|\pm\boldsymbol{\mu}_-|)^{1/2} = \boldsymbol{\mu}_B = \pm\frac{1}{2}\hbar\,\frac{\mathbf{e}_0}{\mathbf{m}_0\mathbf{c}} = \mathbf{S}\,\frac{\mathbf{e}_0}{\mathbf{m}_0\mathbf{c}} \qquad 3.3$$

where: $\mathbf{e}_0/\mathbf{m}_0\mathbf{c}$ is gyromagnetic ratio of Bivacuum fermion, equal to that of the electron.

One may see from (3.3), that the spin of the actual torus, equal to that of the resulting spin of Bivacuum fermion (symmetric or asymmetric), is:

$$\mathbf{S} = \pm\frac{1}{2}\hbar \qquad 3.4$$

Consequently, the permanent absolute value of spin of torus and antitorus is a consequence of 2nd postulate.

It is assumed also in our approach, that the dependence of the *actual inertial* mass ($\mathbf{m}_V^+$)

of torus $\mathbf{V}^+$ of asymmetric Bivacuum fermions ($\mathbf{BVF}_{as}^\uparrow = \mathbf{V}^+\uparrow\uparrow\ \mathbf{V}^-$) on the external translational group velocity ($\mathbf{v}$) follows relativistic mechanics:

$$\pm\ \mathbf{m}_V^+ = \frac{\mathbf{m}_0}{\pm\sqrt{1-(\mathbf{v}/\mathbf{c})^2}}\ = \mathbf{m}\quad \text{(inertial mass)}\qquad\qquad 3.5$$

while the *complementary inertialess* mass ($\mp\mathbf{m}_{\bar{V}}^-$) of antitorus $\mathbf{V}^-$ has the sign, opposite to that of the actual one ($\pm\mathbf{m}_V^+$) the reverse velocity dependence:

$$\mp\ \mathbf{m}_{\bar{V}}^- = \mp\mathbf{m}_0\sqrt{1-(\mathbf{v}/\mathbf{c})^2}\quad \text{(inertialess mass)}\qquad\qquad 3.6$$

The product of actual (inertial) and complementary (inertialess) mass is a constant, equal to the rest mass of particle squared and reflect the *mass compensation principle*. It means, that increasing of mass/energy of the torus is compensated by in-phase decreasing of absolute values of these parameters for antitorus and vice versa:

$$|\pm\mathbf{m}_V^+|\ |\mp\mathbf{m}_{\bar{V}}^-| = \mathbf{m}_0^2\qquad\qquad 3.7$$

Taking (3.7) and (3.1) into account, we get for the product of the *internal* group and phase velocities of positive and negative subquantum particles, forming torus and antitorus, correspondingly:

$$\mathbf{v}_{gr}^{in}\ \mathbf{v}_{ph}^{in} = \mathbf{c}^2\qquad\qquad 3.8$$

A similar symmetric relation is reflecting the *charge compensation principle*:

$$|\mathbf{e}_+|\ |\mathbf{e}_-| = \mathbf{e}_0^2\qquad\qquad 3.9$$

The sum of the actual (positive) and the complementary (negative) total energies of (3.5 and 3.6), i.e. the resulting energy of Bivacuum *fermion* ($\mathbf{BVF}_{as}^\uparrow$) is equal to its doubled external kinetic energy, anisotropic in general case:

$$\left[(\mathbf{m}_V^+ - \mathbf{m}_{\bar{V}}^-)\mathbf{c}^2 = \mathbf{m}_V^+\mathbf{v}^2 = 2T_k = \frac{\mathbf{m}_0\mathbf{v}^2}{\sqrt{1-(\mathbf{v}/\mathbf{c})^2}}\right]_{x,y,z}^i\qquad\qquad 3.10$$

In asymmetric Bivacuum fermions $\left(\mathbf{BVF}_{as}^\uparrow = \mathbf{V}^+\uparrow\uparrow\ \mathbf{V}^-\right)^i$ and Bivacuum antifermions $\left(\mathbf{BVF}_{as}^\downarrow = \mathbf{V}^+\downarrow\downarrow\ \mathbf{V}^-\right)^i$ the actual and complementary torus and antitorus change their place, as well as relativistic dependence of their opposite mass and charge on the external velocity of Bivacuum dipoles ($\mathbf{v}$). Similar exchange of the notions of the actual and complementary torus and antitorus and their relativistic dependence on ($\mathbf{v}$) takes a place for Bivacuum bosons of opposite polarization: ($\mathbf{BVB}^+ = \mathbf{V}^+\uparrow\downarrow\ \mathbf{V}^-)^i$ and $(\mathbf{BVB}^- = \mathbf{V}^+\downarrow\uparrow\ \mathbf{V}^-)^i$. We assume, that the actual mass of asymmetric dipoles of Bivacuum with regular relativistic dependence is always positive (like in conventional consideration of particles and antiparticles) and the uncompensated energy of Bivacuum dipoles is determined by the *absolute* value of their mass symmetry shift.

The resulting energy of asymmetric Bivacuum *antifermion* and negatively polarized Bivacuum boson, the formula (3.10) turns to shape:

$$\left[ (\overline{\mathbf{m}_V^-} - \overline{\mathbf{m}_V^+})\mathbf{c}^2 = \mathbf{m}_V^- \mathbf{v}^2 = 2\mathbf{T}_k = \frac{\mathbf{m}_0 \mathbf{v}^2}{\sqrt{1 - (\mathbf{v/c})^2}} \right]^i_{x,y,z}$$

where in contrast to (3.5) and (3.6), the relativistic dependences of torus and antitorus change their place:

$$\pm \overline{\mathbf{m}_V^-} = \frac{\mathbf{m}_0}{\pm \sqrt{1 - (\mathbf{v/c})^2}} \text{ (inertial mass)}$$

$$\mp \overline{\mathbf{m}_V^+} = \mp \mathbf{m}_0 \sqrt{1 - (\mathbf{v/c})^2} \text{ (inertialess mass)}$$

The fundamental Einstein equation for total energy of particle can be reformed and extended, using eqs. 3.10 and 3.7:

$$\mathbf{E}_{tot} = \mathbf{m}_V^+ \mathbf{c}^2 = \mathbf{mc}^2 = \mathbf{m}_V^- \mathbf{c}^2 + \mathbf{m}_V^+ \mathbf{v}^2 \qquad 3.10a$$

$$or: \ \mathbf{E}_{tot} = \mathbf{m}_V^+ \mathbf{c}^2 = \frac{\mathbf{m}_0^2}{\mathbf{m}_V^+} \mathbf{c}^2 + \mathbf{m}_V^+ \mathbf{v}^2 \qquad 3.10b$$

$$or: \ \mathbf{E}_{tot} = \mathbf{m}_V^+ \mathbf{c}^2 = \sqrt{1 - (\mathbf{v/c})^2}\, \mathbf{m}_0 \mathbf{c}^2 + 2\mathbf{T}_k \qquad 3.10c$$

The ratio of absolute values (3.6) to (3.5), taking into account (3.7), is:

$$\frac{|-\mathbf{m}_V^-|}{\mathbf{m}_V^+} = \frac{\mathbf{m}_0^2}{(\mathbf{m}_V^+)^2} = 1 - \left(\frac{\mathbf{v}}{\mathbf{c}}\right)^2 \qquad 3.11$$

It can easily be transformed to the important formula for resulting external energy of Bivacuum dipoles (3.10).

The opposite shift of symmetry between $\mathbf{V}^+$ and $\mathbf{V}^-$ of two Bivacuum fermions of opposite spins occur due to relativistic effects, accompanied their rotation *side-by-side* as a Cooper pairs $[\mathbf{BVF}^\uparrow \bowtie \mathbf{BVF}^\downarrow]_{as}$ around the common axe. In this case the quantum beats between $\mathbf{V}^+$ and $\mathbf{V}^-$ of $\mathbf{BVF}^\uparrow$ and $\mathbf{BVF}^\downarrow$ can occur in the same phase.

When the external velocity ($\mathbf{v}$) of the external rotation of pair $[\mathbf{BVF}^\uparrow \bowtie \mathbf{BVF}^\downarrow]_{as}$ reach the Golden mean (GM) condition ($\mathbf{v}^2/\mathbf{c}^2 = \phi = 0.618$), this results in origination of *the rest mass*: $\mathbf{m}_0 = |\mathbf{m}_V^+ - \mathbf{m}_V^-|^\phi$ and *elementary charge*: $\mathbf{e}^\phi = |\mathbf{e}_+ - \mathbf{e}_-|$ of opposite sign for sub-elementary fermion: $\left(\mathbf{BVF}_{as}^\uparrow\right)^\phi = \mathbf{F}_\uparrow^+$ and sub-elementary antifermion $\left(\mathbf{BVF}_{as}^\downarrow\right)^\phi = \mathbf{F}_\downarrow^-$ with spatial image of pair of truncated cone of opposite symmetry (section 4.1). The resulting mass/energy, charge and spin of *Cooper pairs* $[\mathbf{F}_\uparrow^+ \bowtie \mathbf{F}_\downarrow^-]$ is zero because of compensation effects.

On the other hand, two adjacent asymmetric Bivacuum fermions and antifermions of *similar* direction of rotation and similar semi-integer spins can not rotate 'side-by-side', like in Cooper pairs: $[\mathbf{BVF}^\uparrow \bowtie \mathbf{BVF}^\downarrow]_{as}$, compensating each other, but only as 'head-to-tail' *complexes* in clockwise or anticlockwise directions:

$$\mathbf{N}^+[\mathbf{BVF}^\uparrow + \mathbf{BVF}^\uparrow]_{as} \quad or \quad \mathbf{N}^-[\mathbf{BVF}^\downarrow + \mathbf{BVF}^\downarrow]_{as} \qquad 3.12$$

In such bosonic configuration, corresponding to integer spin, the energy/ mass, charge and half-integer spin of the Bivacuum dipoles, are the additive values.

As far in primordial Bivacuum the average mass/energy, charge and spin should be zero, it means that the number of 'head-to-tail' pairs of Bivacuum fermions with boson properties is equal to similar bosonic pairs of Bivacuum antifermions: $\mathbf{N}^+ = \mathbf{N}^-$.

In contrast to Bivacuum fermions, which may self-assemble to the doubled virtual microtubules only, the Bivacuum bosons may polymerize also into the mono filaments of two opposite polarization (±), as far it do not contradict the Pauli principle:

$$\sum(\mathbf{BVB^+} = \mathbf{V^+\uparrow\downarrow \ V^-})^i \quad \text{and} \quad \sum(\mathbf{BVB^-} = \mathbf{V^+\downarrow\uparrow \ V^-})^i \qquad 3.13$$

It follows from our model of elementary particles (chapter 5), that the described above opposite symmetry shift of paired *side-by-side* Bivacuum fermions, antifermions and Bivacuum bosons of opposite polarization occur, as a result of their rotation around the common axes with tangential external velocity (**v**).

The 'head-to-tail' associated Bivacuum dipoles may form the straight/linear mono and doubled virtual microtubules, connecting "Sender" and "Receiver" (virtual filaments). Another possible configurations of Bivacuum dipoles self-assembly is a rotating *circles/rings* with perimeter, equal de Broglie wave length of mono dipoles $(\mathbf{BVB^\pm})_{as}$ or their pairs $[\mathbf{BVF^\uparrow} \bowtie \mathbf{BVF^\downarrow}]_{as}$, determined by their actual mass $(\mathbf{m}_V^+)^i$ due to mass symmetry shift $\left(\mathbf{m}_V^+ - \mathbf{m}_V^-\right)^i$, determined, in turn, by their tangential velocity (**v**) of ring rotation:

$$\lambda_{Vir}^i = 2\pi\mathbf{L}_{Vir}^i = \frac{h}{p_{BVB^\pm,BVF}^i} = \frac{h}{\left(\mathbf{m}_V^+ - \mathbf{m}_V^-\right)^i \mathbf{c}} = \frac{hc}{(\mathbf{m}_V^+)^i\mathbf{v}^2} \qquad 3.14$$

This condition corresponds to that of standing de Broglie wave of particle with mass $(\mathbf{m}_V^+)^i$ and tangential velocity (**v**).

For single Bivacuum dipoles ($\mathbf{BVF^\uparrow}$, $\mathbf{BVF^\downarrow}$ and $\mathbf{BVF^\pm})^i$, the conversion of their torus ($\mathbf{V^+}$) or antitorus ($\mathbf{V^-}$) to the actual one depends on direction of Bivacuum dipoles propagation in direction, parallel to the main axes of dipoles rotation. For example, just the *frontier torus* ($\mathbf{V^+}$) of dipole $[\mathbf{V^+}\Updownarrow \mathbf{V^-}]$ as respect to direction of dipole propagation becomes the actual.

In the opposite direction of this dipole propagation with translational velocity ($\vec{\mathbf{v}}$), the antitorus ($\mathbf{V^-}$) turns to the actual one. In the intermediate direction of Bivacuum dipole motion, the probability of torus or antitorus to became actual one, is proportional to (**cos** $\theta$), where $\theta$ is the angle between vectors of dipole velocity ($\vec{\mathbf{v}}$) and vector of its internal symmetry shift $[\mathbf{V^-} \Longrightarrow \mathbf{V^+}]$. In strong electrostatic or gravitational fields tension gradients, the induced vector of Bivacuum dipoles polarization coincides with vector of their external momentum. This means that the probability of the 'frontier' torus or antitorus 'actualization': $P^\pm \sim \mathbf{cos}\,\theta \to 1$, as far $\theta \to 0$.

## 4 The relation between the external and internal parameters of Bivacuum fermions. Quantum roots of Golden mean

The important formula, unifying a lot of internal and external (translational-rotational) parameters of $\mathbf{BVF}_{as}^\updownarrow$, taking into account that the product of internal and external phase and group velocities is equal to light velocity squared:

$$\mathbf{v}_{ph}^{in}\mathbf{v}_{gr}^{in} = \mathbf{v}_{ph}^{ext}\mathbf{v}_{gr}^{ext} = \mathbf{c}^2 \qquad 4.1$$

can be derived from eqs. (3.1 - 3.11):

$$\left(\frac{\mathbf{m}_V^+}{\mathbf{m}_V^-}\right)^{1/2} = \frac{\mathbf{m}_V^+\mathbf{c}^2}{\mathbf{m}_0\mathbf{c}^2} = \frac{\mathbf{v}_{ph}^{in}}{\mathbf{v}_{gr}^{in}} = \left(\frac{\mathbf{c}}{\mathbf{v}_{gr}^{in}}\right)^2 =$$  (4.2)

$$= \frac{\mathbf{L}_V^-}{\mathbf{L}_V^+} = \frac{\mathbf{L}_0^2}{(\mathbf{L}_V^+)^2} = \frac{|\mathbf{e}_+|}{|\mathbf{e}_-|} = \left(\frac{\mathbf{e}_+}{\mathbf{e}_0}\right)^2 = \frac{1}{[1-(\mathbf{v}^2/\mathbf{c}^2)^{ext}]^{1/2}}$$  (4.2a)

where:

$$\mathbf{L}_V^+ = \hbar/(\mathbf{m}_V^+\mathbf{v}_{gr}^{in}) = \mathbf{L}_0[1-(\mathbf{v}^2/\mathbf{c}^2)^{ext}]^{1/4}$$  (4.3)

$$\mathbf{L}_V^- = \hbar/(\mathbf{m}_V^-\mathbf{v}_{ph}^{in}) = \frac{\mathbf{L}_0^2}{\mathbf{L}_V^+} = \frac{\mathbf{L}_0}{[1-(\mathbf{v}^2/\mathbf{c}^2)^{ext}]^{1/4}}$$

$$\mathbf{L}_0 = (\mathbf{L}_V^+\mathbf{L}_V^-)^{1/2} = \hbar/\mathbf{m}_0\mathbf{c} \ - \ Compton \ radius$$  (4.3a)

are the radii of torus ($\mathbf{V}^+$), antitorus ($\mathbf{V}^-$) and the resulting radius of $\mathbf{BVF}_{as}^{\updownarrow} = [\mathbf{V}^+ \Updownarrow \mathbf{V}^-]$, equal to Compton radius, correspondingly.

The *absolute* external velocity of Bivacuum dipoles, squared, as respect to primordial Bivacuum (absolute reference frame), can be expressed, using 4.2 and 4.2a, as a criteria of parameters of torus and antitorus symmetry shift as:

$$\left[\mathbf{v}^2 = \mathbf{c}^2\left(1 - \frac{\mathbf{m}_V^-}{\mathbf{m}_V^+}\right) = \mathbf{c}^2\left(1 - \frac{\mathbf{e}^2}{\mathbf{e}_+^2}\right) = \mathbf{c}^2\left(1 - \frac{\mathbf{S}_+}{\mathbf{S}_-}\right)\right]_{x,y,z}$$  (4.4)

where: $\mathbf{S}_+ = \pi(\mathbf{L}_V^+)^2$ and $\mathbf{S}_- = \pi(\mathbf{L}_V^-)^2$ are the squares of cross-sections of torus and antitorus of Bivacuum dipoles.

*The existence of absolute velocity in our Unified theory (anisotropic in general case) and the Universal reference frame of Primordial Bivacuum, pertinent for Ether concept, is an important difference with Special relativity theory. The light velocity in UT, like sound velocity in the matter, is a function of Bivacuum (primary or secondary) matrix properties.*

The relativistic dependences of the actual charge $\mathbf{e}_+$ and actual mass ($\mathbf{m}_V^+$) on external **absolute** velocity of Bivacuum dipole, following from (4.2a) and (3.5) are:

$$\mathbf{e}_+ = \frac{\mathbf{e}_0}{[1-(\mathbf{v}^2/\mathbf{c}^2)]^{1/4}}$$  (4.5)

$$\mathbf{m}_V^+ = \frac{\mathbf{m}_0}{\sqrt{1-(\mathbf{v}/\mathbf{c})^2}}$$  (4.5a)

The influence of relativistic dependence of *real* particles charge on the resulting charge and electric field density of Bivacuum, which is known to be electrically quasi neutral vacuum/bivacuum, is negligible for two reasons:

1. Densities of positive and negative real charges (i.e. particles and antiparticles) are very small and approximately equal. This quasi-equilibrium of opposite charges is Lorentz invariant;

2. The remnant uncompensated by real antiparticles charges density at any velocities can be compensated totally by virtual antiparticles and asymmetric Bivacuum fermions (BVF) of opposite charges.

The ratio of the actual charge to the actual inertial mass from (4.5 and 4.5a) has also the relativistic dependence:

$$\frac{\mathbf{e}_+}{\mathbf{m}_V^+} = \frac{\mathbf{e}_0}{\mathbf{m}_0}\left[1 - (\mathbf{v}^2/\mathbf{c}^2)\right]^{1/4} \qquad 4.6$$

The decreasing of this ratio with velocity increasing is weaker, than it follows from the generally accepted statement, that charge has no relativistic dependence in contrast to the actual mass $\mathbf{m}_V^+$. The direct experimental investigation of relativistic dependence of this ratio on the external velocity ($\mathbf{v}$) may confirm the validity of our formula (4.6) and general approach.

From eqs. (3.10); (3.13) and (3.13a) we find for mass and charge symmetry shift:

$$\Delta\mathbf{m}_\pm = \mathbf{m}_V^+ - \mathbf{m}_{\overline{V}}^- = \mathbf{m}_V^+\left(\frac{\mathbf{v}}{\mathbf{c}}\right)^2 \qquad 4.7$$

$$\Delta\mathbf{e}_\pm = \mathbf{e}_+ - \mathbf{e}_- = \frac{\mathbf{e}_+^2}{\mathbf{e}_+ + \mathbf{e}_-}\left(\frac{\mathbf{v}}{\mathbf{c}}\right)^2 \qquad 4.7a$$

These mass and charge symmetry shifts determines the relativistic dependence of the *effective* mass and charge of the fermions. In direct experiments only the actual mass ($\mathbf{m}_V^+$) and charge ($\mathbf{e}_\pm$) can be registered. It means that the complementary mass ($\mathbf{m}_{\overline{V}}^-$) and charge are *hidden* for observation.

The ratio of charge to mass symmetry shifts (the *effective* charge and mass ratio) is:

$$\frac{\Delta\mathbf{e}_\pm}{\Delta\mathbf{m}_\pm} = \frac{\mathbf{e}_+^2}{\mathbf{m}_V^+\left(\mathbf{e}_+ + \mathbf{e}_-\right)} \qquad 4.8$$

The mass symmetry shift can be expressed via the squared charges symmetry shift also in the following manner:

$$\Delta\mathbf{m}_\pm = \mathbf{m}_V^+ - \mathbf{m}_{\overline{V}}^- = \mathbf{m}_V^+\frac{\mathbf{e}_+^2 - \mathbf{e}_-^2}{\mathbf{e}_+^2} \qquad 4.8a$$

or using (3.11) this formula turns to:

$$\frac{\mathbf{e}_+^2 - \mathbf{e}_-^2}{\mathbf{e}_+^2} = \frac{\mathbf{v}^2}{\mathbf{c}^2} \qquad 4.9$$

When the mass and charge symmetry shifts of Bivacuum dipoles are small and $|\mathbf{e}_+| + |\mathbf{e}_-| \simeq 2\mathbf{e}_+ \simeq 2\mathbf{e}_0$, we get from 4.7a for variation of charge shift:

$$\Delta\mathbf{e}_\pm = \mathbf{e}_+ - \mathbf{e}_- = \frac{1}{2}\mathbf{e}_0\frac{\mathbf{v}^2}{\mathbf{c}^2} \qquad 4.10$$

The formula, unifying the *internal* and *external* group and phase velocities of asymmetric Bivacuum fermions ($\mathbf{BVF}_{as}^\updownarrow$), derived from (4.2) and (4.2a), is:

$$\left(\frac{\mathbf{v}_{gr}^{in}}{\mathbf{c}}\right)^4 = 1 - \left(\frac{\mathbf{v}}{\mathbf{c}}\right)^2 \qquad 4.11$$

where: $(\mathbf{v}_{gr}^{ext}) \equiv \mathbf{v}$ is the external translational-rotational group velocity of $\mathbf{BVF}_{as}^\updownarrow$.

At the conditions of "Hidden Harmony", meaning the equality of the internal and external rotational group and phase velocities of asymmetric Bivacuum fermions $\mathbf{BVF}_{as}^\updownarrow$:

$$(\mathbf{v}_{gr}^{in})_{V^+}^{rot} = (\mathbf{v}_{gr}^{ext})^{tr} \equiv \mathbf{v} \qquad 4.12$$

$$(\mathbf{v}_{ph}^{in})_{V^-}^{rot} = (\mathbf{v}_{ph}^{ext})^{tr} \qquad 4.12a$$

and introducing the notation:

$$\left(\frac{\mathbf{v}^{in}_{gr}}{\mathbf{c}}\right)^2 = \left(\frac{\mathbf{v}}{\mathbf{c}}\right)^2 = \left(\frac{\mathbf{v}^{in}_{gr}}{\mathbf{v}^{in}_{ph}}\right) = \left(\frac{\mathbf{v}^{ext}_{gr}}{\mathbf{v}^{ext}_{ph}}\right) \equiv \phi \qquad 4.13$$

formula (4.11) turns to a simple quadratic equation:

$$\phi^2 + \phi - 1 = 0, \qquad 4.14$$

which has a few modes : $\quad \phi = \dfrac{1}{\phi} - 1 \quad or : \quad \dfrac{\phi}{(1-\phi)^{1/2}} = 1 \qquad 4.14a$

$$or : \quad \frac{1}{(1-\phi)^{1/2}} = \frac{1}{\phi} \qquad 4.14b$$

The solution of (4.14), is equal to **Golden mean**: $(\mathbf{v}/\mathbf{c})^2 = \phi = 0.618$. *It is remarkable, that the Golden Mean, which plays so important role on different Hierarchic levels of matter organization: from elementary particles to galactics and even in our perception of beauty (i.e. our mentality), has so deep physical roots, corresponding to Hidden Harmony conditions (4.12 and 4.12a). Our theory is the first one, elucidating these roots (Kaivarainen, 1995; 2000; 2005). This important fact points, that we are on the right track.*

The overall shape of asymmetric $\left(\mathbf{BVF}^{\ddagger}_{as} = \left[\mathbf{V}^{+} \Updownarrow \mathbf{V}^{-}\right]\right)^{i}$ is a *truncated cone* (Fig.2) with plane, parallel to the base with radiuses of torus ($L^{+}$) and antitorus ($L^{-}$), defined by eq. (4.3).

### 4.1 The rest mass and charge origination

Using Golden Mean equation in the form (4.14b), we can see, that all the ratios (4.2 and 4.2a) at Golden Mean conditions turn to:

$$\left[\left(\frac{\mathbf{m}^{+}_{V}}{\mathbf{m}^{-}_{V}}\right)^{1/2} = \frac{\mathbf{m}^{+}_{V}}{\mathbf{m}_0} = \frac{\mathbf{v}^{in}_{ph}}{\mathbf{v}^{in}_{gr}} = \frac{\mathbf{L}^{-}}{\mathbf{L}^{+}} = \frac{|\mathbf{e}_{+}|}{|\mathbf{e}_{-}|} = \left(\frac{\mathbf{e}_{+}}{\mathbf{e}_0}\right)^2\right]^{\phi} = \frac{1}{\phi} \qquad 4.15$$

where the actual ($e_+$) and complementary ($e_-$) charges and corresponding mass at GM conditions are:

$$\mathbf{e}^{\phi}_{+} = \mathbf{e}_0/\boldsymbol{\phi}^{1/2}; \qquad \mathbf{e}^{\phi}_{-} = \mathbf{e}_0\boldsymbol{\phi}^{1/2} \qquad 4.16$$

$$(\mathbf{m}^{+}_{V})^{\phi} = \mathbf{m}_0/\boldsymbol{\phi}; \qquad (\mathbf{m}^{-}_{V})^{\phi} = \mathbf{m}_0\boldsymbol{\phi} \qquad 4.16a$$

using (4.16 and 4.16a) it is easy to see, that the difference between the actual and complementary mass at GM conditions is equal to the rest mass:

$$\left[|\boldsymbol{\Delta}\mathbf{m}_{V}|^{\phi} = \mathbf{m}^{+}_{V} - \mathbf{m}^{-}_{V} = \mathbf{m}_0(1/\boldsymbol{\phi} - \boldsymbol{\phi}) = \mathbf{m}_0\right]^{e,\mu,\tau} \qquad 4.17$$

*This is an important result, pointing that just a symmetry shift, determined by the Golden mean conditions, is responsible for origination of the rest mass of sub-elementary particles of each of three generation ($i = e, \mu, \tau$).*

*The same is true for charge origination.* The GM difference between actual and complementary charges, using relation $\boldsymbol{\phi} = (1/\boldsymbol{\phi} - 1)$, determines corresponding minimum charge of sub-elementary fermions or antifermions (at $\mathbf{v}^{ext}_{tr} \to \mathbf{0}$):

$$\phi^{3/2}\mathbf{e}_0 = |\Delta\mathbf{e}_\pm|^\phi = |\mathbf{e}_+ - \mathbf{e}_-|^\phi \equiv |\mathbf{e}|^\phi \qquad 4.18$$

$$\text{where:} \quad (|\mathbf{e}_+||\mathbf{e}_-|) = \mathbf{e}_0^2 \qquad 4.18a$$

The absolute values of charge symmetry shifts for electron, muon and tauon at GM conditions are the same. This result determines the equal absolute values of empirical rest charges of the electron, positron, proton and antiproton. However, the mass symmetry shifts at GM conditions, equal to the rest mass of electrons, muons and tauons are very different.

The ratio of charge to mass symmetry shifts at Golden mean (GM) conditions ($\mathbf{v}_{tr}^{ext} = 0$) is a permanent value for all three electron generations ($e,\ \mu,\ \tau$). The different values of their rest mass are taken into account by postulate III and it consequences of their rest mass and charge relations: $\mathbf{e}_0^\mu = \mathbf{e}_0^e(\mathbf{m}_0^e/\mathbf{m}_0^\mu);\ \ \mathbf{e}_0^\tau = \mathbf{e}_0^e(\mathbf{m}_0^e/\mathbf{m}_0^\tau)$ (see 3.2c) :

$$\left[\ \frac{|\Delta\mathbf{e}_\pm|^\phi}{|\Delta\mathbf{m}_V|^\phi} = \frac{|\mathbf{e}^i|^\phi}{\mathbf{m}_0^e} = \frac{|\mathbf{e}_+|^\phi\phi}{|\mathbf{m}_V^+|^\phi} = \frac{\mathbf{e}_0\phi^{3/2}}{\mathbf{m}_0} = \frac{\mathbf{e}_0\phi^{3/2}}{\mathbf{m}_0^{\mu,\tau}}\ \right]^{e,\mu,\tau} \qquad 4.19$$

where: $(\mathbf{m}_V^+)^\phi = \mathbf{m}_0/\phi$ is the actual mass of unpaired sub-elementary fermion in [C] phase at Golden mean conditions (see next section); $\mathbf{e}_0 \equiv \mathbf{e}_0^e;\ \mathbf{m}_0^e \equiv \mathbf{m}_0$.

Formula (4.19) can be considered as a background of permanent value of gyromagnetic ratio, equal to ratio of magnetic moment of particle to its angular momentum (spin). For the electron it is:

$$\Gamma = \frac{\mathbf{e}_0}{2\mathbf{m}_e\mathbf{c}} \qquad 4.20$$

A huge amount of information, pointing that Golden mean plays a crucial role in Nature, extrapolating similar basic principles of matter formation on higher than elementary particles hierarchical levels, starting from DNA level up to galactics spatial organization, are collected and analyzed in the impressive web site of Dan Winter: http://www.soulinvitation.com/indexdw.html

### 4.2 Quantization of the rest mass/energy and charge of sub-elementary fermions

Formula (3.10), using (4.19), can be transformed to following shape:

$$\mathbf{n}^2 \equiv \left(\frac{\Delta\mathbf{m}_V^+}{\mathbf{m}_0}\right)^2 = \left(\frac{\Delta\mathbf{e}}{\mathbf{e}_0\phi^{3/2}}\right)^2 = \frac{(\mathbf{v}/\mathbf{c})^4}{1 - (\mathbf{v}/\mathbf{c})^2} \qquad 4.21$$

Introducing the definition: $(\mathbf{v}/\mathbf{c})^2 = \mathbf{x}$, eq. 4.21 can be reduced to quadratic equation:

$$\mathbf{x}^2 + \mathbf{n}^2\mathbf{x} - \mathbf{n}^2 = 0 \qquad 4.22$$

The solution of this equation is:

$$\mathbf{x} = \frac{1}{2}\left[-\mathbf{n}^2 + \sqrt{\mathbf{n}^4 + 4\mathbf{n}^2}\right] \qquad 4.23$$

It is easy to calculate, that at $\mathbf{n} = \mathbf{1}$, $\mathbf{n}^2 = \mathbf{1}$ and $\Delta\mathbf{m}_V^+ = \mathbf{m}_0$; $\Delta\mathbf{e} = \mathbf{e}_0\phi^{3/2}$ we have $\mathbf{x}_{n=1} = (\mathbf{v}/\mathbf{c})^2 = \mathbf{0.618} = \phi$.

At $\mathbf{n} = \mathbf{2}$, $\mathbf{n}^2 = \mathbf{4}$ and $\Delta\mathbf{m}_V^+ = 2\mathbf{m}_0$; $\Delta\mathbf{e} = 2\mathbf{e}_0\phi^{3/2}$ we have $\mathbf{x}_{n=2} = \mathbf{0.8284} = \mathbf{1.339}\phi$.

At $\mathbf{n} = \mathbf{3}$, $\mathbf{n}^2 = \mathbf{9}$ and $\Delta\mathbf{m}_V^+ = 3\mathbf{m}_0$; $\Delta\mathbf{e} = 3\mathbf{e}_0\phi^{3/2}$ we have $\mathbf{x}_{n=3} = \mathbf{0.9083} = \mathbf{1.469}\phi$

At $\mathbf{n} = \mathbf{4}$, $\mathbf{n}^2 = \mathbf{16}$ and $\Delta\mathbf{m}_V^+ = 4\mathbf{m}_0$; $\Delta\mathbf{e} = 4\mathbf{e}_0\phi^{3/2}$ we have $\mathbf{x}_{n=4} = \mathbf{0.9442} = \mathbf{1.528}\phi$

### 4.3 The ratio of energies at Golden mean and Dead mean conditions

The known formula, unifying the ratio of phase and group velocity of relativistic de Broglie wave $(\mathbf{v}_{ph}/\mathbf{v}) = (\mathbf{c}^2/\mathbf{v}^2)$ with ratio of its potential energy to kinetic one $(V_B/T_k)$ is:

$$2\frac{\mathbf{v}_{ph}}{\mathbf{v}} - 1 = \frac{\mathbf{V}_B}{\mathbf{T}_k} \qquad\qquad 4.24$$

It is easy to see from (4.24), that at GM condition: $(\mathbf{v}_{ph}/\mathbf{v})^\phi = (\mathbf{c}^2/\mathbf{v}^2)^\phi = 1/\phi$, the ratio:

$$(V_B/T_k)^\phi = 2.236 \quad and \quad [T_k/(T_k + V_B)]^\phi = [T_k/E_B]^\phi = 0.309 \qquad 4.25$$

The Golden mean (GM) conditions for sub-elementary particles, composing free elementary particles are the result of their fast rotation at GM or Compton frequency (section 5):

$$\boldsymbol{\omega}_0^i = \mathbf{m}_0^i \mathbf{c}^2/\hbar \qquad\qquad 4.25a$$

Such spinning of sub-elementary particles in triplets around the common axis (Fig.2), at the Hidden Harmony conditions, when their internal and external group and phase velocities coincide (eq.4.12; 4.12a).

In contrast to Golden mean (4.25), we may introduce here the "**Dead mean**", corresponding to thermal equilibrium. At this conditions any system can be described by the number of independent harmonic oscillators, unable to coupling and self-organization:

$$\left[\frac{\mathbf{V}}{\mathbf{T}_k}\right]^D = \mathbf{1}; \qquad \left[\frac{2\mathbf{T}_k}{\mathbf{E}_B}\right]^D = \left[\frac{\mathbf{T}_k + \mathbf{V}}{\mathbf{E}_B}\right]^D = \mathbf{1} \qquad 4.26$$

The deep natural roots of Golden mean, as a consequence of Hidden Harmony conditions (4.12), leading from our theory, explain the universality of this number ($\phi = 0.618$).

It is demonstrated in our work, that any kind of selected system, able to self-assembly, self-organization and evolution: from atoms to living organisms and from galactics to Universe - are tending to conditions of combinational resonance with virtual pressure waves under the action Tuning Energy (TE) of Bivacuum (section 15).

The less is deviation of ratio of characteristic parameters (dynamic and spatial) of system from $[\phi \equiv Phi]$, the more advanced is evolution of this system. We have to keep in mind that all forms of matter are composed from hierarchic systems of de Broglie waves.

### 4.4 The solution of Dirac monopole problem, following from Unified theory

The Dirac theory, searching for elementary magnetic charges ($g^-$ and $g^+$), symmetric to electric ones ($e^-$ and $e^+$), named **monopoles**, leads to following relation between the magnetic monopole and electric charge of the same signs:

$$\mathbf{ge} = \frac{n}{2}\hbar\mathbf{c} \;\; or: \;\; \mathbf{g} = \frac{n}{2}\frac{\hbar\mathbf{c}}{\mathbf{e}} = \frac{n}{2}\frac{\mathbf{e}}{\boldsymbol{\alpha}} \qquad\qquad 4.27$$

*where : $n = 1, 2, 3$ is the integer number*

where $\boldsymbol{\alpha} = \mathbf{e}^2/\hbar\mathbf{c}$ is the electromagnetic fine structure constant.

It follows from this definition, that minimal magnetic charge (*at* **n = 1**) is as big as $g \cong 67.7e$. The mass of monopole should be huge $\sim 10^{16}\,GeV$. All numerous attempts to reveal such particles experimentally has failed.

Our theory explains this fact in such a way: in contrast to *electric and mass dipoles* symmetry shifts (see 4.17 and 4.18), the symmetry violation between the internal actual $|\boldsymbol{\mu}_+|$ and complementary $|\boldsymbol{\mu}_-|$ *magnetic charges* of elementary fermions is absent because of their permanent values postulated (3.2). The equality of the actual (torus) and complementary (antitorus) magnetic moments of sub-elementary fermions and antifermions:

$$\Delta|\mathbf{\mu}_\pm| = (|\mathbf{\mu}_+| - |\mathbf{\mu}_-|) = \mathbf{0} \qquad\qquad 4.28$$

independent on their external velocity, explains the *absence of magnetic monopoles in Nature.*

The elementary magnetic charge is not a monopole, like electric one (+) or (-). It is a dipole, formed by pair $[\mathbf{F}_\uparrow^+ \bowtie \mathbf{F}_\downarrow^-]$ of triplet $< [\mathbf{F}_\uparrow^+ \bowtie \mathbf{F}_\downarrow^-] + \mathbf{F}_\updownarrow^\pm >^i$ .

## 5  Fusion of elementary particles, as a triplets of sub-elementary fermions at Golden mean conditions

At the Golden Mean (GM) conditions: $(\mathbf{v}/\mathbf{c})^2 = \phi = 0.618$, the Cooper pairs of asymmetric Bivacuum fermions, rotating in opposite direction around the common axis of vorticity, turns to pair of sub-elementary fermion and antifermion with ratio of radiuses of torus and antitorus: $\mathbf{L}^+/\mathbf{L}^- = \pi(\mathbf{L}^+)^2/\pi\mathbf{L}_0^2 = \mathbf{S}^+/\mathbf{S}_0 = \phi$ (see eq. 4.15):

$$[\mathbf{F}_\uparrow^+ \bowtie \mathbf{F}_\downarrow^-] \equiv [\mathbf{BVF}_{as}^\uparrow \bowtie \mathbf{BVF}_{as}^\downarrow]^\phi \qquad\qquad 5.1$$

of opposite charge, spin and energy with common Compton radius. The spatial image of pair $[\mathbf{F}_\uparrow^+ \bowtie \mathbf{F}_\downarrow^-]$ is two identical *truncated cones* of the opposite orientation of planes rotating without slip around common rotation axis (Fig.2).

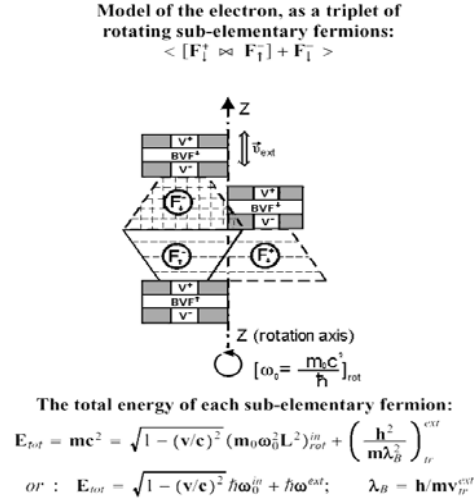

**Model of the electron, as a triplet of rotating sub-elementary fermions:**
$< [\mathbf{F}_\uparrow^+ \bowtie \mathbf{F}_\downarrow^-] + \mathbf{F}_\updownarrow^\pm >$

**The total energy of each sub-elementary fermion:**

$$\mathbf{E}_{tot} = \mathbf{mc}^2 = \sqrt{1 - (\mathbf{v}/\mathbf{c})^2}\ (\mathbf{m}_0\boldsymbol{\omega}_0^2|\mathbf{L}^2)_{in}^{cur} + \left(\frac{\mathbf{h}^2}{\mathbf{m}\boldsymbol{\lambda}_B^2}\right)_{lr}^{cvt}$$

$$or\ :\ \ \mathbf{E}_{tot} = \sqrt{1 - (\mathbf{v}/\mathbf{c})^2}\ \hbar\boldsymbol{\omega}_0^{in} + \hbar\boldsymbol{\omega}^{ext};\ \ \ \ \boldsymbol{\lambda}_B = \mathbf{h}/\mathbf{mv}_{tr}^{ext}$$

**Fig. 2**. Model of the electron, as a triplets $< [\mathbf{F}_\uparrow^+ \bowtie \mathbf{F}_\downarrow^-] + \mathbf{F}_\updownarrow^\pm >^i$ , resulting from fusion of third sub-elementary antifermion $[\mathbf{F}_\updownarrow^-]$ to sub-elementary antifermion $[\mathbf{F}_\uparrow^-]$ with opposite spin in rotating pair $[\mathbf{F}_\uparrow^+ \bowtie \mathbf{F}_\uparrow^-]$. The velocity of rotation of unpaired sub-elementary $[\mathbf{F}_\updownarrow^-]$ around the same axis of common rotation axis of pair provide the similar rest mass ($\mathbf{m}_0$) and absolute charge $|\mathbf{e}^\pm|$, as have the paired $[\mathbf{F}_\uparrow^+$ and $\mathbf{F}_\downarrow^-]$. Three effective anchor $(\mathbf{BVF}^\updownarrow = [\mathbf{V}^+ \Updownarrow \mathbf{V}^-])_{anc}$ in the vicinity of sub-elementary particles base, participate in recoil effects, accompanied their $[\mathbf{C} \rightleftharpoons \mathbf{W}]$ pulsation and modulation of Bivacuum pressure waves $(\mathbf{VPW}_q^\pm)$. The recoil effects of paired $[\mathbf{F}_\uparrow^+ \bowtie \mathbf{F}_\downarrow^-]$ totally compensate each other and the relativistic mass change of triplets is determined only by the anchor Bivacuum fermion $(\mathbf{BVF}^\updownarrow)_{anc}$ of the unpaired sub-elementary fermion $\mathbf{F}_\updownarrow^\pm >$.

The fusion of asymmetric sub-elementary fermions and antifermions of $e$, $\mu$ and $\tau$ generations $\left[\mathbf{F}_\updownarrow^\pm \equiv \left(\mathbf{BVF}_{as}^\uparrow\right)^\phi\right]^{e,\mu,\tau}$ (Fig.2) to triplets results in corresponding electrons/positrons, muons/antimuons and protons/antiprotons origination

$$< [\mathbf{F}_{\uparrow}^{+} \bowtie \mathbf{F}_{\downarrow}^{-}]_{x,y} + \mathbf{F}_{\updownarrow}^{\pm} >_{z}^{e,\mu,p} \qquad 5.2$$

This fusion becomes possible at the Golden mean (GM) conditions, stimulated by the resonant exchange interaction with basic $(\mathbf{q = 1})$ Bivacuum virtual pressure waves $(\mathbf{VPW}_{q=1}^{\pm})$. In the case protons it is accompanied by the energy release and gluons origination, equal in sum to the mass defect, as far the mass of tauons is bigger, than the mass of the proton. In section 12.5 it will be proved, that stabilization of the electron/positron triplets is possible without e-gluons exchange. The centrifugal force, generated by rotation of pair $[\mathbf{F}_{\uparrow}^{+} \bowtie \mathbf{F}_{\downarrow}^{-}]_{x,y}^{e}$ can be compensated by the Coulomb attraction between $\mathbf{F}_{\uparrow}^{+}$ and $\mathbf{F}_{\downarrow}^{-}$.

Similar consideration of *muons* with mass 0.106 GeV/c$^2$ (about 200 times bigger, than electron's) reveals that the centrifugal force, generated by fast rotation of pair $[\mathbf{F}_{\uparrow}^{+} \bowtie \mathbf{F}_{\downarrow}^{-}]_{x,y}^{\mu}$ around common axis exceeds strongly the Coulomb attraction between sub-elementary fermions of corresponding lepton generation $(\mathbf{F}_{\uparrow}^{+})^{\mu}$ and $(\mathbf{F}_{\downarrow}^{-})^{\mu}$. This makes the triplet structure of $\mathbf{\mu}$ −electron unstable even at Golden mean conditions. The experimental life-time of *muon* is $2.19703{\times}10^{-6}s$. The life-time of *tauon* with mass 1.7771 GeV/c$^2$ is even much shorter $2.95{\times}10^{-13}$ s. We suppose the reason of low stability of $\tau$ −electron, is that, in contrast to electron and muon, it represents just a monomeric form of asymmetric Bivacuum fermion at GM conditions $\left[ \mathbf{F}_{\updownarrow}^{\pm} \equiv \left( \mathbf{BVF}_{as}^{\updownarrow} \right)^{\phi} \right]^{\tau}$. The fusion of these sub-elementary fermions to protons and neutrons stabilize the structure of these triplets.

It was demonstrated theoretically, that the vortical structures at certain conditions self-organizes into vortex crystals (Jin and Dubin, 2000).

The fusion of triplets is accompanied by 'switching on' the resonant exchange interaction of $\mathbf{CVC}_{q=1}^{\pm}$ with Bivacuum virtual pressure waves $(\mathbf{VPW}_{q=1}^{\pm})^{i}$ of fundamental frequency $(\omega_0 = \mathbf{m}_0 \mathbf{c}^2/\hbar)^{e,\mu,\tau}$ in the process of $[\mathbf{corpuscle(C)} \rightleftharpoons \mathbf{wave(W)}]$ transitions of elementary particles. The *triplets* of elementary particles and antiparticles formation (Fig.2) is a result of fusion of third sub-elementary fermion (antifermion) $[\mathbf{F}_{\updownarrow}^{\pm}]$ with one of sub-elementary fermion (antifermion) of rotating pair $[\mathbf{F}_{\uparrow}^{+} \bowtie \mathbf{F}_{\downarrow}^{-}]$ of the opposite spins. The opposite spins means that their $[\mathbf{C} \rightleftharpoons \mathbf{W}]$ pulsations are counterphase and these two sub-elementary particles are spatially compatible (see section 9). The velocity of rotation of unpaired sub-elementary fermion $[\mathbf{F}_{\downarrow}^{-}]$ around the same axis of common rotation axis of pair (Fig.2) provide the similar mass and charge $|e^{\pm}|$, as have the paired $[\mathbf{F}_{\uparrow}^{+} \text{ and } \mathbf{F}_{\downarrow}^{-}]$ because of similar symmetry shift.

Let us analyze the rotational dynamics of unpaired $\mathbf{F}_{\updownarrow}^{\pm} >^{e,\mu,\tau} = [\mathbf{V}^{+}\Updownarrow \; \mathbf{V}^{-}]^{as}$ in triplets (Fig.2) just after fusion to triplet at GM conditions in the absence of the external translational motion of triplet.

Its properties are the result of participation in two rotational process simultaneously:

1) rotation of asymmetric $\mathbf{F}_{\updownarrow}^{\pm} >^{e,\mu,\tau}$ around its own axis (Fig.2) with spatial image of truncated cone with resulting radius:

$$\mathbf{L}_{\mathbf{BVF}_{as}}^{\phi} = \hbar/|\mathbf{m}_{V}^{+} + \mathbf{m}_{V}^{-}|^{\phi}\mathbf{c} = \hbar/[\mathbf{m}_0(1/\phi + \phi)\mathbf{c}] = \hbar/2.236\,\mathbf{m}_0\mathbf{c} = \mathbf{L}_0/2.23 \qquad 5.3$$

2) rolling of this truncated cone of $\mathbf{F}_{\updownarrow}^{\pm} >^{e,\mu,\tau}$ around the another axis, common for pair of sub-elementary particles $[\mathbf{F}_{\uparrow}^{+} \bowtie \mathbf{F}_{\downarrow}^{-}]$ (Fig.2) inside of a larger vorticity with bigger radius, equal to *Compton radius*:

$$\mathbf{L}_{\mathbf{BVF}_{as}^{\updownarrow} \bowtie \mathbf{BVF}_{as}^{\updownarrow}}^{\phi} = \hbar/|\mathbf{m}_{V}^{+} - \mathbf{m}_{V}^{-}|^{\phi}\mathbf{c} = \hbar/\mathbf{m}_0\mathbf{c} = \mathbf{L}_0 \qquad 5.4$$

with Golden mean angular frequency:

$$\left(\boldsymbol{\omega}_{\mathbf{v},\bar{\mathbf{v}}}^{i}\right)_{rot}^{\phi} = \frac{\mathbf{c}}{\mathbf{L}_0} = \omega_0 = \frac{\mathbf{m}_0^i \mathbf{c}^2}{\hbar} \qquad 5.4a$$

The ratio of radius of $\left(\mathbf{BVF}_{as}^{\uparrow}\right)^{\phi} \equiv \mathbf{F}_{\updownarrow}^{\pm} >$ to radius of pairs $[\mathbf{F}_{\uparrow}^{+} \bowtie \mathbf{F}_{\downarrow}^{-}]$ at GM conditions is equal to the ratio of potential energy ($\mathbf{V}$) to kinetic energy ($\mathbf{T}_k$) of relativistic de Broglie wave (wave B) at GM conditions. This ratio is the same, as in known formula for relativistic wave B $\left(\frac{\mathbf{V}}{\mathbf{T}_k} = 2\frac{\mathbf{v}_{ph}}{\mathbf{v}_{gr}} - 1\right)$:

$$\frac{\mathbf{L}_{\mathbf{BVF}_{as}^{\uparrow} \bowtie \mathbf{BVF}_{as}^{\downarrow}}^{\phi}}{\mathbf{L}_{\mathbf{BVF}_{as}}^{\phi}} = \frac{|\mathbf{m}_V^{+} + \mathbf{m}_V^{-}|^{\phi}}{|\mathbf{m}_V^{+} - \mathbf{m}_V^{-}|^{\phi}} = \left(\frac{\mathbf{V}}{\mathbf{T}_k}\right)^{\phi} = 2\left(\frac{\mathbf{v}_{ph}}{\mathbf{v}_{gr}}\right)^{\phi} - 1 = 2,236 \qquad 5.5$$

where the potential and kinetic energy of asymmetric Bivacuum dipoles, forming triplets - elementary fermions are equal correspondingly to:

$$\mathbf{V} = \frac{1}{2}|\mathbf{m}_V^{+} + \mathbf{m}_V^{-}|\mathbf{c}^2 \qquad 5.5a$$

$$\mathbf{T}_k = \frac{1}{2}|\mathbf{m}_V^{+} - \mathbf{m}_V^{-}|\mathbf{c}^2 = \frac{1}{2}\mathbf{m}_V^{+}\mathbf{v}^2 \qquad 5.5b$$

This result is a good evidence in proof of our expressions for total energy of sub-elementary particle, as a sum of internal potential and rotational kinetic energies (see section 7, eqs. 7.1 - 7.3).

*The triplets of the electrons and muons* of the same or opposite spin state are the result of fusion of sub-elementary particles of $e$ or $\mu$ − leptons generation, correspondingly:

$$\mathbf{e}^{-} \equiv < [\mathbf{F}_{\uparrow}^{-} \bowtie \mathbf{F}_{\downarrow}^{+}] + \mathbf{F}_{\updownarrow}^{-} >^{e,\mu} \qquad 5.6$$

$$\mathbf{e}^{+} \equiv < [\mathbf{F}_{\uparrow}^{-} \bowtie \mathbf{F}_{\downarrow}^{+}] + \mathbf{F}_{\updownarrow}^{+} >^{e,\mu} \qquad 5.7$$

with mass, charge and spins, determined by uncompensated/unpaired sub-elementary particle: $\mathbf{F}_{\uparrow}^{+} >^{e,\mu}$.

### 5.1 Correlation between new model of hadrons and conventional quark model of protons and neutrons in Standard Model

The *proton* ($Z = +1$; $S = \pm 1/2$) *is constructed* by the same principle as the electron (Fig.2). It is a result of fusion of pair of sub-elementary fermion and antifermion of $\boldsymbol{\tau}$ −generation $< [\mathbf{F}_{\uparrow}^{-} \bowtie \mathbf{F}_{\downarrow}^{+}]_{S=0}$ and one unpaired sub-elementary fermion $\left(\mathbf{F}_{\updownarrow}^{+}\right)_{S=\pm 1/2}^{\tau}$, accompanied by huge energy release, corresponding to mass defect: $\mathbf{\Delta E} \sim (m^{\tau} - m^{p})\mathbf{c}^2$. These three components of proton correspond to three quarks: $\left(\mathbf{F}_{\updownarrow}^{+}\right)_{S=\pm 1/2}^{p} \sim \mathbf{q}^{+}$ and antiquarks $\left(\mathbf{F}_{\updownarrow}^{-}\right)_{S=\pm 1/2}^{p} \sim \mathbf{q}^{-}$.

The difference with quark model is that we do not need to use the notion of fractional charge in our model of proton with spin $S = \pm 1/2$:

$$\mathbf{p} \equiv < [\mathbf{F}_{\uparrow}^{-} \bowtie \mathbf{F}_{\downarrow}^{+}]_{S=0} + (\mathbf{F}_{\uparrow}^{+})_{S=\pm 1/2} >^{\tau} \qquad 5.9$$

$$or : \mathbf{p} \sim \left\langle [\mathbf{q}^{-} \bowtie \mathbf{q}^{+}]_{S=0} + (\mathbf{q}^{+})_{S=\pm 1/2} \right\rangle \qquad 5.9a$$

$$or : \mathbf{p} \sim \left\langle [\boldsymbol{\tau}^{-} \bowtie \boldsymbol{\tau}^{+}]_{S=0} + (\boldsymbol{\tau}^{+})_{S=\pm 1/2} \right\rangle \qquad 5.9b$$

The charges, spins and mass/energy of sub-elementary particles and antiparticles in pairs

$[\mathbf{F}_\uparrow^- \bowtie \mathbf{F}_\uparrow^+]^\tau$ compensate each other. The resulting properties of protons (**p**) are determined by unpaired/uncompensated sub-elementary particle $\mathbf{F}_\uparrow^+ >^\tau$ of heavy $\tau$ −electrons generation, taking into account the mass defect after fusion.

The *neutron* ($Z = 0$; $S = \pm 1/2$) can be presented as:

$$\mathbf{n} \; \equiv \; < [\mathbf{F}_\uparrow^- \bowtie \mathbf{F}_\downarrow^+]_{S=0}^\tau + [(\mathbf{F}_\uparrow^+)^\tau \bowtie (\mathbf{F}_\downarrow^-)^e]_{S=\pm 1/2} > \qquad 5.10$$

$$or : \mathbf{n} \sim [\mathbf{q}^+ \bowtie \mathbf{q}^-]_{S=0}^\tau + \left(\mathbf{q}_\uparrow^0\right)_{S=\pm 1/2}^{\tau e} \qquad 5.10a$$

$$or : \mathbf{n} \sim [\tau^+ \bowtie \tau^-]_{S=0}^\tau + ([\tau_\uparrow^+]^\tau \bowtie [\mathbf{F}_\downarrow^-]^e) \qquad 5.10b$$

where: the neutral quark $\left(\mathbf{q}_\uparrow^0\right)_{S=\pm 1/2}^{\tau e}$ is introduced, as a metastable complex of positive sub-elementary $\tau$ −fermion $\left(\mathbf{F}_\uparrow^+\right)^\tau$ with negative electron's $\mathbf{e}^- \equiv \; < [\mathbf{F}_\uparrow^+ \bowtie \mathbf{F}_\downarrow^-] + \mathbf{F}_\uparrow^- >^e$ sub-elementary fermion of opposite charge $[\mathbf{F}_\downarrow^-]^e$:

$$\left(\mathbf{q}_\uparrow^0\right)_{S=\pm 1/2}^{\tau e} = \left([\mathbf{q}_\uparrow^+] \bowtie [\mathbf{F}_\downarrow^-]^e\right) \qquad 5.11$$

This means that the positive charge of unpaired heavy sub-elementary particle $(\mathbf{F}_\uparrow^+)^\tau$ in neutron (**n**) is compensated by the charge of the light sub-elementary fermion $(\mathbf{F}_\downarrow^-)^e$. In contrast to charge, the spin of unpaired $(\mathbf{F}_\uparrow^+)^\tau$ is not compensated (totally) by spin of $(\mathbf{F}_\downarrow^-)^e$ in neutrons, because of strong mass and angular momentum difference in conditions of the $(\mathbf{F}_\downarrow^-)^e$ confinement.

Another possible explanation of zero charge of the neutron is a such configuration of its three sub-elementary fermions with structure,

$$\langle [\mathbf{F}_\uparrow^+ \bowtie \mathbf{F}_\downarrow^-]_W \bowtie (\mathbf{F}_\uparrow^-)_C \rangle_n \rightleftharpoons \langle [\mathbf{F}_\uparrow^+ \bowtie \mathbf{F}_\downarrow^-]_C \bowtie (\mathbf{F}_\uparrow^-)_W \rangle_n \qquad 5.11a$$

providing the recoilless $\mathbf{C} \rightleftharpoons \mathbf{W}$ pulsation of all three sub-elementary fermions, like in Mössbauer effect (see section 8.10).

Different superpositions of three sub-elementary fermions, like different combinations of three interlocing Borromean rings (symbol, popular in Medieval Italy) can be responsible for different properties of the electrons, protons and neutrons.

The mass of $\tau$- electron, equal to that of $\tau$−positron is: $\mathbf{m}_{\tau^\pm} = 1782(3)$ MeV, the mass of the regular electron is: $\mathbf{m}_{e^\pm} = 0,511003(1)$ MeV and the mass of $\mu$ − electron is: $\mathbf{m}_{\mu^\pm} = 105,6595(2) \, MeV$.

For the other hand, the mass of proton and neutron are correspondingly:
$\mathbf{m_p} = 938,280(3)$ MeV and $\mathbf{m_n} = 939,573 \,(3)$ MeV. They are about two times less, than the mass of $\tau$- electron, equal, in accordance to our model, to mass of its unpaired sub-elementary fermion $(\mathbf{F}_\uparrow^+)^\tau$. This difference characterize the energy of neutral massless *gluons* (exchange bosons), stabilizing the triplets of protons and neutrons. In the case of neutrons this difference is a bit less (taking into account the mass of $[\mathbf{F}_\downarrow^-]^e$), providing, however, much shorter life-time of isolated neutrons (918 sec.), than that of protons ($>10^{31}$ years).

In accordance to our *hadrons* models, each of three quarks (sub-elementary fermions of $\tau$ − generation) in **protons** and **neutrons** can exist in 3 states (*red*, *green* and *blue*), but not simultaneously:

1. The *red* state of **quark/antiquark** means that it is in corpuscular [C] phase;
2. The *green* state of **quark/antiquark** means that it is in wave [W] phase;
3. The *blue* state means that **quark/antiquark** $(\mathbf{F}_\uparrow^+)^\tau$ is in the transition [C]⟺[W] state.
The 8 different combinations of the above defined states of 3 quarks of protons and neutrons correspond to *8 gluons colors*, stabilizing the these hadrons. The triplets of quarks are stabilized by the emission $\rightleftharpoons$ absorption of cumulative virtual clouds ($\mathbf{CVC}^{\pm}$) in the process of quarks [$\mathbf{C} \rightleftharpoons \mathbf{W}$] pulsation.

The known experimental values of life-times of $\boldsymbol{\mu}$ and $\boldsymbol{\tau}$ electrons, corresponding in accordance to our model, to monomeric asymmetric sub-elementary fermions $\left(\mathbf{BVF}_{as}^{\updownarrow}\right)^{\mu,\tau}$, are equal only to $2,19 \cdot 10^{-6}s$ and $3,4 \cdot 10^{-13}s$, respectively. We assume here, that stability of monomeric sub-elementary particles/antiparticles of $\mathbf{e}$, $\boldsymbol{\mu}$ and $\boldsymbol{\tau}$ generations, strongly increases, as a result of their fusion in triplets, possible at Golden mean conditions.

The well known example of weak interaction, like $\beta - decay$ of the neutron to proton, electron and $\mathbf{e}$ −antineutrino:

$$\mathbf{n} \rightarrow \mathbf{p} + \mathbf{e}^- + \overline{\mathbf{v}}_e \qquad\qquad 5.12$$

$$or : \left\langle [\mathbf{q}^+ \bowtie \overline{\mathbf{q}}^-] + \left(\mathbf{q}_\uparrow^0\right)_{S=\pm 1/2}^{\boldsymbol{\tau e}} \right\rangle \rightarrow ([\mathbf{q}^+ \bowtie \overline{\mathbf{q}}^-] + \mathbf{q}^+) + \mathbf{e}^- + \overline{\mathbf{v}}_e \qquad 5.12a$$

is in accordance with our model of elementary particles and theory of neutrino (section 8.4).

The sub-elementary fermion of $\tau -$ generation in composition of proton or neutron can be considered, as a quark and the sub-elementary antifermion, as antiquark:

$$\left(\mathbf{F}_\uparrow^+\right)^\tau \sim \mathbf{q}^+ \quad and \quad \left(\mathbf{F}_\updownarrow^-\right)^\tau \sim \overline{\mathbf{q}}^- \qquad\qquad 5.13$$

In the process of $\beta$ −decay of neutron (5.12 and 5.11) the unpaired negative sub-elementary fermion $[\mathbf{F}_\uparrow^-]^e$ in (5.11) forms a complex - triplet (electron) with complementary virtual Cooper pair $[\mathbf{F}_\uparrow^- \bowtie \mathbf{F}_\downarrow^+]_{S=0}^e$ from the vicinal to neutron polarized Bivacuum:

$$[\mathbf{F}_\uparrow^-]^e + [\mathbf{F}_\uparrow^- \bowtie \mathbf{F}_\downarrow^+]_{S=0}^e \rightarrow \mathbf{e}^- \qquad\qquad 5.14$$

If we accept the explanation of zero charge of neutron, as a result of total compensation of recoil dynamics in the process of correlated $\mathbf{C} \rightleftharpoons \mathbf{W}$ pulsation of three of its sub-elementary fermions, then $\beta$ −decay can be considered as conversion of such specific configuration of neutron (5.11a) to another configuration, pertinent for proton:

$$\left\langle [\mathbf{F}_\uparrow^+ \bowtie \mathbf{F}_\downarrow^-] \bowtie (\mathbf{F}_\uparrow^-) \right\rangle_n \rightarrow \mathbf{p} + \mathbf{e}^- + \overline{\mathbf{v}}_e \qquad\qquad 5.14a$$

where the configuration of proton is described by (5.9). This conversion is accompanied by excitation of vicinal virtual electron($\widetilde{\mathbf{e}}^-$) and its transition to the real pair [electron + antineutrino] $\mathbf{e}^- + \overline{\mathbf{v}}_e$.

The energy of 8 gluons, corresponding to different superposition of $[\mathbf{CVC}^+ \bowtie \mathbf{CVC}^-]_{S=0,1}$, emitted and absorbed with in-phase $[\mathbf{C} \rightleftharpoons \mathbf{W}]$ pulsation of pair [quark + antiquark] in triplets (5.9 - 5.9b):

$$[\mathbf{F}_\uparrow^+ \bowtie \mathbf{F}_\downarrow^-]_{S=0,1}^\tau = [\mathbf{q}^+ + \widetilde{\mathbf{q}}^-]_{S=0,1} \qquad\qquad 5.15$$

is about 50% of energy/mass of quarks and antiquarks ($\tau$ sub-elementary fermions and antifermions).

These 8 gluons, responsible for strong interaction, can be presented as a following combinations of transitions states of $\tau -$ sub-elementary fermions (quarks $q_2$ and $q_3$) and antifermion (antiquark $\widetilde{q}_1$), corresponding to two spin states of proton ($S = \pm 1/2\, \hbar$) of unpaired quark.

For its spin state: $S_{q_3} = +1/2\, \hbar$ we have following 4 transition combinations of triplets transition states, corresponding to four types of gluons:

$$1)\quad \left\langle \left([C \to W]^{S=1/2}_{\overline{q}_1} \;\bowtie\; [C \to W]^{S=-1/2}_{q_2}\right) + [C \to W]^{S=1/2}_{q_3} \right\rangle \qquad 5.16$$

$$2)\quad \left\langle \left([W \to C]^{S=1/2}_{\overline{q}_1} \;\bowtie\; [W \to C]^{S=-1/2}_{q_2}\right) + [C \to W]^{S=1/2}_{q_3} \right\rangle \qquad 5.16a$$

$$3)\quad \left\langle \left([C \to W]^{S=1/2}_{\overline{q}_1} \;\bowtie\; [C \to W]^{S=-1/2}_{q_2}\right) + [W \to C]^{S=1/2}_{q_3} \right\rangle \qquad 5.16b$$

$$4)\quad \left\langle \left([W \to C]^{S=1/2}_{\overline{q}_1} \;\bowtie\; [W \to C]^{S=-1/2}_{q_2}\right) + [W \to C]^{S=1/2}_{q_3} \right\rangle \qquad 5.16c$$

and for the opposite spin state of unpaired quark: $S_{q_3} = -1/2\,\hbar$ we have also 4 transition states combinations, representing another four types of gluons:

$$5)\quad \left\langle \left([C \to W]^{S=1/2}_{\overline{q}_1} \;\bowtie\; [C \to W]^{S=-1/2}_{q_2}\right) + [C \to W]^{S=-1/2}_{q_3} \right\rangle \qquad 5.17$$

$$6)\quad \left\langle \left([W \to C]^{S=1/2}_{\overline{q}_1} \;\bowtie\; [W \to C]^{S=-1/2}_{q_2}\right) + [C \to W]^{S=-1/2}_{q_3} \right\rangle \qquad 5.17a$$

$$7)\quad \left\langle \left([C \to W]^{S=1/2}_{\overline{q}_1} \;\bowtie\; [C \to W]^{S=-1/2}_{q_2}\right) + [W \to C]^{S=-1/2}_{q_3} \right\rangle \qquad 5.17b$$

$$8)\quad \left\langle \left([W \to C]^{S=1/2}_{\overline{q}_1} \;\bowtie\; [W \to C]^{S=-1/2}_{q_2}\right) + [W \to C]^{S=-1/2}_{q_3} \right\rangle \qquad 5.17c$$

One of our versions of elementary particle fusion have some similarity with thermonuclear *fusion* and can be as follows. The rest mass of *isolated* sub-elementary fermion/antifermion *before* fusion of the electron or proton, is equal to the rest mass of unstable muon or tauon, correspondingly. The 200 times decrease of muons mass is a result of mass defect, equal to the binding energy of triplets: electrons or positrons. It is provided by origination of electronic *e-gluons* and release of the huge amount of excessive kinetic (thermal) energy, for example in form of high energy photons or *e-neutrino* beams.

In protons, as a result of fusion of three $\tau$ −electrons/positrons, the contribution of hadron *h-gluon* energy to mass defect is only about 50% of their mass. However, the absolute hadron fusion energy yield is higher, than that of the electrons and positrons.

*Our hypothesis of stable electron/positron and hadron fusion from short-living $\mu$ and $\tau$ - electrons, as a precursor of electronic and hadronic quarks, correspondingly, can be verified using special collider.*

In accordance to our Unified Theory, there are two different mechanisms of stabilization of the electron and proton structures in form of triplets of sub-elementary fermions/antifermions of the reduced $\mu$ and $\tau$ generations, correspondingly, preventing them from exploding under the action of self-charge:

1. Each of sub-elementary fermion/antifermion, representing asymmetric pair of torus ($\mathbf{V}^+$) and antitorus ($\mathbf{V}^-$), as a charge, magnetic and mass dipole, is stabilized by the Coulomb, magnetic and gravitational attraction between torus and antitorus;

2. The stability of triplet, as a whole, is provided by the exchange of Cumulative Virtual Clouds (CVC$^+$ and CVC$^-$) between three sub-elementary fermions/antifermions in the process of their $[\mathbf{C} \rightleftharpoons \mathbf{W}]$ pulsation. In the case of proton and neutron, the 8 transition states corresponds to 8 *h-gluons* of hadrons, responsible for strong interaction. In the case of the electron or positron, the stabilization of triplets is realized by 8 lighter *e-gluons*. The process of internal energy exchange of pairs $[\mathbf{F}_{\uparrow}^- \bowtie \mathbf{F}_{\downarrow}^+]^{e,p}_{S=0,1}$ with unpaired sub-elementary fermion in triplets of hadrons is accompanied also by the energy exchange with external Bivacuum medium. It is resulted in modulation of positive and negative virtual pressure waves $[\mathbf{VPW}^+ \bowtie \mathbf{VPW}^-]$ of Bivacuum, generating the Virtual Replica Multiplication of nucleons (see chapter 13). The feedback reaction between Bivacuum and elementary particles is also existing.

### 5.2 Possible structure of mesons, $W^{\pm}$ and $Z^0$ bosons of electroweak interaction

By definition of Standard Model, the *mesons* are a family of subatomic particles (about 140) that participate in strong interactions and have masses intermediate between leptons and baryons. When the mass of such particles, formed by quarks like baryons, exceeds the mass of baryons (proton, neutron, lambda and omega), they named *bosonic hadrons*. It is generally assumed, that they are composed of a quark and an antiquark. They are bosons with spin, equal to zero or 1 and possible charge: 0, +1 and -1.

In our approach (see 5.15) the pairs of sub-elementary fermions of $\tau$ or $\mu$ generations $[\mathbf{F}_\uparrow^- + \mathbf{F}_\downarrow^+]_{S=0,1}^{\tau,\mu} = [\mathbf{q}^- + \mathbf{q}^+]_{S=0,1}^{\tau,\mu}$ (see 5.6 - 5.9a), have a properties of *mesons,* as a neutral [quark + antiquark] pair with bosonic integer spin. However these sub-elementary fermions are not symmetric necessarily, like $[\mathbf{F}_\uparrow^- \bowtie \mathbf{F}_\downarrow^+]_{S=0,1}^\tau$ of triplets. The coherent cluster of such pairs - from one to four pairs: $(\mathbf{n}[\mathbf{q}^+ + \mathbf{q}^-])_{S=0,1,2,3,4}$ can provide the experimentally revealed integer spins of mesons - from zero to four.

We assume also that some of experimentally revealed charged mesons, like *pions ($\pi^+$),* standing for nucleons interaction, may represent the intermediate bosonic state of spin exchange process between sub-elementary fermion and antifermion of muon generation $(\mathbf{BVB})_{S=0}^{z=+1}$:

$$[\mathbf{F}_\uparrow^- + \mathbf{F}_\downarrow^+]^\mu \rightleftharpoons \left[\mathbf{F}_\uparrow^- \rightleftharpoons (\mathbf{BVB})_{S=0}^{z=\pm 1} \rightleftharpoons \mathbf{F}_\downarrow^+\right]^\mu \qquad 5.18$$

If *pion* with mass (0.140 GeV/c$^2$), is the intermediate state between muon and antimuon, indeed, this explains the decay of pion and antipion on muon (antimuon) and muonic neutrino (antineutrino):

$$(\mathbf{BVB})_{S=0}^{z=\pm 1} \rightarrow \boldsymbol{\mu}^\pm + \boldsymbol{\nu}_\mu(\overline{\boldsymbol{\nu}}_\mu) \qquad 5.19$$

The negatively charged *kaon ($K^-$)* and antikaon *($\overline{K}^+$)* with mass (0.494 GeV/c$^2$) about 5 times bigger than that of muon (0.106 GeV/c$^2$), can represent a small cluster of the odd number of Bivacuum bosons of $\mu$ − generation, like:

$$[2(\mathbf{BVB}^+ \bowtie \mathbf{BVB}^-) + \mathbf{BVB}^\pm]^{z=\pm 1} \qquad 5.19a$$

The neutral heavy B-zero meson ($\mathbf{B}^0$) with mass (5.279 GeV/c$^2$) and eta-c meson (2.980 GeV/c$^2$) can be a clusters of *even* number of Bivacuum bosons of $\tau$ − generation of opposite symmetry shift, compensating the opposite charges of each other in pairs.

The interrelation between muon and the electron follows from two decay reactions of muon and antimuon:

$$\boldsymbol{\mu}^- \rightarrow \mathbf{e}^- + \overline{\boldsymbol{\nu}}_e + \boldsymbol{\nu}_\mu \qquad 5.20$$

$$\boldsymbol{\mu}^+ \rightarrow \mathbf{e}^+ + \boldsymbol{\nu}_e + \overline{\boldsymbol{\nu}}_\mu \qquad 5.20a$$

In terms of our Unified theory (UT), the neutrino and antineutrino are stable carriers of the excessive Bivacuum dipoles mass/energy symmetry shifts - positive ($\boldsymbol{\nu}_{e,\mu}$) or negative ($\overline{\boldsymbol{\nu}}_{e,\mu}$) see section 8.4.

The existence of heavy charged ($\mathbf{W}^+$, $\mathbf{W}^-$) and neutral ($\mathbf{Z}^0$) force carriers bosons with integer spin $\mathbf{0}$, $\mathbf{1}$, $\mathbf{2}$... and mass: $\left(80.4; \ 80.4 \ and \ 91.187\right) \ GeV/c^2$, correspondingly, introduced in electroweak theory is confirmed experimentally. These bosons complex structure differs strongly from that of photons. This author suggest, that the charged bosons $\mathbf{W}^+$, $\mathbf{W}^-$ are the 'rings' constructed from the *odd* number of asymmetric Bivacuum bosons of $\tau$ − generation of opposite symmetry shift and charge and the neutral bosons ($\mathbf{Z}^0$) from the *even* number of paired Bivacuum bosons $(\mathbf{BVB^+} \bowtie \mathbf{BVB^-})^{\tau}_{as}$, compensating the charges of each other. These heavy bosons belongs to class of very unstable particles, named *resonances*, as far their decay/disassembly is related to strong interaction. Their life times $\tau = \hbar/\Gamma$ interrelated with *width of resonance* ($\Gamma$) are very short $\sim 2 \times 10^{-25}$ s.

The rotating around common axes $\mathbf{BVB^+}$ and $\mathbf{BVB^-}$ forming virtual microtubules has a positive and negative charge and mass symmetry shift, corresponding to Golden mean condition $(\mathbf{v}^2/\mathbf{c}^2) = \phi = 0.618$. These dipoles interact *side-by-side* in the same pairs and by *head-to-tail* principle when forming doubled microtubules from adjacent pairs:

$$\mathbf{n} \times (\mathbf{BVB^+} \bowtie \mathbf{BVB^-})^{\tau}_{S=0,1,..} = \mathbf{n} \times \left[ (\mathbf{V^+}{\uparrow}{\downarrow} \, \mathbf{V^-})^i \bowtie (\mathbf{V^+}{\downarrow}{\uparrow} \, \mathbf{V^{-i}}) \right]^{\tau}_{S=0,1,..} \qquad 5.21$$

We suppose, that these pairs polymerize in ring structures, different from that of photon and providing the uncompensated mass of such rotating virtual rings, equal to mass of $W^{\pm}$ and $Z^0$ bosons. The positive and negative charges in each pair $(\mathbf{BVB^+} \bowtie \mathbf{BVB^-})^{\tau}_{S=0,1,..}$ compensate each other and the resulting charge of the 'ring' is equal to charge ($\mathbf{e^{\pm}}$) of one excessive unpaired $(\mathbf{BVB^+})^{\tau}_{S=0,1,..}$ or $(\mathbf{BVB^-})^{\tau}_{S=0,1,..}$.

It is possible to evaluate the velocity of bosonic 'ring' rotation, taking its mass, equal to: $\mathbf{M}_{W^{\pm}} = 80.4$ GeV/c$^2$ and the ring radius, equal to Compton radius of neutron: $\mathbf{L_n} = \hbar/\mathbf{m}_n c$, the region of electroweak interaction action. Then using the obtained earlier formula (3.14) for de Broglie radius of Bivacuum dipoles circulation, we get the condition for bosonic 'ring' ($\mathbf{L}^{W^{\pm}}_{\mathbf{Vir}}$):

$$\mathbf{L}^{W^{\pm}}_{\mathbf{Vir}} = \frac{\hbar \mathbf{c}}{\mathbf{M}_{W^{\pm}} \mathbf{v}^2} = \frac{\hbar}{\mathbf{m}_n \mathbf{c}} = \mathbf{L_n} \qquad 5.22$$

where the mass of neutron $\mathbf{m}_n = 0.940$ GeV/c$^2$.

From this formula we may get the velocity of 'ring' rotation:

$$\mathbf{v} = \mathbf{c} \times \left( \frac{\mathbf{m}_n}{\mathbf{M}_{W^{\pm}}} \right)^{1/2} = \mathbf{c} \times \mathbf{0.1081} \qquad 5.23$$

The corresponding velocity for $\mathbf{Z}^0$ boson is very close to that. We may see, that rotation of these ring - shape bosons is nonrelativistic. However, it becomes equal to light velocity, at the assumption, that radius of heavy bosons is determined by their Compton radius.

Otherwise, the heavy bosons and other *resonances* can be considered as the intermediate - gluonic state, when the asymmetric Bivacuum boson and antiboson with zero charge, but opposite polarization, exchange their cumulative virtual clouds, being simultaneously in the wave [W] phase. In this case the equality (5.21) turns to:

$$\mathbf{n} \times (\mathbf{BVB^+} \bowtie \mathbf{BVB^-})^{\tau}_{S=0,1,..} \overset{\mathbf{C} \to \mathbf{W}}{\Rightarrow} \mathbf{n} \times (\mathbf{CVC^+} \bowtie \mathbf{CVC^-})^{\tau}_{S=0,1,..} \qquad 5.24$$

The proposed approach to analysis of hadrons and mesons intrinsic features can be developed further to explain the general roots of all know elementary particles, taking into account their duality of sub-elementary fermions of all three generation and combination of their different states. It looks that it is possible to do without strong contradictions with Standard model. However our theory explains the origination of mass of elementary particles without Higgs field and corresponding bosons, not detected experimentally.

## 6 Total, potential and kinetic energies of elementary de Broglie waves

The total energy of sub-elementary particles of triplets of the electrons or protons $< [\mathbf{F}^-_{\uparrow} \bowtie \mathbf{F}^+_{\downarrow}]_{S=0} + (\mathbf{F}^{\pm}_{\updownarrow})_{S=\pm 1/2} >^{e,p}$ we can present in three modes, as a sum of total potential $\mathbf{V}_{tot}$ and kinetic $\mathbf{T}_{tot}$ energies, including the internal and external contributions:

$$\mathbf{E}_{tot} = \mathbf{V}_{tot} + \mathbf{T}_{tot} = \frac{1}{2}(\mathbf{m}_V^+ + \mathbf{m}_V^-)\mathbf{c}^2 + \frac{1}{2}(\mathbf{m}_V^+ - \mathbf{m}_V^-)\mathbf{c}^2 \qquad 6.1$$

$$\mathbf{E}_{tot} = \mathbf{m}_V^+\mathbf{c}^2 = \frac{1}{2}\mathbf{m}_V^+(2\mathbf{c}^2 - \mathbf{v}^2) + \frac{1}{2}\mathbf{m}_V^+\mathbf{v}^2 \qquad 6.1a$$

$$\mathbf{E}_{tot} = 2\mathbf{T}_k(\mathbf{v}/\mathbf{c})^2 = \frac{1}{2}\mathbf{m}_V^+\mathbf{c}^2[1 + \mathbf{R}^2] + \frac{1}{2}\mathbf{m}_V^+\mathbf{v}^2 \qquad 6.1b$$

where: $\mathbf{R} = \mathbf{m}_0/\mathbf{m}_V^+ = \sqrt{1 - (\mathbf{v}/\mathbf{c})^2}$ is the dimensionless relativistic factor; $\mathbf{v}$ is the external translational - rotational velocity of particle; $\mathbf{m}_V^+$ and $\mathbf{m}_V^-$ are the *absolute* masses of torus and antitorus of dipoles.

One may see, that $\mathbf{E}_{tot} \to \mathbf{m}_0\mathbf{c}^2$ at $\mathbf{v} \to \mathbf{0}$ and $\mathbf{m}_V^+ \to \mathbf{m}_0$.

Taking into account that the kinetic and potential energies of dipoles are defined by (5.5b and 5.5a):

$$\mathbf{T}_{tot} = \frac{1}{2}(\mathbf{m}_V^+ - \mathbf{m}_V^-)\mathbf{c}^2 = \frac{1}{2}\mathbf{m}_V^+\mathbf{v}^2 \qquad 6.1c$$

$\mathbf{T}_{tot}^W = \frac{1}{2}(\mathbf{m}_V^+ - \mathbf{m}_V^-)\mathbf{c}^2 = \mathbf{T}_{tot}^C = \frac{1}{2}\mathbf{m}_V^+\mathbf{v}^2$ and $\mathbf{c}^2 = \mathbf{v}_{gr}\mathbf{v}_{ph}$, where $\mathbf{v}_{gr} \equiv \mathbf{v}$, then dividing the left and right parts of (6.1 and 6.1a) by $\frac{1}{2}\mathbf{m}_V^+\mathbf{v}^2$, we get:

$$2\frac{\mathbf{c}^2}{\mathbf{v}^2} - 1 = 2\frac{\mathbf{v}_{ph}}{\mathbf{v}_{gr}} - 1 = \frac{(\mathbf{m}_V^+ + \mathbf{m}_V^-)\mathbf{c}^2}{\mathbf{m}_V^+\mathbf{v}^2} = \frac{\mathbf{m}_V^+ + \mathbf{m}_V^-}{\mathbf{m}_V^+ - \mathbf{m}_V^-} \qquad 6.2$$

Comparing formula (6.2) with known relation for relativistic de Broglie wave for ratio of its potential and kinetic energy (Grawford, 1973), we get the confirmation of our definitions of potential and kinetic energies of elementary particle in (6.1):

$$2\frac{\mathbf{v}_{ph}}{\mathbf{v}_{gr}} - 1 = \frac{\mathbf{V}_{tot}}{\mathbf{T}_{tot}} = \frac{(\mathbf{m}_V^+ + \mathbf{m}_V^-)\mathbf{c}^2}{(\mathbf{m}_V^+ - \mathbf{m}_V^-)\mathbf{c}^2} \qquad 6.3$$

In Golden mean conditions, necessary for triplet fusion, the ratio $(\mathbf{V}_{tot}/\mathbf{T}_{tot})^\phi = (1/\phi + \phi) = 2.236$.

In the case of symmetric primordial Bivacuum fermions $[\mathbf{BVF}^\uparrow \bowtie \mathbf{BVF}^\downarrow]$ and bosons $\mathbf{BVB}^\pm$ the absolute values of their energy/masses of their torus and antitorus are equal: $\mathbf{m}_V^+\mathbf{c}^2 = \mathbf{m}_V^-\mathbf{c}^2 = \mathbf{m}_0\mathbf{c}^2(\frac{1}{2} + \mathbf{n})$ (eq.1.1). This means that their kinetic energy is zero and total energy is determined by the value of potential energy:

$$\mathbf{E}_{tot} = \mathbf{V}_{tot} = \frac{1}{2}(\mathbf{m}_V^+ + \mathbf{m}_V^-)\mathbf{c}^2 = \frac{1}{2}\mathbf{m}_0\mathbf{c}^2(1 + 2\mathbf{n}) \qquad 6.3a$$

It is a half of the energy gap between torus and antitorus of Bivacuum dipoles (eq.1.3). The bigger is the potential energy of Bivacuum, the bigger is frequency of virtual pressure waves ($\mathbf{VPW}_{q>1}^\pm$). It will be shown in chapters 14 and 19 of this paper, that the forced resonance of $\mathbf{VPW}_{q>1}^\pm$ with $[\mathbf{corpuscle(C)} \rightleftharpoons \mathbf{wave(W)}]$ pulsation of elementary particles accelerate them and is a source of energy for overunity devices. The idea, that the potential energy of vacuum, as a sum of absolute values of its positive and negative energies, can be used as a source of 'free' energy for overunity devices was discussed also by Frolov (2003) and much earlier by Gustav Naan (1964).

In general case the total potential ($\mathbf{V}_{tot}$) and kinetic ($\mathbf{T}_{tot}$) energies of sub-elementary fermions and their increments can be presented as:

$$\mathbf{V}_{tot}^{W} = \frac{1}{2}(\mathbf{m}_{V}^{+} + \mathbf{m}_{V}^{-})\mathbf{c}^2 = \mathbf{V}_{tot}^{C} = \frac{1}{2}\mathbf{m}_{V}^{+}(2\mathbf{c}^2 - \mathbf{v}^2) = \frac{1}{2}\frac{\hbar\mathbf{c}}{\mathbf{L}_{\mathbf{V}_{tot}}} \geqslant \mathbf{V}_{tot}^{\phi}; \qquad 6.4$$

$$\Delta\mathbf{V}_{tot} = \Delta\mathbf{m}_{V}^{+}\mathbf{c}^2 - \Delta\mathbf{T}_{tot} = -\frac{1}{2}\frac{\hbar\mathbf{c}}{\mathbf{L}_{\mathbf{V}_{tot}}}\frac{\Delta\mathbf{L}_{\mathbf{V}_{tot}}}{\mathbf{L}_{\mathbf{V}_{tot}}} = -\mathbf{V}_p\frac{\Delta\mathbf{L}_{\mathbf{V}_{tot}}}{\mathbf{L}_{\mathbf{V}_{tot}}} \qquad 6.4a$$

where: the characteristic velocity of potential energy, squared, is related to the group velocity of particle ($\mathbf{v}$), as $\mathbf{v}_p^2 = \mathbf{c}^2(2 - \mathbf{v}^2/\mathbf{c}^2)$ and the characteristic *curvature of potential energy* of elementary particles is:

$$\mathbf{L}_{\mathbf{V}_{tot}} = \frac{\hbar}{(\mathbf{m}_{V}^{+} + \mathbf{m}_{V}^{-})\mathbf{c}} \leqslant \mathbf{L}_0^{\phi} \quad \text{at } \left(\frac{\mathbf{v}_{tot}}{\mathbf{c}}\right)^2 \geqslant \phi \qquad 6.4b$$

The total kinetic energy of unpaired sub-elementary fermion of triplets includes the internal vortical dynamics and external translational one, which determines their de Broglie wave length, ($\mathbf{\lambda}_B = 2\pi\mathbf{L}_{\mathbf{T}_{ext}}$) :

$$\mathbf{T}_{tot} = \frac{1}{2}|\mathbf{m}_{V}^{+} - \mathbf{m}_{V}^{-}|\mathbf{c}^2 = \frac{1}{2}\mathbf{m}_{V}^{+}\mathbf{v}^2 = \frac{1}{2}\frac{\hbar\mathbf{c}}{\mathbf{L}_{\mathbf{T}_{tot}}} \geqslant \mathbf{T}_{tot}^{\phi}; \qquad 6.5$$

$$\Delta\mathbf{T}_{tot} = \mathbf{T}_{tot}\frac{1 + \mathbf{R}^2}{\mathbf{R}^2}\frac{\Delta\mathbf{v}}{\mathbf{v}} = -\mathbf{T}_k\frac{\Delta\mathbf{L}_{\mathbf{T}_{tot}}}{\mathbf{L}_{\mathbf{T}_{tot}}} \qquad 6.5a$$

where the characteristic *curvature of kinetic energy* of sub-elementary particles in triplets is:

$$\mathbf{L}_{\mathbf{T}_{tot}} = \frac{\hbar}{(\mathbf{m}_{V}^{+} - \mathbf{m}_{V}^{-})\mathbf{c}} \leqslant \mathbf{L}_0^{\phi} \quad \text{at } \left(\frac{\mathbf{v}_{tot}}{\mathbf{c}}\right)^2 \geqslant \phi \qquad 6.5b$$

It is important to note, that in compositions of triplets $< [\mathbf{F}_{\uparrow}^{-} \bowtie \mathbf{F}_{\downarrow}^{+}]_{S=0} + (\mathbf{F}_{\updownarrow}^{\pm})_{S=\pm 1/2} >^{e,p}$ the *minimum* values of *total* potential and kinetic energies and the *maximum* values of their characteristic curvatures correspond to that, determined by Golden mean conditions (see eqs. 5.3 and 5.4). In our formulas above it is reflected by corresponding inequalities. In accordance to our theory, the Golden mean conditions determine a threshold for triplets fusion from sub-elementary fermions.

The increment of total energy of elementary particle is a sum of total potential and kinetic energies increments:

$$\Delta\mathbf{E}_{tot} = \Delta\mathbf{V}_{tot} + \Delta\mathbf{T}_{tot} = -\mathbf{V}_{tot}\frac{\Delta\mathbf{L}_{\mathbf{V}_{tot}}}{\mathbf{L}_{\mathbf{V}_{tot}}} - \mathbf{T}_{tot}\frac{\Delta\mathbf{L}_{\mathbf{T}_{tot}}}{\mathbf{L}_{\mathbf{T}_{tot}}} \qquad 6.6$$

In the process of corpuscle-wave pulsation $[\mathbf{C} \rightleftharpoons \mathbf{W}]$ (section 7) at the permanent velocity $\mathbf{v} = const$, the total energy is also permanent and it its increment is zero: $\Delta\mathbf{E}_{tot} = 0$. This means that the oscillation of potential and kinetic energy in (6.6), accompanied $[\mathbf{C} \rightleftharpoons \mathbf{W}]$ pulsation should be opposite by value and compensating each other:

$$-\mathbf{V}_{tot}\frac{\Delta\mathbf{L}_{\mathbf{V}_{tot}}}{\mathbf{L}_{\mathbf{V}_{tot}}} \overset{\mathbf{C} \rightleftharpoons \mathbf{W}}{\rightleftharpoons} \mathbf{T}_{tot}\frac{\Delta\mathbf{L}_{\mathbf{T}_{tot}}}{\mathbf{L}_{\mathbf{T}_{tot}}} \qquad 6.6a$$

The well known Dirac equation for energy of a free relativistic particle, following also from Einstein relativistic formula (3.5), can be easily derived from (6.1a), multiplying its left and right part on $\mathbf{m}_{V}^{+}\mathbf{c}^2$ and using introduced mass compensation principle (3.7):

$$\mathbf{E}_{tot}^2 = (\mathbf{m}_{V}^{+}\mathbf{c}^2)^2 = (\mathbf{m}_0\mathbf{c}^2)^2 + (\mathbf{m}_{V}^{+})^2\mathbf{v}^2\mathbf{c}^2 \qquad 6.6b$$

where: $\mathbf{m}_0^2 = \left| \mathbf{m}_V^+ \, \mathbf{m}_V^- \right|$ and the actual inertial mass of torus of unpaired sub-elementary fermion in triplets is equal to regular mass of particle: $\mathbf{m}_V^+ = \mathbf{m}_0$.

Dividing the left and right parts of (6.6b) to $\mathbf{m}_V^+ \mathbf{c}^2$, we may present the total energy of an elementary de Broglie wave, as a difference between *doubled kinetic energy, representing the Maupertuis function (*$2\mathbf{T_k}$*)* and *Lagrange function (*$\mathcal{L} = \mathbf{T}_k - \mathbf{V}$*)* contributions, in contrast to sum of *total potential and kinetic* energies (6.1):

$$\mathbf{E}_{tot} = \mathbf{m}_V^+ \mathbf{c}^2 = (\mathbf{m}_V^+ - \mathbf{m}_V^-)\mathbf{c}^4/\mathbf{v}^2 = \qquad 6.7$$

$$= \frac{\mathbf{m}_0}{\mathbf{m}_V^+}(\mathbf{m}_0\mathbf{c}^2)_{rot}^{in} + (\mathbf{m}_V^+\mathbf{v}^2) \qquad 6.7a$$

$$\mathbf{E}_{tot} = \mathbf{V} + \mathbf{T_k} = \left[ \mathbf{R}(\mathbf{m}_0\mathbf{c}^2)_{rot}^{in} + \tfrac{1}{2}(\mathbf{m}_V^+\mathbf{v}^2) \right] + \tfrac{1}{2}(\mathbf{m}_V^+\mathbf{v}^2) \qquad 6.8$$

$$\mathbf{E}_{tot} = 2\mathbf{T}_k - \mathcal{L}, \quad \text{where} \quad -\mathcal{L} = \mathbf{V} - \mathbf{T}_k = \mathbf{R}(\mathbf{m}_0\mathbf{c}^2)_{rot}^{in} \qquad 6.8a$$

$$\mathbf{E}_{tot} = \mathbf{m}_V^+\mathbf{c}^2 = \mathbf{h}\mathbf{v}_{C \rightleftharpoons W} = \mathbf{R}\,(\mathbf{m}_0\boldsymbol{\omega}_0^2\mathbf{L}_0^2)_{rot}^{in} + \left[ \frac{\mathbf{h}^2}{\mathbf{m}_V^+\lambda_{\mathbf{B}}^2} \right] \qquad 6.8b$$

where: $\mathbf{R} \equiv \sqrt{1 - (\mathbf{v}/\mathbf{c})^2}$ is relativistic factor, dependent on the *external* translational velocity ($\mathbf{v}$) of sub-elementary fermion in composition of triplet; $\mathbf{m}_V^+ = \mathbf{m}_0/\mathbf{R} = \mathbf{m}$ is the actual inertial mass of sub-elementary fermion; the external translational de Broglie wave length is: $\lambda_{\mathbf{B}} = \frac{h}{\mathbf{m}_V^+\mathbf{v}}$ and $\mathbf{v}_{C \rightleftharpoons W}$ is the resulting frequency of corpuscle - wave pulsation (see next section).

We can easily transform formula (6.8) to a mode, including the internal rotational parameters of sub-elementary fermion, necessary for the rest mass and charge origination:

$$\mathbf{E}_{tot} = \mathbf{R}\,(\mathbf{m}_0\boldsymbol{\omega}_0^2\mathbf{L}_0^2)_{rot}^{in} + [(\mathbf{m}_V^+ - \mathbf{m}_V^-)\mathbf{c}^2]_{tr} \qquad 6.9$$

where: $\mathbf{L}_0 = \hbar/\mathbf{m}_0\mathbf{c}$ is the Compton radius of sub-elementary particle; $\boldsymbol{\omega}_0 = \mathbf{m}_0\mathbf{c}^2/\hbar$ is the angular Compton frequency of sub-elementary fermion rotation around the common axis in a triplet (Fig.2).

For potential energy of a sub-elementary fermion, we get from (6.8), taking into account, that $(\mathbf{m}_V^+\mathbf{v}^2)_{tr}^{ext} = 2\mathbf{T}_k$ and $\mathbf{E}_{tot} = \mathbf{V} + \mathbf{T}_k$ :

$$\mathbf{V} = \mathbf{E}_{tot} - \tfrac{1}{2}(\mathbf{m}_V^+\mathbf{v}^2) = \mathbf{R}(\mathbf{m}_0\mathbf{c}^2)_{rot}^{in} + \tfrac{1}{2}(\mathbf{m}_V^+\mathbf{v}^2)_{tr} \qquad 6.9a$$

The difference between potential and kinetic energies, as analog of Lagrange function, from (4.9a) is:

$$-\mathcal{L} = \mathbf{V} - \mathbf{T_k} = \mathbf{V}_{tot} - \tfrac{1}{2}(\mathbf{m}_V^+\mathbf{v}^2)_{tr} = \mathbf{R}(\mathbf{m}_0\mathbf{c}^2)_{rot}^{in} \qquad 6.9b$$

It follows from (6.9 - 6.9b), that at $\mathbf{v}_{tr}^{ext} \to \mathbf{c}$, the *total* relativistic factor, involving both the external and internal translational - rotational dynamics of sub-elementary fermions in triplets: $\mathbf{R} \equiv \sqrt{1 - (\mathbf{v}_{tr}/\mathbf{c})^2} \to 0$ and the rest mass contribution to total energy of sub-elementary particle also tends to zero: $\mathbf{R}(\mathbf{m}_0\mathbf{c}^2)_{rot}^{in} \to 0$. Consequently, the total potential and kinetic energies tend to equality $\mathbf{V}_{tot} \to \mathbf{T}_{tot}$, and the Lagrangian to zero. *This is a conditions for harmonic oscillations of the photon, propagating in unperturbed Bivacuum.*

The important formula for doubled external kinetic energy (Maupertuis function) can be derived from (4.8), taking into account that the relativistic relation between the actual and rest mass is $\mathbf{m}_V^+ = \mathbf{m}_0/\mathbf{R}$ :

$$2\mathbf{T}_k = \mathbf{m}_V^+\mathbf{v}^2 = \mathbf{m}_V^+\mathbf{c}^2 - \mathbf{R}\,\mathbf{m}_0\mathbf{c}^2 = \frac{\mathbf{m}_0\mathbf{c}^2}{\mathbf{R}}(1^2 - \mathbf{R}^2)\ \ or:$$ 6.10

$$2\mathbf{T}_k = \frac{\mathbf{m}_0\mathbf{c}^2}{\mathbf{R}}(1 - \mathbf{R})(1 + \mathbf{R}) = (1 + \mathbf{R})[\mathbf{m}_V^+\mathbf{c}^2 - \mathbf{m}_0\mathbf{c}^2]$$ 6.10a

This formula is a background of the introduced in section 9 notion of *Tuning energy* of Bivacuum Virtual Pressure Waves (**VPW**$^\pm$).

From the formula (3.6), describing dependence of *inertialess* mass $\mathbf{m}_V^-$ of antitorus ($\mathbf{V}^-$) on the external velocity of Bivacuum dipole or unpaired sub-elementary fermion in triplets $\mathbf{m}_V^- = \mathbf{m}_0\sqrt{1 - (\mathbf{v}/\mathbf{c})^2}$, we get:

$$(\mathbf{m}_V^-\mathbf{c}^2)^2 = (\mathbf{m}_0\mathbf{c}^2)^2 - \mathbf{m}_0^2\mathbf{v}^2\mathbf{c}^2$$ 6.11

The difference between 6.6b and 6.11 can be easily transformed to product of kinetic and potential energies of asymmetric Bivacuum dipole (see 5.5a and 5.5b):

$$(\mathbf{m}_V^+\mathbf{c}^2)^2 - (\mathbf{m}_V^-\mathbf{c}^2)^2 = [(\mathbf{m}_V^+)^2 + \mathbf{m}_0^2]\mathbf{v}^2\mathbf{c}^2$$ 6.12

$$(\mathbf{m}_V^+\mathbf{c}^2 - \mathbf{m}_V^-\mathbf{c}^2)(\mathbf{m}_V^+\mathbf{c}^2 + \mathbf{m}_V^-\mathbf{c}^2) = [(\mathbf{m}_V^+)^2 + \mathbf{m}_0^2]\mathbf{v}^2\mathbf{c}^2$$ 6.12a

$$\mathbf{T}_k\mathbf{V} = \frac{1}{4}[(\mathbf{m}_V^+)^2 + \mathbf{m}_0^2]\mathbf{v}^2\mathbf{c}^2$$ 6.12b

We got the new important formula, expressing the product of kinetic and potential energy of asymmetric Bivacuum dipole or unpaired sub-elementary fermion in triplets ($\mathbf{T}_k\mathbf{V}$) via its actual inertial ($\mathbf{m}_V^+$), the rest mass ($\mathbf{m}_0$) and external velocity ($\mathbf{v}$). As far the kinetic energy of asymmetric dipole like the unpaired sub-elementary fermion of triplet is $\mathbf{T}_k = \mathbf{m}_V^+\mathbf{v}^2/2$, the potential energy from 6.12b can be calculated from the known empirical data:

$$\mathbf{V} = \frac{1}{2}[\mathbf{m}_V^+ + \mathbf{m}_0^2/\mathbf{m}_V^+]\mathbf{c}^2$$ 6.13

Our expressions (6.1 - 6.13) are more general, than the conventional ones, as far they take into account the properties of both poles (actual and complementary) of Bivacuum dipoles and subdivide the total energy of particle to the internal and external or to kinetic and potential ones.

## 7. The dynamic mechanism of corpuscle-wave duality

It is generally accepted, that the manifestation of corpuscle - wave duality of a particle is dependent on the way in which the observer interacts with a system. However, the mechanism of duality, as a background of quantum physics, is still obscure.

It follows from our theory, that the [corpuscle (C) $\rightleftharpoons$ wave (W)] duality represents modulation of the internal (hidden) quantum beats frequency between the asymmetric 'actual' (torus) and 'complementary' (antitorus) states of sub-elementary fermions or antifermions by the external - empirical de Broglie wave frequency of these particles, equal to beats frequency of the *'anchor'* Bivacuum fermion (Kaivarainen, 2005). The [C] phase of each sub-elementary fermions of triplets $< [\mathbf{F}_\uparrow^+ \bowtie \mathbf{F}_\downarrow^-] + \mathbf{F}_\updownarrow^\pm >^i$ of elementary particles, like electrons and protons, exists as a mass and an electric and magnetic asymmetric dipole. The total energy, charge and spin of fermion, moving in space with velocity ($\mathbf{v}$) is determined by the unpaired sub-elementary fermion ($\mathbf{F}_\updownarrow^\pm$)$_z$, as far the energy, charge, spin of paired ones in $[\mathbf{F}_\uparrow^+ \bowtie \mathbf{F}_\downarrow^-]_{x,y}$ of triplets compensate each other.

The [$\mathbf{C} \rightarrow \mathbf{W}$] transition of each of sub-elementary fermions in triplets is a result of two stages superposition.

*The 1st stage* of transition is a reversible dissociation of charged sub-elementary fermion in [C] phase $(\mathbf{F}_{\updownarrow}^{\pm})_{\mathbf{C}}^{\mathbf{e}^{\pm}}$ to charged Cumulative Virtual Cloud $(\mathbf{CVC}^{\pm})_{\mathbf{F}_{\updownarrow}^{\pm}}^{\mathbf{e}^{\pm}-\mathbf{e}_{anc}^{\pm}}$ of subquantum particles and the *'anchor'* Bivacuum fermion with internal frequency $(\boldsymbol{\omega}^{in})^{i}$ (eq. 7.4c):

$$(\mathbf{I}): \quad \left[ (\mathbf{F}_{\updownarrow}^{\pm})_{\mathbf{C}}^{\mathbf{e}^{\pm}} \xLeftrightarrow{\text{Recoil/Antirecoil}} \left( \mathbf{BVF}_{anc}^{\updownarrow} \right)_{\mathbf{C}}^{\mathbf{e}_{anc}^{\pm}} + (\mathbf{CVC}^{\pm})_{\mathbf{F}_{\updownarrow}^{\pm}}^{\mathbf{e}^{\pm}-\mathbf{e}_{anc}^{\pm}} \right]^{i} \qquad 7.1$$

where notations $\mathbf{e}^{\pm}$, $\mathbf{e}_{anc}^{\pm}$ and $\mathbf{e}_{\mathbf{CVC}^{\pm}} = \mathbf{e}^{\pm} - \mathbf{e}_{anc}^{\pm}$ mean, correspondingly, the total charge, the anchor charge and their difference, equal to charge of $\mathbf{CVC}^{\pm}$. Between the uncompensated charge and uncompensated mass of Bivacuum dipoles the direct correlation is existing (eq.4.6).

*The 2nd stage* of $[\mathbf{C} \rightarrow \mathbf{W}]$ transition is a reversible dissociation of the anchor Bivacuum fermion $(\mathbf{BVF}_{anc}^{\updownarrow})^{i} = [\mathbf{V}^{+} \updownarrow \mathbf{V}^{-}]_{anc}^{i}$ to symmetric and neutral $(\mathbf{BVF}^{\updownarrow})^{i}$ and the anchor cumulative virtual cloud $(\mathbf{CVC}^{\pm})_{\mathbf{BVF}_{anc}^{\updownarrow}}$, with charge $\mathbf{e}_{anc}^{\pm}$ and frequency $(\boldsymbol{\omega}_{B}^{ext})_{tr}$, equal to the empirical frequency of de Broglie wave of particle (eq. 7.4):

$$(\mathbf{II}) \; : \; \left( \mathbf{BVF}_{anc}^{\updownarrow} \right)_{\mathbf{C}}^{\mathbf{e}_{anc}^{\pm}} \xLeftrightarrow{\text{Recoil/Antirecoil}} \left[ (\mathbf{BVF}^{\updownarrow})^{0} + (\mathbf{CVC}^{\pm})_{\mathbf{BVF}_{anc}^{\updownarrow}}^{\mathbf{e}_{anc}^{\pm}} \right]_{W}^{i} \qquad 7.2$$

The 2nd stage takes a place if $(\mathbf{BVF}_{anc}^{\updownarrow})^{i}$ is asymmetric only in the case of nonzero external translational - rotational velocity of particle. The beats frequency of $(\mathbf{BVF}_{anc}^{\updownarrow})^{e,p}$ is equal to that of the empirical de Broglie wave frequency: $\boldsymbol{\omega}_{B} = \hbar/(\mathbf{m}_{\updownarrow}^{+}\mathbf{L}_{B}^{2})$. The higher is the external kinetic energy of fermion, the higher is frequency $\boldsymbol{\omega}_{B}$. The frequency of the 2st stage oscillations modulates the internal frequency of $[\mathbf{C} \rightleftharpoons \mathbf{W}]$ pulsation: $(\boldsymbol{\omega}^{in})^{i} = \mathbf{R} \, \boldsymbol{\omega}_{0}^{i} = \mathbf{R} \, \mathbf{m}_{0}^{i}\mathbf{c}^{2}/\hbar$, related to contribution of the rest mass energy to the total energy of the de Broglie wave (Kaivarainen, http://arxiv.org/abs/physics/0103031).

The $[\mathbf{C} \rightleftharpoons \mathbf{W}]$ pulsations of unpaired sub-elementary fermion $\mathbf{F}_{\updownarrow}^{\pm} >$, of triplets of the electrons or protons $< [\mathbf{F}_{\uparrow}^{+} \bowtie \mathbf{F}_{\downarrow}^{-}] + \mathbf{F}_{\updownarrow}^{\pm} >^{e,p}$ are in counterphase with the in-phase pulsation of paired sub-elementary fermion and antifermion, modulating Bivacuum virtual pressure waves $(\mathbf{VPW}^{\pm})$ :

$$[\mathbf{F}_{\uparrow}^{+} \bowtie \mathbf{F}_{\downarrow}^{-}]_{W}^{e,p} \xLeftrightarrow{\mathbf{CVC}^{+}+\mathbf{CVC}^{-}} [\mathbf{F}_{\uparrow}^{+} \bowtie \mathbf{F}_{\downarrow}^{-}]_{\mathbf{C}}^{e,p} \qquad 7.3$$

The basic frequency of $[\mathbf{C} \rightleftharpoons \mathbf{W}]$ pulsation of particle in the state of rest, corresponding to Golden mean conditions, $(\mathbf{v}^{in}/\mathbf{c})^{2} = \mathbf{0,618} = \boldsymbol{\phi}$, is equal to that of the 1st stage frequency (5.1) at zero external translational velocity $(\mathbf{v}_{tr}^{ext} = 0; \; \mathbf{R} = \mathbf{1})$. This frequency is the same as the basic Bivacuum virtual pressure waves $(\mathbf{VPW}_{q=1}^{\pm})$ and virtual spin waves $(\mathbf{VirSW}_{q=1}^{S=\pm 1/2})$ frequency (1.7 and 1.10a): $[\boldsymbol{\omega}_{q=1} = \mathbf{m}_{0}\mathbf{c}^{2}/\hbar]^{i}$.

The empirical parameters of de Broglie wave of elementary particle are determined by asymmetry of the torus and antitorus of the *anchor* Bivacuum fermion $(\mathbf{BVF}_{anc}^{\updownarrow})^{e,p} = [\mathbf{V}^{+} \updownarrow \mathbf{V}^{-}]_{anc}^{e,p}$ (Fig.2) and the frequency of its reversible dissociation to symmetric $(\mathbf{BVF}^{\updownarrow})^{i}$ and the anchor cumulative virtual cloud $(\mathbf{CVC}_{anc}^{\pm})$ – stage $(\mathbf{II})$ of duality mechanism (7.2). The dimensions of $\mathbf{CVC}_{anc}^{\pm}$, *i.e.* the Wave phase of $(\mathbf{BVF}_{anc}^{\updownarrow})^{e,p}$ are determined by the empirical de Broglie wave length and can be much bigger than dimension of the *anchor* Bivacuum fermion in Corpuscular phase, close to Compton length.

The total energy, charge and spin of triplets - fermions, moving in space with external translational velocity $(\mathbf{v}_{tr}^{ext})$ is determined by the unpaired sub-elementary fermion $(\mathbf{F}_{\updownarrow}^{\pm})_{z}$, as far the paired ones in $[\mathbf{F}_\uparrow^+ \bowtie \mathbf{F}_\downarrow^-]_{x,y}$ of triplets compensate each other. From (6.9; 6.9a and 6.9b) it is easy to get:

$$\mathbf{E}_{tot} = \mathbf{m}_V^+ \mathbf{c}^2 = \hbar\omega_{\mathbf{C} \rightleftharpoons \mathbf{W}} = \mathbf{R}(\hbar\omega_0)_{rot}^{in} + (\hbar\omega_B^{ext})_{tr} = \mathbf{R}(\mathbf{m}_0\mathbf{c}^2)_{rot}^{in} + (\mathbf{m}_V^+ \mathbf{v}_{tr}^2)^{ext} \qquad 7.4$$

$$\mathbf{E}_{tot} = \mathbf{m}_V^+ \mathbf{c}^2 = -\mathscr{L} + 2\mathbf{T}_k = \mathbf{R}(\mathbf{m}_0\omega_0^2\mathbf{L}_0^2)_{\mathbf{rot}}^{in} + \left(\frac{\mathbf{h}^2}{\mathbf{m}_V^+ \boldsymbol{\lambda}_B^2}\right) \qquad 7.4a$$

$$\mathbf{E}_{tot} = \mathbf{V} + \mathbf{T_k} = \left[\mathbf{R}(\mathbf{m}_0\mathbf{c}^2)_{rot}^{in} + \frac{1}{2}(\mathbf{m}_V^+ \mathbf{v}_{tr}^2)\right] + \frac{1}{2}(\mathbf{m}_V^+ \mathbf{v}_{tr}^2) \qquad 7.4b$$

$$or: \quad \mathbf{E}_{tot} = \mathbf{m}_V^+ \mathbf{c}^2 = \mathbf{V} + \mathbf{T_k} = \frac{1}{2}(\mathbf{m}_V^+ + \mathbf{m}_V^-)\mathbf{c}^2 + \frac{1}{2}(\mathbf{m}_V^+ - \mathbf{m}_V^-)\mathbf{c}^2 \qquad 7.4c$$

where: $\mathbf{R} = \sqrt{1 - (\mathbf{v}/\mathbf{c})^2}$ is the relativistic factor; $\mathbf{v} \equiv \mathbf{v}_{tr}^{ext}$ is the external translational group velocity; $\boldsymbol{\lambda}_B = h/\mathbf{m}_V^+ \mathbf{v} = 2\pi\mathbf{L}_B$ is the external translational de Broglie wave length; the actual inertial mass is $\mathbf{m}_V^+ = \mathbf{m} = \mathbf{m}_0/\mathbf{R}$; $\mathbf{L}_0^i = \hbar/\mathbf{m}_0^i \mathbf{c}$ is a Compton radius of the elementary particle.

It follows from our approach, that the fundamental phenomenon of **corpuscle − wave** duality (Fig.3) is a result of modulation of the primary - carrying frequency of the internal $[\mathbf{C} \rightleftharpoons \mathbf{W}]^{in}$ pulsation of individual sub-elementary fermions (*1st stage*):

$$(\boldsymbol{\omega}^{in})^i = \mathbf{R}\omega_0^i = \mathbf{R} = \sqrt{1 - (\mathbf{v}/\mathbf{c})^2} \;\; \mathbf{m}_0^i \mathbf{c}^2/\hbar \qquad 7.4d$$

by the frequency of the external empirical de Broglie wave of triplet:
$\omega_B^{ext} = \mathbf{m}_V^+ \mathbf{v}_{ext}^2/\hbar = 2\pi\mathbf{v}_{ext}/\mathbf{L}_B$, equal to angular frequency of $[\mathbf{C} \rightleftharpoons \mathbf{W}]_{anc}$ pulsation of the anchor Bivacuum fermion $(\mathbf{BVF}_{anc}^{\updownarrow})^i$ (*2nd stage*).

The contribution of this external translational dynamics to the total one is determined by asymmetry of the *anchor* $(\mathbf{BVF}_{anc}^{\updownarrow})^i = [\mathbf{V}^+ \Updownarrow \mathbf{V}^-]_{anc}^i$ of particle, i.e. by second terms in (7.4) and (7.4a):

$$2\mathbf{T}_k = (\hbar\omega_B)_{tr} = \left(\frac{\mathbf{h}^2}{\mathbf{m}_V^+ \boldsymbol{\lambda}_B^2}\right)_{\mathbf{tr}} = [(\mathbf{m}_V^+ - \mathbf{m}_V^-)\mathbf{c}^2]_{tr} \qquad 7.5$$

$$= (\mathbf{m}_V^+ \mathbf{v}^2)_{tr} = (\mathbf{m}_V^+ \omega_B^2\mathbf{L}_B^2)_{\mathbf{rot}} = \frac{\mathbf{p}_B^2}{\mathbf{m}_V^+} \qquad 7.5a$$

This contribution is increasing with particle acceleration and tending to light velocity. At $\mathbf{v} \rightarrow \mathbf{c}$, and $\mathbf{R} \rightarrow 0$ :

$$2\mathbf{T}_k = (\mathbf{m}_V^+ \mathbf{v}^2)_{tr}^{ext} \rightarrow \mathbf{m}_V^+ \mathbf{c}^2 = \mathbf{E}_{tot} = \mathbf{V} + \mathbf{T}_k \qquad 7.5b$$

$$or \quad \mathbf{V} = \mathbf{T}_k = \frac{1}{2}\mathbf{m}_V^+ \mathbf{c}^2 = \frac{1}{2}\hbar\omega_{\mathbf{C} \rightleftharpoons \mathbf{W}} \qquad 7.5c$$

For example, the equality of the averaged potential and kinetic energies of sub-elementary fermions and antifermions should take a place for photon (fig.4).

The 1st stage of particle duality (7.1) is a consequence of the rest mass influence on propagation of fermions. In the case of bosons, like photons, propagating with light velocity, the contribution of the rest mass and 1st stage to process is negligible as it follows from eq.(7.4). *The mechanism of photon duality is determined by the 2nd stage only (7.2), determined by dynamics of the anchor Bivacuum fermion.* In general case the process of $[\mathbf{C} \rightleftharpoons \mathbf{W}]$ pulsation is accompanied by reversible conversion of rotational energy of elementary particles in [C] phase to their translational energy in [W] phase (see section 7.1).

The double kinetic energy of sub-elementary particle can be expressed via electromagnetic fine structure constant $\alpha = e^2/(\hbar c)$, electric charge squared, frequency of $[\mathbf{C} \rightleftharpoons \mathbf{W}]$ pulsation $\omega_{\mathbf{C} \rightleftharpoons \mathbf{W}} = \mathbf{m}_V^+ \mathbf{c}^2/\hbar$ and de Broglie wave length, equal to that of cumulative virtual cloud $CVC^\pm$ : $\mathbf{L}_B = \mathbf{L}_{CVC} = \hbar/\mathbf{m}_V^+ \mathbf{v}$ :

$$2\mathbf{T}_k = \frac{\hbar^2 \mathbf{c}^2}{\mathbf{m}_V^+ \mathbf{c}^2 \mathbf{L}_{CVC}^2} = \frac{1}{\alpha} \frac{e^2}{\mathbf{L}_{CVC}^2} \frac{\mathbf{c}}{\omega_{\mathbf{C} \rightleftharpoons \mathbf{W}}} = \frac{1}{\alpha} \frac{e^2}{\mathbf{L}_{CVC}^2} \mathbf{L}_{res} \qquad 7.6$$

where the resulting curvature of de Broglie wave is: $\mathbf{L}_{res} = \frac{\mathbf{c}}{\omega_{\mathbf{C} \rightleftharpoons \mathbf{W}}}$ .

In contrast to *external* translational contribution of triplets, the *internal* rotational-translational contribution of individual unpaired sub-elementary fermions, described by the Lagrange function, is tending to zero at the same conditions:

$$-\mathcal{L} = \mathbf{V} - \mathbf{T_k} = \mathbf{R}(\hbar\omega_0)_{rot}^{in} = \hbar\omega^{in} = \mathbf{R}(\mathbf{m}_0\omega_0\mathbf{L}_0^2)_{rot}^{in} \to 0 \quad \text{at } \mathbf{v} \to \mathbf{c} \qquad 7.6a$$

as far as $\mathbf{v} \to \mathbf{c}$, the $\mathbf{R} = \sqrt{1 - (\mathbf{v}/\mathbf{c})^2} \to 0$.

For a regular nonrelativistic electron the carrier frequency is $\omega^{in} = R\omega_0^e \sim 10^{21} s^{-1} >> \omega_B^{ext}$. However, for relativistic case at $\mathbf{v} \to \mathbf{c}$, the situation is opposite: $\omega_B^{ext} >> \omega^{in}$ at $\omega^{in} \to 0$.

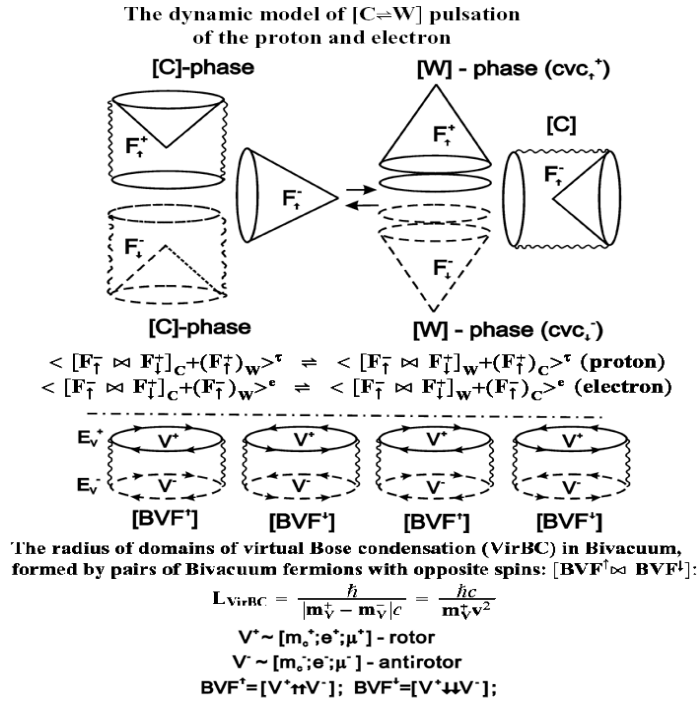

**Fig.3**. Dynamic model of $[\mathbf{C} \rightleftharpoons \mathbf{W}]$ pulsation of triplets of sub-elementary fermions/antifermions (the reduced by fusion to triplets $\mu$ and $\tau$ electrons) composing, correspondingly, electron and proton $< [\mathbf{F}_\uparrow^+ \bowtie \mathbf{F}_\downarrow^-] + \mathbf{F}_\uparrow^\pm >^{e,p}$ . The pulsation of the pair $[\mathbf{F}_\uparrow^- \bowtie \mathbf{F}_\uparrow^+]$, modulating virtual pressure waves of Bivacuum ($VPW^+$ and $VPW^-$), is counterphase to pulsation of unpaired sub-elementary fermion/antifermion $\mathbf{F}_\uparrow^\pm >$.

The properties of the *anchor* Bivacuum fermion $\mathbf{BVF}_{anc}^\updownarrow$ where analyzed (Kaivarainen, 2005), at three conditions:
1. The external translational velocity ($\mathbf{v}$) is zero;
2. The external translational velocity corresponds to Golden mean ($\mathbf{v} = \mathbf{c}\phi^{1/2}$);
3. The relativistic case, when $\mathbf{v} \sim \mathbf{c}$.

Under nonrelativistic conditions ($\mathbf{v} \ll \mathbf{c}$), the de Broglie wave (modulation) frequency is low: $2\pi(\nu_B)_{tr} \ll (\omega^{in} = \mathbf{R}\omega_0)$. However, in relativistic case ($\mathbf{v} \sim \mathbf{c}$), the modulation frequency of the 'anchor' ($\mathbf{BVF}_{anc}^{\updownarrow}$), equal to that of de Broglie wave, can be higher, than the internal one : $2\pi(\nu_B)_{tr} \geq \omega^{in}$.

The paired sub-elementary fermion and antifermion of $[\mathbf{F}_{\uparrow}^{-} \bowtie \mathbf{F}_{\uparrow}^{+}]_{S=0}$ also have the 'anchor' Bivacuum fermion and antifermion ($\mathbf{BVF}_{anc}^{\updownarrow}$), similar to that of unpaired. However, the opposite energies of their $[\mathbf{C} \rightleftharpoons \mathbf{W}]$ pulsation compensate each other in accordance with proposed model.

If we proceed from the assumption that the total energy of the corpuscular and wave phase of each sub-elementary fermion do not change in the process of $[\mathbf{C} \rightleftharpoons \mathbf{W}]$ pulsation of sub-elementary fermions $\mathbf{E}_{tot}^{\mathbf{C} \rightleftharpoons \mathbf{W}} = 0$ in the inertial system ($\mathbf{v} = const$), then, from (4.6 and 4.6a) we get, that the oscillations of potential and kinetic energy should be opposite and compensating each other:

$$\Delta \mathbf{E}_{tot}^{\mathbf{C} \rightleftharpoons \mathbf{W}} = \Delta \mathbf{V_{tot}} + \Delta \mathbf{T_{tot}} = 0 \qquad 7.7$$

$$or : \quad -\mathbf{V}_{tot} \frac{\Delta \mathbf{L}_{\mathbf{V}_{tot}}}{\mathbf{L}_{\mathbf{V}_{tot}}} \overset{\mathbf{C} \rightleftharpoons \mathbf{W}}{\rightleftharpoons} \mathbf{T}_{tot} \frac{\Delta \mathbf{L}_{\mathbf{T}_{tot}}}{\mathbf{L}_{\mathbf{T}_{tot}}} \qquad 7.7a$$

Let us analyze what happens with contributions of the Lagrange function and doubled kinetic energy (Maupertuis function) to the permanent total energy of particle in the process of $[\mathbf{C} \rightleftharpoons \mathbf{W}]$ pulsation in the rest state condition. When the external translational velocity of particle is zero ($\mathbf{v} = 0 = const$ and $\mathbf{R} = \mathbf{1}$) and symmetry shift of sub-elementary fermions in [C] phase is determined only by the relative rotation of the paired $[\mathbf{F}_{\uparrow}^{-} \bowtie \mathbf{F}_{\uparrow}^{+}]$ around common axes with internal rotational-translational velocity, determined by Golden Mean $(\mathbf{v}^{in}/\mathbf{c})^2 = \phi = 0.618$. For the opposite counterphase increments pulsation of the Lagrange function and doubled kinetic energy we get:

$$\Delta \mathcal{L} = \Delta\left[ \mathbf{R}(\mathbf{m}_0\mathbf{c}^2)_{rot}^{in} \right] = \Delta[\mathbf{R}(\mathbf{m}_V^{+} - \mathbf{m}_{\bar{V}})^{\phi}\mathbf{c}^2] = \Delta\left( \mathbf{R}\, \mathbf{m}_0\omega_0^2\mathbf{L}_0^2 \right) \qquad 7.8$$

$$\Delta 2\mathbf{T_{tot}} = \Delta\left( \frac{\mathbf{h}^2}{\mathbf{m}_V^{+}\lambda_B^2} \right) = \Delta(\mathbf{m}_V^{+}\omega_{CVC}^2\mathbf{L}_{CVC}^2) = \Delta\left( \frac{1}{\alpha}\, \frac{\mathbf{e}^2}{\mathbf{L}_{CVC}^2}\, \frac{\mathbf{c}}{\omega_{\mathbf{C} \rightleftharpoons \mathbf{W}}} \right) \qquad 7.8a$$

$$\Delta \mathcal{L} \overset{\mathbf{C} \rightleftharpoons \mathbf{W}}{\rightleftharpoons} -\Delta 2\mathbf{T_{tot}} = -\Delta \frac{\mathbf{p}_0^2}{\mathbf{m}_0} \qquad 7.8b$$

where $\mathbf{L}_{CVC} = \mathbf{L}_0 = \hbar/\mathbf{m}_0\mathbf{c}$ is a radius cumulative virtual cloud with charge, squared: $\mathbf{e}^2 = \mathbf{e}_{+}\mathbf{e}_{-}$; $\omega_{\mathbf{C} \rightleftharpoons \mathbf{W}} = \mathbf{m}_V^{+}\mathbf{c}^2/\hbar$ is the resulting frequency of $[\mathbf{C} \rightleftharpoons \mathbf{W}]$ pulsation; $\mathbf{p}_0 = \mathbf{m}_0\mathbf{c}$.

The decreasing of $\Delta \mathcal{L} = \mathcal{L}$ to zero ( $\Delta \mathcal{L} \to 0$) as a result of $\mathbf{C} \to \mathbf{W}$ transition, due equalizing of torus and antitorus energies and masses: $\mathbf{m}_V^{+} = \mathbf{m}_{\bar{V}} = \mathbf{m}_0$, is accompanied by the Cumulative Virtual Cloud ($CVC^{\pm}$) emission and increasing of its energy from zero to $\Delta(\mathbf{m}_V^{+}\omega_{CVC}^2\mathbf{L}_{CVC}^2) = \left( \frac{1}{\alpha}\, \frac{\mathbf{e}^2}{\mathbf{L}_{CVC}^2}\, \frac{\mathbf{c}}{\omega_{\mathbf{C} \rightleftharpoons \mathbf{W}}} \right)$.

The linear dimension of [C] phase of the triplets is determined by their Compton radius. For the Wave phase, the configuration of triplets may change and they 'jump' from the Corpuscular spatial state to another one in form of Cumulative Virtual Cloud ($CVC^{\pm}$). We named this jumping process from the one Bivacuum fermion to another as the 'Kangaroo effect'. These $[\mathbf{C} \rightleftharpoons \mathbf{W}]$ pulsation in the process of particle propagation in space occur without dissipation in superfluid matrix of Bivacuum in the absence of external fields or matter.

The linear dimension of the Wave phase of the electron in nonrelativistic condition $0 < \mathbf{v}_{tr}^{ext} \ll \mathbf{c}$   $\lambda_B = \mathbf{h}/\mathbf{m}_V^{+}\mathbf{v}_{tr}^{ext}$ can be much bigger, than that [C] phase, determined by Compton length of particle: $\lambda_0 = h/\mathbf{m}_0\mathbf{c}$ $(\lambda_B > \lambda_0)$.

The counterphase oscillations of momentum $(\Delta\mathbf{p})$ and dimensions $(\Delta\mathbf{x})$ in the process of $[\mathbf{C} \rightleftharpoons \mathbf{W}]$ pulsation of elementary particles (fig.3) is reflected by the uncertainty principle:

$$\Delta\mathbf{p}\,\Delta\mathbf{x} \geq \hbar/\mathbf{2} \qquad\qquad 7.9$$

The decreasing of momentum uncertainty $\Delta\mathbf{p} \rightarrow \mathbf{0}$ in the Wave [W] phase is accompanied by the increasing of the *effective* de Broglie wave length: $\Delta\mathbf{x} \rightarrow \lambda_B$ and vice verse.

Taking the differential of de Broglie wave length, it is easy to get:

$$\lambda_B = \mathbf{h}/\mathbf{p}_{tr}^{ext} \quad \rightarrow \quad \frac{\Delta\lambda_B}{\lambda_B} = -\frac{\Delta\mathbf{p}}{\mathbf{p}} \qquad\qquad 7.9a$$

In conditions, when $\Delta\lambda_B = \lambda_B$ we have $-\Delta\mathbf{p} = \mathbf{p}$. The de Broglie wave length characterize the dimension of cumulative virtual cloud, positive for particles or negative for antiparticles $(\mathbf{CVC}^{\pm})$ in their [W] phase and momentum $\mathbf{p} = \mathbf{m}_V^+\mathbf{v}_{tr}^{ext}$ characterize the corpuscular [C] phase.

The other presentation of uncertainty principle reflects the counterphase oscillation of the kinetic energy and time for free particle in process of $[\mathbf{C} \rightleftharpoons \mathbf{W}]$ pulsation:

$$\Delta\mathbf{T}_k\,\Delta\mathbf{t} \geq \hbar/\mathbf{2} \qquad\qquad 7.10$$

This kind of counterphase energy-time pulsation is in accordance with our theory of time (section 7.1).

The wave function for de Broglie wave of particle, moving in direction $\mathbf{x}$ with certain momentum:

$$\mathbf{p} = \mathbf{m}_V^+\mathbf{v}_{tr}^{ext} = \hbar/\mathbf{L}_B = \hbar\mathbf{k} \qquad\qquad 7.10a$$

is described by the wave function in conventional mode:

$$\boldsymbol{\Psi}(\mathbf{x},\mathbf{t}) = \mathbf{C}\exp\left[\frac{i}{\hbar}(\mathbf{px} - \mathbf{Et})\right] = \mathbf{C}\exp\left[i\left(\frac{\mathbf{x}}{\mathbf{L}_B} - \boldsymbol{\omega}_\mathbf{B}\mathbf{t}\right)\right] \qquad 7.11$$

where: $\mathbf{C}$ is a permanent complex number.

The module of the wave function squared: $|\boldsymbol{\Psi}|^2 = \boldsymbol{\Psi}^*\boldsymbol{\Psi} = \mathbf{C}^*\mathbf{C} = \mathbf{const}$ is independent on $\mathbf{x}$. This means that the probability to find a particle with permanent $\mathbf{p}$ is equal in any space volume (or it can be localized everywhere). This contradicts the experimental data.

The Quantum Mechanics solve this contradiction assuming the idea of Shrödinger, that particle represents the 'wave packet' with big number of de Broglie waves with different $\mathbf{p} = \hbar\mathbf{k}$, localized in a small interval $\Delta\mathbf{p}$. The amplitude of all this number of de Broglie waves in the packet with spatial dimension $\Delta\mathbf{x} = \lambda_B$ add to each other because of close phase. For the other hand, at the $\Delta\mathbf{x} >> \lambda_B$ they damper out each other because of phase difference.

The wave packet model can be explained, using our eq.7.4 for nonrelativistic particles: $\mathbf{v} << \mathbf{c}$ and $\mathbf{R} = \sqrt{1 - (\mathbf{v}/\mathbf{c})^2} \sim 1$. For this case, the carrying internal frequency of $\mathbf{C} \rightleftharpoons \mathbf{W}$ pulsation (5.4c) is much higher, than the external translational de Broglie wave modulation frequency (7.5): $\boldsymbol{\omega}^{in} >> \boldsymbol{\omega}_B^{ext}$. The wave packet, consequently, in this case, is formed by the waves, generated by the internal $[\mathbf{C} \rightleftharpoons \mathbf{W}]^{in}$ dynamics, corresponding to *zitterbewegung* (Shrödinger, 1930). However, the wave packet concept itself do not explain the mechanism of $\mathbf{C} \rightleftharpoons \mathbf{W}$ duality.

Our dynamic corpuscle - wave $[\mathbf{C} \rightleftharpoons \mathbf{W}]$ duality theory suggests another possible explanation of the uncertainty principle realization, as a counterphase pulsation of momentum and position, energy and time, described above by eqs. 7.9 and 7.9a. The $\mathbf{C} \to \mathbf{W}$ transition is accompanied by conversion of real mass to virtual one, presented by cumulative virtual cloud $\mathbf{CVC^{\pm}}$. As far the energies of both phase $[\mathbf{C}]$ and $[\mathbf{W}]$ are equal, it makes possible to apply the relativistic mechanics to both of them.

### 7.1  The dynamic model of pulsing photon

The model of a photon with integer spin (boson), resulting from fusion (annihilation) of pairs of triplets: electron + positron (see Fig.2), are presented by Fig.4:

$$< \left[\mathbf{F}^-_\uparrow \bowtie \mathbf{F}^+_\downarrow\right]_{S=0} + \left(\mathbf{F}^-_\uparrow + \mathbf{F}^+_\uparrow\right)_{S=\pm1} + \left[\mathbf{F}^-_\uparrow \bowtie \mathbf{F}^+_\downarrow\right]_{S=0} > \qquad 7.11a$$

Two side pairs represent a Cooper pairs with zero spin. The central pair $\left(\mathbf{F}^-_\uparrow + \mathbf{F}^+_\uparrow\right)_{S=\pm1}$ have the uncompensated integer spin and energy ($\mathbf{E}_{ph} = \mathbf{h}\nu_{ph}$). This structure determines the properties of photon.

Usually the photon originate, as a result of excitation and fusion of three pairs of asymmetric Bivacuum fermions and antifermions - one of *secondary anchor site* of photon (7.46), in the process of transition of the excited state of atom or molecule to the ground state.

There are *two possible ways* to make the rotation of adjacent sub-elementary fermion and sub-elementary antifermion compatible. One of them is interaction 'side-by-side', like in the 1st and 3d pairs of (7.11a). In such a case of Cooper pairs, they are rotating in opposite directions and their angular momenta (spins) compensate each other, turning the resulting spin of such a pair to zero. The resulting energy and charge of such a pair of sub-elementary particle and antiparticle is also zero, because their symmetry shifts with respect to Bivacuum is exactly opposite, compensating each other.

The other way of compatibility is interaction 'head-to-tail', like in a central pair of sub-elementary fermions of 7.11a. In this configuration they rotate in the *same direction* and the sum of their spins is: $\mathbf{s} = \pm\mathbf{1}\hbar$. The energy of this pair is a sum of the *absolute values of the energies of sub-elementary fermion and antifermion*, as far their resulting symmetry shift is a sum of the symmetry shifts of each of them.

In such a case, pertinent for photon, its total energy is interrelated with photon frequency ($\mathbf{\nu}_{ph}$) can be presented as:

$$\mathbf{E}_{ph} = \mathbf{h}\nu_{ph} = \left[(\mathbf{m}^+_V - \mathbf{m}^-_V)\mathbf{c}^2\right]^{\mathbf{F}^+_\uparrow}_{(\mathbf{F}^+_\uparrow + \mathbf{F}^-_\uparrow)} + \left[|-\mathbf{m}^-_V|-\mathbf{m}^+_V|\mathbf{c}^2\right]^{\mathbf{F}^-_\uparrow}_{(\mathbf{F}^+_\uparrow + \mathbf{F}^-_\uparrow)} \qquad 7.12$$

$$or: \quad \mathbf{E}_{ph} = \mathbf{h}\nu_{ph} \cong \left[\mathbf{m}^-_V\mathbf{c}^2\right]^{\mathbf{F}^+_\uparrow}_{(\mathbf{F}^+_\uparrow + \mathbf{F}^-_\uparrow)} + \left[|-\mathbf{m}^-_V|\mathbf{c}^2\right]^{\mathbf{F}^-_\uparrow}_{(\mathbf{F}^+_\uparrow + \mathbf{F}^-_\uparrow)} = 2\left[\mathbf{m}^\pm_V\mathbf{c}^2\right]^{\mathbf{F}^\pm_\uparrow}_{(\mathbf{F}^+_\uparrow + \mathbf{F}^-_\uparrow)} \qquad 7.12a$$

In accordance to our theory (see eqs. 7.4 and 7.4a), the rest mass contribution to energy of sub-elementary fermion $\mathbf{R}[\mathbf{m}_0\mathbf{c}^2]^{\mathbf{F}^+_\uparrow}_{(\mathbf{F}^+_\uparrow + \mathbf{F}^-_\uparrow)}$ and that of sub-elementary antifermion $\left[\mathbf{R}\mathbf{m}_0\mathbf{c}^2\right]^{\mathbf{F}^-_\uparrow}_{(\mathbf{F}^+_\uparrow + \mathbf{F}^-_\uparrow)}$ in *symmetric pairs* are tending to zero: $\mathbf{R} = \sqrt{1 - (\mathbf{v}_{tr}/\mathbf{c})^2} \to 0$, when the external *translational* group velocity of the whole particle is tending to light velocity $\mathbf{v} \to \mathbf{c}$. At these conditions the masses/energies of complementary torus of sub-elementary fermion $(\mathbf{m}^-_V\mathbf{c}^2)^{\mathbf{F}^+_\uparrow}_{(\mathbf{F}^+_\uparrow + \mathbf{F}^-_\uparrow)} = \mathbf{m}_0\sqrt{1 - (\mathbf{v}_{tr}/\mathbf{c})^2}$ and that of complementary sub-elementary antifermion: $(\mathbf{m}^+_V\mathbf{c}^2)^{\mathbf{F}^-_\uparrow}_{(\mathbf{F}^+_\uparrow + \mathbf{F}^-_\uparrow)} = \mathbf{m}_0\sqrt{1 - (\mathbf{v}_{tr}/\mathbf{c})^2}$ are also close to zero; $\mathbf{\nu}_{ph} = E_{ph}/h$ is the photon frequency, equal to frequency of quantum beats between the actual states of asymmetric pair of $\mathbf{F}^+_\uparrow$ and $\mathbf{F}^-_\uparrow$ in photon.

The energy of photon in Corpuscular phase is a sum of energy of tori of asymmetric sub-elementary fermion and antifermion. Equal to this energy, the energy of the Wave phase $(E_{ph})_W$ is determined by the energy of two corresponding cumulative virtual clouds $\varepsilon_{CVC^+} + \varepsilon_{CVC^-}$ :

$$(\mathbf{E}_{ph})_\mathbf{W} = \varepsilon_{CVC^+} + \varepsilon_{CVC^-} = \mathbf{h}\nu_{ph} = \frac{hc}{\lambda_{ph}}; \qquad\qquad 7.13$$

$$(\mathbf{E}_{ph})_\mathbf{C} = (\mathbf{E}_{ph})_\mathbf{W} = \mathbf{h}\nu_{ph} = \mathbf{m}_{ph}\mathbf{c}^2 = 2\mathbf{m}_\mathbf{V}^+\mathbf{c}^2 = \frac{2\mathbf{m}_0(\mathbf{L}_0\boldsymbol{\omega}_0)^2}{\sqrt{1 - \left(\frac{\mathbf{L}_{ph}^C\boldsymbol{\omega}_{rot}}{\mathbf{c}}\right)^2}} \qquad 7.13a$$

where: $\mathbf{L}_0 = \hbar/\mathbf{m}_0\mathbf{c}$; $\boldsymbol{\omega}_0 = \mathbf{m}_0\mathbf{c}^2/\hbar$ are the Compton radius and angle frequency; $\mathbf{L}_{ph}^C$ is the radius of photon rotation in corpuscular phase (fig.4); $\omega_{rot}$ is the angle frequency of photon rotation around the direction of its propagation;

$$\mathbf{m}_{ph} = (\mathbf{m}_{\bar{V}}^+ + |{-}\mathbf{m}_{\bar{V}}^-|) = 2\mathbf{m}_\mathbf{V}^+ = \ 2|{-}\mathbf{m}_{\bar{V}}^-| = \ \mathbf{h}\nu_{ph}/\mathbf{c}^2 = \ \frac{\mathbf{h}}{\mathbf{c}\lambda_{ph}} \qquad 7.14$$

is the effective mass of photon; $\lambda_{ph} = \mathbf{c}/\nu_{ph}$ is the photon wave length.

The energy of photon can be presented as a sum of potential $\mathbf{V} = (\mathbf{m}_{\bar{V}}^+ + |\mathbf{m}_{\bar{V}}^-|)\mathbf{c}^2$ and kinetic $\mathbf{T}_k = (\mathbf{m}_{\bar{V}}^+ - |\mathbf{m}_{\bar{V}}^-|)\mathbf{c}^2$ energies of uncompensated central pair of sub-elementary fermions:

$$(\mathbf{E}_{ph})_\mathbf{C} = \mathbf{m}_{ph}\mathbf{c}^2 = \mathbf{V} + \mathbf{T}_k = (\mathbf{m}_{\bar{V}}^+ + |\mathbf{m}_{\bar{V}}^-|)\mathbf{c}^2 + (\mathbf{m}_{\bar{V}}^+ - |\mathbf{m}_{\bar{V}}^-|)\mathbf{c}^2 \qquad 7.14a$$

We suppose that potential energy of photon or elementary fermion, like electron or proton stands for electric component of photon and kinetic - for its magnetic field energy.

The mechanism of photon duality is determined by the 2nd stage only (7.2), determined by dynamics of the anchor Bivacuum fermion. In general case the process of $[\mathbf{C} \rightleftharpoons \mathbf{W}]$ pulsation is accompanied by reversible conversion of rotational energy of elementary particles in [C] phase (eq.7.13a) to their translational energy in [W] phase (eq.7.13).

It follows from our model (fig.4), that the *minimum* value of the photon effective mass and energy, necessary for splitting of photon to [*electron + positron*] pair in strong fields is equal to the sum of *absolute values* of rest mass/energy of central pair of sub-elementary fermion and antifermion: $(\mathbf{E}_{ph})_C = 2\mathbf{m}_0\mathbf{c}^2$ with positive or negative integer spins $(\pm 1)$ : $(\mathbf{F}_\uparrow^+ + \mathbf{F}_\uparrow^-)_{S=+1}$ or sub-elementary antifermions $(\mathbf{F}_\downarrow^+ + \mathbf{F}_\downarrow^-)_{S=-1}$. This consequence of our model is in accordance with available experimental data.

**Model of photon, as a double
[electron + positron] rotating structure:**
$$< 2[\mathbf{F}_\downarrow^+ \bowtie \mathbf{F}_\uparrow^-] + (\mathbf{F}_\downarrow^- + \mathbf{F}_\uparrow^+) >_{S=\pm1}$$

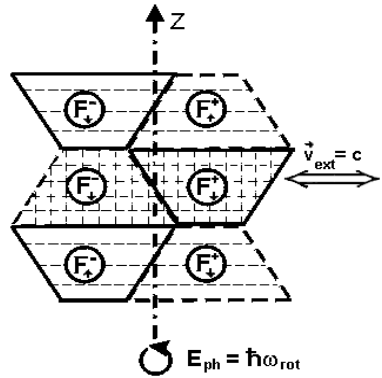

**Fig.4** Model of photon $< 2[\mathbf{F}_\uparrow^- \bowtie \mathbf{F}_\downarrow^+]_{S=0} + (\mathbf{F}_\uparrow^- + \mathbf{F}_\downarrow^+)_{S=\pm1} >$, as result of fusion of electron and positron-like triplets $< [\mathbf{F}_\uparrow^- \bowtie \mathbf{F}_\downarrow^-] + \mathbf{F}_\uparrow^+ >$ of sub-elementary fermions , presented on Fig.2. The resulting symmetry shift of such structure is equal to zero, providing the absence or very close to zero rest mass of photon and its propagation in primordial Bivacuum with light velocity or very close to it in the asymmetric secondary Bivacuum.

We may see, that it has axially symmetric configurations in respect to the directions of rotation and propagation, which are normal to each other. These configurations periodically change in the process of sub-elementary fermions and antifermions correlated [*Corpuscle* $\leftrightharpoons$ *Wave*] pulsations in composition of photon (Fig.4). The volume of sextet of sub-elementary fermions in Corpuscular [C] phase is equal to volume, occupied by 6 asymmetric pairs of torus ($\mathbf{V}^+$) and antitorus ($\mathbf{V}^-$) with geometry of truncated cones and bases: $\mathbf{S}_{V^+} = \pi\mathbf{L}_{V^+}^2$;   $\mathbf{S}_{V^-} = \pi\mathbf{L}_{V^-}^2$ (Korn and Korn, 1968):

$$\mathbf{V_C} = 6\mathbf{d}\,\pi(\mathbf{L}_{V^+}^2 + \mathbf{L}_{V^+}\mathbf{L}_{V^-} + \mathbf{L}_{V^-}^2) \qquad 7.15$$

where: $\mathbf{d}$ is the height of truncated cone (eq.1.4); the radiuses of Compton bases $\mathbf{L}_{V^+}$ and $\mathbf{L}_{V^-}$ and their squares $\mathbf{S}_{V^+}$ and $\mathbf{S}_{V^-}$ of the electron's torus and antitorus can be calculated, using eqs. 4.3 and 4.3a.

For the simple case, when the radiuses of torus of sub-elementary fermion and antitorus in paired sub-elementary antifermion in photons are close: $\mathbf{L}_{V^+} \simeq \mathbf{L}_{V^-} \simeq \mathbf{L}_0^e$, then 7.15 turns to: $\mathbf{V_C^0} \simeq 18\,\mathbf{d}\,\pi\mathbf{L}_0^2$.

The volume of Wave phase of photon in general case is much bigger, than that [C] phase. It can be evaluated as a 3D standing wave:

$$\mathbf{V_W} = \frac{3}{8\pi}\lambda_{ph}^3 = \frac{3}{8\pi}\left(\frac{\mathbf{c}}{\mathbf{v}_{ph}}\right)^3 \qquad 7.16$$

The energy density in [C] phase is much higher, than that of [W] phase as far the volume is much less and the energies are equal:

$$\varepsilon_\mathbf{C} = \frac{\mathbf{E_C}}{\mathbf{V_C}} = \frac{\mathbf{m}_V^+\mathbf{v}_{gr}^2}{18\,\mathbf{d}\,\pi\mathbf{L}_0^2} >> \frac{8\pi\,\mathbf{h}\mathbf{v}_{ph}}{3\,\lambda_{ph}^3} = \frac{\mathbf{E}_W}{\mathbf{V}_W} = \varepsilon_\mathbf{W} \qquad 7.17$$

The expanded Wave phase in contrast to compact Corpuscular phase represents a big number ($\mathbf{N}_{BVF}$) of Bivacuum fermions and antifermions in the volume of wave [W] phase $\mathbf{V}_W$ with resulting symmetry shift and uncompensated energy:

$$\mathbf{c}^2 \int\limits_0^{\mathbf{m}_{ph}} [(\mathbf{m}_V^+ - \mathbf{m}_V^-)]_\mathbf{W}\mathbf{d}\Delta\mathbf{m}_V^\pm = [\mathbf{m_{ph}}\mathbf{c}_{\mathbf{gr}}^2]_\mathbf{C} = \mathbf{h}\mathbf{c}^2/(\mathbf{v}_{ph}\lambda_{ph}^2)_W = \mathbf{h}\mathbf{v}_{ph} \qquad 7.18$$

For photon in primordial symmetric Bivacuum its group and phase velocities are equal: $\mathbf{v}_{gr} = \mathbf{v}_{ph} = c$. This means that the average kinetic and potential energies are also equal: $\mathbf{T}_k = \mathbf{V}_p$. In the process of $\mathbf{C} \rightleftharpoons \mathbf{W}$ pulsation the rotational-translational local kinetic energy of photon: $\mathbf{m}_0\omega_0\mathbf{L}_0^2 = \mathbf{m}_0\,\mathbf{v}_{gr}\mathbf{v}_{ph}$ in [C] phase turns to non-local symmetry shift of Bivacuum dipoles in volume of [W] phase.

The clockwise and counter clockwise rotation of photons in [C] phase around the z-axis (fig.2) stands for two possible polarizations of photon.

The asymmetric pair [actual torus ($\mathbf{V}^+$) + complementary antitorus ($\mathbf{V}^-$)] of sub-elementary fermion has a spatial image of truncated cone (Fig.3 and Fig.4). Using vector analysis, the energy of Cumulative Virtual Cloud ($\mathbf{CVC}^\pm$), equal to energy of quantum beats between the torus and antitorus, can be expressed via internal group and phase velocity fields of sub-quantum particles and antiparticles, composing torus and antitorus: $\mathbf{v}^+$ and $\mathbf{v}^-$, with radiuses $L^+$ and $L^-$:

$$\mathbf{E}_{CVC} = \mathbf{E}_W = \mathbf{n}\,\hbar\boldsymbol{\omega}_{C*W}^+ = \mathbf{n}\,\hbar(\boldsymbol{\omega}_V^+ - \boldsymbol{\omega}_{\bar{V}}^-) = \tfrac{1}{2}\hbar\Big[\,rot\,\mathbf{v}^+ - rot\,\mathbf{v}^-\,\Big] \qquad 7.18a$$

where: $\mathbf{n}$ is the unit-vector, common for both: torus and antitorus of sub-elementary fermion ($\mathbf{F}_{\updownarrow}^{\pm}$); $\boldsymbol{\omega}_{C*W} = 2\pi\mathbf{v}_{ph} = \mathbf{n}\,\hbar(\boldsymbol{\omega}_V^+ - \boldsymbol{\omega}_{\bar{V}}^-)$ is a frequency of quantum beats between actual and complementary states of $\mathbf{F}_{\updownarrow}^{\pm}$.

It is assumed here, that all of subquantum particles/antiparticles, forming actual and complementary torus and antitorus of [C] phase of sub-elementary fermion have the same angular frequency: $\boldsymbol{\omega}_V^+$ and $\boldsymbol{\omega}_{\bar{V}}^-$, correspondingly.

### 7.2 The correlated dynamics of pairs of sub-elementary fermions and antifermions of the opposite and similar spins

We define the energy, as the ability of system to perform a work. In this definition the energy of asymmetric Bivacuum fermions and antifermions is *always positive*, independently of sign of symmetry shift between the mass and charge of torus ($\mathbf{V}^+$) and antitorus ($\mathbf{V}^-$), if they are spatially separated.

If the adjacent asymmetric Bivacuum fermions and antifermions of the opposite spins (i.e. rotating in opposite direction), contacting with each other 'side-by-side', form Cooper pairs $[\mathbf{BVF}^{\uparrow}\bowtie\mathbf{BVF}^{\downarrow}]_{as}$, are pulsing in the same phase between the actual and complementary states, their energy, charge and spin compensate each other.

On the other hand, if the adjacent asymmetric Bivacuum fermion and antifermion of the same spin (i.e. direction of rotation) form 'head-to-tail' complexes, they are spatially compatible only in the case if their pulsation are not in-phase. It will be shown in section 9, that Pauli repulsion between fermions of the same spin due to superposition of their cumulative virtual clouds $\mathbf{CVC}^+$ and $\mathbf{CVC}^-$ is absent, if their emission $\rightleftharpoons$ absorption in the process of [$\mathbf{C} \rightleftharpoons \mathbf{W}$] pulsation are *counter-phase*. It is true also for pair of sub-elementary fermion and antifermion $(\mathbf{F}_{\updownarrow}^- + \mathbf{F}_{\updownarrow}^+)_{S=\pm 1}$, like in photon (Fig.4). In case of this configuration and dynamics the total spin and energy of pair is a sum of spins and *absolute energies* of $\mathbf{F}_{\updownarrow}^-$ and $\mathbf{F}_{\updownarrow}^+$ eqs.(7.13-7.13b).

### 7.3 Spatial images of sub-elementary particles in [C] and [W] phase

The spatial images of torus $[V^+]$ and antitorus $[V^-]$ of asymmetric sub-elementary fermion in [C] phase, reflecting the energy distribution of the actual and complementary states of sub-elementary fermions, can be analyzed in terms of wave numbers. For this end we analyze the basic equations for actual and complementary energy of Bivacuum fermions, squared, leading from (3.5 and 3.6):

$$(\mathbf{E}_V^+)^2 = \left(\mathbf{m}_V^+\mathbf{c}^2\right)^2 = (\mathbf{m}_0\mathbf{c}^2)^2 + \left(\mathbf{m}_V^+\mathbf{v}\right)^2\mathbf{c}^2 \qquad 7.19$$

$$(\mathbf{E}_{\bar{V}}^-)^2 = \left(\mathbf{m}_{\bar{V}}^-\mathbf{c}^2\right)^2 = (\mathbf{m}_0\mathbf{c}^2)^2 - (\mathbf{m}_0\mathbf{v})^2\mathbf{c}^2 \qquad 7.19a$$

These equations can be transformed to following combinations of wave numbers squared:

for actual torus $[V^+]$ :  $\left(\dfrac{\mathbf{m}_V^+\,\mathbf{c}}{\hbar}\right)^2 - \left(\dfrac{\mathbf{m}_V^+\,\mathbf{v}}{\hbar}\right)^2 = \left(\dfrac{\mathbf{m}_0\,\mathbf{c}}{\hbar}\right)^2$ $\qquad$ 7.20

for complementary antitorus $[V^-]$ :  $\left(\dfrac{\mathbf{m}_V^-\,\mathbf{c}}{\hbar}\right)^2 + \left(\dfrac{\mathbf{m}_0\,\mathbf{v}}{\hbar}\right)^2 = \left(\dfrac{\mathbf{m}_0\,\mathbf{c}}{\hbar}\right)^2$ $\qquad$ 7.20a

The spatial image of energy distribution of the *actual* torus $[\mathbf{V}^+]$, defined by equation (7.20), is described by *equilateral hyperbola* (Fig.5a):

$$[\mathbf{V}^+] : \quad X_+^2 - Y_+^2 = a^2 \qquad\qquad 7.21$$

where: $X_+ = (k_V^+)_{tot} = m_V^+ c/\hbar; \qquad Y_+ = m_V^+ \mathbf{v}/\hbar; \quad a = m_0 c/\hbar$

The spatial image of *complementary* antitorus $[\mathbf{V}^-]$ (7.20a) corresponds to *circle* (Fig. 5b), described by equation:

$$[\mathbf{V}^-] : \quad X_-^2 + Y_-^2 = R^2 \qquad\qquad 7.22$$

where: $X_- = (k_V^-)_{tot} = m_V^- c/\hbar; \qquad Y_- = (k_0)_{kin} = m_0 \mathbf{v}/\hbar$.

The radius of complementary circle: $R = k_0 = m_0 c/\hbar$ is equal to the axis length of equilateral hyperbola: $R = a$ of actual $[\mathbf{V}^+]$ state. In fact this circle is a spatial image of the complementary torus of asymmetric $\mathbf{BVF}^{\updownarrow}$ sub-elementary particle or antiparticle ($F_{\updownarrow}^\pm$).

*A spatial image of sub-elementary fermion* $[\mathbf{F}_{\updownarrow}^\pm]$ *in corpuscular* $[C]$ *phase* is a correlated asymmetric pair: *[actual torus + complementary antitorus]* with radiuses of their cross sections, defined, correspondingly, as $(L^+)$ and $(L^-)$:

$\left[ L^+ = \dfrac{\hbar}{\mathbf{m}_V^+ \mathbf{v}_{gr}^{in}} \right]^i$ and $\left[ L^- = \dfrac{-\hbar}{-\mathbf{m}_V^- \mathbf{v}_{ph}^{in}} \right]^i$

*the resulting Compton radius of vorticity of* $[\mathbf{F}_{\updownarrow}^\pm]$ *is* : $\left[ L_0 = (L^+ L^-)^{1/2} = \dfrac{\hbar}{m_0 c} \right]^i$ $\qquad$ 7.23 where: $m_V^+$ and $m_V^-$ are actual (inertial) and complementary (inertialess) effective mass of torus and antitorus of sub-elementary particle, correspondingly; $m_0 = (m_V^+ m_V^-)^{1/2}$ is the rest mass of sub-elementary particle; $\mathbf{v}_{gr}^{in}$ and $\mathbf{v}_{ph}^{in}$ are the internal group and phase velocities, characterizing collective motion (circulation) of subquantum particles and antiparticles, forming actual vortex and complementary torus (Fig.5 a, b).

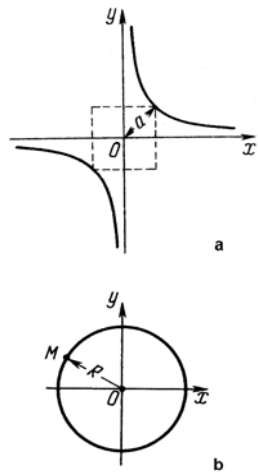

**Fig**. **5a**. Equilateral hyperbola, describing the energy distribution for the actual torus corpuscular $[\mathbf{V}^+]$ of sub-elementary fermion (positive energy region) and sub-elementary

antifermion (negative energy region). This asymmetrically excited *torus* is responsible also for inertial mass ($\mathbf{m}_V^+$), the internal actual magnetic moment ($\boldsymbol{\mu}_+^{in}$) and actual electric charge component ($\mathbf{e}_+$) of sub-elementary fermion (Kaivarainen, 2001a; 2004);

**Fig. 5b**. Circle, describing the energy distribution for the *complementary* state [$\mathbf{V}^-$] of antitorus of sub-elementary fermion. This state is responsible for inertialess mass ($\mathbf{m}_V^-$), the internal complementary magnetic moment ($\boldsymbol{\mu}^{in}$) and complementary component ($\mathbf{e}_-$) of elementary charge. Such antitorus is general for Bivacuum fermions (BVF$_{\updownarrow}^{\pm}$) and Bivacuum bosons (BVB$^{\pm}$).

The [Wave] phase of sub-elementary fermions in form of *cumulative virtual cloud (CVC)* is a result of quantum beats between the actual and complementary torus and antitorus of [Corpuscular] phase of elementary wave B. Consequently, the spatial image of $\mathbf{CVC}^{\pm}$ energy distribution can be considered as a geometric difference between energetic surfaces of the actual [$\mathbf{V}^+$] and complementary [$\mathbf{V}^-$] states of Fig 5a and Fig.5b.

After subtraction of left and right parts of (7.20 and 7.20a) and some reorganization, we get the energetic *spatial image of the [Wave] phase or* [$\mathbf{CVC}^{\pm}$], as a geometrical difference of equilateral hyperbola and circle:

$$\frac{(m_V^+)^2}{m_0^2} + \frac{(m_V^-)^2}{m_0^2} \frac{c^2}{\mathbf{v}^2} - \frac{(m_V^+)^2}{m_0^2} \frac{c^2}{\mathbf{v}^2} = -1 \qquad 7.24$$

This equation in dimensionless form describes the *parted (two-cavity) hyperboloid* (Fig.6):

$$\frac{x^2}{a^2} + \frac{y^2}{b^2} - \frac{z^2}{c^2} = -1 \qquad 7.25$$

The ($c$) is a real semi-axe; $a$ and $b-$ the imaginary ones.

*A spatial image of the wave [W] phase (Fig.6), in form of cumulative virtual cloud (CVC$^{\pm}$) of subquantum particles, is a parted hyperboloid* (Kaivarainen, 2001a).

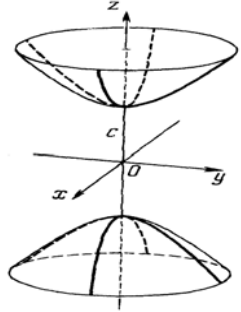

**Fig. 6**. The parted (two-cavity) hyperboloid is a spatial image of twin cumulative virtual cloud [$\mathbf{CVC}^+$ **and CVC**$^-$], corresponding to [Wave] phase of sub-elementary fermion (positive cavity) and sub-elementary antifermion (negative cavity). It may characterize also the twofold CVC$^+$ and CVC$^-$ of positive and negative energy, corresponding to [W] phase of pair (sub-elementary fermion + sub-elementary antifermion) pairs [$\mathbf{F}_{\uparrow}^- \bowtie \mathbf{F}_{\downarrow}^+$], as a general symmetric part of the triplets of electron, positron, photon, proton and neutron (see Figs. 2 and 3).

*For the external observer, the primordial Bivacuum* looks like a isotropic system of 3D double cells (Bivacuum fermions) with shape of pair of donuts of positive and negative energy, separated by energetic gap (see eq.1.4). There are three kinds of like virtual dipoles with three Compton radiuses, corresponding to the rest mass of three *electron's* generation: $i = e, \mu, \tau$ and the external group velocity, equal to zero ($\mathbf{v}_{gr}^{ext} \equiv \mathbf{v} = 0$). The absence of translational dynamics of Bivacuum dipoles provide their zero external momentum and the conditions of virtual Bose condensation, related directly to Bivacuum nonlocal properties (section 1.3). The dimensions of Bivacuum dipoles (radius of two donuts and gap between them) are pulsing in a course of virtual clouds ($\mathbf{VC}^{\pm}$) emission ⇌ absorption.

The following reversible energy oscillations of the positive actual torus ($V^+$) and negative complementary antitorus ($V^-$), accompanied the [Corpuscle ⇌ Wave] transitions of asymmetric sub-elementary fermions of elementary particles.

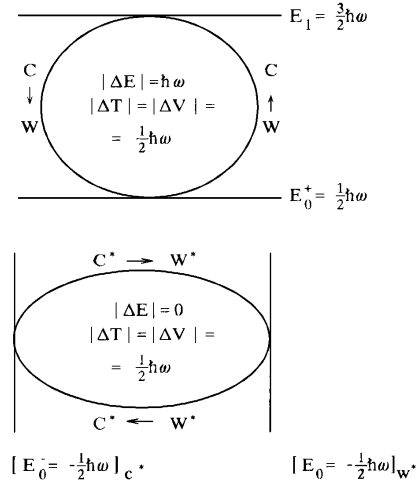

**Fig. 7.** The in-phase oscillation of the total energy [$\mathbf{E_1} \rightleftharpoons \mathbf{E_0^+}$] of the actual state (upper fig.) and the symmetry oscillation [|$\mathbf{T} - \mathbf{V}|_C \rightleftharpoons$ |$\mathbf{T} - \mathbf{V}|_W$] of the complementary state (down) during [$\mathbf{C} \rightleftharpoons \mathbf{W}$] transitions of [vortex + torus] dipole of sub-elementary particles.

### 7.4 New interpretation of Shrödinger equation and general shape of wave function, describing both the external and internal dynamics of elementary particle

The stationary Shrödinger equation can be easily derived from universal for homogeneous medium wave equation:

$$\nabla^2 \Phi(r,t) - \frac{1}{\mathbf{v}^2} \frac{\partial \Phi(r,t)}{\partial t^2} = 0 \qquad 7.26$$

where $\Phi(r,t)$ is the wave amplitude (scalar), depending distance from source (r) and time (t) in the process of its propagation with permanent velocity (**v**). One of possible form of time and space dependent wave function is like (7.11):

$$\Phi(r,t) = \mathbf{C} \exp\left[ i\left( \frac{\mathbf{X}}{\mathbf{L}_B} - \boldsymbol{\omega}_\mathbf{B} \mathbf{t} \right) \right] = \mathbf{C} \exp\left( i \frac{\mathbf{X}}{\mathbf{L}_B} \right) \exp(-i\boldsymbol{\omega}_\mathbf{B}\mathbf{t}) \qquad 7.26a$$

In the case of harmonic dependence of the wave amplitude on time with angle frequency $\omega$, it can be presented as:

$$\Phi(r,t) = \Phi(r) \exp(-i\omega t) \qquad 7.27$$

Putting 7.27 to 7.26, we get the following equation.

$$\nabla^2 \Phi^{m,e}(r) + \mathbf{k}^2 \Phi^{m,e}(r) = 0 \qquad 7.28$$

where **k** is a wave number ($\mathbf{k} = \omega/\mathbf{v} = 2\boldsymbol{\pi}/(\mathbf{vT}) = 2\boldsymbol{\pi}/\boldsymbol{\lambda} = 1/\mathbf{L}$).

The conversion of (7.28) to form describing corpuscle-wave duality can be done using de Broglie relations:

$$\mathbf{k} = \mathbf{p}/\hbar = 2\pi/\mathbf{L}_B; \qquad \mathbf{L}_B = \hbar/\mathbf{p} \qquad\qquad 7.29$$

$$\mathbf{k}^2 = \mathbf{p}^2/\hbar^2 = (2\pi/\mathbf{L}_B)^2 = 1/\lambda_B^2 \qquad\qquad 7.29a$$

in stationary conditions, when the total energy of de Broglie wave, equal to sum of its external kinetic ($\mathbf{T}_k$) and potential ($\mathbf{V}$) energies, is time-independent, like in standing waves, for example:

$$\mathbf{E} = \mathbf{T}_k + \mathbf{V} = \frac{\mathbf{p}^2}{2\mathbf{m}} + \mathbf{V} = \mathbf{const} \qquad\qquad 7.30$$

$$or : \ \mathbf{p}^2 = 2\mathbf{m}(\mathbf{E} - \mathbf{V}) \qquad\qquad 7.30a$$

The de Broglie wave number squared from 7.29a and 7.30a is

$$\mathbf{k}^2 = (2\mathbf{m}/\hbar)(\mathbf{E} - \mathbf{V}) \qquad\qquad 7.31$$

Combining 7.31 with 7.28, we get the *stationary* Shrödinger equation:

$$\nabla^2 \Phi(r) + (2\mathbf{m}/\hbar)(\mathbf{E} - \mathbf{V})\Phi(r) = 0 \qquad\qquad 7.32$$

It has solutions for continuous wave function, existing as *eigenfunctions* only at certain discreet *eigenvalues* of energy ($\mathbf{E}_n$). It was shown by Shrödinger, that spectra of these energies of the electron in potential electric field ($\mathbf{V}$) describes correctly the absorption spectra of hydrogen atoms.

The time-dependent form of Shrödinger equation includes the time and space dependent wave function, like (7.26a):

$$\Phi(\mathbf{r}, t) = \Phi(\mathbf{r})\exp(-i\mathbf{E}t/\hbar) = \mathbf{C}\exp\left(i\frac{\mathbf{X}}{\mathbf{L}_B}\right)\exp(-i\boldsymbol{\omega_B}\mathbf{t}) \qquad\qquad 7.33$$

The corresponding equation can be presented as:

$$-\frac{\hbar}{i}\frac{\partial \Phi(\mathbf{r}, t)}{\partial t} = \left(-\frac{\hbar}{2\mathbf{m}}\nabla^2 + \mathbf{V}\right)\Phi(\mathbf{r}, t) \qquad\qquad 7.34$$

The inertial mass in 7.34, in accordance to our Unified theory, is equal to the actual mass of unpaired/uncompensated sub-elementary fermion of elementary particle: $\mathbf{m} = \mathbf{m}_V^+$.

The properties of stationary wave function $\Phi(\mathbf{r})$ and time-dependent $\Phi(\mathbf{r}, t)$ should be the same, i.e. they are *continuous, single-valued and finitesimal*. The product of wave function with its *complex conjugate* function, characterize the density of probability of particle location in this point of space at certain time moment:

$$\Phi(\mathbf{r}, t)\Phi^*(\mathbf{r}, t) = |\Phi(\mathbf{r}, t)|^2 \qquad\qquad 7.35$$

In solutions of Shrödinger equation the certain eigenvalues of energy ($\mathbf{E}_n$) corresponds to eigenfunctions ($\Phi_n$), describing *anchor sites (primary and secondary)* of elementary particles in their corpuscular [C] phase.

It follows from our theory of wave-corpuscle duality, that de Broglie wave length ($\lambda_B = 2\pi\mathbf{L}_B$) and its frequency ($\boldsymbol{\omega_B}$), as a crucial parameters of wave function (7.33), are determined by properties of the *anchor Bivacuum fermions* of uncompensated sub-elementary fermions of the electron or proton or bosons, like photon.

From eqs.7.4, 7.4a and 7.5 we can see, that the *external* de Broglie wave frequency ($\boldsymbol{\omega}_B^{ext}$) and wave number ($\mathbf{k}_B$) of particle can be expressed via *internal* ($\boldsymbol{\omega}_0^{in}$), *total* ($\boldsymbol{\omega_{C \rightleftharpoons W}}$) frequencies and corresponding energies as:

$$\boldsymbol{\omega}_B^{ext} = \frac{1}{\hbar}[(\mathbf{m}_V^+ - \mathbf{m}_V^-)_{anc}^{ext}\mathbf{c}^2]_{tr} = \boldsymbol{\omega}_{\mathbf{C}\rightleftharpoons\mathbf{W}} - \mathbf{R}\boldsymbol{\omega}_0^{in} \qquad 7.36$$

$$or: \quad \mathbf{k}_B = \frac{1}{\mathbf{L}_B} = \frac{\mathbf{c}}{\hbar}\,[\mathbf{m}_V^+(\mathbf{m}_V^+ - \mathbf{m}_V^-)]^{1/2} = \frac{\mathbf{c}}{\hbar}[\mathbf{m}_V^+(\mathbf{m}_V^+ - \mathbf{R}\,\mathbf{m}_0)]_{tr}^{1/2} \qquad 7.37$$

where relativistic factor: $\mathbf{R} = \sqrt{1 - (\mathbf{v}/\mathbf{c})^2}$ is dependent on the external translational group velocity ($\mathbf{v}$); $\mathbf{m}_V^+ = \mathbf{m}_0/\mathbf{R}$; $\quad \mathbf{m}_V^- = \mathbf{R}\,\mathbf{m}_0$.

At $\mathbf{v} \to \mathbf{c}$, the $\mathbf{R} \to 0$, the rest mass contribution decreases and $\boldsymbol{\omega}_B^{ext} \to \boldsymbol{\omega}_{\mathbf{C}\rightleftharpoons\mathbf{W}}$ and $\mathbf{k}_B \to (\mathbf{m}_V^+\mathbf{c}/\hbar)$.

The mass and charge symmetry shifts of asymmetric Bivacuum fermions and antifermions are interrelated (eqs. 4.7- 4.8):

$$\Delta\mathbf{m}_V^+ = (\mathbf{m}_V^+ - \mathbf{m}_V^-) = \mathbf{m}_V^+\Big(\frac{\mathbf{v}}{\mathbf{c}}\Big)^2 \qquad 7.38$$

$$\Delta\mathbf{e}_\pm = (\mathbf{e}_+ - \mathbf{e}_-) = \frac{\Delta\mathbf{m}_V^+\,\mathbf{e}_+^2}{\mathbf{m}_V^+(\mathbf{e}_+ + \mathbf{e}_-)} = \Big(\frac{\mathbf{v}}{\mathbf{c}}\Big)^2\frac{\mathbf{e}_+^2}{\mathbf{e}_+ + \mathbf{e}_-} \qquad 7.38a$$

where the *actual* charge ($\mathbf{e}_+$), in accordance to eq.4.5, has the following relativistic dependence on the external velocity of Bivacuum dipoles:

$$\mathbf{e}_+ = \frac{\mathbf{e}_0}{[1 - \mathbf{v}^2/\mathbf{c}^2]^{1/4}} \qquad 7.38b$$

The complementary charge ($\mathbf{e}_-$) can be calculated from the earlier obtained relation (eq. 4.18a): $|\mathbf{e}_+\mathbf{e}_-| = \mathbf{e}_0^2$.

Using the relations above, we may present the dimensionless coefficient of wave function (C) in (7.33), as a maximum symmetry shift of the *anchor* Bivacuum fermion, reduced to the rest mass ($\mathbf{m}_0$) and rest charge ($\mathbf{e}_0$):

$$\mathbf{C_m} = \Delta\mathbf{m}_V^+/\sqrt{2}\,\mathbf{m}_0 = (\mathbf{m}_V^+ - \mathbf{m}_V^-)/\sqrt{2}\,\mathbf{m}_0 = \frac{\mathbf{m}_V^+}{\sqrt{2}\,\mathbf{m}_0}\Big(\frac{\mathbf{v}}{\mathbf{c}}\Big)^2 \qquad 7.39$$

$$\mathbf{C_e} = \Delta\mathbf{e}_\pm/\sqrt{2}\,\mathbf{e}_0 = (\mathbf{e}_+ - \mathbf{e}_-)/\sqrt{2}\,\mathbf{e}_0 = \Big(\frac{\mathbf{v}}{\mathbf{c}}\Big)^2\frac{\mathbf{e}_+^2/\sqrt{2}\,\mathbf{e}_0}{\mathbf{e}_+ + \mathbf{e}_-} \qquad 7.39a$$

We assume here, that as far the complementary mass and charge are undetectable directly and we may consider them as imaginary ones: $i\mathbf{m}_V^-$ and $i\mathbf{e}_-$. Consequently, using 7.36; 7.37 and 7.39, we may present the wave function (7.33) and its complex conjugate in terms of Bivacuum dipoles symmetry shifts for understanding the mechanism of particle internal dynamics and its propagation in space:

$$\Phi(\mathbf{r},t) = \mathbf{C}\exp\Big(i\frac{\mathbf{X}}{\mathbf{L}_B}\Big)\exp(-i\boldsymbol{\omega_B}t); \qquad \Phi^*(\mathbf{r},t) = \mathbf{C}^*\exp\Big(-i\frac{\mathbf{X}}{\mathbf{L}_B}\Big)\exp(i\boldsymbol{\omega_B}t) \qquad 7.40$$

$$\Phi(\mathbf{r},t) = \frac{\mathbf{m}_V^+ - i\mathbf{m}_V^-}{\sqrt{2}\,\mathbf{m}_0}\exp\Big[i\frac{\mathbf{X}}{\hbar}\mathbf{c}\,[\mathbf{m}_V^+(\mathbf{m}_V^+ - i\mathbf{m}_V^-)]^{1/2}\Big]\exp\Big\{-i\frac{1}{\hbar}[(\mathbf{m}_V^+ - i\mathbf{m}_V^-)\mathbf{c}^2]_{tr}\mathbf{t}\Big\} = \qquad 7.40a$$

$$\Phi(\mathbf{r},t) = \frac{\mathbf{m}_V^+ - i\mathbf{R}\,\mathbf{m}_0}{\sqrt{2}\,\mathbf{m}_0}\exp\Big[i\frac{\mathbf{X}}{\hbar}\mathbf{c}\,[\mathbf{m}_V^+(\mathbf{m}_V^+ - i\mathbf{R}\,\mathbf{m}_0)]^{1/2}\Big]\exp\Big\{-i\frac{1}{\hbar}[(\mathbf{m}_V^+ - i\mathbf{R}\,\mathbf{m}_0)\mathbf{c}^2]_{tr}\mathbf{t}\Big\} \qquad 7.40b$$

$$\Phi^*(\mathbf{r},t) = \frac{\mathbf{m}_V^+ + i\mathbf{m}_V^-}{\sqrt{2}\,\mathbf{m}_0}\exp\Big[i\frac{\mathbf{X}}{\hbar}\mathbf{c}\,[\mathbf{m}_V^+(\mathbf{m}_V^+ + i\mathbf{m}_V^-)]^{1/2}\Big]\exp\Big\{-i\frac{1}{\hbar}[(\mathbf{m}_V^+ + i\mathbf{m}_V^-)\mathbf{c}^2]_{tr}\mathbf{t}\Big\} = \qquad 7.41$$

$$\Phi^*(\mathbf{r},t) = \frac{\mathbf{m}_V^+ + i\mathbf{R}\,\mathbf{m}_0}{\sqrt{2}\,\mathbf{m}_0}\exp\Big[i\frac{\mathbf{X}}{\hbar}\mathbf{c}\,[\mathbf{m}_V^+(\mathbf{m}_V^+ + i\mathbf{R}\,\mathbf{m}_0)]^{1/2}\Big]\exp\Big\{-i\frac{1}{\hbar}[(\mathbf{m}_V^+ + i\mathbf{R}\,\mathbf{m}_0)\mathbf{c}^2]_{tr}\mathbf{t}\Big\} \qquad 7.41a$$

From 7.40b and 7.41a it follows, that at $\mathbf{v} = \mathbf{c}$ and $\mathbf{R} = \mathbf{0}$ these wave functions turn to that, describing *photons* with effective mass $\mathbf{m}_V^+ = \hbar\omega/\mathbf{c}^2$; and frequency $\boldsymbol{\omega} = \frac{1}{\hbar}[\mathbf{m}_V^+\mathbf{c}^2]_{tr}$.

$$[\Phi(\mathbf{r},t) = \Phi^*(\mathbf{r},t)]_{ph} = \frac{\mathbf{m}_V^+}{\sqrt{2}\,\mathbf{m}_0} \exp\!\left[i\frac{\mathbf{x}}{\hbar}\mathbf{m}_V^+\mathbf{c}\right]\exp\!\left\{-i\frac{1}{\hbar}[\mathbf{m}_V^+\mathbf{c}^2]_{tr}\mathbf{t}\right\} \qquad 7.42$$

where: $\mathbf{m}_V^+\mathbf{c}^2 = \mathbf{h}\nu_{ph}$ is the photon energy.

The product of the conventional forms of complex conjugate wave functions (7.40) gives the space and time independent pre-exponential coefficient squared: $|\Phi(\mathbf{r},t)|^2 = \mathbf{C}^*\mathbf{C} = const.$

From product of 7.40b and 7.41a we get the new general formula for density of probability of particle in [C] phase location, dependent on space and time $|\Phi(\mathbf{r},t)|^2$:

$$|\Phi(\mathbf{r},t)|^2 = \Phi(\mathbf{r},t)\Phi^*(\mathbf{r},t) = \qquad\qquad 7.43$$
$$= \frac{(\mathbf{m}_V^+)^2 + (\mathbf{m}_V^-)^2}{2\mathbf{m}_0^2}\exp\!\left[i\frac{\sqrt{2}\,\mathbf{x}}{\mathbf{L}_C}\right]\exp\!\left\{-i\,2\omega_{C\rightleftharpoons W}\mathbf{t}\right\}$$

where the resulting frequency of $\mathbf{C} \rightleftharpoons \mathbf{W}$ pulsation of uncompensated sub-elementary fermions: $\boldsymbol{\omega}_{C\rightleftharpoons W} = \mathbf{m}_V^+\mathbf{c}^2/\hbar$ and $\mathbf{L}_C = \hbar/\mathbf{m}_V^+\mathbf{c}$ is the characteristic dimension of elementary particle in [C] phase.

The resulting energy of this state is characterized by the length of hypotenuse of triangle with adjacent cathetus squared:

$$\mathbf{E}_{\mathbf{V}^+\Uparrow\mathbf{V}^-}^{Res} = \mathbf{m}_{\mathbf{V}^+\Uparrow\mathbf{V}^-}\mathbf{c}^2 = \sqrt{(\mathbf{m}_V^+)^2 + (\mathbf{m}_V^-)^2}\;\mathbf{c}^2 \qquad\qquad 7.44$$

It is important to point out, that in state of rest, when the external translational velocity of elementary particle is zero ($\mathbf{v} = \mathbf{0}$), the real and complementary mass are equal to the rest mass: $\mathbf{m}_V^+ = \mathbf{m}_V^- = \mathbf{m}_0$, the external de Broglie wave length tends to infinity ($\lambda_B = 2\pi L_B = \infty$) and its frequency to zero ($\omega_B = 0$), the wave function, described by *conventional* expression (7.26a) becomes equal to coefficient $\mathbf{C}$. This coefficient itself, as a square root of pre-exponential factor $\mathbf{C} = \sqrt{\frac{(\mathbf{m}_V^+)^2 + (\mathbf{m}_V^-)^2}{2\mathbf{m}_0^2}}$ at these conditions is equal to $\mathbf{C} = \mathbf{1}$. The corresponding density of probability describing only the external properties of particle $\mathbf{C}^2 = \mathbf{1}$ is a permanent value, independent on space and time.

However, the general expression of density of probability (7.43) of particle location in selected point of space-time, when its external translational velocity is equal to zero ($\mathbf{v}^{ext} = \mathbf{0}$), following *from our theory*, turns to:

$$|\Phi(\mathbf{r},t)|^2 = \exp\!\left(i\sqrt{2}\,\frac{\mathbf{x}}{\mathbf{L}_0}\right)\exp(-i2\boldsymbol{\omega}_0\mathbf{t}) \qquad\qquad 7.45$$

where the Compton wave length and frequency of particle are equal, correspondingly, to:

$$\mathbf{L}_0 = \frac{\mathbf{c}}{\omega_0} = \frac{\hbar}{\mathbf{m}_0\mathbf{c}} \quad\text{and}\quad \boldsymbol{\omega}_0 = \frac{\mathbf{m}_0\mathbf{c}^2}{\hbar} \qquad\qquad 7.45a$$

We can see, that the general expression of density probability of particle in [C] phase location (7.45), in contrast to conventional, the permanent one, is *oscillating* due to internal $[\mathbf{C} \rightleftharpoons \mathbf{W}]_{in}$ pulsation of sub-elementary fermions, rotating around common axes, as presented in Fig.1 and Fig.3. At fixed coordinate ($\mathbf{x}$), the probability of particle in [C] phase location is dependent on time, i.e. phase of pulsation. At fixed time ($\mathbf{t}$) this probability is dependent on coordinate of particle in [C] phase.

### 7.5 The mechanism of free particle propagation in space

The propagation of elementary particles, like triplets-fermions $< [\mathbf{F}_\uparrow^+ \bowtie \mathbf{F}_\downarrow^-] + \mathbf{F}_\uparrow^\pm >^{e,p}$ or sextets - bosons $< 2[\mathbf{F}_\uparrow^- \bowtie \mathbf{F}_\downarrow^+]_{S=0} + (\mathbf{F}_\updownarrow^- + \mathbf{F}_\updownarrow^+)_{S=\pm1} >^{ph}$ throw the 'empty' Bivacuum or throw perturbed Bivacuum in the volume of condensed matter, transparent for these particles, can be considered as a **two stage process**:

**Stage I**: It corresponds to elementary particle state, when the unpaired/uncompensated sub-elementary fermions $\mathbf{F}_\updownarrow^\pm >^{e,p}$ or $(\mathbf{F}_\updownarrow^- + \mathbf{F}_\updownarrow^+)_{S=\pm1} >^{ph}$ are in [C] phase and compensated each other in pairs $[\mathbf{F}_\uparrow^+ \bowtie \mathbf{F}_\downarrow^-]$ are in [W] phase. This stage is accompanied by excitation of elastic waves in Bivacuum matrix, representing reversible Bivacuum dipoles symmetry shifts, provided by the external translational momentum of uncompensated sub-elementary fermions in [C] phase. The stage I stands for *kinetic* energy and momentum transmission to big number of *secondary anchor sites* of elementary particle in matrix, using Bivacuum *nonlocal* properties. At the same stage the wave [W] phase of symmetric pairs $[\mathbf{F}_\uparrow^- \bowtie \mathbf{F}_\downarrow^+]_{S=0}$ simultaneously transfer the *potential* energy to the *secondary anchor sites*. The properties and locations of the anchor sites corresponds to particle's eigenfunctions and corpuscular eigen states dependent on de Broglie wave length of the particle. The mechanism of the instant momentum and energy transmission, responsible for *anchor sites* can be realized via bundles of Virtual Guides (see section 14).

The eigenfunctions, characterizing *anchor sites* are alternative, i.e. incompatible with each other - *orthogonal*. It means, that only one of many may be occupied by Cumulative Virtual Cloud ($\mathbf{CVC}^\pm$) of particle in the process of its propagation throw Bivacuum (*stage II*).

The energy and charge conservation law demands, that in the absence of external fields, the resulting energy of all activated anchor sites should be zero. It is possible, if we assume that all *secondary anchor sites* (**AS**) are composed from two or three pairs of conjugated and correlated Cooper pairs of asymmetric Bivacuum fermions with energy, spin and charge compensating each other:

$$\mathbf{AS} = \sum^N 3[\mathbf{BVF}_\pm^\uparrow \bowtie \mathbf{BVF}_\updownarrow^\downarrow]_n \qquad 7.46$$

The opposite asymmetry of Bivacuum fermions and antifermions, forming virtual Cooper pairs, is provided by their rotation around common basic axis. Such anchor sites are proper for absorption of Cumulative Virtual Clouds ($CVC^\pm$) of the electrons, positrons and photons in their [W] phase.

**Stage II**: Corresponds to particle state, when the unpaired/uncompensated sub-elementary fermions $\mathbf{F}_\updownarrow^\pm >^{e,p}$ or $(\mathbf{F}_\updownarrow^- + \mathbf{F}_\updownarrow^+)_{S=\pm1} >^{ph}$ are in expanded [W] phase, representing cumulative virtual cloud ($\mathbf{CVC}^\pm$), modulated by de Broglie wave of particles, determined by properties of its *primary anchor site*. The symmetric pairs $[\mathbf{F}_\uparrow^+ \bowtie \mathbf{F}_\downarrow^-]$ on this stage II are in the compact [C] phase.

The jumps of the triplets (fermions) or sextet (photons) with *group velocity* of wave packet to one of prepared in previous **stage I** *secondary anchor sites* occur on this stage. The properties of secondary anchor site can change after complex formation with particle, however without violation of energy conservation and energy dissipation.

The most probable distance of such 'jump' is determined by de Broglie wave length of particle ($\lambda_B = h/\mathbf{p}$), equal to that of cumulative virtual cloud (CVC) of uncompensated sub-elementary fermions and the most probable direction of jump coincide with particle momentum in its [C] phase. However the new location of particle, as only one of many possible, is not rigidly predetermined and the 'jumps' can be considered as the stochastic process. The described mechanism of elementary particles propagation in space can be named "the kangaroo effect".

*The principle of superposition in quantum mechanics has the same formal expression as the waves superposition in classical mechanics:*

$$\Phi(\mathbf{r}, t) = c_1 \Phi(\mathbf{r}, t)_1 + c_2 \Phi(\mathbf{r}, t)_2 + \ldots c_n \Phi(\mathbf{r}, t)_n \qquad 7.47$$

where: $c_n$ are arbitrary complex numbers; $\Phi(\mathbf{r}, t)_n$ is wave function, describing different and alternative/orthogonal ($n$) states of quantum system. In accordance to our theory these quantum states correspond to multiple *secondary anchor sites* of moving in space particle.

However, in contrast to state/wave superposition of classical systems, in quantum system any state is not the result of 'mixing' of other states, but always the alternative or *orthogonal*, i.e. only one state of many allowed can be realized. It is so-called collapsing of the wave function.

Our description of the 'anchor' sites is in accordance with interpretation of wave function as a cohomological measure of quantum vorticity by Kiehn (1989, 1998). An exact complex mapping of the wave function has been found, which, when followed by a separation into real and imaginary parts, transforms the two dimensional Schrödinger equation for a charged particle interacting with an electromagnetic field into two partial differential systems. The first partial differential system is exactly the evolutionary equation for the vorticity of a compressible, viscous two dimensional Navie-Stokes fluid. The second system is related to the Beltrami equation defining a minimal surface in terms of the kinetic and potential energy. *The absolute square of the wave function is exactly the vorticity distribution in a fluid. This distribution corresponds to distribution of secondary anchor sites in our model of particle propagation (7.46 and 7.47).* This interpretation of the wave function offers an alternative to the Copenhagen dogma.

## 8. The nature of electrostatic, magnetic and gravitational interaction, based on Unified theory

### 8.1 Electromagnetic dipole radiation as a consequence of charge oscillation

The [*emission* ⇌ *absorption*] of photons in a course of elementary fermions - triplets $< [\mathbf{F}_\uparrow^- \bowtie \mathbf{F}_\downarrow^+]_{S=0} + (\mathbf{F}_\uparrow^+)_{S=\pm 1/2} >^{e,\tau}$ vibrations can be described by known mechanism of the electric dipole radiation ($\varepsilon_{EH}$), induced by charge acceleration ($a$), following from Maxwell equations (Berestetsky, et. al.,1989):

$$\varepsilon_{EH} = \frac{2e^2}{3c^3} a^2 \qquad 8.1$$

The resulting frequency of $[C \rightleftharpoons W]$ pulsation of each of three sub-elementary fermions in triplets is a sum of internal frequency contribution ($\mathbf{R}\,\omega_0^{in}$) and the external frequency ($\omega_B$) of de Broglie wave from (7.4):

$$[\omega_{C \rightleftharpoons W} = \mathbf{R}\,\omega_0^{in} + \omega_B]^i \qquad 8.2$$

where: $\mathbf{R} = \sqrt{1 - (\mathbf{v}/\mathbf{c})^2}$ is relativistic factor.

The acceleration can be related only with external translational dynamics which determines the empirical de Broglie wave parameters of particles. Acceleration is a result of alternating change of the charge deviation from the position of equilibrium:

$\Delta\lambda_B(\mathbf{t}) = (\lambda_B^i - \lambda_0) \sin \omega_B \mathbf{t}$ with de Broglie wave frequency of triplets: $\omega_B = \hbar/(m_V^+ L_B^2)$, where $L_B = \hbar/m_V^+ \mathbf{v}$. It is accompanied by oscillation of the instant de Broglie wave length $(\boldsymbol{\lambda}_B^t)$.

The acceleration of charge in the process of $\mathbf{C} \leftrightharpoons \mathbf{W}$ pulsation of the anchor $\mathbf{BVF}_{anc}^{\updownarrow}$ can be expressed as:

$$\mathbf{a} = \boldsymbol{\omega}_B^2 \Delta \boldsymbol{\lambda}_B(\mathbf{t}) \qquad\qquad 8.3$$

$$\mathbf{a} = \boldsymbol{\omega}_B^2 (\boldsymbol{\lambda}_B^t - \boldsymbol{\lambda}_0) \sin \boldsymbol{\omega}_B \mathbf{t} \qquad\qquad 8.4$$

where: $\boldsymbol{\lambda}_B^t = 2\boldsymbol{\pi} \mathbf{L}_B^t$ is the instant de Broglie wave length of the particle and $\boldsymbol{\lambda}_0 = \mathbf{h}/\mathbf{m}_0 \mathbf{c}$ is the Compton length of triplet.

The intensity of dipole radiation of pulsing $\mathbf{BVF}_{anc}^{\updownarrow}$ from 8.2 and 8.4 is:

$$\boldsymbol{\varepsilon}_{EM} = \frac{2}{3c^3} \boldsymbol{\omega}_B^4 \ (\mathbf{d}_E^t)^2 \qquad\qquad 8.5$$

where the oscillating electric dipole moment is: $\mathbf{d}_E^t = \mathbf{e}(\boldsymbol{\lambda}_B^t - \boldsymbol{\lambda}_0)$.

Consequently, in accordance with our model of duality, the EM dipole radiation is due to modulation of the frequency of $\mathbf{C} \leftrightharpoons \mathbf{W}$ pulsation of three sub-elementary fermions of the electron or proton by $[\mathbf{C} \leftrightharpoons \mathbf{W}]_{anc}$ frequency of anchor Bivacuum fermions $\mathrm{BVF}_{anc}^{\updownarrow}$, related to thermal vibrations of elementary particles. These vibrations are are accompanied by creation of secondary *anchor sites* (**AS**), described in previous section (eq.7.46). When the accelerations and final kinetic energy of elementary charges are big enough for resonant interaction with basic Bivacuum virtual pressure waves $[\mathbf{VPW}^+ \bowtie \mathbf{VPW}^-]_{q=1}$, the **AS from virtual excitations transform to photons** (Fig.4 of this paper).

The electromagnetic field, is a result of correlated Corpuscle - Wave pulsation of group of such transformed photons and their fast rotation in opposite directions with angle velocity ($\omega_{rot}$), equal to $[\mathbf{C} \leftrightharpoons \mathbf{W}]$ pulsation frequency of sub-elementary fermions and antifermions, forming photons. The superposition of clockwise or anticlockwise direction of photon's rotation as respect to direction of their propagation, determines their polarization.

### 8.2 Different kind of Bivacuum dipoles symmetry perturbation by dynamics of elementary particles, as a background of fields origination

In the process of $[\mathbf{C} \leftrightharpoons \mathbf{W}]$ pulsation of sub-elementary particles in triplets $< [\mathbf{F}_\uparrow^+ \bowtie \mathbf{F}_\downarrow^-] + \mathbf{F}_\updownarrow^\pm >^{e,p}$ the reversibility of [local (internal) $\Leftrightarrow$ distant (external)] symmetry compensation effects stand for the energy conservation law. The *local* symmetry effects pertinent for the [C] phase of particles. They are confined in the volume of sub-elementary fermions and stabilized by the Coulomb, magnetic and gravitational attraction between opposite charges and mass of asymmetric torus and antitorus of sub - elementary fermions. The attraction forces between two sub-elementary fermions in pairs $[\mathbf{F}_\uparrow^+ \bowtie \mathbf{F}_\downarrow^-]$ are balanced by centrifugal force of their axial rotation around common axes. The axis of triplet rotation is strictly related, in accordance to our model, with its spin and direction of translational propagation. It is supposed, that like magnetic field force lines, this rotation follows the *right hand screw rule* and is responsible for *magnetic field* origination. The total energy of triplet, the angular frequency of its rotation and the velocity of its translational propagation in space are interrelated (see eqs. 6.8 and 6.8b).

The $[\mathbf{C} \to \mathbf{W}]$ transitions of unpaired/uncompensated $\mathbf{F}_\updownarrow^\pm >^{e,p}$ of elementary particles are accompanied by the *diverging effects - translational and rotational (angular)*, accompanied by distant elastic deformation of Bivacuum matrix, shifting the corresponding symmetry (charge and spin equilibrium) of Bivacuum dipoles.

The reverse [$\mathbf{W} \to \mathbf{C}$] transition represents the *converging effect*. The latter is accompanied by getting back the energy, *diverged* in previous phase and restoration of the unpaired sub-elementary fermion and the whole triplet *local/enfolded* asymmetric properties.

The [*divergence* ⇌ *convergence*] of mass/energy, charge and spin equilibrium shifts in surrounding medium of Bivacuum dipoles (BVF$^\uparrow$ and BVF$^\downarrow$) in form of spherical elastic waves, are induced by [$\mathbf{C} \rightleftharpoons \mathbf{W}$] pulsations of triplets and accompanied *recoil* ⇌ *antirecoil* effects. These effects are generated by *unpaired* positive sub-elementary fermion $\mathbf{F}_{\updownarrow}^+ >$ of triplets $< [\mathbf{F}_{\updownarrow}^+ \bowtie \mathbf{F}_{\downarrow}^-] + \mathbf{F}_{\updownarrow}^\pm >^{e,p}$ . They are opposite for particles and antiparticles.

Corresponding *charge symmetry shifts* between torus and antitorus of Bivacuum dipoles are dependent on distance ($R$) from pulsing triplets, as ($\vec{r}/R$). The induced by such mechanism attraction and assembly of Bivacuum dipoles can be accompanied by formation of Cooper pairs [$\mathbf{BVF}_+^\uparrow \bowtie \mathbf{BVF}_-^\downarrow$] in space between remote $\mathbf{F}_{\updownarrow}^+ >$ and $\mathbf{F}_{\updownarrow}^- >$ of different triplets. The *attraction* between elementary particles of opposite charges is a result of Bivacuum tendency to minimize the uncompensated symmetry shift and charge density by formation of Cooper pairs from $\mathbf{BVF}_\pm^\updownarrow$. This compensation effect is increasing with with decreasing the separation between charges ($R \to 0$). The corresponding ordering of Cooper pairs, like bundles of virtual microfilaments stands for *electrostatic field and its 'force lines* origination. The Coulomb *repulsion* between similar charges is consequence of decreasing the resulting Bivacuum asymmetry of the same sign (positive or negative) in space between them by increasing the separation between these charges ($R \to \infty$).

The electrostatic field tension, produced by charged particles, is proportional to their kinetic energy ($\alpha \mathbf{T}_k^{\mathbf{F}_{\updownarrow}^\pm >^{e,p}}$). It can be expressed via gradients of charge symmetry shift of Bivacuum dipoles of surrounding medium, interrelated also mass symmetry shift and the external kinetic energy of dipoles:

$$\mathbf{E}_E = -grad \, |e_+ - e_-|_{BVF} = -grad \, |\mathbf{m}_V^+ - \mathbf{m}_V^-|\mathbf{c}_{BVF}^2 \sim \alpha \mathbf{T}_k^{\mathbf{F}_{\updownarrow}^\pm >^{e,p}} \qquad 8.5a$$

$$\alpha \mathbf{T}_k^{\mathbf{F}_{\updownarrow}^\pm >^{e,p}} = \alpha \frac{1}{2}|\mathbf{m}_V^+ - \mathbf{m}_V^-|\mathbf{c}^2 = \alpha \frac{1}{2}\mathbf{m}_V^+ \mathbf{v}^2$$

where: $\alpha = \mathbf{e}^2/\hbar\mathbf{c}$ is electromagnetic fine structure constant.
The validity of 8.5a will be presented in the next section.

The direction of fast rotation of pairs of sub-elementary fermion and antifermion [$\mathbf{F}_\uparrow^+ \bowtie \mathbf{F}_\downarrow^-$] of triplets $< [\mathbf{F}_\uparrow^+ \bowtie \mathbf{F}_\downarrow^-] + \mathbf{F}_{\updownarrow}^\pm >^{e,p}$ of opposite charges - clockwise or anticlockwise and unpaired $\mathbf{F}_{\updownarrow}^\pm >$ is dependent on direction of triplets propagation. The rotational motion is pertinent for [C] phase of [$\mathbf{F}_\uparrow^+ \bowtie \mathbf{F}_\downarrow^-$] and is absent for their [W] phase. Consequently, their [$\mathbf{C} \rightleftharpoons \mathbf{W}$] pulsation, counterphase to pulsation of $\mathbf{F}_{\updownarrow}^\pm >$ should induce the oscillation of spin equilibrium shift between Bivacuum fermions and antifermions of clockwise and anticlockwise rotation [$\mathbf{BVF}^\uparrow \leftrightharpoons \mathbf{BVB}^\pm \rightleftharpoons \mathbf{BVF}^\downarrow$] to the left or right. The sign of shift is dependent on direction of triplets propagation.

The shift of spin equilibrium in Bivacuum is accompanied by disassembly of Cooper pairs:

$$\mathbf{n}_{\updownarrow}[\mathbf{BVF}^\uparrow \bowtie \mathbf{BVF}^\downarrow] \to \mathbf{n}_\uparrow \mathbf{BVF}^\uparrow + \mathbf{n}_\downarrow \mathbf{BVF}^\downarrow \qquad 8.5a$$

In the absence of magnetic field the densities of Bivacuum fermions and antifermions are equal to each other $\mathbf{n}_{\updownarrow} = \mathbf{n}_\uparrow = \mathbf{n}_\downarrow$ and all of them compensate each other spins.

Let's assume, that the *increasing* of $\mathbf{BVF}^\uparrow$ density ($\mathbf{n}_\uparrow$) and corresponding *decreasing* of $\mathbf{BVF}^\downarrow$ density ($\mathbf{n}_\downarrow$) corresponds to the *North (N) magnetic pole* formation. The opposite to that, Bivacuum dipoles densities shifts stands for South (S) pole formation, i.e. when $\mathbf{n}_\downarrow$ is *increasing* and $\mathbf{n}_\uparrow$ *decreasing*:

$$\mathbf{N\ pole}:\quad \mathbf{n}_\uparrow > \mathbf{n}_\downarrow \qquad\qquad 8.5b$$

$$\mathbf{S\ pole}:\quad \mathbf{n}_\downarrow > \mathbf{n}_\uparrow$$

The *attraction between opposite poles* **N** and **S** reflects the tendency of $\mathbf{BVF}^\uparrow$ and $\mathbf{BVF}^\downarrow$ of the excessive density to form stable Cooper pairs, equalizing the symmetry shift between densities of Bivacuum dipoles of opposite spins:

$$\left[\ \text{attraction:}\ (\mathbf{n}_\uparrow \mathbf{BVF}^\uparrow)^{\mathbf{N}} \bowtie (\mathbf{n}_\downarrow \mathbf{BVF}^\downarrow)^{\mathbf{S}}\ \right]$$

For the other hand, the *repulsion between similar magnetic poles* is a consequence of Pauli principle of spatial incompatibility of two fermions (real or virtual) of the same spins (see section 9):

$$\text{repulsion:}\ (\mathbf{n}_\uparrow \mathbf{BVF}^\uparrow)^{\mathbf{N}} \Leftrightarrow (\mathbf{n}_\uparrow \mathbf{BVF}^\uparrow)^{\mathbf{N}} \qquad\qquad 8.5c$$

$$\text{repulsion:}\ (\mathbf{n}_\downarrow \mathbf{BVF}^\downarrow)^{\mathbf{S}} \Leftrightarrow (\mathbf{n}_\downarrow \mathbf{BVF}^\downarrow)^{\mathbf{S}}$$

The magnetic attraction and repulsion between Bivacuum dipoles is most effective, when $\mathbf{n}_\uparrow \simeq \mathbf{n}_\downarrow$ and is increasing with their densities.

*Consequently, just the equilibrium shift between Bivacuum fermions and antifermions of opposite spins, depending on direction of current and rotation of triplets, stands for the pole and intensity of curled magnetic field origination around current.*

The thermal motion of conducting electrons in metals or ions in plasma became more ordered in electric current, increasing correspondingly the magnetic cumulative effects due to increasing of probability and number of triplets, rotating in the same plane and direction. The bigger is velocity and kinetic energy of triplets, the faster is their rotation and bigger magnetic field tension, excited by this rotation:

$$\mathbf{H} = \mathbf{grad}\ (\mathbf{K}_{BVF\uparrow \rightleftharpoons BVF\downarrow}) = (\overrightarrow{r}/R)\mathbf{K}_{BVF\uparrow \rightleftharpoons BVF\downarrow} \ \sim \alpha \mathbf{T}_k^{\mathbf{F}_i^\pm > e,p} \qquad 8.5d$$

$$\alpha \mathbf{T}_k^{\mathbf{F}_i^\pm > e,p} \ = \alpha \frac{1}{2}|\mathbf{m}_V^+ - \mathbf{m}_V^-|\mathbf{c}^2 \ = \alpha \frac{1}{2}\mathbf{m}_V^+ \mathbf{v}^2 \ = \alpha \frac{1}{2}\mathbf{m}_V^+ \boldsymbol{\omega}_{\mathbf{I}}^2 \mathbf{L}_T^2$$

The pulsation of *potential energy* of sub-elementary fermions, in contrast to that of kinetic one, is determined by the sum of absolute energies of their torus and antitorus: $\mathbf{V} = \frac{1}{2}(\mathbf{m}_V^+ + \mathbf{m}_V^-)\mathbf{c}^2$. Consequently, the amplitude of this kind of energy pulsation is independent on the charge of fermion.

The potential energy oscillation of each of paired sub-elementary fermions $[\mathbf{F}_\uparrow^+ \bowtie \mathbf{F}_\downarrow^-]$ of triplets have similar but opposite effect on excitation of $(\mathbf{V}^+)$ and $(\mathbf{V}^-)$ of surrounding $\mathbf{BVF}^\updownarrow = [\mathbf{V}^+ \Updownarrow \mathbf{V}^-]$, equal to unpaired one by absolute value.

The excitation of positive and negative virtual pressure waves (VPW$_q^+$ and VPW$_q^-$) by the **recoil** $\rightleftharpoons$ **antirecoil** effects, accompanied the $[\mathbf{C} \rightleftharpoons \mathbf{W}]$ pulsation of *potential* energy of sub-elementary fermions of elementary particles is a background of *gravitational field* in accordance to our theory, independently on charge. The influence of the in-phase recoil/antirecoil effects of pulsing $[\mathbf{F}_\uparrow^+ \bowtie \mathbf{F}_\downarrow^-]$ on the probability of excitation of positive and negative virtual pressure waves (VPW$^+$ and VPW$^-$) in Bivacuum by torus $(\mathbf{V}^+)$ and antitorus $(\mathbf{V}^-)$ of Bivacuum dipoles $\mathbf{BVF}^\updownarrow = [\mathbf{V}^+ \Updownarrow \mathbf{V}^-]$ is equal by absolute value to increment. It is determined by corresponding potential energy oscillation of unpaired $\mathbf{F}_i^\pm >$ of triplets.

**It is possible to present the given above explanation of the Coulomb, magnetic and gravitational fields nature in more formal way**. The total energies of $[\mathbf{C} \rightarrow \mathbf{W}]$ and $[\mathbf{W} \rightarrow \mathbf{C}]$ transitions of particles we present using general formula (6.1): $\mathbf{E}_{tot} = \mathbf{V}_{tot} + \mathbf{T}_{tot}$. However, here we take into account the *diverging $\rightleftharpoons$ converging* effects, accompanied $[\mathbf{C} \rightleftharpoons \mathbf{W}]$ transitions and reversible transformation of the *internal* - local (Loc) gravitational, Coulomb and magnetic potentials to the *external* - distant (Dis) Bivacuum perturbation, stimulated by these transitions. For the end of energy conservation it is assumed, that the local and distant energy increments are opposite by sign and compensate each other. The distant *diverging $\rightleftharpoons$ converging* effects, in contrast to local *emission $\rightleftharpoons$ absorption* of $\mathbf{CVC}^{\pm}$, can be described in terms of *recoil (Rec) $\rightleftharpoons$ antirecoil (ARec)* effects.

The $[\mathbf{C} \rightarrow \mathbf{W}]$ transition, accompanied by three kinds of *diverging* effects, can be described as:

$$\mathbf{E}^{C \rightarrow W} = \mathbf{m}_V^+ \mathbf{c}^2 = \mathbf{V}_{tot} + [(\mathbf{E}_G)_{\text{Rec}}^{Loc} - (\mathbf{E}_G)_{\text{Rec}}^{Dist}] + \qquad 8.6$$

$$+ \, \mathbf{T}_{tot} + [(\mathbf{E}_E)_{\text{Rec}}^{Loc} - (\mathbf{E}_E)_{\text{Rec}}^{Dist}]_{tr} + \qquad 8.6a$$

$$+ \, [(\mathbf{E}_H)_{\text{Rec}}^{Loc} - (\mathbf{E}_H)_{\text{Rec}}^{Dist}]_{rot} \qquad 8.6b$$

In the process of the reverse $[\mathbf{W} \rightarrow \mathbf{C}]$ *converging* transition the unpaired sub-elementary fermion $\mathbf{F}_{\updownarrow}^{\pm} >$ of triplet $< [\mathbf{F}_{\uparrow}^- \bowtie \mathbf{F}_{\downarrow}^+]_{S=0} + (\mathbf{F}_{\updownarrow}^{\pm})_{S=\pm 1/2} >$ gets back the *diverged* in previous phase *antirecoil* energy due to elastic properties of Bivacuum, turning its symmetry shift from the distant to local one of opposite energy:

$$\mathbf{E}^{W \rightarrow C} = \mathbf{m}_V^+ \mathbf{c}^2 = \mathbf{V}_{tot} + [-(\mathbf{E}_G)_{A\text{Rec}}^{Loc} + (\mathbf{E}_G)_{A\text{Rec}}^{Dist}] + \qquad 8.7$$

$$+ \, \mathbf{T}_{tot} + [-(\mathbf{E}_E)_{A\text{Rec}}^{Loc} + (\mathbf{E}_E)_{A\text{Rec}}^{Dist}]_{tr} + \qquad 8.7a$$

$$+ \, [-(\mathbf{E}_H)_{A\text{Rec}}^{Loc} + (\mathbf{E}_H)_{A\text{Rec}}^{Dist}]_{rot} + \qquad 8.7b$$

where:

$$\mathbf{V}_{tot}^W = \frac{1}{2}(\mathbf{m}_V^+ + \mathbf{m}_V^-)\mathbf{c}^2 = \mathbf{V}_{tot}^C = \frac{1}{2}\mathbf{m}_V^+ \mathbf{c}^2 [\mathbf{2} - (\mathbf{v}/\mathbf{c})^2]$$

is a total potential energy of each sub-elementary fermion of triplet (6.4) in the wave and corpuscular phase, non equal to zero at $\mathbf{v} = \mathbf{0}$;

$$\mathbf{T}_{tot}^W = \frac{1}{2}(\mathbf{m}_V^+ - \mathbf{m}_V^-)\mathbf{c}^2 = \mathbf{T}_{tot}^C = \frac{1}{2}\mathbf{m}_V^+ \mathbf{v}^2$$

is its total kinetic energy, equal to zero at the external velocity $\mathbf{v}_{ext} = 0$ (6.5).

The reversible conversions of the *localized potential energy* $\pm (\mathbf{V}_G)_{\text{Rec},A\text{Rec}}^{Loc}$ to the distant one $\mp (\mathbf{V}_G)_{\text{Rec},A\text{Rec}}^{Dist}$, accompanied the *recoil $\rightleftharpoons$ antirecoil* effects, induced by $[\mathbf{C} \rightleftharpoons \mathbf{W}]$ pulsation of unpaired sub-elementary fermion of triplets at $\mathbf{v}_{ext} = 0$, i.e. when its mass symmetry shift is equal to the rest mass, can be evaluated quantitatively. The increment of these oscillation are equal to difference of potential energies of $(\mathbf{F}_{\updownarrow}^{\pm})_{S=\pm 1/2} >$, corresponding to Golden mean conditions $(\mathbf{v}_{in}/\mathbf{c})^2 = \mathbf{\phi} = \mathbf{0.618}$, and energy of symmetric Bivacuum fermion with zero mass symmetry shift $\mathbf{V}_0 = \frac{1}{2}(\mathbf{m}_V^+ + \mathbf{m}_V^-)_0 \mathbf{c}^2 = \mathbf{m}_0 \mathbf{c}^2$ :

$$\Delta \mathbf{V}_{(\mathbf{F}_{\updownarrow}^{\pm})_{S=\pm 1/2}}^{\mathbf{VPW}^{\pm}} = \mathbf{V}_{[\mathbf{C}]} - \mathbf{V}_0 = \frac{1}{2}(\mathbf{m}_V^+ + \mathbf{m}_V^-)^{\phi}\mathbf{c}^2 - \mathbf{m}_0 \mathbf{c}^2 = 0.118\,\mathbf{m}_0 \mathbf{c}^2 \qquad 8.7c$$

where: $(\mathbf{m}_V^+)^{\phi} = \mathbf{m}_0/\mathbf{\phi} = 1.618\mathbf{m}_0$; $\quad (\mathbf{m}_V^-)^{\phi} = \mathbf{\phi}\mathbf{m}_0 = 0.618\mathbf{m}_0$.

The conversions between local and distant Bivacuum perturbations, related to potential energy oscillation, are mediated by Virtual Pressure Waves ($\textbf{VPW}^+$ and $\textbf{VPW}^-$).

Pulsations of unpaired $\left(\textbf{F}_{\updownarrow}^{\pm}\right)_{S=\pm 1/2}$ are interrelated with those of paired ones $[\textbf{F}_{\uparrow}^+ \bowtie \textbf{F}_{\downarrow}^-]$. The latter excite the positive $\textbf{VPW}^+$ and negative $\textbf{VPW}^-$ spherical virtual pressure waves, propagating in space with light velocity and energy:

$$\textbf{V}_{[\textbf{F}_{\uparrow}^+ \bowtie \textbf{F}_{\downarrow}^-]}^{\textbf{VPW}^+ + \textbf{VPW}^-} = \left|\Delta\textbf{V}^{\textbf{VPW}^+}\right| + |\Delta\textbf{V}^{\textbf{VPW}^-}| = 0.236\,\textbf{m}_0\textbf{c}^2 \qquad 8.7d$$

The pulsation of potential energy 8.7c and 8.7d of unpaired and paired sub-elementary fermions are counterphase.

They are a consequence of transitions of torus $\textbf{V}^+$ and antitorus $\textbf{V}^-$ of surrounding Bivacuum dipoles $\textbf{BVF}^{\updownarrow} = [\textbf{V}^+ \Updownarrow \textbf{V}^-]$ between the excited and ground states. The increasing of particle external translational velocity is accompanied by its relativistic mass ($\textbf{m}_\textbf{V}^{\mp}$) and potential energy increasing.

The $\pm(\textbf{E}_E)_{\text{Rec},\,Arec}^{Loc} = \mp(\textbf{E}_E)_{\text{Rec},\,A\,\text{Rec}}^{Dist}$ in $(8.6 - 8.7b)$ are the local and distant electrostatic potential oscillations, equal to each other.

The $\pm(\textbf{E}_H)_{\text{Rec},\,Arec}^{Loc} = \mp(\textbf{E}_H)_{\text{Rec},A\,\text{Rec}}^{Dist}$ are the local and distant magnetic potentials oscillations, equal to each other.

These [local $\rightleftharpoons$ distant] reversible interconversions, exciting the electric and magnetic fields, are the result of $[\textbf{C} \leftrightarrows \textbf{W}]$ pulsations and [*emission $\rightleftharpoons$ absorption*] of CVC$^{\pm}$ of sub-elementary fermions of triplets, determined by increments of translational and rotational momentum of CVC$^{\pm}$, correspondingly.

The residual momentum, kinetic energy and charge of the *anchor* Bivacuum fermion after emission of CVC$^{\pm}$ by unpaired/uncompensated sub-elementary fermion in the rest conditions ($\textbf{v}_{ext} = 0$) is equal to zero: $\textbf{T}_0 = \frac{1}{2}(\textbf{m}_\textbf{V}^+ - \textbf{m}_{\bar{\textbf{V}}}^-)_0\textbf{c}^2 = 0$ in contrast to the rest potential energy $\textbf{V}_0 = \textbf{m}_0\textbf{c}^2$ (8.7c):

$$\Delta\textbf{T}_{\left(\textbf{F}_{\updownarrow}^{\pm}\right)_{S=\pm 1/2}}^{\textbf{CVC}^{\pm}} = \textbf{T}_{[\textbf{C}]} - \textbf{T}_0 = \textbf{T}_{[\textbf{C}]} = \textbf{T}_{[\textbf{W}]} \qquad 8.7e$$

Let us consider in more detail the interconversions of the *internal - local* and the *external - distant* gravitational, Coulomb and magnetic interactions of charged elementary fermions, like electron or proton.

### 8.3 The new approach to quantum gravity and antigravity

The unified right parts of eqs. (8.6) and (8.7), describing the excitation of *gravitational waves*, represented by small part of potential energy of positive and negative virtual pressure waves ($\textbf{VPW}^+$ and $\textbf{VPW}^-$) with frequency, equal to frequency $[\textbf{C} \rightleftharpoons \textbf{W}]$ pulsation of unpaired sub-elementary fermions, equal to frequency of **recoil $\rightleftharpoons$ antirecoil** vibrations. These waves excitation is a result of corresponding oscillation of *potential energy* of unpaired $\textbf{F}_{\updownarrow}^{\pm} >^{e,p}$ of triplets $< [\textbf{F}_{\uparrow}^+ \bowtie \textbf{F}_{\downarrow}^-] + \textbf{F}_{\updownarrow}^{\pm} >^{e,p}$, correlated with similar vibrations of paired $[\textbf{F}_{\uparrow}^+ \bowtie \textbf{F}_{\downarrow}^-]$:

$$\overline{\textbf{V}}_{tot}^{\textbf{C}\rightleftharpoons\textbf{W}} = \textbf{V}_{tot} \pm \left[(\Delta\textbf{V}_G)_{[\textbf{C}]}^{Loc} - (\Delta\textbf{V}_G)_{[\textbf{W}]}^{Dist}\right] = \textbf{V}_{tot} \qquad 8.8$$

where: $(\Delta\textbf{V}_G)_{[\textbf{C}]}^{Loc} = (\textbf{V}_{[\textbf{C}]} - \textbf{V}_0)^{Loc}$; $(\Delta\textbf{V}_G)_{[W]}^{Dis} = (\textbf{V}_{[W]} - \textbf{V}_0)^{Dis}$ are the local and distant increments of part of potential energy oscillation in [C] and [W] phase of sub-elementary fermions of elementary particles, determined by reversible **recoil $\rightleftharpoons$ antirecoil** effects.

The general formula for fluctuation of total *potential energy,* accompanied the $[\textbf{C} \rightleftharpoons \textbf{W}]$ pulsation of unpaired sub-elementary fermion, can be presented in similar way as

8.7c:

$$\Delta \mathbf{V}^{\mathbf{VPW}^{\pm}}_{\left(\mathbf{F}^{\pm}_{\updownarrow}\right)_{S=\pm 1/2}} = \mathbf{V}_{[C]} - \mathbf{V}_0 = \frac{1}{2}(\mathbf{m}^+_V + \mathbf{m}^-_V)\mathbf{c}^2 - \mathbf{m}_0\mathbf{c}^2 = \qquad 8.8a$$

$$= \frac{1}{2}\frac{\hbar\mathbf{c}}{\mathbf{L}_V} - \frac{\hbar\mathbf{c}}{\mathbf{L}_0} = \frac{\hbar\mathbf{c}}{2}\left(\frac{1}{\mathbf{L}_V} - \frac{1}{\mathbf{L}_0}\right) = \qquad 8.8b$$

$$= \frac{1}{2}\frac{\mathbf{m}_0\mathbf{c}^2}{\mathbf{R}}\left[2 - (\mathbf{v}/\mathbf{c})^2\right] - \mathbf{m}_0\mathbf{c}^2 \qquad 8.8c$$

where the curvature, characterizing potential energy of asymmetric sub-elementary fermion is defined as: $\mathbf{L}_V = \hbar/[(\mathbf{m}^+_V + \mathbf{m}^-_V)\mathbf{c}]$ and the $\mathbf{L}_0 = \hbar/[\mathbf{m}_0\mathbf{c}]$ is a curvature, characterizing the potential energy of symmetric Bivacuum fermion, equal to Compton radius.

Taking into account, that $1 - (\mathbf{v}/\mathbf{c})^2 = \mathbf{R}^2$ we easily get from 8.8c the following expression for the total amplitude of sub-elementary fermion potential energy oscillation:

$$\Delta \mathbf{V}^{\mathbf{VPW}^{\pm}}_{\left(\mathbf{F}^{\pm}_{\updownarrow}\right)_{S=\pm 1/2}} = \frac{1}{2}\frac{\mathbf{m}_0\mathbf{c}^2}{\mathbf{R}}[\mathbf{R}^2 - 2\mathbf{R} + 1] \qquad 8.8d$$

This potential energy increment of Virtual pressure waves, generated by elementary particle pulsation, turns to zero, when the solution of quadratic equation is zero: $\mathbf{R}^2 - 2\mathbf{R} + 1 = 0$. It is easy to see, that this happens at $\mathbf{R} = 1$, i.e. when the elementary particle is in rest state condition: $\mathbf{v} = \mathbf{0}$.

The more detailed presentation of 8.8 is:

$$\overline{\mathbf{V}}^{\mathbf{C}\rightleftharpoons\mathbf{W}}_{tot} = \frac{1}{2}(\mathbf{m}^+_V + \mathbf{m}^-_V)\mathbf{c}^2 \pm \frac{\mathbf{r}}{\mathscr{r}}\left\{ \begin{array}{l} \left[\mathbf{G}\frac{(\mathbf{m}^+_V\mathbf{m}^-_V)}{\mathbf{L}_V} - \mathbf{G}\frac{\mathbf{m}^2_0}{\mathbf{L}_0}\right]^{Loc} - \\ -\left[\frac{1}{2}\left(\frac{\mathbf{m}^i_0}{\mathbf{M}_{Pl}}\right)^2(\mathbf{m}^+_V + \mathbf{m}^-_V)\mathbf{c}^2 - \left(\frac{\mathbf{m}^i_0}{\mathbf{M}_{Pl}}\right)^2\mathbf{m}_0\mathbf{c}^2\right]^{Dist} \end{array} \right\} \qquad 8.9$$

The local *internal* gravitational interaction between the opposite mass poles of the mass-dipoles of unpaired sub-elementary fermions (antifermions) $\left(\mathbf{F}^{\pm}_{\updownarrow}\right)_{S=\pm 1/2}$ turns reversibly to the *external* distant one. The corresponding dynamic equilibrium between the *diverging* and *converging* flows of potential energy, following $[\mathbf{C} \rightleftharpoons \mathbf{W}]$ pulsation and corresponding **recoil** $\rightleftharpoons$ **antirecoil** effects can be described as:

$$(\mathbf{V}_G)_{\mathbf{F}^+_{\uparrow}\bowtie\mathbf{F}^-_{\downarrow}} = \frac{\mathbf{r}}{\mathscr{r}}\left[\mathbf{G}\frac{|\mathbf{m}^+_V\mathbf{m}^-_V|}{\mathbf{L}_V} - \mathbf{G}\frac{\mathbf{m}^2_0}{\mathbf{L}_0}\right]^{Loc}_{\mathbf{F}^+_{\uparrow}\bowtie\mathbf{F}^-_{\downarrow}} \overset{\substack{\text{Recoil}\\\mathbf{C}\rightarrow\mathbf{W}}}{\underset{\substack{\mathbf{W}\rightarrow\mathbf{C}\\\text{Antirecoil}}}{\rightleftharpoons}} \frac{\mathbf{r}}{\mathscr{r}}\left[(\beta\,\mathbf{m}^+_V\mathbf{c}^2(2 - \mathbf{v}^2/\mathbf{c}^2) - \beta^i\mathbf{m}_0\mathbf{c}^2)\right]^{Dist}_{\mathbf{F}^+_{\uparrow}\bowtie\mathbf{F}^-_{\downarrow}} \quad 8.10$$

where: $\mathbf{L}_V = \hbar/(\mathbf{m}^+_V + \mathbf{m}^-_V)\mathbf{c}$ is a characteristic curvature of potential energy (4.4b); $\mathbf{M}^2_{Pl} = \hbar\mathbf{c}/\mathbf{G}$ is a Plank mass; $\frac{\mathbf{r}}{\mathscr{r}}$ is ratio of unitary vector to distance from particle; $\mathbf{m}^2_0 = |\mathbf{m}^+_V\,\mathbf{m}^-_V|$ is a rest mass squared; $\beta^i = \left(\frac{\mathbf{m}^i_0}{\mathbf{M}_{Pl}}\right)^2$ is the introduced earlier dimensionless gravitational fine structure constant (Kaivarainen, 1995-2005). For the electron $\beta^e = 1.739 \times 10^{-45}$ and $\sqrt{\beta^e} = \frac{\mathbf{m}^e_0}{\mathbf{M}_{Pl}} = 0.41 \times 10^{-22}$.

The effective velocity of particle's *recoil* $\rightleftharpoons$ *antirecoil* process, responsible for excitation of gravitational waves squared $(\mathbf{v}^2_G)_{eff}$, can be introduced from the right part of (8.10) as

$$\beta\,\mathbf{m}^+_V\mathbf{c}^2(2 - \mathbf{v}^2/\mathbf{c}^2) = \beta\,(\mathbf{m}^+_V + \mathbf{m}^-_V)\mathbf{c}^2 = \mathbf{m}^+_V(\mathbf{v}^2_G)_{eff}$$

in form:

$$(\mathbf{v}_G^2)_{eff} = \beta \, \mathbf{c}^2 (2 - \mathbf{v}^2/\mathbf{c}^2) \qquad\qquad 8.10a$$

This effective recoil velocity, providing excitation of gravitational waves ($\mathbf{VPW}^+$ and $\mathbf{VPW}^-$)$_G$ is decreasing up to $(\mathbf{v}_G^2)_{eff}^{min} = \beta \, \mathbf{c}^2$ at $\mathbf{v} = \mathbf{c}$, like in the case of photons or neutrino, and increasing up two times $(\mathbf{v}_G^2)_{eff}^{max} = 2\beta \, \mathbf{c}^2$ at $\mathbf{v} = \mathbf{0}$, i.e. in primordial Bivacuum dipoles.

At the Golden mean conditions, when $(\mathbf{v}^2/\mathbf{c}^2) = 0.618 = \phi$, we get from (8.10a) the reduced value of characteristic gravitational velocity of zero-point oscillation, of elementary particles in state of rest:

$$\frac{(\mathbf{v}_G^2)_{eff}^{\phi}}{\mathbf{c}^2} = 1.382 \, \beta$$

In triplets $< [\mathbf{F}_{\uparrow}^+ \bowtie \mathbf{F}_{\downarrow}^-] + \mathbf{F}_{\updownarrow}^{\pm} >^{e,p}$ the contribution of symmetric pair $[\mathbf{F}_{\uparrow}^+ \bowtie \mathbf{F}_{\downarrow}^-]$ pulsation to gravitation field energy is the additive function of energies of their cumulative virtual clouds energies: $\boldsymbol{\varepsilon}_{CVC^+}^{\mathbf{F}_{\uparrow}^+}$ and $\boldsymbol{\varepsilon}_{CVC^-}^{\mathbf{F}_{\downarrow}^-}$ :

$$(\mathbf{V}_G)_{<[\mathbf{F}_{\uparrow}^+\bowtie\mathbf{F}_{\downarrow}^-]+\mathbf{F}_{\updownarrow}^{\pm}>} = \frac{\mathbf{r}}{r}\left[\mathbf{G}\frac{|\mathbf{m}_{\mathbf{V}}^+\mathbf{m}_{\mathbf{V}}^-|}{\mathbf{L_V}}\right]_{\mathbf{F}_{\uparrow}^+\bowtie\mathbf{F}_{\downarrow}^-}^{Loc} \underset{\mathbf{W}\to\mathbf{C}}{\overset{\mathbf{C}\to\mathbf{W}}{\rightleftarrows}} \frac{\mathbf{r}}{r}\left[\beta^i\left(\frac{1}{2}\mathbf{m}_{\mathbf{V}}^+\mathbf{c}^2(2-\mathbf{v}^2/\mathbf{c}^2)-\mathbf{m}_0\mathbf{c}^2\right)\right]_{\mathbf{F}_{\uparrow}^+\bowtie\mathbf{F}_{\downarrow}^-}^{Dist}$$

$$or: \; (\mathbf{V}_G)_{<[\mathbf{F}_{\uparrow}^+\bowtie\mathbf{F}_{\downarrow}^-]+\mathbf{F}_{\updownarrow}^{\pm}>} = \beta^i\left(\boldsymbol{\varepsilon}_{CVC^+}^{\mathbf{F}_{\uparrow}^+}+\boldsymbol{\varepsilon}_{CVC^-}^{\mathbf{F}_{\downarrow}^-}\right)+\beta^i\boldsymbol{\varepsilon}_{CVC^{\pm}}^{\mathbf{F}_{\updownarrow}^{\pm}} \; \sim \qquad 8.10ab$$

$$or: (\mathbf{V}_G)_{<[\mathbf{F}_{\uparrow}^+\bowtie\mathbf{F}_{\downarrow}^-]+\mathbf{F}_{\updownarrow}^{\pm}>} = [\mathbf{VPW}_q^+ \bowtie \mathbf{VPW}_q^-]_G^{[\mathbf{F}_{\uparrow}^+\bowtie\mathbf{F}_{\downarrow}^-]} + [\mathbf{VPW}_q^+ \bowtie \mathbf{VPW}_q^-]_G^{\mathbf{F}_{\updownarrow}^{\pm}}$$

where: $\mathbf{m}_{\mathbf{V}}^+\mathbf{c}^2(2-\mathbf{v}^2/\mathbf{c}^2) = (\mathbf{m}_{\mathbf{V}}^+ + \mathbf{m}_{\mathbf{V}}^-)\mathbf{c}^2$

The excitation of the *external* - distant spherical virtual pressure waves of positive and negative energy: $\mathbf{VPW}_q^+$ and $\mathbf{VPW}_q^-$ is a result of pair of torus and antitorus energy beats, accompanied $[\mathbf{C} \leftrightarrows \mathbf{W}]$ counterphase pulsation of unpaired $\mathbf{F}_{\updownarrow}^{\pm} >^{e,p}$ and paired sub-elementary fermions $[\mathbf{F}_{\downarrow}^- \bowtie \mathbf{F}_{\uparrow}^+]_{S=0}$ with equal by absolute values energy.

It is important to note, that the energy of introduced gravitational field does not depend on charge of triplet, determined by unpaired sub-elementary fermion of triplets $< [\mathbf{F}_{\uparrow}^- \bowtie \mathbf{F}_{\downarrow}^+]_{S=0} + (\mathbf{F}_{\updownarrow}^{\pm})_{S=\pm1/2} >$, in contrast to electrostatic and magnetic field.

*It follows from our approach, that the gravitational energy is pertinent even for 'empty' primordial Bivacuum in the absence of matter and fields or when their influence is negligible.* This phenomena can be responsible for the attraction effect of '*cold dark matter*' of the Universe. The primordial Bivacuum dipoles are symmetric and their absolute mass/energies, charges and magnetic moments are equal:

$$\mathbf{m}_{\mathbf{V}}^+ = \mathbf{m}_{\mathbf{V}}^- = \mathbf{m}_0$$

The in-phase fluctuations of torus and antitorus of equal and opposite energy, compensating each other, can be presented as:

$$(\mathbf{E}_G)_{\mathbf{F}_{\uparrow}^+\bowtie\mathbf{F}_{\downarrow}^-} = \frac{\mathbf{r}}{r}\beta^i(\mathbf{m}_{\mathbf{V}}^+ + \mathbf{m}_{\mathbf{V}}^-)^i\mathbf{c}^2 = \frac{\mathbf{r}}{r}\beta^i\mathbf{m}_0\mathbf{c}^2(1+2\mathbf{n}) \qquad 8.11$$

$$or: (\mathbf{E}_G)_{\mathbf{F}_{\uparrow}^+\bowtie\mathbf{F}_{\downarrow}^-} = \frac{\mathbf{r}}{r}\beta^i\hbar\boldsymbol{\omega}_0(1+2\mathbf{n}) \; \sim [\mathbf{VPW}_q^+ \bowtie \mathbf{VPW}_q^-]_G \qquad 8.11a$$

Consequently, the *cold dark matter* phenomena can be a consequence of simultaneous excitation of huge number of Bivacuum dipoles, symmetric as respect to positive and negative energy, in virtual domains of nonlocality (see section 1.3). The energy, proportional to $V \sim (\mathbf{m}_V^+ + \mathbf{m}_V^-)\mathbf{c}^2$ considered in our theory as the potential one of Bivacuum dipoles, in contrast to kinetic one: $\mathbf{T}_k \sim (\mathbf{m}_V^+ - \mathbf{m}_V^-)\mathbf{c}^2$. The bigger is quantum number of Bivacuum dipoles excitation ($\mathbf{n}$), the higher is frequency of virtual pressure waves $[\mathbf{VPW}_q^+ \bowtie \mathbf{VPW}_q^-]_G$, responsible for gravitational field.

From the proposed here mechanism of gravitation and similar values of ($\mathbf{m}_V^+$) in the left and right parts of eq. (8.10) follows *the equality of gravitational and inertial mass*. The *inertia* itself can be defined, as a *resistance to additional symmetry shift* between the actual and complementary masses/energy of sub-elementary fermions ($\mathbf{m}_V^+$ and $\mathbf{m}_V^-$) of elementary particles and surrounding Bivacuum dipoles, accompanied positive and negative particles acceleration. Consequently, the inertia follows from generalized Le Chatelier's Principle, which this author formulate, as a resistance of any system, containing sub-elementary fermions of elementary particles in state of dynamic equilibrium, to additional symmetry shift, accompanied particles acceleration.

### 8.4 The hydrodynamic mechanism of gravitational attraction and repulsion

In accordance to our hypothesis (Kaivarainen, 1995; 2000; 2005), the mechanism of gravitational attraction and repulsion is similar to Bjerknes attraction/repulsion between pulsing spheres in liquid medium of Bivacuum. The dependence of Bjerknes force on distance between centers of pulsing objects is quadratic: $\mathbf{F}_{Bj} \sim 1/r^2$:

$$\mathbf{F}_G = \mathbf{F}_{Bj} = \frac{1}{\mathbf{r}^2}\pi\rho_G\mathbf{R}_1^2\mathbf{R}_2^2\mathbf{v}^2\cos\beta \qquad 8.12$$

where $\rho_G$ is density of liquid, i.e. virtual density of secondary Bivacuum. It is determined by Bivacuum dipoles (BVF$^{\updownarrow}$ and BVB$^{\pm}$) symmetry shift; $\mathbf{R}_1$ and $\mathbf{R}_2$ radiuses of pulsing/gravitating spheres; $\mathbf{v}$ is velocity of spheres surface oscillation (i.e. velocity of $\mathbf{VPW}_q^{\pm}$, excited by $[\mathbf{C} \rightleftharpoons \mathbf{W}]$ pulsation of elementary particles, which can be assumed to be equal to light velocity: $\mathbf{v} = \mathbf{c}$); $\beta$ is a phase shift between pulsation of spheres or system of coherent elementary particles.

It is important to note, that on the big enough distances the *attraction* may turn to *repulsion*. The latter effect, depending on the phase shift of coherent $[\mathbf{C} \rightleftharpoons \mathbf{W}]$ pulsation of interacting remote triplets ($\beta$), can explain the revealed *acceleration of the Universe expansion*. The corresponding *antigravitation* energy or *negative pressure energy (dark energy)*, is about 70% of the total Universe energy.

The possibility of artificial phase shift of $[\mathbf{C} \rightleftharpoons \mathbf{W}]$ pulsation of coherent elementary particles of any object may (for example by magnetic field) may change its gravitational attraction to repulsion and vice versa. The volume and radius of pulsing spheres ($R_1$ and $R_2$) in such approach is determined by sum of volume of hadrons, composing gravitating systems in solid, liquid, gas or plasma state. The gravitational attraction or repulsion is a result of increasing or decreasing of virtual pressure of subquantum particles between interacting systems as respect to its value outside them. This model can serve as a background for new *quantum gravity theory*.

The effective radiuses of gravitating objects $\mathbf{R}_1$ and $\mathbf{R}_2$ can be calculated from the effective volumes of the objects:

$$\mathbf{V}_{1,2} = \frac{4}{3}\pi\mathbf{R}_{1,2}^3 = \mathbf{N}_{1,2}\frac{4}{3}\pi\mathbf{L}_{p,n}^3 \qquad 8.12a$$

where: $\mathbf{N}_{1,2} = \mathbf{M}_{1,2}/\mathbf{m}_{p,n}$ is the number of protons and neutrons in gravitating bodies with mass $\mathbf{M}_1$ and $\mathbf{M}_2$; $\mathbf{m}_{p,n}$ is the mass of proton and neutron; $\mathbf{L}_{p,n} = \hbar/\mathbf{m}_{p,n}\mathbf{c}$ is the Compton radius of proton and neutron.

From (8.12a) we get for effective radiuses:

$$\mathbf{R}_{1,2} = \left(\frac{\mathbf{M}_{1,2}}{\mathbf{m}_{p,n}}\right)^{1/3} \mathbf{L}_{p,n} = \left(\frac{\mathbf{M}_{1,2}}{\mathbf{m}_{p,n}}\right)^{1/3} \frac{\hbar}{\mathbf{m}_{p,n}\mathbf{c}} \qquad 8.12b$$

Putting this to (8.12) we get for gravitational interaction between two macroscopic objects, each of them formed by atoms with coherently pulsing protons and neutrons:

$$\mathbf{F}_G = \frac{1}{\mathbf{r}^2}\pi\boldsymbol{\rho}_{Bv}\frac{(\mathbf{M}_1\mathbf{M}_2)^{2/3}}{\mathbf{m}_{p,n}^{4/3}}\left(\frac{\hbar}{\mathbf{m}_{p,n}}\right)^4\frac{1}{\mathbf{c}^2} \qquad 8.13$$

Equalizing this formula with Newton's one: $\mathbf{F}_G^N = \frac{1}{\mathbf{r}^2}\mathbf{G}(\mathbf{M}_1\mathbf{M}_2)$, we get the expression for gravitational constant:

$$\mathbf{G} = \pi\frac{\boldsymbol{\rho}_G}{\sqrt[3]{\mathbf{M}_1\mathbf{M}_2}}\frac{\hbar^2/\mathbf{c}^2}{\sqrt[3]{\mathbf{m}_{p,n}^{16}}} \qquad 8.14$$

The condition of gravitational constant permanency from (8.14), is the anticipated from our theory interrelation between the mass of gravitating bodies $\sqrt[3]{\mathbf{M}_1\mathbf{M}_2}$ and the virtual density $\boldsymbol{\rho}_G$ of secondary Bivacuum, determined by Bivacuum fermions symmetry shift and excitation in gravitational field:

$$\mathbf{G} = \mathbf{const}, \quad if \quad \frac{\boldsymbol{\rho}_G}{\sqrt[3]{\mathbf{M}_1\mathbf{M}_2}} = \mathbf{const} \qquad 8.14a$$

where, taking into account (8.10):

$$\sqrt[3]{\mathbf{M}_1\mathbf{M}_2} \ \sim \boldsymbol{\rho}_G \ = \frac{\frac{1}{2}\left(\frac{\mathbf{m}_0}{\mathbf{M}_{Pl}}\right)^2(\mathbf{m}_V^+ + \mathbf{m}_{\bar{V}}^-)}{\frac{3}{4}\pi\mathbf{L}_V^3} \ = \frac{2}{3}\frac{\left(\frac{\mathbf{m}_0}{\mathbf{M}_{Pl}}\right)^2\mathbf{m}_V^+(2 - \mathbf{v}^2/\mathbf{c}^2)}{\pi\mathbf{L}_V^3} \qquad 8.15$$

assuming, that the radius/curvature of Bivacuum fermion, characterizing it s potential energy, is:

$$\mathbf{L}_V = \frac{\hbar}{(\mathbf{m}_V^+ + \mathbf{m}_{\bar{V}}^-)\mathbf{c}} \qquad 8.16$$

we get for reduced gravitational density:

$$\boldsymbol{\rho}_G = \frac{2}{3}\frac{1}{\pi\hbar^3}\left(\frac{\mathbf{m}_0}{\mathbf{M}_{Pl}}\right)^2(\mathbf{m}_V^+ + \mathbf{m}_{\bar{V}}^-)^4\mathbf{c}^3 \qquad 8.16a$$

we may see from (8.16) that the bigger is potential energy of Bivacuum: $\mathbf{V} = \frac{1}{2}(\mathbf{m}_V^+ + \mathbf{m}_{\bar{V}}^-)\mathbf{c}^2$ the bigger is gravitational density and corresponding interaction.

### 8.5  Possible nature of neutrino and antineutrino

Following from our approach to elementary particles formation (chapter 5), the neutrino (antineutrino) of three lepton generation ($i = \mathbf{e}, \boldsymbol{\mu}, \boldsymbol{\tau}$) can be considered, as a stable neutral fermion, formed by pair of asymmetric charged Bivacuum fermion (antifermion) and asymmetric Bivacuum antiboson (boson) of zero spin and opposite energy and charge $(\mathbf{BVB}^\pm)^i \equiv [\mathbf{V}^+ \uparrow\downarrow \mathbf{V}^-]$, compensating that of
$\left(\mathbf{BVF}_-^{S=1/2} \ \equiv [\mathbf{V}^+ \uparrow\uparrow \mathbf{V}^-] \ or \ \mathbf{BVF}_+^{S=-1/2} \equiv [\mathbf{V}^+ \downarrow\downarrow \mathbf{V}^-]\right)^i$:

$$(\mathbf{\nu})^i \sim [\mathbf{BVF}_{\mp}^{\uparrow} \Leftrightarrow \overline{\mathbf{BVB}^{\pm}}]^i \qquad \text{8.16b}$$

$$(\overline{\mathbf{\nu}})^i \sim [\overline{\mathbf{BVF}_{\mp}^{\downarrow}} \Leftrightarrow \mathbf{BVB}^{\pm}]^i \qquad \text{8.16c}$$

These two Bivacuum dipoles are rotating as respect to each other 'side-by-side' principle.

Their relativistic mass/energy and charge symmetry shifts are close to Golden mean conditions:

$$|m_V^+ - m_V^-|_{\mathbf{BVF}_{\mp}^{\uparrow}} c^2 \cong \mathbf{m}_0^i c^2 \qquad \text{8.17}$$

$$|m_V^+ - m_V^-|_{\mathbf{BVB}^{\pm}}^i c^2 \cong \mathbf{m}_0^i c^2 \qquad \text{8.17a}$$

This asymmetric pair is *rotating* around main common axes with Golden Mean angular frequency and tangential velocity squared: $\mathbf{v}^2 = \mathbf{\phi} c^2$, providing corresponding symmetry shift and frequency of $[C \rightleftharpoons W]$ pulsation of each of $[\mathbf{BVF}_{\mp}^{S=\pm 1/2} \Leftrightarrow \mathbf{BVB}^{\pm}]^i$ pairs (eq.5.4a):

$$(\mathbf{\omega}_{\mathbf{v},\overline{\mathbf{v}}}^i)_{rot}^{\phi} = \frac{\mathbf{c}}{L_0} = \mathbf{\omega}_0 = \frac{\mathbf{m}_0^i \mathbf{c}^2}{\hbar} = \mathbf{\omega}_{C \rightleftharpoons W}^i \qquad \text{8.18}$$

The rotating Cooper pairs (neutrinos) propagate in direction parallel to rotation axis, with light velocity or very close to that, like the photons, because of their quasi-ideal symmetry as respect to Bivacuum. The in-phase $[C \rightleftharpoons W]$ beats between the actual and complementary states of these Bivacuum dipoles $[\mathbf{BVF}_{\mp}^{S=\pm 1/2} \Leftrightarrow \mathbf{BVB}^{\pm}]^i$ almost totally compensate each other energy/mass and charge. The latter means that this pair interact with matter as the neutral particle.

The spin/spirality of neutrino is positive and that of antineutrino - negative. The stability of elementary particles is provided in general case by the resonant energy exchange interaction of their sub-elementary particles with basic Bivacuum virtual pressures waves of corresponding generation:

$$[\mathbf{VPW}^+ \bowtie \mathbf{VPW}^-]_{q=1}^i \qquad \text{8.18a}$$

in the process of particles $[C \rightleftharpoons W]$ pulsation. The *internal* Coulomb attraction between opposite charges of $[\mathbf{BVF}_{\mp}^{S=\pm 1/2}$ and $\mathbf{BVB}^{\pm}]^i$ of neutrino also stabilize their structure, like in the case of photons (see section 12.3).

The frequency of beats between asymmetric and symmetric states of pairs $[\mathbf{BVF}_{\mp}^{S=\pm 1/2} \Leftrightarrow \mathbf{BVB}^{\pm}]^i$, equal to neutrino frequency, is determined by slight difference in the energy of sub-elementary fermion $(\mathbf{BVF}_{\mp}^{S=\pm 1/2})^i$ and sub-elementary antiboson $(\mathbf{BVB}^{\pm})^i$ in pairs. This energy difference for each lepton generation is defined by gravitational potential of corresponding electron generation:

$$\mathbf{\omega}_{\mathbf{v},\overline{\mathbf{v}}}^i = \left| \mathbf{\omega}_{C \rightleftharpoons W}^{\mathbf{BVF}_{\mp}^{S=\pm 1/2}} - \mathbf{\omega}_{C \rightleftharpoons W}^{\mathbf{BVB}^{\pm}} \right|^i = \frac{\left| \mathbf{E}_{\mathbf{BVF}_{\mp}^{S=\pm 1/2}}^i - \mathbf{E}_{\mathbf{BVB}^{\pm}}^i \right|}{\hbar} = \mathbf{\beta}^i \frac{(\mathbf{m}_0^i/\phi)\mathbf{c}^2}{\hbar} \qquad \text{8.19}$$

where the gravitational fine structure constant is different for each lepton generation:

$$\mathbf{\beta}^i = \left( \frac{\mathbf{m}_0^i}{\mathbf{M}_{Pl}} \right)^2 \qquad \text{8.19a}$$

where: $(\mathbf{m}_0^i/\phi) = (\mathbf{m}_V^+)_{e,\mu,\tau}^{\phi}$ are the actual mass of the electrons or positrons of three generation at Golden mean conditions, participating in a weak interaction, following by corresponding neutrino and antineutrino emission.

The mass/energy of each of three generation of neutrino can be estimated from (8.19 and 8.19a) as:

$$\mathbf{m}_{\nu,\bar{\nu}}^i = \frac{\hbar\omega_{\nu,\bar{\nu}}^i}{\mathbf{c}^2} = \frac{1}{\phi}\frac{(\mathbf{m}_0^i)^3}{\mathbf{M}_{Pl}^2} = 1.618\frac{(\mathbf{m}_0^i)^3}{\mathbf{M}_{Pl}^2} \qquad 8.19b$$

Corresponding mass evaluations fit the currently existing ones in form of inequalities, i.e. mass of the electron neutrino is less than $1\times10^{-8}$ Ge/c$^2$, mass of muon neutrino is less than 0.0002 Ge/c$^2$ and mass of the tau neutrino - less, than 0.02 Ge/c$^2$. Good description of neutrino properties could be found at: http://en.wikipedia.org/wiki/Neutrino.

It is important to mention, that in accordance of our formula for total energy of relativistic particle (7.4) at $\mathbf{v} \simeq \mathbf{c}$, the relativistic factor $\mathbf{R} = \sqrt{1-(\mathbf{v}/\mathbf{c})^2} \simeq 0$, its total energy is determined by its kinetic energy. For neutrino in general case:

$$\mathbf{E}_{\nu,\bar{\nu}} = \beta\,\mathbf{m}_{\overline{\tau}}^+\mathbf{c}^2 = \beta\big[\mathbf{R}(\mathbf{m}_0\mathbf{c}^2)_{rot}^{in} + (\mathbf{m}_{\overline{\tau}}^+\mathbf{v}_{tr}^2)^{ext}\big] \simeq \beta[(\mathbf{m}_{\overline{\tau}}^+\mathbf{v}_{tr}^2)^{ext}] = \beta\,2T_k \qquad 8.19c$$

The spatially delocalized asymmetry and spin of neutrino and antineutrino compensates the local mass/energy asymmetry and the angular momentum, accompanied the origination of positrons or electrons of three generation in different reactions of weak interaction. This compensating energy and spin asymmetry/shift is assumed to be positive for electrons and negative for positrons of all three generation the triplets for $e^\pm$ and $\mu^\pm$ generations $< [\mathbf{F}_{\uparrow}^- \bowtie \mathbf{F}_{\downarrow}^+] + (\mathbf{F}_{\downarrow}^\pm)>^{e,\mu}$ and monomeric $(\mathbf{F}_{\downarrow}^\pm)>^\tau$ for tauons.

*Neutrino oscillation* between different lepton flavor (electron, muon, or tau) follows from experimental data. For example, the solution of the *solar neutrino problem*, as a major discrepancy between measurements of the neutrinos flowing through the Earth and theoretical models of the solar interior needs the neutrino oscillation. The probability of measuring a particular flavor for a neutrino varies periodically as it propagates. In accordance to our model of neutrino these interconversions can be a result of simultaneous reversible excitation of pair $(\mathbf{v})^i$ ~$[\mathbf{BVF}_{\overline{\tau}}^\uparrow \Leftrightarrow \overline{\mathbf{BVB}^\pm}]^e$ from it ground state with minimum energy of torus and antitorus to their certain excited states, corresponding to muon and tau neutrinos $[\mathbf{BVF}_{\overline{\tau}}^\uparrow \Leftrightarrow \overline{\mathbf{BVB}}^\pm]^{\mu,\tau}$. Consequently, the neutrino oscillation between different generations can be a result of *absorbtion* or *emission* by one type of neutrino the high frequency pair of standing Bivacuum virtual pressure waves (8.18a) of corresponding generation $[\mathbf{VPW}^+ \bowtie \mathbf{VPW}^-]_{q>>1}^{\mu,\tau}$. These neutrino oscillations:

$$(\mathbf{v})^e \rightleftharpoons (\mathbf{v})^\mu \rightleftharpoons (\mathbf{v})^\tau$$

do not violate the energy conservation due to compensation of positive and negative Bivacuum energies (see eq. 1.8 from section 1.2 and the next section).

### 8.6 The background of energy conservation law

The law of energy conservation for elementary particles, as a sum of their kinetic and potential energies in wave and corpuscular phase can be reformulated in terms of our Unified theory. The additivity of different forms of energy means the additivity of Bivacuum dipoles torus and antitorus energy difference (i.e. forms of kinetic energy) and sum of their absolute values (forms of potential energy). These energy conservation quantum roots are illustrated for one sub-elementary particle case by eqs.(6.1 and 6.1a).

The reversible conversion of the localized asymmetry of sub-elementary fermions of elementary particles to spatially delocalized asymmetry in huge number of Bivacuum dipoles around these particles in the process of their $[\mathbf{C} \rightleftharpoons \mathbf{W}]$ pulsation, is a general phenomena. This idea of dynamic equilibrium between *diverging* energy, charge and angular momentum (spin) in the process of $\mathbf{C} \to \mathbf{W}$ transition, responsible for fields origination, and *converging* process of matter formation: $\mathbf{W} \to \mathbf{C}$, can be formulated as follows:

**The total sum of local (corpuscular) and non-local (wave/field) kinetic and potential energies**,

**responsible for Matter and Bivacuum interconversions and interaction is zero**:

$$\left[ \begin{array}{l} \frac{1}{\mathbf{Z}} \sum\limits^{\infty} \mathbf{P}_k \mathbf{c}^2 \Big[ \Delta(\mathbf{m}_V^+ - \mathbf{m}_V^-) + \Delta(\mathbf{m}_V^+ + \mathbf{m}_V^-) \Big]_k^W \\ + \; \frac{1}{\mathbf{Z}} \sum\limits^{\infty} \mathbf{P}_j \Big[ \Delta(\mathbf{m}_V^+ \mathbf{v}^2) + \Delta \mathbf{m}_V^+ \mathbf{c}^2 (2 - \frac{\mathbf{v}^2}{\mathbf{c}^2}) \Big]_j^C \end{array} \right] = \mathbf{0} \qquad 8.20$$

where: $\mathbf{Z} = \sum\limits^{\infty} \mathbf{P}_k + \sum\limits^{\infty} \mathbf{P}_j$ is the total partition function, i.e. sum of probabilities of all possible transitions of energy in the Universe, including interconversions of fields and matter.

In the process of $[\mathbf{C} \rightleftharpoons \mathbf{W}]$ pulsation we have following transitions of kinetic energy:

$$\Delta \mathbf{T}_k = \Delta(\mathbf{m}_V^+ - \mathbf{m}_V^-)\mathbf{c}^2 \; \rightleftharpoons \Delta(\mathbf{m}_V^+ \mathbf{v}^2) \qquad 8.20a$$

and following transitions of potential energy:

$$\Delta \mathbf{V} = \Delta(\mathbf{m}_V^+ + \mathbf{m}_V^-)\mathbf{c}^2 \; \rightleftharpoons \; \Delta \mathbf{m}_V^+ \mathbf{c}^2 (2 - \frac{\mathbf{v}^2}{\mathbf{c}^2})_j \qquad 8.20b$$

$\mathbf{m}_V^+$ and $\mathbf{m}_V^-$ are the actual and complementary mass of torus ($\mathbf{V}^+$) and antitorus ($\mathbf{V}^-$) of each Bivacuum dipoles and elementary particle in the Universe.

Such matter - fields energy interconversions in the Universe, as consequence of proposed in this work duality mechanism, can be considered, as a background for the energy conservation law.

### 8.7 The mechanism of electrostatic and magnetic field origination

It is demonstrated, that the charge symmetry and spin equilibrium shift oscillation in Bivacuum matrix in form of spherical elastic waves, provide the electric and magnetic fields origination. These excitations are the consequence of reversible $[diverging \rightleftharpoons converging]$ of Cumulative Virtual Clouds ($\mathbf{CVC}^\pm$), accompanied the $[Corpuscle \rightleftharpoons Wave]$ pulsation of sub-elementary fermions/antifermions of triplets and their fast rotation. The tendency of asymmetric Bivacuum fermions and antifermions of *opposite* spins and charge shifts to formation of *Cooper pairs* $[\mathbf{BVF}^\uparrow \bowtie \mathbf{BVF}^\downarrow]$ is responsible for Coulomb attraction and the Pauli and electric repulsion between Bivacuum dipoles of *similar* spins and charge shift stands for Coulomb repulsion. Consequently, the electric field formation is a result of *internal shift* of charge equilibrium in each Bivacuum dipole.

The magnetic field and $\mathbf{N}$ or $\mathbf{S}$ poles origination is a result of shift of equilibrium $[\mathbf{BVF}^\uparrow \rightleftharpoons \mathbf{BVB}^\pm \rightleftharpoons \mathbf{BVF}^\downarrow]$ to the left or right, correspondingly, depending on clockwise or anticlockwise rotation of triplets, correlated with direction of their propagation and charge. The magnetic poles attraction or repulsion is also dependent on possibility of *Cooper pairs* of Bivacuum dipoles assembly or disassembly. However, this process is independent on internal symmetry shifts of Bivacuum dipoles, responsible for electric field.

*Let us consider the origination of electrostatic and magnetic field* in more formalized way. The unified *right* parts of eqs. (8.6 - 8.6b) can be subdivided to translational *(electrostatic)* and rotational (*magnetic*) contributions, determined by corresponding degrees of freedom of Cumulative Virtual Cloud ($\mathbf{CVC}^{\pm}_{tr,rot}$):

$$\overline{\mathbf{T}}^{\mathbf{C \rightleftharpoons W}}_{tot} = \mathbf{T}_{tot} \pm \left[ (\mathbf{E}_E)^{Loc}_{[C]} - (\mathbf{E}_E)^{Dist}_{[W]} \right]_{tr} \pm \left[ (\mathbf{E}_H)^{Loc}_{[C]} - (\mathbf{E}_H)^{Dist}_{[W]} \right]_{rot} \qquad 8.21$$

where the most probable total kinetic energy of particle can be expressed via its mass symmetry shift ($\mathbf{m}^+_V - \mathbf{m}^-_V$) or actual inertial mass ($\mathbf{m}^+_V$) and external velocity ($\mathbf{v}$):

$$\mathbf{T}_{tot} = \frac{1}{2}(\mathbf{m}^+_V - \mathbf{m}^-_V)\mathbf{c^2} = \frac{1}{2}\mathbf{m}^+_V\mathbf{v^2} \qquad 8.21a$$

Formula (8.21) reflects the fluctuations of the most probable total kinetic energy, accompanied [$\mathbf{C} \rightleftharpoons \mathbf{W}$] pulsation of unpaired sub-elementary fermion, responsible for linear - electrostatic and curled - magnetic fields origination. In more detailed form the eq. (8.21) can be presented as:

$$\overline{\mathbf{T}}^{\mathbf{C \rightleftharpoons W}}_{tot} = \frac{1}{2}(\mathbf{m}^+_V - \mathbf{m}^-_V)\mathbf{c^2} \pm \left\{ \left[ \frac{|\mathbf{e}_+\mathbf{e}_-|}{\mathbf{L}_T} \right]^{Loc} - \left[ \frac{\mathbf{e^2}}{\hbar\mathbf{c}}(\mathbf{m}^+_V - \mathbf{m}^-_V)\mathbf{c^2} \right]^{Dist}_{tr} \right\} \qquad 8.22$$

$$\pm \left\{ \left[ \mathbf{K_H}\frac{|\boldsymbol{\mu}_+\boldsymbol{\mu}_-|}{\mathbf{L}_T} \right]^{Loc} - \left[ \mathbf{K_H}\frac{\boldsymbol{\mu}^2_0}{\hbar\mathbf{c}}(\mathbf{m}^+_V - \mathbf{m}^-_V)\mathbf{c^2} \right]^{Dist}_{rot} \right\} \qquad 8.23a$$

The **Loc** $\rightleftharpoons$ **Dist** oscillation of electrostatic *translational* contributions, in-phase with [$C \rightleftharpoons W$] pulsation and [*recoil* $\rightleftharpoons$ *antirecoil*] effects energetically compensate each other. Taking into account the obtained relation between mass and charge symmetry shifts (4.8a): $\mathbf{m}^+_V - \mathbf{m}^-_V = \mathbf{m}^+_V\frac{\mathbf{e}^2_+ - \mathbf{e}^2}{\mathbf{e}^2_+}$ they can be described as:

$$\left[ \frac{|\mathbf{e}_+\mathbf{e}_-|}{\mathbf{L}_T} \right]^{Loc} \underset{\mathbf{W \to C}}{\overset{\mathbf{C \to W}}{\rightleftharpoons}} \left[ \boldsymbol{\alpha}\left( \mathbf{m}^+_V\mathbf{c^2}\frac{\mathbf{e}^2_+ - \mathbf{e^2}}{\mathbf{e}^2_+} \right) \right]^{Dist}_{tr} \qquad 8.24$$

where: $\mathbf{L}_T = \hbar/(\mathbf{m}^+_V - \mathbf{m}^-_V)\mathbf{c}$ is a characteristic curvature of kinetic energy (6.5b); $|\mathbf{e}_+\mathbf{e}_-| = \mathbf{e}^2_0$ is a rest charge squared; $\alpha = \mathbf{e}^2/\hbar\mathbf{c}$ is the well known dimensionless electromagnetic fine structure constant.

The right part of (8.24) taking into account that: $\mathbf{e}^2_+ - \mathbf{e}^2_- = (\mathbf{e}_+ - \mathbf{e}_-)(\mathbf{e}_+ + \mathbf{e}_-)$ characterizes the electric dipole moment of triplet, equal to that of unpaired sub-elementary fermion $\left( \mathbf{F}^{\pm}_{\updownarrow} \right)$.

The local *internal* Coulomb interaction between opposite and asymmetric charges of torus and antitorus of unpaired sub-elementary fermions (antifermions) $\left( \mathbf{F}^{\pm}_{\updownarrow} \right)_{S=\pm 1/2}$ turn reversibly to the *external* electric field due to elastic $\left[ \textit{diverging} \rightleftharpoons \textit{converging} \right]$ effects, induced by $\mathbf{C} \rightleftharpoons \mathbf{W}$ pulsation of $\left( \mathbf{F}^{\pm}_{\updownarrow} \right)_{S=\pm 1/2}$.

### 8.8 The factors, responsible for Coulomb interaction between elementary particles

There are three factors, which determines the attraction or repulsion between opposite or similar elementary charges, correspondingly. They are provided by $\left[ \textit{diverging} \rightleftharpoons \textit{converging} \right]$ effects, including the *recoil* $\rightleftharpoons$ *antirecoil* effects, induced by [$\mathbf{C} \rightleftharpoons \mathbf{W}$] pulsation and *emission* $\rightleftharpoons$ *absorption* of positive or negative cumulative virtual clouds CVC$^+$ or CVC$^-$ of the unpaired sub-elementary fermion $\left( \mathbf{F}^{\pm}_{\updownarrow} \right)_{S=\pm 1/2}$ of triplets.

These factors are listed below:

1. The opposite or similar Bivacuum dipoles charge symmetry shifts, providing their

attraction or repulsion, correspondingly;

2. Assembly or disassembly of Bivacuum fermions and antifermions of opposite or similar charge symmetry shifts, correspondingly;

3. The different conditions for standing waves formation by virtual pressure waves of the opposite (± and ∓) or similar (± and ±) by sign energy:

$$\left[\mathbf{VPW}^{\pm} + \mathbf{VPW}^{\mp}\right] \text{- standing waves}$$

$$\left[\mathbf{VPW}^{\pm} + \mathbf{VPW}^{\pm}\right] \text{- no standing waves}$$

These virtual pressure waves are excited by corresponding cumulative virtual clouds - opposite or similar by the energy and angular momentum:

$$\left[\mathbf{CVC}^{\pm} + \mathbf{CVC}^{\mp}\right] \text{or} \left[\mathbf{CVC}^{\pm} + \mathbf{CVC}^{\pm}\right]$$

*The 1st factor* is a basic one. The asymmetry of torus ($V^+$) and antitorus ($V^-$) of Bivacuum dipoles means their ability to beats, accompanied by *emission⇌absorption* of Virtual Clouds ($\mathbf{VC}^{\pm}$) of the opposite or similar energy. The **attraction** between opposite charges is a consequence of exchange interaction between Bivacuum fermions ($BVF^{\pm}$) with opposite by sign $[\mathbf{VC}^+ \bowtie \mathbf{VC}^-]$, following by decreasing of the resulting symmetry shift of Bivacuum. The less is separation between real charged particles, the more is symmetry shift of $\mathbf{BVF}^{\pm}$ in space between them and more effective is the exchange interaction, stimulating the attraction between opposite charges. The attraction decreases with distance between charges (R) as ($\mathbf{r}$/R), where $\mathbf{r}$ is radius vector between charges.

The **repulsion** between similar charges is also due to superposition of $\mathbf{VC}^{\pm}$ of similar sign decreases with distance increasing between charges. Both of these processes are the consequence of energy conservation law, formulated as eq. 8.20, involving tendency of the Bivacuum symmetry increments to zero.

*The 2nd factor - the assembly of Bivacuum dipoles of opposite charge* is a consequence of the 1st one as a result of exchange of $\mathbf{VC}^{\pm}$ between $BVF^+$ and $BVF^-$ of opposite charge and their assembly in virtual Cooper pairs:

$$\left\{\left(\mathbf{BVF}^{\updownarrow}_+ = [\mathbf{V}^+ \uparrow\uparrow \mathbf{V}^-]_{S=+1/2}\right) \bowtie (\mathbf{BVF}^{\updownarrow}_- = [\mathbf{V}^+ \downarrow\downarrow \mathbf{V}^-]_{S=-1/2})\right\} \qquad 8.25$$

induced by the unpaired sub-elementary fermions of triplets of opposite charge. The *flip-flop* spin exchange also is possible in these Cooper pairs.

The linear polymerization of such pairs by *"head to tail"* principle is possible in space between $\left(\mathbf{F}^{\pm}_{\updownarrow}\right)_{S=\pm1/2}$ of triplets of opposite charges, like. electron and positron.

Such virtual microtubules, composed from Cooper pairs $\sum(\mathbf{BVF}^{\updownarrow}_+ \bowtie (\mathbf{BVF}^{\updownarrow}_-)$ are responsible for the 'force lines' origination between the opposite distant charges.

In space between *similar* charge the probability of virtual Cooper pairs (8.25) disassembly increases due to repulsion between similar charges of Bivacuum fermions of the same charge symmetry shift. This effect also decreases with distance (R) between charges as ($\mathbf{r}$/R), where $\mathbf{r}$ is unitary radius vector.

*The 3d factor* is determined by interaction of positive and negative subquantum particles density oscillation, representing virtual pressure waves: $\mathbf{VPW}^+$ and $\mathbf{VPW}^-$ is directly interrelated with 2nd one. Its effect on attraction or repulsion of charges also can be explained in terms of tending of system: $\left[\text{Charges} + \text{Bivacuum}\right]$ to minimum symmetry shift and energy density in space between charges in accordance to energy conservation law in form of eq. 8.20.

### 8.9 The magnetic field origination

*The oscillation of magnetic dipole radiation contribution* in the process of $[\mathbf{C} \leftrightarrows \mathbf{W}]$ pulsations of sub-elementary fermions between local and distant modes do not accompanied by magnetic moments symmetry change, but only by the oscillation of separation between torus and antitorus of $BVF^{\updownarrow}$ : $\mathbf{L}_T = \hbar/(\mathbf{m}_V^+ - \mathbf{m}_{\bar{V}})\mathbf{c}$ and rotational energy of $CVC^{\pm}$ [*emitted $\rightleftarrows$ absorbed*] in the process of $[\mathbf{C} \leftrightarrows \mathbf{W}]$ pulsation.

It can be described as:

$$\left[ \mathbf{K}_H^i \frac{|\mathbf{\mu}_+\mathbf{\mu}_-|}{\mathbf{L}_T} \right]_{[C]}^{Loc} \overset{\mathbf{C}\to\mathbf{W}}{\underset{\mathbf{W}\to\mathbf{C}}{\rightleftarrows}} \left[ \mathbf{K}_H^i \frac{\mathbf{\mu}_0^2}{\hbar\mathbf{c}} (\mathbf{m}_V^+ - \mathbf{m}_{\bar{V}})\mathbf{c}^2 \right]_{[W]}^{Dis} \qquad 8.26$$

$$or : \left[ \mathbf{K}_H^i \frac{|\mathbf{\mu}_+\mathbf{\mu}_-|}{\mathbf{L}_T} \right]_{[C]}^{Loc} \overset{\mathbf{C}\to\mathbf{W}}{\underset{\mathbf{W}\to\mathbf{C}}{\rightleftarrows}} \left[ \mathbf{K}_H^i \frac{\mathbf{\mu}_0^2}{\hbar\mathbf{c}} \mathbf{m}_V^+ \mathbf{\omega}_T^2 \mathbf{L}_T^2 \right]_{[W]}^{Dis} \qquad 8.26a$$

where: $\frac{\mathbf{\mu}_0^2}{\hbar\mathbf{c}} = \gamma$ is the magnetic fine structure constant, introduced in our theory. The magnetic conversion coefficient $\mathbf{K}_H$ we find from the equality of the electrostatic and magnetic energy contributions, determined by recoil$\rightleftarrows$antirecoil effects:

$$\mathbf{E}_E = \mathbf{T}_{rec} = \frac{1}{2} \frac{\mathbf{e}^2}{\hbar\mathbf{c}} (\mathbf{m}_V^+ - \mathbf{m}_{\bar{V}})\mathbf{c}^2 = \frac{1}{2} \mathbf{K}_H^i \frac{\mathbf{\mu}_0^2}{\hbar\mathbf{c}} \mathbf{m}_V^+ \mathbf{\omega}_T^2 \mathbf{L}_T^2 = \mathbf{E}_H \qquad 8.27$$

These equality is a consequence of *equal probability* of energy distribution between translational (electrostatic) and rotational (magnetic) independent degrees of freedom of an unpaired sub-elementary fermion and its cumulative virtual cloud ($CVC^{\pm}$) in conditions of zero-point oscillation. This becomes evident for the limiting case of photon in vacuum. The sum of these two contributions is equal to

$$\mathbf{E}_H + \mathbf{E}_E = \alpha \ \mathbf{m}_V^+ \mathbf{v}_{\mathbf{res}}^2 = \alpha \ \mathbf{m}_V^+ (\mathbf{L}_{ph}\mathbf{\omega}_{ph})^2 \qquad 8.28$$

where $\mathbf{v}_{\mathbf{res}}$ is a resulting *recoil$\rightleftarrows$antirecoil* vibration velocity; $\mathbf{L}_{ph} = \lambda_{ph}/2\pi$ is a radius of photon gyration; $\mathbf{\omega}_{ph}$ is the angle frequency of gyration.

From the above conditions it follows, that:

$$\mathbf{K}_H \frac{\mathbf{\mu}_0^2}{\hbar\mathbf{c}} = \mathbf{K}_H \frac{\hbar\mathbf{e}_0^2}{4\mathbf{m}_0^2\mathbf{c}^3} = \frac{\mathbf{e}_0^2}{\hbar\mathbf{c}} \qquad 8.29$$

where $\mathbf{\mu}_0^2 = |\mathbf{\mu}_+\mathbf{\mu}_-| = \left(\frac{1}{2}\mathbf{e}_0 \frac{\hbar}{\mathbf{m}_0\mathbf{c}}\right)^2$ is the Bohr magneton.

The introduced *magnetic conversion coefficient* can be obtained from 8.29 as:

$$\mathbf{K}_H^{e,p} = \left( \frac{\mathbf{m}_0^{e,p}\mathbf{c}}{\hbar/2} \right)^2 = \left( \frac{2}{\mathbf{L}_0^{e,p}} \right)^2 \qquad 8.30$$

where $\mathbf{L}_0^{e,p} = \hbar/\mathbf{m}_0^{e,p}\mathbf{c}$ is the Compton radius of the electron or proton.

*Origination of magnetic field* can be a result of dynamic equilibrium shift between Bivacuum fermions and Bivacuum antifermions to the left or right, corresponding to the North or South poles:

$$\mathbf{BVF}_{S=+1/2}^{\uparrow} \rightleftarrows \mathbf{BVB}_{S=0}^{\pm} \rightleftarrows \mathbf{BVF}_{S=-1/2}^{\downarrow} \qquad 8.31$$

accompanied by corresponding shift of equilibrium between Bivacuum bosons of opposite polarization:

$$\langle \mathbf{BVB}^+ \equiv [\mathbf{V}^+ \uparrow \downarrow \ \mathbf{V}^-]\rangle_{S=0} \ \rightleftharpoons \ \langle [\mathbf{V}^+ \downarrow \uparrow \ \mathbf{V}^-] \equiv \mathbf{BVB}^-\rangle_{S=0} \qquad 8.32$$

and clockwise or anticlockwise circulation in the plane, normal to direction of charged particles propagation in the current and dependent on this direction and sign of charge.

We assume, that the leftward shift of the equilibrium (8.31) corresponds to North (**N**) magnetic pole formation and the rightward - to South (**S**) pole. The attraction between opposite magnetic poles is determined by tendency of Bivacuum fermions of opposite spins to formation of virtual Cooper pairs (8.25).

In contrast to *linear* Virtual microtubules, formed by Cooper pairs of Bivacuum fermions, responsible for electrostatic interaction, the magnetic field is determined by system of *closed/axial* system of virtual microtubules around the direction of current, formed by Bivacuum fermions: $\mathbf{BVF}^{\uparrow}_{S=+1/2}$ and $\mathbf{BVF}^{\downarrow}_{S=-1/2}$ and difference between positive (**VirP**$^+$) and negative (**VirP**$^-$) virtual pressure because of mass and charge symmetry shifts in these dipoles and difference in their density:

$$\mathbf{\Delta VirP}^{\pm}(R) = \frac{\mathbf{r}}{R}|\mathbf{VirP}^+ - \mathbf{VirP}^-| \sim \frac{\mathbf{r}}{R}\left|\mathbf{n}_+\mathbf{BVF}^{\uparrow}_{S=+1/2} - \mathbf{n}_-\mathbf{BVF}^{\downarrow}_{S=-1/2}\right| \qquad 8.32a$$

$$or : \mathbf{\Delta VirP}^{\pm} \ \sim \frac{\mathbf{r}}{R}|\mathbf{n}_+\mathbf{VC}^+ - \mathbf{n}_-\mathbf{VC}^-| = \frac{\mathbf{r}}{R}\mathbf{n}_+\left|\mathbf{VC}^+ - \frac{\mathbf{n}_-}{\mathbf{n}_+}\mathbf{VC}^-\right|$$

where: **r** is the unitary vector; $R$ is a distance from electric current to certain 'ring' of Bivacuum dipoles; $\mathbf{n}_+$ and $\mathbf{n}_-$ are the densities of $\mathbf{BVF}^{\uparrow}_{S=+1/2}$ and $\mathbf{BVF}^{\downarrow}_{S=-1/2}$; $\mathbf{VC}^+$ and $\mathbf{VC}^-$ are positive and negative virtual clouds, emitted $\rightleftharpoons$ absorbed in the process of transitions between asymmetric and symmetric states of $\mathbf{BVF}^{\uparrow}_{S=+1/2}$ and $\mathbf{BVF}^{\downarrow}_{S=-1/2}$, correspondingly; $\mathbf{K}_{BVF^{\uparrow} \rightleftharpoons BVF^{\downarrow}} = \frac{\mathbf{BVF}^{\downarrow}_{S=-1/2}}{\mathbf{BVF}^{\uparrow}_{S=+1/2}} = \frac{\mathbf{n}_-}{\mathbf{n}_+}$ is the equilibrium constant (see eq. 8.33).

The magnetic field origination is related to asymmetric properties of unpaired sub-elementary fermion $\left(\mathbf{F}^{\pm}_{\updownarrow}\right)_{S=\pm 1/2}$> of moving triplets < $[\mathbf{F}^-_{\uparrow} \bowtie \mathbf{F}^+_{\downarrow}]_{S=0} + \left(\mathbf{F}^{\pm}_{\updownarrow}\right)_{S=\pm 1/2}$> and fast rotation of uncompensated $\mathbf{CVC}^{\pm}$ and *pairs* of charge and magnetic dipoles $[\mathbf{F}^-_{\uparrow} \bowtie \mathbf{F}^+_{\downarrow}]_{S=0}$ in plane, normal to *directed* motion of triplets, i.e. current. This statement is in accordance with empirical fact, that the magnetic field can be exited only by the electric current: $\overrightarrow{\mathbf{j}} = \mathbf{n} \ e\overrightarrow{\mathbf{v}}_j$, i.e. *directed motion* of the charged particles.

The resulting effect of rotation of uncompensated cumulative virtual clouds ($\mathbf{CVC}^{\pm}$) of many of the electrons of current in plane, normal to current direction and axis of $\left(\mathbf{F}^{\pm}_{\updownarrow}\right)_{S=\pm 1/2}$> and paired $\mathbf{CVC}^{\pm}$ rotation is determined by the *hand screw rule* and induce the circular structure formation around $\overrightarrow{\mathbf{j}}$ in Bivacuum. These axisymmetric closed structures are the result of assembly of Bivacuum dipoles of opposite spins in Cooper pairs. If these dipoles have opposite charges, the probability of Cooper pairs formation increases. The rotation velocity of these axial structures, formed by Cooper pairs, representing the force lines of magnetic field is due to symmetry shift between mass and charge of torus and antitorus of $\left[\mathbf{BVF}^{\uparrow}_+ \bowtie \mathbf{BVF}^{\downarrow}_-\right]$ in accordance with (4.2 and 4.2a). This asymmetry of dipoles is dependent on the distance (R) from current as $(\overrightarrow{r}/R)$.

The unpaired sub-elementary fermions and antifermions of the opposite charges in elementary particles have the opposite influence on symmetry shift between torus and antitorus, interrelated with their opposite influence on the direction of the $[\mathbf{BVF}^{\uparrow}_{S=+1/2} \ \rightleftarrows \ \mathbf{BVB}^{\pm}_{S=0} \ \rightleftarrows \ \mathbf{BVF}^{\downarrow}_{S=-1/2}]$ equilibrium shift.

The equilibrium constant between Bivacuum fermions of opposite spins, characterizing their uncompensated magnetic moment, we introduce, using (3.11), as function of the external translational velocity of $\mathbf{BVF}^{\updownarrow}$:

$$\mathbf{K}_{BVF\uparrow \rightleftharpoons BVF\downarrow} = \frac{\mathbf{BVF}^{\uparrow}_{S=-1/2}}{\mathbf{BVF}^{\downarrow}_{S=+1/2}} = \frac{\mathbf{n}_-}{\mathbf{n}_+} = \exp\left[-\frac{\alpha(\mathbf{m}_V^+ - \mathbf{m}_V^-)}{\mathbf{m}_V^+}\right] =$$

$$= \exp\left[-\alpha\frac{\mathbf{v}^2}{\mathbf{c}^2}\right] = \exp\left[-\frac{\omega_T^2\mathbf{L}_T^2}{\mathbf{c}^2}\right] \qquad 8.33$$

The Bivacuum dipoles with equilibrium constants $\mathbf{K}_{BVF\uparrow \rightleftharpoons BVF\downarrow}$ of the same values, have the axial distribution with respect to the current vector ($\mathbf{j}$) of charges. The conversion of Bivacuum fermions or Bivacuum antifermions to Bivacuum bosons ($\mathbf{BVB}^{\pm} = \mathbf{V}^+ \Updownarrow \mathbf{V}^-$) with different probabilities ($\mathbf{P}^{\uparrow}$ and $\mathbf{P}^{\downarrow}$):

$$\mathbf{BVF}^{\uparrow}_{S=+1/2} \xrightarrow{\mathbf{P}^{\uparrow}} \langle \mathbf{BVB}^+ \equiv [\mathbf{V}^+\uparrow\downarrow \mathbf{V}^-] \rangle$$

$$\mathbf{BVF}^{\downarrow}_{S=-1/2} \xrightarrow{\mathbf{P}^{\downarrow}} \langle \mathbf{BVB}^- \equiv [\mathbf{V}^+\downarrow\uparrow \mathbf{V}^-] \rangle$$

may provide an increasing or decreasing of the equilibrium constant $\mathbf{K}_{BVF\uparrow \rightleftharpoons BVF\downarrow}$. The corresponding sign of probability difference: $\Delta\mathbf{P} = \mathbf{P}^{\uparrow} - \mathbf{P}^{\downarrow}$ is dependent on the direction of current, related in-turn with direction of paired sub-elementary fermions $[\mathbf{F}_{\uparrow}^- \bowtie \mathbf{F}_{\downarrow}^+]_{S=0}$ and uncompensated $\mathbf{CVC}^{\pm}$ circulation of unpaired sub-elementary fermion of triplet.

*The magnetic field tension* can be presented as a gradient of the constant of equilibrium:

$$\mathbf{H} = \mathbf{grad}(\mathbf{K}_{BVF\uparrow \rightleftharpoons BVF\downarrow}) = (\vec{r}/R)\mathbf{K}_{BVF\uparrow \rightleftharpoons BVF\downarrow} \qquad 8.34$$

The chaotic thermal velocity of the 'free' conductivity electrons in metals and ions at room temperature is very high even in the absence of current, and follows Maxwell-Boltzmann distribution:

$$\mathbf{v}_T = \sqrt{\frac{\mathbf{kT}}{\mathbf{m}_V^+}} \sim 10^7 \ cm/s \qquad 8.35$$

It proves, that not the acceleration, but the ordering of the electrons translational and rotational dynamics in space, provided by current, is a main reason of the curled magnetic field excitation. In contrast to conventional view, the electric current itself is not a *primary*, but only a *secondary* reason of magnetic field origination, as the charges translational and rotational dynamics ordering or 'vectorization factor'.

## 8.10 Interpretation of the Maxwell displacement current, based on Bivacuum model

The magnetic field origination in Bivacuum can be analyzed also from more conventional point of view.

Let us analyze the 1st Maxwell equation, interrelating the circulation of vector of magnetic field tension $\mathbf{H}$ along the closed contour $\mathbf{L}$ with the conduction current ($\mathbf{j}$) and *displacement current* $\mathbf{j}_d = \frac{1}{4\pi}\frac{\partial \mathbf{E}_{BVF}}{\partial t}$ through the surface, limited by $\mathbf{L}$ :

$$\oint_{\mathbf{L}} \mathbf{H}\,dl = \frac{4\pi}{c}\int_{\mathbf{S}}\left(\mathbf{j} + \frac{1}{4\pi}\frac{\partial \mathbf{E}_{BVF}}{\partial t}\right)d\mathbf{s} \qquad 8.36$$

where ($\mathbf{s}$) is the element of surface, limited with contour ($l$).

The existence of the displacement current: $\mathbf{j}_d = \frac{1}{4\pi}\frac{\partial \mathbf{E}}{\partial t}$ is in accordance with our model of Bivacuum the result of oscillating virtual dipoles ($\mathbf{BVF}^{\ddagger}$ and $\mathbf{BVB}^{\pm}$) continuum.

In condition of *primordial* Bivacuum of the ideal virtual dipoles symmetry (i.e. in the absence of matter and fields) the charges of torus and antitorus totally compensate each other. However, even in primordial symmetric Bivacuum the oscillations of distance between torus and antitorus of Bivacuum dipoles, following energy gap oscillation, is responsible for *displacement current*. This alternating current generates corresponding *displacement magnetic field*:

$$H_d = \frac{4\pi}{c} \int_{\mathbf{S}} \frac{1}{4\pi} \frac{\partial \mathbf{E}_{BVF}}{\partial t} d\mathbf{s} \qquad 8.36a$$

Corresponding virtual dipole oscillations are the consequence of the in-phase transitions of $\mathbf{V}^+$ and $\mathbf{V}^-$ between the excited and ground states, compensating each other. These transitions are accompanied by spontaneous emission and absorption of positive and negative virtual pressure waves: $\mathbf{VPW}^+$ and $\mathbf{VPW}^-$. The excitation of such transitions and $\mathbf{VPW}^{\pm}_{q=1,2,3}$ for example by pulsing electric field, like one, accompanied discharge in condensers, should influence on gravitational effects (see paragraph 8.3) and interaction of Bivacuum with pulsing elementary particles.

The displacement current and corresponding displacement magnetic field can be enhanced as result of feedback reaction by presence of pulsing particles and their thermal fluctuations.

### 8.11 New kind of current in secondary Bivacuum, additional to displacement one. Velocity of zero-point oscillation, providing the Coulomb and gravitational interactions. Physical sense of electric charge

This additional current is a consequence of vibrations of $\mathbf{BVF}^{\updownarrow}$, induced by recoil-antirecoil effects, accompanied $[\mathbf{C} \leftrightharpoons \mathbf{W}]$ transitions of unpaired sub-elementary fermion of triplets $< [\mathbf{F}^-_{\uparrow} \bowtie \mathbf{F}^+_{\uparrow}]_{S=0} + (\mathbf{F}^+_{\uparrow})_{S=\pm 1/2} >^{e,p}$ It can be also a consequence of Bivacuum dipoles perturbations, induced by relativistic translational propagation of particles in Bivacuum.

The corresponding elastic deformations of Bivacuum fermions ($\mathbf{BVF}^{\updownarrow}$) $\equiv [\mathbf{V}^+ \Updownarrow \mathbf{V}^-]$ are followed by small charge-dipole symmetry zero-point oscillations ($\mathbf{v}^{ext} = 0$) with amplitude, determined by the most probable resulting translational - rotational recoil velocity ($\mathbf{v}_{rec}$). At conditions $\mathbf{e}_+ \simeq \mathbf{e}_- \simeq \mathbf{e}_0$ and $|\mathbf{e}_+ - \mathbf{e}_-| << \mathbf{e}_0$, i.e. at small perturbations of torus and antitorus: $\mathbf{V}^+$ and $\mathbf{V}^-$ we have for the charge symmetry shift oscillation amplitude:

$$\Delta \mathbf{e}_{\pm} = \mathbf{e}_+ - \mathbf{e}_- = \frac{1}{2} \mathbf{e}_0 \frac{\mathbf{v}_{rec}^2}{\mathbf{c}^2} \qquad 8.37$$

The resulting most probable recoil kinetic energy and velocity, standing for electromagnetism (8.27), can be defined as:

$$\mathbf{T}_{rec} = \frac{1}{2} \mathbf{E}_{el} = \frac{1}{2} \alpha (\mathbf{m}_{\mathbf{V}}^+ - \mathbf{m}_{\mathbf{V}}^-) \mathbf{c}^2 = \frac{1}{2} \alpha \, \mathbf{m}_{\mathbf{V}}^+ \mathbf{v}_{\mathbf{res}}^2 \qquad 8.38$$

$$\mathbf{v}_{rec}^2 = \alpha \, \mathbf{v}_{\mathbf{res}}^2 \qquad 8.38a$$

Using interrelation between the mass and charge symmetry shifts (4.8a), formula (8.38) for recoil kinetic energy can be presented as:

$$\mathbf{T}_{rec} = \frac{1}{2} \alpha \, \mathbf{m}_{\mathbf{V}}^+ \mathbf{v}_{\mathbf{res}}^2 = \frac{1}{2} \alpha \, \mathbf{m}_{\mathbf{V}}^+ \mathbf{c}^2 \frac{\mathbf{e}_+^2 - \mathbf{e}_-^2}{\mathbf{e}_+^2} \qquad 8.38b$$

In presence of matter and fields, when primordial Bivacuum turns to secondary one, composed from Bivacuum dipoles of small asymmetry: $\mathbf{e}_+ \simeq \mathbf{e}_- \simeq \mathbf{e}_0$, we may assume,

that:

$$\mathbf{e}_+^2 - \mathbf{e}_-^2 = (\mathbf{e}_+ + \mathbf{e}_-)(\mathbf{e}_+ - \mathbf{e}_-) \simeq 2\mathbf{e}_0(\mathbf{e}_+ - \mathbf{e}_-)$$

and right part of (8.38b) turns to formula, interrelating external kinetic energy of asymmetric Bivacuum dipoles with their charge symmetry shift:

$$\mathbf{T}_{rec} = \frac{1}{2}\alpha\,\mathbf{m}_\mathbf{V}^+\mathbf{v}_{\mathbf{res}}^2 = \alpha\,\mathbf{m}_\mathbf{V}^+\mathbf{c}^2\,\frac{\mathbf{e}_+ - \mathbf{e}_-}{\mathbf{e}_0} \qquad 8.38c$$

As far formula (8.24) can be applied not only for sub-elementary fermions, but also for asymmetric Bivacuum fermions, our formula (8.38c) reflects electromagnetic oscillation of Bivacuum dipoles, generated by their kinetic energy oscillation. It will be shown in chapter 20, that thrust, accompanied the condenser electric discharge in Biefeld -Brown and Podkletnov - Modanese effect is a result of force and excessive momentum origination due to collective coherent Bivacuum dipoles polarization/asymmetry jump.

The minimum value of recoil velocity, corresponding to zero *external* translational velocity of triplets, like electrons, positrons and protons, can be evaluated from internal velocity of sub-elementary fermions, determined by Golden mean conditions $(\mathbf{v}_{\mathbf{res}}/\mathbf{c})^2 = \phi = 0.61803398$ (see chapter 4), can be considered as a *velocity of zero-point oscillations of elementary particles*:

$$(\mathbf{v}_{rec}^2)^{\min} \equiv (\mathbf{v}_0^2)_{HE}^{\min} = \alpha\phi\,\mathbf{c}^2 \qquad 8.39$$

$$or: \frac{(\mathbf{v}_{rec}^2)^{\min}}{\mathbf{c}^2} = \alpha\phi \qquad 8.39a$$

where: $\alpha = e^2/\hbar c = 0,0072973506$; $\alpha\phi = (\mathbf{v}_{rec}^2)^{\min}/\mathbf{c}^2 = 4.51 \cdot 10^{-3}$.
*The physical sense of the electric charge follows from 8.38 in form:*

$$\mathbf{E}_{el} = \frac{1}{2}\frac{\mathbf{e}^2}{\hbar c}(\mathbf{m}_\mathbf{V}^+ - \mathbf{m}_\mathbf{V}^-)\mathbf{c}^2 = \frac{1}{2}\frac{\mathbf{e}^2}{\hbar c}\mathbf{m}_\mathbf{V}^+\mathbf{v}^2 \qquad 8.39b$$

$$\mathbf{v}_{rec}^2 \equiv (\mathbf{v}_0^2)_{HE} = \frac{1}{\hbar c}\left(\frac{\mathbf{e}}{\mathbf{Q}}\mathbf{v}\right)\left(\frac{\mathbf{e}}{\mathbf{Q}}\mathbf{v}\right) \qquad 8.39c$$

The product $\hbar c \equiv \mathbf{Q}^2$ is the total elementary charge squared and the ratio: $\mathbf{e}_\pm/Q$ is the relative charge of sub-elementary fermions. This means that the relative electric charge can be considered as the *recoil factor,* which interrelate the external group velocity of particle (**v**) and the velocity of its *recoil⇌antirecoil* vibrations of elementary charge ($\mathbf{v}_{rec}$), its mass/energy symmetry shift: $\pm(\mathbf{m}_\mathbf{V}^+ - \mathbf{m}_\mathbf{V}^-)\mathbf{c}^2$ and the energy of electric field, representing the Bivacuum matrix perturbation, generated by this charge vibrations.

The alternating *recoil current ($j_{rec}^{EH}$)*, additional to that of Maxwell *displacement current ($j_d$,)* existing in presence of charged particles even in the absence of conducting current (**j = 0**) is equal to product of (8.37) and square root of (8.39). At Golden mean conditions $(\mathbf{v}/\mathbf{c})^2 = \phi$ this new *recoil current,* following from our approach, is:

$$(\mathbf{j}_{rec}^\phi)^{EH} = (\Delta\mathbf{e}_\pm)^\phi(\mathbf{v}_{rec})^{\min} = \frac{1}{2}\alpha^{1/2}\phi^{3/2}\,\mathbf{e}_0\mathbf{c} \qquad 8.40$$

Corresponding gravitational contribution of recoil velocity, related to the increment of the elastic recoil vibration of potential energy of particle (8.10) is much smaller, as far $\beta << \alpha$:

$$\mathbf{V}_{rec} = \tfrac{1}{2}\beta(\mathbf{m_V^+} + \mathbf{m_{\bar V}})\mathbf{c}^2 = \tfrac{1}{2}\beta\,\mathbf{m_{\bar V}^+}\mathbf{c}^2(2 - \mathbf{v}^2/\mathbf{c}^2) \qquad 8.41$$

The zero-point recoil/antirecoil velocity squared, providing the potential energy of particle recoil/antirecoil oscillation at GM conditions $(\mathbf{v}^2/\mathbf{c}^2)^\phi = \mathbf{0.618} = \phi$ is:

$$(\mathbf{v}_0^2)_G = \beta\mathbf{c}^2(2 - \phi); \quad (\mathbf{v}_0^2)_G/\mathbf{c}^2 = \beta(2 - \phi) \qquad 8.42$$
$$(\mathbf{v}_0)_G = \mathbf{c}\beta^{1/2}(2 - \phi)^{1/2} = 1,446 \cdot 10^{-12}\ cm/s$$

Consequently, the Maxwell equation (8.36) can be modified, taking into account the EH recoil current, as

$$\oint_{\mathbf{L}} \mathbf{H}\,dl = \frac{4\pi}{c}\int_{\mathbf{S}}\left(\mathbf{j} + \frac{1}{4\pi}\frac{\partial\mathbf{E}}{\partial t} + \mathbf{j}_{rec}^{EH}\right)d\mathbf{s} = \mathbf{I}_{tot} \qquad 8.43$$

where: $\mathbf{I}_{tot}$ is the total current throw the surface ($\mathbf{S}$).

We have to note, that $\mathbf{j}_{rec}^{EH}$ is nonzero not only in the vicinity of particles, but as well in any remote space regions of Bivacuum, perturbed by electric and magnetic potentials. This consequence of our theory coincides with the extended electromagnetic theory of Bo Lehnert (2004, 2004a), also considering current in vacuum, additional to displacement one.

In accordance with the known Helmholtz theorem, each kind of vector field ($\mathbf{F}$), tending to zero at infinity, can be presented, as a sum of the gradient of some scalar potential ($\phi$) and a rotor of vector potential ($\mathbf{A}$):

$$\mathbf{F} = \mathbf{grad}\,\varphi + \mathbf{rot}\,\mathbf{A} \qquad 8.43a$$

The scalar and vector potentials are convenient to use for description of electromagnetic field, i.e. photon properties. They are characterized by the interrelated translational and rotational degrees of freedom, indeed.

To explain the *ability of secondary Bivacuum to keep the average (macroscopic) mass and charge equal to zero,* we have to postulate, that the mass and charge symmetry shifts oscillations of Bivacuum fermions and antifermions, forming virtual Cooper pairs:

$$(\mathbf{BVF}^\uparrow)_{S=+1/2}^\pm \equiv [\mathbf{V}^+ \uparrow\uparrow \mathbf{V}^-]^\pm \;\bowtie\; [\mathbf{V}^+ \downarrow\downarrow \mathbf{V}^-]^\mp \;\equiv\; (\mathbf{BVF}^\downarrow)_{S=-1/2}^\mp \qquad 8.44$$

are opposite by sign, but equal by the absolute value. Consequently, the polarized secondary Bivacuum (i.e. perturbed by matter and field) can be considered, as a *plasma of the in-phase oscillating virtual dipoles (BVF)* of opposite resulting charge and mass/energy.

### 8.12 The mechanisms, increasing the refraction index of Bivacuum

By definition, the *torus* is a figure, formed by rotation of a circle with maximum radius, corresponding to minimum quantum number ($\mathbf{n} = \mathbf{0}$, see 1.1a) $\mathbf{L}_{\mathbf{V}^\pm}^i = \frac{2\hbar}{\mathbf{m}_0^i\mathbf{c}}$, around the axis, shifted from the center of the circle at the distance $\pm\mathbf{\Delta L}_{EH,G}$. The electromagnetic (*EH*) and gravitational (*G*) vibrations of positions $(\pm\mathbf{\Delta L}_{EH,G})_{V^\pm}$ of the big number of recoiled $\mathbf{BVF}_{rec}$, induced by the elastic *recoil$\leftrightharpoons$antirecoil* deformations of Bivacuum matrix, are accompanied by vibrations of square and volume of torus ($\mathbf{V}^+$) and antitorus ($\mathbf{V}^-$) of perturbed Bivacuum dipoles: $(\mathbf{BVF}_{rec}^\updownarrow)^i = [\mathbf{V}^+ \updownarrow \mathbf{V}^-]_{rec}^i$. The electromagnetic and gravitational increments of square ($\mathbf{\Delta S}_{\mathbf{V}^\pm}^{E,G}$) and volume ($\mathbf{\Delta V}_{\mathbf{V}^\pm}^{E,G}$) of toruses and antitoruses of $(\mathbf{BVF}_{rec}^\updownarrow)^i$, as a consequence of their center vibrations can be presented, correspondingly, as:

$$\Delta \mathbf{S}_{\mathbf{V}^{\pm}}^{EH,G} = 4\pi^2 |\Delta \mathbf{L}_{EH,G}|_{V^{\pm}}^{EH,G} \cdot \mathbf{L}_{\mathbf{V}^{\pm}} \qquad 8.45$$

$$\Delta \mathbf{V}_{\mathbf{V}^{\pm}}^{EH,G} = 4\pi^2 |\Delta \mathbf{L}_{EH,G}|_{V^{\pm}}^{EH,G} \cdot \mathbf{L}_{\mathbf{V}^{\pm}}^2 \qquad 8.45a$$

At conditions of zero-point oscillations, corresponding to Golden Mean (GM), when the ratio $(\mathbf{v}_0/\mathbf{c})^2 = \boldsymbol{\phi}$ and external translational velocity ($\mathbf{v}$) is zero, the maximum shifts of center of secondary Bivacuum dipoles *in vicinity of pulsing elementary particles* due to electromagnetic and gravitational recoil-antirecoil (zero-point) vibrations are, correspondingly:

$$(\Delta \mathbf{L}_{\mathbf{EH}}^{\mathbf{i}})_{V^{\pm}}^{\phi} = \left( \boldsymbol{\tau}_{C \rightleftharpoons W}^{\phi} \, \mathbf{v}_{EH}^{\phi} \right)^i = \frac{\hbar}{\mathbf{m}_0^i \mathbf{c}} (\alpha \phi)^{1/2} = 0,067 \, (\mathbf{L}_{\mathbf{V}^{\pm}}^i) \qquad 8.46$$

$$(\Delta \mathbf{L}_{\mathbf{G}}^{\mathbf{i}})_{V^{\pm}}^{\phi} = \left( \boldsymbol{\tau}_{C \rightleftharpoons W}^{\phi} \, \mathbf{v}_{G}^{\phi} \right)^i = \frac{\hbar}{\mathbf{m}_0^i \mathbf{c}} \beta^{1/2} (2 - \phi)^{1/2} = 3,27 \cdot 10^{-23} \, (\mathbf{L}_{\mathbf{V}^{\pm}}^i) \qquad 8.46a$$

where: the recoil $\rightleftharpoons$ antirecoil oscillation period is $\left[ \boldsymbol{\tau}_{C \rightleftharpoons W}^{\phi} = 1/\boldsymbol{\omega}_{C \rightleftharpoons W}^{\phi} = \hbar/\mathbf{m}_0^i \mathbf{c}^2 \right]^i$; the recoil$\rightleftharpoons$antirecoil most probable velocity of zero-point oscillations, which determines the electrostatic and magnetic fields is: $\mathbf{v}_{EH}^{\phi} = \mathbf{c}(\alpha \phi)^{1/2} = 0.201330447 \times 10^8 \, \mathrm{m \, s^{-1}}$ and $(\alpha \phi)^{1/2} = 0,067$ the corresponding zero-point velocity, which determines gravitational field is: $\mathbf{v}_G^{\phi} = \mathbf{c} \beta_e^{1/2} (2 - \phi)^{1/2} = 1,446 \cdot 10^{-12} \, \mathrm{m \, s^{-1}}$ and $\beta_e^{1/2}(2 - \phi)^{1/2} = 0,48 \cdot 10^{-22}$.

The dielectric permittivity of Bivacuum and corresponding refraction index, using our theory of refraction index of matter (Kaivarainen, 1995; 2001), can be presented as a ratio of volume of Bivacuum fermions and bosons in symmetric *primordial* Bivacuum ($\mathbf{V_{pr}}$) to their volume in *secondary* Bivacuum: $\mathbf{V_{sec}} = \mathbf{V}_{BVF} - (\mathbf{r}/r)\Delta \mathbf{V}_{\mathbf{BVF}_{rec}}^{E,G}$, perturbed by matter and fields. The secondary Bivacuum is optically more dense, if we assume that the volume, occupied by Bivacuum fermion torus and antitorus, is excluded for photons. The Coulomb and gravitational potentials and the related excluded volumes of perturbed Bivacuum fermions/antifermions decline with distance ($r$) as:

$$(\vec{\mathbf{r}}/r)\Delta \mathbf{V}_{\mathbf{BVF}_{rec}}^{EH} \quad \text{and} \quad (\vec{\mathbf{r}}/r)\Delta \mathbf{V}_{\mathbf{BVF}_{rec}}^{G}$$

where: ($r$) is a distance from the charged and/or gravitating particle and $\vec{\mathbf{r}}$ is the unitary radius vector. Taking all this into account, we get for permittivity of secondary Bivacuum:

$$\boldsymbol{\varepsilon} = \mathbf{n}^2 = \left( \frac{\mathbf{c}}{\mathbf{v}_{EH,G}} \right)^2 = \frac{N \mathbf{V_{pr}}}{N \mathbf{V_{sec}}} =$$

$$= \frac{\mathbf{V}_{BVF}}{\mathbf{V}_{BVF} - (\mathbf{r}/r)\Delta \mathbf{V}_{\mathbf{BVF}_{rec}}^{EH,G}} = \frac{1}{(1 - \mathbf{r}/r)\Delta \mathbf{V}_{\mathbf{BVF}_{rec}}^{EH,G}/\mathbf{V}_{BVF})} \qquad 8.47$$

$$\mathbf{n}^2 = \frac{1}{1 - (\mathbf{r}/r) \, 3\pi |\Delta \mathbf{L}|_{V^{\pm}}^{EH,G} \cdot \mathbf{L}_{\mathbf{V}^{\pm}}} \qquad 8.47a$$

where: the velocity of light propagation in asymmetric secondary Bivacuum of higher virtual density, than in primordial one, is notated as: $\mathbf{v}_{EH,G} = \mathbf{c}_{EH,G}$; the volume of primordial Bivacuum fermion is $\mathbf{V}_{BVF} = (4/3)\pi \mathbf{L}_{V^{\pm}}^3$ and its increment in secondary Bivacuum: $\Delta \mathbf{V}_{\mathbf{BVF}_{rec}}^{E,G} = \Delta \mathbf{V}_{\mathbf{V}^{\pm}}^{E,G}$ (8.45a).

($\mathbf{r}/r$) is a ratio of unitary radius-vector to distance between the source of $[\mathbf{C} \rightleftharpoons \mathbf{W}]$ pulsations (elementary particle) and perturbed by the electrostatic, magnetic and gravitational potential $\mathbf{BVF}_{rec}^{EH,G}$.

Putting (8.46) into formula (8.46a) we get for the refraction index of Bivacuum and relativistic factor ($\mathbf{R}_E$) in the vicinity of charged elementary particle (electron, positron or proton, antiproton) the following expression:

$$\left[ \varepsilon = \mathbf{n}^2 = \left( \frac{\mathbf{c}}{\mathbf{c}_{EH}} \right)^2 \right]_E = \frac{1}{1 - (\mathbf{r}/r)\, 3\pi(\alpha\phi)^{1/2}} \lesssim 2.71 \qquad 8.48$$

where:  $1 \lesssim \mathbf{n}^2 \lesssim 2,71$ is tending to 1 at $r \to \infty$.
The Coulomb relativistic factor:

$$\mathbf{R}_{EH} = \sqrt{1 - \frac{(\mathbf{c}_{EH})^2}{\mathbf{c}^2}} = \sqrt{(\mathbf{r}/r)\, 0,631} \lesssim (\mathbf{r}/r)^{1/2}\, 0.794 \qquad 8.49$$

$0 \lesssim \mathbf{R}_E \lesssim 0,794$ is tending to zero at $r \to \infty$.

In similar way, using (8.46a) and (8.47a), for the refraction index of Bivacuum and the corresponding relativistic factor ($\mathbf{R}_G$) of gravitational vibrations of Bivacuum fermions ($\mathbf{BVF}^{\updownarrow}$) in the vicinity of pulsing elementary particles at zero-point conditions, we get:

$$\left[ \varepsilon = \mathbf{n}^2 = \left( \frac{\mathbf{c}_G}{\mathbf{c}} \right)^2 \right]_G = \frac{1}{1 - (\mathbf{r}/r)3\pi(\beta^e)^{1/2}(2-\phi)^{1/2}} \gtrsim 1 \qquad 8.50$$

where $(\beta^e)^{1/2}(2 - \phi)^{1/2} = 0.48 \times 10^{-22}$.
The gravitational relativistic factor:

$$\mathbf{R}_G = \sqrt{1 - \left( \frac{\mathbf{c}_G}{\mathbf{c}} \right)^2} = \sqrt{(\mathbf{r}/r)\, 0,48 \cdot 10^{-22}} \lesssim (\mathbf{r}/r)^{1/2}\, 0,69 \cdot 10^{-11} \qquad 8.51$$

Like in previous case, the Bivacuum refraction index, increased by gravitational potential, is tending to its minimum value: $\mathbf{n}^2 \to 1$ at the increasing distance from the source: $r \to \infty$.

The charge - induced refraction index increasing of secondary Bivacuum, in contrast to the mass - induced one, is independent of lepton generations of Bivacuum dipoles ($e, \mu, \tau$).

The formulas (8.48) and (8.50) for Bivacuum dielectric permittivity and refraction index near elementary particles, perturbed by their Coulomb and gravitational potentials, point out that bending and scattering probability of photons on charged particles is much higher, than that on neutral particles with similar mass.

We have to point out, that the *light velocity* in conditions:
$[\mathbf{n}^2_{EH,G} = \mathbf{c}/\mathbf{v}_{EH,G} = \mathbf{c}/\mathbf{c}_{EH,G}] > 1$ is not longer a scalar, but a vector, determined by the gradient of Bivacuum fermion symmetry shift:

$$grad\, \Delta|\mathbf{m}^+_V - \mathbf{m}^-_V|_{EH,G}\, \mathbf{c}^2 = grad\, \Delta(\mathbf{m}^+_V \mathbf{v}^2) \qquad 8.52$$

and corresponding gradient of torus and antitorus equilibrium constant increment:
$\Delta \mathbf{K}_{\mathbf{V}^+\updownarrow\mathbf{V}^-} = \mathbf{1} - \mathbf{m}^-_V / \mathbf{m}^+_V = (\mathbf{c}_{EH,G}/\mathbf{v})^2$:

$$grad[\Delta \mathbf{K}_{\mathbf{V}^+\updownarrow\mathbf{V}^-} = \mathbf{1} - \mathbf{m}^-_V / \mathbf{m}^+_V] = \qquad 8.53$$

$$= grad\left( \frac{\mathbf{c}_{EH,G}}{\mathbf{c}} \right)^2 = grad\, \frac{1}{\mathbf{n}^2} \qquad 8.53a$$

The other important consequence of: $[\mathbf{n}^2]_{E,G} > 1$ is that *the contributions of the rest mass energy of photons and neutrino (Kaivarainen, 2005) to their total energy is not zero*, as far the electromagnetic and gravitational relativistic factors ($\mathbf{R}_{EH,G}$) are greater than zero. It follows from the basic formula for the total energy of de Broglie wave (the photon in our

case):

$$E_{tot} = m_V^+ c^2 = \hbar \omega_{C \leftrightharpoons W} = R(m_0 c^2)_{rot}^{in} + (\hbar \omega_B^{ext})_{tr} \qquad 8.54$$

where the gravitational relativistic factor of electrically neutral objects:
$R_G = \sqrt{(r/r)3{,}08 \cdot 10^{-22}} \lesssim (r/r)^{1/2} \, 1.75 \times 10^{-11}$.

This consequence is also consistent with a theory of the photon and neutrino, developed by Bo Lehnert (2004a).

We can see, that in conditions of *primordial* Bivacuum, when $r \to \infty$, the $n_{EH,G} \to 1$, $R_{EH,G} \to 0$ and the contribution of the rest mass energy $R(m_0 c^2)_{rot}^{in}$ tends to zero. At these limiting conditions the frequency of photon [*Corpuscle* $\leftrightharpoons$ *Wave*] pulsation is equal to the frequency of the photon as a wave:

$$E_{ph} = \hbar \omega_{C \leftrightharpoons W} = \hbar \omega_{ph} = h \frac{c}{\lambda_{ph}} \qquad 8.55$$

The results of our analysis explain the bending of light beams, under the influence of strong gravitational potential in another way, than by Einstein's general theory of relativity. A similar idea of polarizable vacuum and it permittivity variations has been developed by Dicke (1957), Fock (1964) and Puthoff (2001), as a background of 'vacuum engineering'.

For the spherically symmetric star or planet it was shown using Dicke model (Dicke, 1957), that the dielectric constant $K$ of polarizable vacuum is given by the exponential form:

$$K = \exp(2GM/rc^2) \qquad 8.56$$

where $G$ is the gravitational constant, $M$ is the mass, and $r$ is the distance from the mass center.

For comparison with expressions derived by conventional General Relativity techniques, it is sufficient a following approximation of the formula above (Puthoff, 2001):

$$K \approx 1 + \frac{2GM}{rc^2} + \frac{1}{2}\left(\frac{2GM}{rc^2}\right)^2 \qquad 8.57$$

Our approach propose the concrete mechanism of Bivacuum optical density increasing near charged and gravitating particles, inducing light beams bending.

The increasing of the excluded for photons volume of toruses and antitoruses due to their rotations and vibrations, enhance the refraction index of Bivacuum and decrease the light velocity near gravitating and charged objects. The nonzero contribution of the rest mass energy to photons and neutrino energy is a consequence of the enhanced refraction index of secondary Bivacuum and corresponding decreasing of the effective light velocity. The latter can be revealed by small shift of Doppler effect in EM radiation of the probe in gravitational field. The 'Pioneer anomaly' (Turushev et al., 2005) is a good example of such phenomena.

### 8.13 Application of angular momentum conservation law for evaluation of curvatures of electric and gravitational potentials

From the formulas of total energy of $[W]$ phase of unpaired sub-elementary fermion (8.17) of triplet $< [F_{\uparrow}^- \bowtie F_{\downarrow}^+]_{S=0} + (F_{\updownarrow}^\pm)_{S=\pm 1/2} >^{e,\tau}$ we can find out the relation between the sum of internal and external angular momentum of $CVC$, including the electric and gravitational increments of $CVC$ of $[W]$ phase for the one side, and a sum of corresponding *recoil* angular momentums, for the other.

For the end of convenience, this expression can be subdivided to the internal $[\mathbf{M}_0^{in}]$ (zero-point) and external $[\mathbf{M}_\lambda^{ext}]$ contributions to the total angular momentum $[\mathbf{M}_{tot}]$:

$$\mathbf{M}_{tot} = \mathbf{M}_0^{in} + \mathbf{M}_\lambda^{ext} \qquad 8.58$$

It follows from the law of angular momentum conservation, that the angular momentums of Cumulative virtual cloud (**CVC**) and the recoil (*rec*) angular momentums, accompanied $[\mathbf{C} \to \mathbf{W}]$ transitions of sub-elementary fermions, should be equal:

$$\mathbf{M}_0^{in} = [\mathbf{R}\,\alpha\mathbf{m}_0\mathbf{c}\mathbf{L}_\mathbf{E}^0 + \mathbf{R}\,\beta\mathbf{m}_0\mathbf{c}\,\mathbf{L}_\mathbf{G}^0]_{rec} = [\mathbf{R}\,\mathbf{m}_0\mathbf{c}\,\mathbf{L}_0 - \mathbf{R}\,\alpha\mathbf{m}_0\mathbf{c}\,\mathbf{L}_0 - \mathbf{R}\,\beta\mathbf{m}_0\mathbf{c}\,\mathbf{L}_0]_{\mathbf{CVC}} \qquad 8.59$$

where the internal momentum of elementary particle at Golden mean (zero-point) conditions:

$$\mathbf{p}_0^{in} = \mathbf{m}_0\mathbf{c} = |\mathbf{m}_V^+ - \mathbf{m}_V^-|^\phi c = (\mathbf{m}_V^+\mathbf{v}^2)^\phi/\mathbf{c} \qquad 8.60$$

$$\mathbf{L}_0 = \hbar/\mathbf{m}_0\mathbf{c} \qquad \text{Compton radius} \qquad 8.60a$$

and the external contribution to angular momentum:

$$\mathbf{M}_\lambda^{ext} = [\alpha\mathbf{m}_V^+\mathbf{v}\mathbf{L}_\mathbf{E}^{ext} + \beta\mathbf{m}_V^+\mathbf{v}\mathbf{L}_\mathbf{G}^{ext}]_{rec} = [\mathbf{m}_V^+\mathbf{v}\,\mathbf{L}_B - \alpha\mathbf{m}_V^+\mathbf{v}\,\mathbf{L}_B - \beta\mathbf{m}_V^+\mathbf{v}\,\mathbf{L}_B]_{\mathbf{CVC}} \qquad 8.61$$

where the external momentum of particle is directly related to its de Broglie wave length ($\lambda_B = 2\pi\mathbf{L}_B = \mathbf{h}/\mathbf{m}_V^+\mathbf{v}$):

$$\mathbf{p}^{ext} = \mathbf{m}_V^+\mathbf{v} = \mathbf{h}/\lambda_B = \frac{\hbar}{\mathbf{L}_B} \qquad 8.62$$

The sum of zero-point and angular momentums is:

$$\mathbf{M}_{tot} = \alpha(\mathbf{R}\,\mathbf{m}_0\mathbf{c}\mathbf{L}_\mathbf{E}^0 + \mathbf{m}_V^+\mathbf{v}\mathbf{L}_\mathbf{E}^{ext})_{rec} + \beta(\mathbf{R}\,\mathbf{m}_0\mathbf{c}\,\mathbf{L}_\mathbf{G}^0 + \mathbf{m}_V^+\mathbf{v}\mathbf{L}_\mathbf{G}^{ext})_{rec} = \qquad 8.63$$

$$= \mathbf{R}\,\mathbf{m}_0\mathbf{c}\,\mathbf{L}_0 + \mathbf{m}_V^+\mathbf{v}\,\mathbf{L}_B - \alpha(\mathbf{R}\,\mathbf{m}_0\mathbf{c}\,\mathbf{L}_0 + \mathbf{m}_V^+\mathbf{v}\,\mathbf{L}_B)_{\mathbf{CVC}} - \beta(\mathbf{R}\,\mathbf{m}_0\mathbf{c}\,\mathbf{L}_0 + \mathbf{m}_V^+\mathbf{v}\,\mathbf{L}_B)_{\mathbf{CVC}}$$

The minimum space curvatures, related to electromagnetism, corresponding to zero-point longitudinal recoil effects, accompanied $[C \rightleftharpoons W]$ pulsation, can be find out from (8.59), reflecting the angular momentum conservation law, as:

$$\mathbf{L}_\mathbf{E}^0 = \frac{\mathbf{L}_0}{\alpha}(1 - \alpha - 2\beta) \cong \mathbf{L}_0\left(\frac{1}{\alpha} - 1\right) = a_B - \mathbf{L}_0 = 136,036\,\mathbf{L}_0 \qquad 8.64$$

$$\beta <<< \alpha = 0,0072973506 \cong 1/137$$

We can see, that the space curvature, characteristic for electric potential of the electron at Golden Mean (zero-point) conditions ($\mathbf{L}_\mathbf{E}^0$) is very close to the radius of the *1st Bohr orbit* ($a_B$) in hydrogen atom:

$$a_B = \frac{1}{\alpha}\mathbf{L}_0 = 137,036\,\mathbf{L}_0 = 0.5291 \cdot 10^{-10}\,m \qquad 8.65$$

In similar way we can find from (8.59) zero-point Bivacuum curvature, determined by elementary particle gravitational potential:

$$\mathbf{L}_\mathbf{G}^0 = \frac{\lambda_G^0}{2\pi} = \frac{\mathbf{L}_0}{\beta}(1 - 2\alpha - \beta) \cong \frac{\mathbf{L}_0}{\beta^{e,p}} \qquad 8.66$$

where: $\beta^e = (m_0^e/M_{Pl})^2 = 1.7385 \cdot 10^{-45}$;   $\beta^p = (m_0^p/M_{Pl})^2 = 5.86 \cdot 10^{-39}$   are introduced in our theory gravitational fine structure constant, different for electrons and protons; $M_{Pl} = (\hbar c / G)^{1/2} = 2.17671 \cdot 10^{-8} \, kg$ is a Plank mass; $m_0^e = 9.109534 \cdot 10^{-31} \, kg$ is a rest mass of the electron; $m_0^p = 1.6726485 \cdot 10^{-27} \, kg = m_0^e \cdot 1.8361515 \cdot 10^3 \, kg$ is a rest mass of proton.

The length of one light year is $9.46 \cdot 10^{15} \, m$. The gravitational curvature radius of proton from (8.66), equal to $(\mathbf{L}_G^0)^p = a_G^p = 3.58 \cdot 10^{22} m$. may have the same importance in cosmology, like the electromagnetic curvature of the electron, equal to 1st orbit radius of the hydrogen atom: $a_B = 0.5291 \cdot 10^{-10} m$ in atomic physics. For comparison with $a_G^p$, the characteristic distance between galactics in their groups and clusters is in range: $(0.3 - 1.5) \cdot 10^{22} m$. The radius of Local group of galactics, like Milky way, Andromeda galaxy and Magellan clouds, equal approximately to $3 \cdot 10^6$ light years. The radius of Vigro cluster of galactics is also close to $a_G^p$.

*Let us consider now the curvature of electric potential, determined by the external dynamics* of the charged particle and its de Broglie wave length from (8.61):

$$\mathbf{L}_\mathbf{E}^{ext} = \frac{\mathbf{L}_B}{\alpha}(1 - \alpha - 2\beta) \cong \mathbf{L}_B\left(\frac{1}{\alpha} - 1\right) = 136,036 \, \mathbf{L}_B = 136,036 \, \frac{\lambda_B}{2\pi} \qquad 8.67$$

In most common nonrelativistic conditions the de Broglie wave length of elementary particle is much bigger than it its Compton length ($\mathbf{L}_B = \frac{\lambda_B}{2\pi} = \frac{1}{2\pi}\frac{\mathbf{h}}{\mathbf{mv}} >> \mathbf{L}_0 = \frac{\hbar}{\mathbf{m}_0\mathbf{c}}$) and, consequently, the effective external radius of Coulomb potential action is much bigger, than the minimum internal one: $\mathbf{L}_\mathbf{E}^{ext} >> \mathbf{L}_\mathbf{E}^0$.

Similar situation is valid for external gravitational potential curvature from (8.61):

$$\mathbf{L}_\mathbf{G}^{ext} = \frac{\lambda_G}{2\pi} = \frac{\mathbf{L}_B}{\beta}(1 - 2\alpha - \beta) = \mathbf{L}_B\left(\frac{1}{\beta} - \frac{2\alpha}{\beta} - 1\right) \cong \frac{\mathbf{L}_B}{\beta} \qquad 8.68$$

### 8.14 Curvatures of Bivacuum domains of nonlocality, corresponding to zero-point electromagnetic and gravitational potentials of elementary particles

Let us analyze the length of coherence (de Broglie waves), determined by zero-point vibrations velocity, accompanied the recoil effects of unpaired and paired sub-elementary fermions of triplets $< [\mathbf{F}_\uparrow^- \bowtie \mathbf{F}_\downarrow^+]_{S=0} + (\mathbf{F}_\updownarrow^\pm)_{S=\pm 1/2} >^{e,p}$, equal to radius of Bivacuum domain of nonlocality. It is assumed that the *translational external* velocity of triplets is zero ($\mathbf{v}_{tr}^{ext} = 0$).

The corresponding curvatures are related to electromagnetic and gravitational potential of pulsing elementary particle of any of (i) generation:

$$\left(L_E^\phi\right)_{VirBC} = \frac{\hbar}{\left(m_{BVF}^\phi\right)(\mathbf{v}_0)_E} = \frac{\hbar}{\left(m_{BVF}^\phi\right)\mathbf{c}(\alpha\phi)^{1/2}} \qquad 8.69$$

$$\left(L_G^\phi\right)_{VirBC} = \frac{\hbar}{\left(m_{BVF}^\phi\right)(\mathbf{v}_0)_G} = \frac{\hbar}{\left(m_{BVF}^\phi\right)\mathbf{c}[\beta(2-\phi)]^{1/2}} \qquad 8.69a$$

where zero-point velocities: $(\mathbf{v}_0)_{HE} = \mathbf{c}(\alpha\phi)^{1/2}$ and $(\mathbf{v}_0)_G = \mathbf{c}[\beta(2-\phi)]^{1/2}$ are defined by (8.39) and (8.42).

The uncompensated masses of BVF, due to mass symmetry shifts, induced by electromagnetic and gravitational vibrations can be evaluated as:

$$\left(\mathbf{m}^{\phi}_{BVF}\right)_E = \left(|\mathbf{m}^+_V| - |-\mathbf{m}^-_V|\right)^{\phi}_E = \left[\mathbf{m}^+_V (\mathbf{v}/\mathbf{c})^2\right]^{\phi}_E = \frac{\mathbf{m}_0\boldsymbol{\alpha}\boldsymbol{\phi}}{\sqrt{1 - \boldsymbol{\alpha}\boldsymbol{\phi}}} \simeq \mathbf{m}_0\boldsymbol{\alpha}\boldsymbol{\phi} \qquad 8.70$$

$$\left(\mathbf{m}^{\phi}_{BVF}\right)_G = \left[\mathbf{m}^+_V (\mathbf{v}/\mathbf{c})^2\right]^{\phi}_G = \frac{\mathbf{m}_0\boldsymbol{\beta}(2 - \boldsymbol{\phi})}{\sqrt{1 - \boldsymbol{\beta}(2 - \boldsymbol{\phi})}} \cong \mathbf{m}_0\boldsymbol{\beta}(2 - \boldsymbol{\phi}) \qquad 8.70a$$

Putting 8.70 and 8.70a into 8.69 and 8.69a, we get radiuses of vortices of BVF$^{\uparrow}$ and BVF$^{\downarrow}$, determined by their recoil ⇌ antirecoil longitudinal vibrations, induced by zero-point [$\mathbf{C} \rightleftharpoons \mathbf{W}$] pulsations of unpaired sub-elementary fermions of triplets - elementary particles, like electrons, protons and neutrons:

$$\left(L^{\phi}_E\right)_{VirBC} = \frac{\hbar(\sqrt{1 - \boldsymbol{\alpha}\boldsymbol{\phi}})}{m_0\mathbf{c}(\boldsymbol{\alpha}\boldsymbol{\phi})^{3/2}} \simeq \frac{L_0}{(\boldsymbol{\alpha}\boldsymbol{\phi})^{3/2}} \qquad 8.71$$

$$\left(L^{\phi}_G\right)_{VirBC} = \frac{\hbar\sqrt{1 - \boldsymbol{\beta}(2 - \boldsymbol{\phi})}}{m_0\mathbf{c}[\boldsymbol{\beta}(2 - \boldsymbol{\phi})]^{3/2}} \simeq \frac{L_0}{[\boldsymbol{\beta}(2 - \boldsymbol{\phi})]^{3/2}} \qquad 8.71a$$

These vortices of two very different radiuses represent standing circular virtual waves. In accordance to our theory, they characterize the regions of virtual Bose condensation, representing the domains of nonlocality.

## 9. Pauli principle: How it works?

Let us consider the reasons why the Pauli principle "works" for fermions and do not work for bosons. In accordance to our model of elementary particles, the numbers of sub-elementary fermions and sub-elementary antifermions, forming bosons, like photons (Fig.4), are equal. Each of sub-elementary fermion and sub-elementary antifermion in symmetric pairs [$\mathbf{F}^+_{\updownarrow} + \mathbf{F}^-_{\updownarrow}$] of bosons can pulsate between their [C] and [W] states in-phase ($S = 0$) or counterphase ($S = \pm 1\hbar$). In both cases the positive and negative subquantum particles, forming $\mathbf{CVC}^+$ and $\mathbf{CVC}^-$ do not overlap, as far they are in realms of opposite energy.

For the other hand, the numbers of sub-elementary particles and sub-elementary antiparticles in composition of fermions (i.e. triplets < [$\mathbf{F}^+_{\uparrow} \bowtie \mathbf{F}^-_{\downarrow}$] + $\mathbf{F}^{\pm}_{\updownarrow} >^i$ are not equal to each other. Consequently, the $\mathbf{CVC}^+$ and $\mathbf{CVC}^-$ of sub-elementary fermions and antifermions in triplets do not compensated each other. It leads to the external oscillations of Bivacuum subquantum particles density in the process of [$\mathbf{C} \rightleftharpoons \mathbf{W}$] pulsation, which can be uncompensated also.

*In the framework of our model, Pauli repulsion effect between fermions with the same spin states and energy, i.e. the same phase and frequency of* [$\mathbf{C} \rightleftharpoons \mathbf{W}$] *pulsation, is similar to the effect of excluded volume.*

This effect is provided by spatial incompatibility of two cumulative virtual clouds: $\mathbf{CVC}^{\pm}_{\uparrow}$ and $\mathbf{CVC}^{\pm}_{\updownarrow}$ of the *anchor* Bivacuum fermions of unpaired sub-elementary particles of triplets, emitted in the same moment of time in the same volume. The latter is a case, if the distance between $\mathbf{CVC}^{\pm}_{\uparrow}$ and $\mathbf{CVC}^{\pm}_{\updownarrow}$ is equal or less, than space of their superposition [$\mathbf{CVC}^{\pm}_{\uparrow} + \mathbf{CVC}^{\pm}_{\updownarrow}$], determined by doubled de Broglie wave length of triplets: $\boldsymbol{\lambda}_B = \boldsymbol{\lambda}_{CVC} = \mathbf{h}/\mathbf{m}^+_{\uparrow}\mathbf{v}^{ext}$.

**Let us analyze this situation in more detail.**

The average *external* translational kinetic energy ($\overline{\mathbf{T}}^{\mathbf{C} \rightleftharpoons \mathbf{W}}_{tot}$) of fermions [$\mathbf{F}^+_{\uparrow} \bowtie \mathbf{F}^-_{\downarrow}$] + $\mathbf{F}^{\pm}_{\updownarrow} >^i$ is:

$$\overline{\mathbf{T}}_{tot}^{\mathbf{C} \rightleftharpoons \mathbf{W}} = \mathbf{T}_{tot} \pm \left[ (\mathbf{E}_E)_{[C]}^{Loc} - (\mathbf{E}_E)_{[W]}^{Dist} \right]_{tr} \qquad 9.1$$

It involves opposite by sign oscillation of local/internal Coulomb *potential* interaction in [C] phase $\left[ \frac{|\mathbf{e}_+\mathbf{e}_-|}{\mathbf{L}_T} \right]^{Loc}$ of the *anchor* Bivacuum fermion of $\mathbf{F}_{\updownarrow}^{\pm} >^i$, transforming to distant *kinetic* recoil perturbation of Bivacuum matrix $\alpha[(\mathbf{m}_V^+ - \mathbf{m}_V^-)\mathbf{c}^2]^{Dis}$, representing electric field, in the process of $[\mathbf{C} \rightleftharpoons \mathbf{W}]$ pulsation:

$$\mathbf{E}_{E[C]}^{Loc} = \left[ \frac{|\mathbf{e}_+\mathbf{e}_-|}{\mathbf{L}_T} \right]^{Loc} = \alpha[\mathbf{m}_V^+\boldsymbol{\omega}_B^2\mathbf{L}_B^2]^{Loc} \overset{[\mathbf{C} \rightleftharpoons \mathbf{W}]}{\rightleftharpoons} \alpha[(\mathbf{m}_V^+ - \mathbf{m}_V^-)\mathbf{c}^2]^{Dis} = \alpha[\mathbf{m}_V^+\mathbf{v}^2]^{Dis} = (\mathbf{E}_E)_{[W]}^{Dist} \quad 9.2$$

The energy of the *anchor* site of unpaired $\mathbf{F}_{\updownarrow}^{\pm} >^i$ in [W] phase of triplet, equal to external energy of de Broglie wave:

$$\mathbf{E}_{anc} = \mathbf{E}_B = (\mathbf{E}_E)_{[W]}^{Dist} + \mathbf{T}_k^{CVC^\pm} \qquad 9.3$$

can be presented as a sum of energy of electric field, equal to recoil energy:

$$(\mathbf{E}_E)_{[W]}^{Dist} = \alpha[(\mathbf{m}_V^+ - \mathbf{m}_V^-)\mathbf{c}^2]^{Dis} = \alpha[\mathbf{m}_V^+\mathbf{v}^2]^{Dis} \sim \boldsymbol{\varepsilon}_C \; (electric\; field\; energy) \qquad 9.4$$

and real energy of $CVC^\pm$, equal to maximum kinetic energy of cumulative virtual cloud $\frac{\mathbf{h}^2}{\mathbf{m}_V^+\boldsymbol{\lambda}_B^2} = \mathbf{m}_V^+\mathbf{v}^2$ minus recoil energy:

$$\mathbf{T}_k^{CVC^\pm} = \frac{\mathbf{h}^2}{\mathbf{m}_V^+\boldsymbol{\lambda}_B^2} - \alpha[\mathbf{m}_V^+\mathbf{v}^2] = \mathbf{m}_V^+\mathbf{v}^2(1 - \alpha) \sim \boldsymbol{\varepsilon}_P \qquad 9.5$$

The Coulomb repulsion ($\boldsymbol{\varepsilon}_C$) between two similar elementary charge is determined by electric field energy (9.4). For the other hand, the Pauli repulsion ($\boldsymbol{\varepsilon}_P$) between these charges, as a fermions, pulsing in the same phase and frequency on the distance, *close to de Broglie wave length*: $\boldsymbol{\lambda}_B = h/\mathbf{m}_V^+\mathbf{v}$ is dependent on real energy of $\mathbf{CVC}^\pm$ (9.5).

The ratio between Pauli and Coulomb repulsion energies between two similar fermions on the distances about or less, than de Broglie wave length of these charges ($\boldsymbol{\lambda}_B$) is equal to ratio of 9.5 and 9.4:

$$\frac{\boldsymbol{\varepsilon}_P}{\boldsymbol{\varepsilon}_C} = \frac{1 - \alpha}{\alpha} = \frac{1}{\alpha} - 1 \simeq 136 \qquad 9.6$$

We can see, that it is close to reverse value of electromagnetic fine structure constant: $\mathbf{1/\alpha} \simeq 137$.

This means, that on these distances, comparable with linear dimensions of $CVC^\pm$ usually much bigger than Compton length of charges: $\boldsymbol{\lambda}_B >> (\mathbf{L}_0 = \hbar/\mathbf{m}_0\mathbf{c})$, the Pauli nonelectromagnetic repulsion is more than hundred times bigger, than Coulomb interaction.

Pauli repulsion regulate the counterphase $[\mathbf{C} \rightleftharpoons \mathbf{W}]$ pulsation in a system of two sub-elementary fermions: $\mathbf{F}_\downarrow^-$ and $\mathbf{F}_\uparrow^-$ of the electron $< [\mathbf{F}_\uparrow^+ \bowtie \mathbf{F}_\downarrow^-] + \mathbf{F}_\uparrow^- >^i$ or two sub-elementary antifermions $\mathbf{F}_\uparrow^+$ and $\mathbf{F}_\downarrow^+$ of the positron $< [\mathbf{F}_\uparrow^+ \bowtie \mathbf{F}_\downarrow^-] + \mathbf{F}_\downarrow^+ >^i$, because their $\mathbf{CVC}^\pm$ do not overlap in in the same space in the same time.

Fore the other hand, the $[\mathbf{C} \rightleftharpoons \mathbf{W}]$ dynamics of sub-elementary fermion and sub-elementary antifermion ($\mathbf{F}_\uparrow^+$ and $\mathbf{F}_\downarrow^-$), localized in opposite energetic realms of Bivacuum, can be in-phase, as well as counterphase, because the $\mathbf{CVC}^+$ and $\mathbf{CVC}^-$ do not overlap in both cases. These conditions may occur in the process of $[\mathbf{C} \rightleftharpoons \mathbf{W}]$ pulsation of sub-elementary fermions, composing elementary bosons, like photons, and complex bosons - neutral atoms. In these two situations the effect of excluded volume is absent and fermions are spatially compatible. The mechanism, proposed, explains the absence of the Pauli repulsion in systems of Bosons and Cooper pairs, making possible their Bose condensation.

### 9.1 Spatial compatibility of sub-elementary fermions of the same charge and opposite spins

We postulate in our model, that $[C \Leftrightarrow W]$ pulsation of paired sub-elementary fermion and antifermion $[F_\uparrow^+ \bowtie F_\downarrow^-]$ of opposite spins in composition of the electron $< [F_\uparrow^+ \bowtie F_\downarrow^-] + \mathbf{F}_\uparrow^-$ > or positron $[(F_\uparrow^+ \bowtie F_\downarrow^-) + \mathbf{F}_\downarrow^+ >]$ are counterphase with pulsation of unpaired $\mathbf{F}_\uparrow^\pm$ > (see the upper part of Fig. 8).

In the case of counterphase $[C \Leftrightarrow W]$ pulsations of paired $[F_\downarrow^+]^{(1)}$ and unpaired $\mathbf{F}_\uparrow^\pm >^{(2)}$ with *opposite* spins, but similar charges, localized in the *same* energy realm, they are *spatially compatible*, as far their corpuscular [C] and wave [W] phase are realized alternatively in different semi-periods. Consequently, the Pauli repulsion, described above, is absent.

The example of such compatible pairs in composition of the electron or positron is presented on (Fig.8).

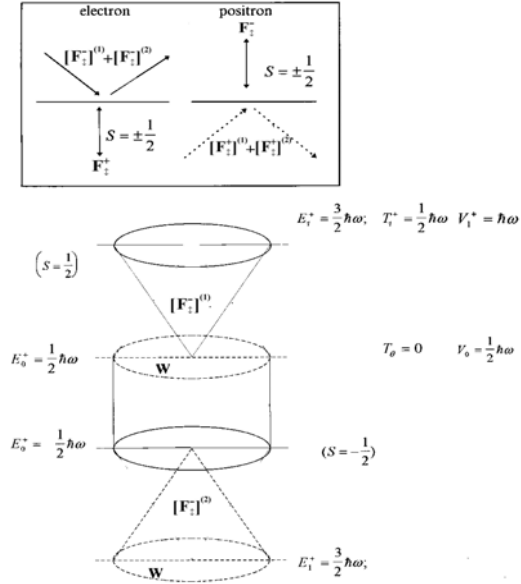

**Fig. 8**. Schematic representation of pair of a *spatially compatible* sub-elementary antifermions of the *electron* $< \left[ F_\uparrow^+ \bowtie F_\downarrow^- \right] + \mathbf{F}_\uparrow^- >$, with opposite half-integer spins: $\mathbf{F}_\uparrow^- >$ and $F_\downarrow^-$ and same charge ($e^-$), energy and frequency of $[\mathbf{C} \Leftrightarrow \mathbf{W}]$ pulsation. The *counterphase* $[\mathbf{C} \rightleftharpoons \mathbf{W}]$ transitions of two sub-elementary antifermions with opposite spins: $\mathbf{F}_\uparrow^- >$ and $F_\downarrow^-$ neutralize the both - Pauli and electromagnetic repulsion between them.

In the electron $< [F_\uparrow^+ \bowtie F_\downarrow^-]_{S=0} + (\mathbf{F}_\uparrow^-)_{S=1/2} >$, the resulting spin and charge is determined by unpaired and uncompensated spin of $(\mathbf{F}_\downarrow^-)_{S=\pm 1/2} >$. The actual inertial mass $(m_V^+)$ and energy of the electron also is determined by this unpaired/uncompensated sub-elementary fermion.

The dynamics of sub-elementary fermions of positron $[(F_\uparrow^+ \bowtie F_\downarrow^-) + \mathbf{F}_\downarrow^+ >]$ is similar to that of electron, determined, however, by unpaired sub-elementary antifermion $(\mathbf{F}_\downarrow^+)_{S=\pm 1/2} >$.

The process of the triplets of sub-elementary fermions spin state inversion needs $720^0$ not $360^0$. It will be explained in the next section.

## 9.2 *The double turn ($720^0$) of magnetic field, as a condition of the fermions spin state reversibility*

It is known fact, that the total rotating cycle for spin of the electrons or positrons is not $360^0$, but $720^0$, i.e. *double turn* by external magnetic field of special configuration, is necessary to return elementary fermions to starting state (Davies, 1985). The correctness of any new model of elementary particles should be testified by its ability to explain this nontrivial fact.

We may propose *three* possible explanations, using our model of the electrons, positrons, protons and antiprotons, as a triplets of sub-elementary fermions/antifermions.

Let us analyze them on example of the electron:

$$< [F_\uparrow^+ \bowtie F_\downarrow^-] + F_\uparrow^- >^e \qquad\qquad 9.5$$

**1**. We may assume, that the direction of external magnetic field rotation acts *only on unpaired* sub-elementary fermion, as asymmetric [torus ($V^-$) + antitorus ($V^+$)] pair: $F_\uparrow^- = (V^- \upuparrows V^+)_{as}$, if the resulting magnetic moment of pair $[F_\uparrow^+ \bowtie F_\downarrow^-]$ is zero and the pair do not interact with external magnetic field at all. In such conditions the 1st $360^0$ turn of external **H** field change the direction of rotation of one of two toruses rotation to the opposite one: $V^- \uparrow \to V^- \downarrow$, transforming sub-elementary fermion to sub-elementary boson: $[F_\uparrow^- \equiv (V^- \upuparrows V^+)] \overset{360^0}{\to} [B^- \equiv (V^- \downuparrows V^+)]$. One more $360^0$ turn of the external magnetic field converts this sub-elementary boson and the triplet (9.5) to starting condition. The total cycle for unpaired $F_\uparrow^- >$ of triplet can be presented as:

$$\textbf{(I)} \quad [F_\uparrow^- > \ \equiv (V^- \upuparrows V^+)] \overset{360^0}{\to} [B^- \equiv (V^- \downuparrows V^+)] \overset{360^0}{\to} [F_\uparrow^- > \ \equiv (V^- \upuparrows V^+)] \qquad 9.6$$

**2**. *The second possible explanation* of double $720^0$ turn may be a consequence of following two stages, involving origination of pair of sub-elementary bosons ($B^\pm \bowtie B^\pm$) from *pair* of sub-elementary fermions, as intermediate stage and two full turns ($2 \cdot 360^0$) of unpaired sub-elementary fermion:

$$\textbf{(II)} \quad < [F_\uparrow^+ \bowtie F_\downarrow^-] + F_\uparrow^- > \overset{360^0}{\to} < [B^\pm \bowtie B^\pm] + F_\uparrow^- > \overset{360^0}{\to} < [F_\uparrow^+ \bowtie F_\downarrow^-] + F_\uparrow^- > \qquad 9.7$$

Both of these mechanisms are not very probable, because they involve the action of external magnetic field on single or paired sub-elementary bosons with zero spin and, consequently, zero magnetic moment.

**3**. *The most probable third mechanism* avoids such strong assumption. The external rotating **H** field interact in two stage manner ($2 \cdot 360^0$) only with sub-elementary fermions/antifermions, changing their spins. However this mechanism demands that the angle of spin rotation of sub-elementary particle and antiparticles of neutral pairs $[F_\uparrow^+ \bowtie F_\downarrow^-]$ are the additive parameters. It means that turn of resulting spin of *pair* on $360^0$ includes reorientation spins of each $F_\uparrow^+$ and $F_\downarrow^-$ only on $180^0$. Consequently, the full spin turn of pair $[F_\uparrow^+ \bowtie F_\downarrow^-]$ resembles that of Mobius transformation.

The spin of unpaired sub-elementary fermion $F_\uparrow^- >$, in contrast to paired ones, makes a *full turn* each $360^0$, i.e. twice in $720^0$ cycle:

$$< [(F_\uparrow^+)_x \bowtie (F_\downarrow^-)_y] + (F_\uparrow^-)_z > \overset{360^0}{\to} < [(F_\downarrow^+)_x \overset{180^0+180^0}{\bowtie} (F_\uparrow^-)_y] + (F_\uparrow^-)_z > \to \qquad 9.8$$

$$\overset{360^0}{\to} < [(F_\uparrow^+)_x \bowtie (F_\downarrow^-)_y] + (F_\uparrow^-)_z >$$

The difference between the intermediate - 2nd stage and the original one in (9.8) is in opposite spin states of paired sub-elementary particle and antiparticle:

$$[(F_\uparrow^+)_x \bowtie (F_\downarrow^-)_y] \overset{360^0}{\to} [(F_\downarrow^+)_x \overset{180^0+180^0}{\bowtie} (F_\uparrow^-)_y]$$ 9.9

Because of Pauli repulsion (see previous section) between two sub-elementary fermions of the same spin state $(F_\uparrow^-)_y$ and $(\mathbf{F}_\uparrow^-)_z >$, in intermediate state of (9.8), the corresponding triplet configuration has deformed - stretched configuration, different from original and final ones.

In the latter - equilibrium configurations of triplet, the $[\mathbf{C} \rightleftharpoons \mathbf{W}]$ pulsation of unpaired sub-elementary fermion $(\mathbf{F}_\uparrow^-)_z >$ and paired $(F_\downarrow^-)_y$ is counterphase and spatially compatible due to the absence of Pauli repulsion.

One more known "strange" experimental result can be explained by our dynamic model of triplets of elementary particles. The existence in triplets paired in-phase pulsating sub-elementary fermions (9.9) with opposite charge, representing double electric dipoles (i.e. double charge), can be responsible for *two times stronger magnetic field*, generated by electron, as compared with those, generated by rotating sphere with single charge $|e^-|$.

### 9.3. *Bosons as a coherent system of sub-elementary and elementary fermions*

The spatial image of sub-elementary boson is a superposition of **strongly correlated** sub-elementary fermions with opposite charges and spin states with properties of Cooper pairs. In general case the elementary bosons are composed from the *integer* number of such pairs.

Bosons have zero or integer spin $(0, 1, 2\dots)$ in the $\hbar$ units, in contrast to the half integer spins of fermions. In general case, bosons with $S = 1$ include: photons, gluons, mesons and boson resonances, phonons, pairs of elementary fermions with opposite spins, atoms and molecules.

**We subdivide bosons into elementary and complex bosons**:

1. *Elementary bosons* (like photons), composed from equal number of *sub-elementary* fermions and antifermions, moving with light velocity in contrast to complex bosons, like atoms;

2. *Complex bosons*, represent a coherent system of *elementary* fermions (electrons and nucleons), like neutral atoms and molecules.

Formation of stable *complex* bosons from elementary fermions with different actual masses: $(\mathbf{m}_V^+)_1 \neq (\mathbf{m}_V^+)_2$ is possible due to their electromagnetic attraction, like in *proton + electron* pairs in atoms. It may occur, if the length of their waves B are the same and equal to distance between them. These conditions may be achieved by difference in their external group velocities, adjusting the momentums to the same value:

$$\mathbf{L}_1 = \hbar/(\mathbf{m}_V^+\mathbf{v})_1 = \mathbf{L}_2 = \hbar/(\mathbf{m}_V^+\mathbf{v})_2\dots = \mathbf{L}_n = \hbar/(\mathbf{m}_V^+\mathbf{v})_n$$ 9.10

$$at: \quad \mathbf{v}_1/\mathbf{v}_n = (\mathbf{m}_V^+)_n/(\mathbf{m}_V^+)_1$$

The mentioned above conditions are the base for assembly of complex bosons, unified in the volume of 3D standing waves of fermions of the opposite or same spins.

The **hydrogen atom**, composing from two fermions: electron and proton is a simplest example of complex bosons. The heavier atoms also follow the same principle of self-organization.

The elementary boson, such as photon, represents dynamic superposition of two triplets of sub-elementary fermions and antifermions, corresponding to electron and positron structures. Such composition determines the resulting external charge of photon, equal to zero and the value of photon's spin: $J = +1, 0$ or -1.

Stability of all types of *elementary* particles: bosons and fermions (electrons, positrons etc.) is a result of superposition/exchange of cumulative virtual clouds [$\mathbf{CVC}^+ \bowtie \mathbf{CVC}^-$] with gluon properties, emitted and absorbed in the process of in-phase [$C \rightleftharpoons W$] pulsations of paired sub-elementary particles and sub-elementary antiparticles [$F_\uparrow^+ \bowtie F_\downarrow^-$] (Fig.9).

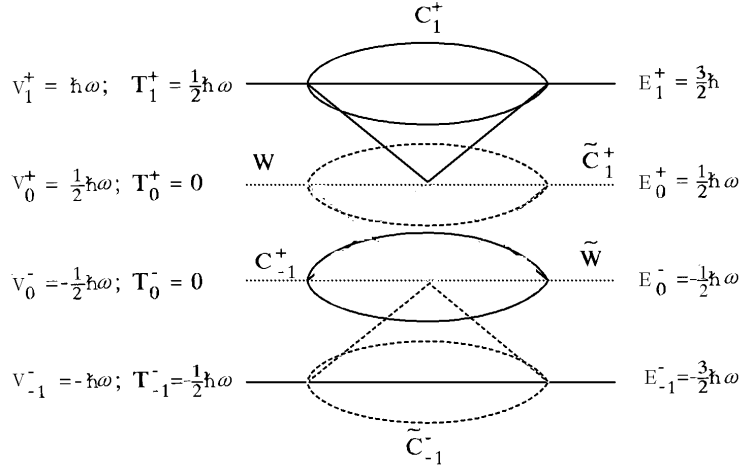

**Fig**. 9. Schematic representation of symmetric pair of the in-phase pulsing sub-elementary fermion and sub-elementary antifermion [$\mathbf{F}_\uparrow^+ \bowtie \mathbf{F}_\downarrow^-$] with boson properties. The $\mathbf{F}_\uparrow^+$ and $\mathbf{F}_\downarrow^-$, pulsing in-phase between the corpuscle and wave states compensate the mass, spin and charge of each other. Such a pair is a neutral component of elementary particles, like electrons, positrons, protons, neutrons, etc. *Properties of symmetric pair of* [$\mathbf{F}_\uparrow^+ \bowtie \mathbf{F}_\downarrow^-$]: resulting electric charge is zero; resulting magnetic charge is zero; resulting spin: $S_{[\mathbf{F}^+ + \mathbf{F}^-]} = \pm 1, 0$.

The neutral symmetric pairs of $\tau$ generations [$\mathbf{F}_\uparrow^- \bowtie \mathbf{F}_\downarrow^+$]$_{S=0,1}^{\tau, \mu}$, forming part of triplets - protons have a properties of *mesons,* as a neutral [quark + antiquark] pairs with integer spin. The coherent cluster of such pairs - from one to four pairs: $(\mathbf{n}[\mathbf{q}^+ \bowtie \widetilde{\mathbf{q}}^-])_{S=0,1,2,3,4}$ can provide the experimentally revealed integer spins of mesons - from zero to four.

## 10  The Mystery of Sri Yantra Diagram

In accordance to ancient archetypal ideas, geometry and numbers describe the fundamental energies in course of their dance - dynamics, transitions. For more than ten millenniums it was believed that the famous Tantric diagram-Sri Yantra contains in hidden form the basic functions active in the Universe (Fig. 10).

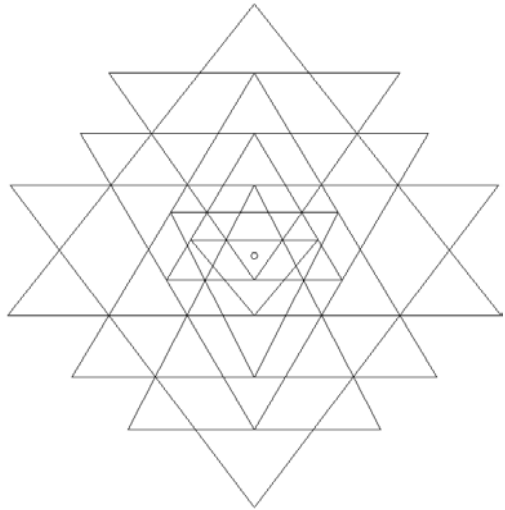

**Fig**. **10**. The Sri Yantra diagram is composed from nine triangles. Four of them are pointed up and five down.
In another way this diagram can be considered as superposition of:
a) the set of pairs of cones of opposite apex, corresponding to torus and antitorus of asymmetric Bivacuum fermions in [C] phase in different excitation states (see Fig. 11a) and
b) the set of diamonds, corresponding to [W] phase of corresponding excitation states of Bivacuum fermions (dashed lines).
 Author is grateful to P. Flanagan for submitting of Sri Yantra diagram with precise coordinates of most important points, making possible its quantitative analysis.

Triangle is a symbol of a three-fold Nature. The Christian trinity, the symbol of God may be represented by triangle. The symbol of trinity is coherent to our idea of *triplets* of sub-elementary particles and antiparticles, as elementary particles. In Buddhism-Hindu triangle with *apex up* is a symbol of God-male and that with *apex down* is a symbol of God-female.

For millenniums it was believed, that Sri Yantra diagram represents geometric language, containing encrypted information about the principles of matter formation.

Let us analyze this diagram, using notions of our theory of elementary particles origination from Bivacuum dipoles and the mechanism of corpuscle - wave duality.

First of all, the ratio 5:4 between positive and negative energy states may reflect the primordial asymmetry of torus and antitorus of Bivacuum dipoles, as a condition of matter origination.

We may see also, that Sri Yantra diagram contains the information about duality of sub-elementary fermions, forming elementary particles, i.e. their discrete corpuscular [C] and wave [W] phases. The diagram at **Fig.10** can be considered as a superposition of:

a) set of pairs of cones of opposite apex, corresponding to asymmetric torus and antitorus of asymmetric Bivacuum fermions in [C] phase in different excitation states (see Fig. 11a, where the diameters of bases of pairs of cones correspond to diameters of torus and antitorus of Bivacuum fermions) and

b) set of diamonds, corresponding to [W] phase of Bivacuum fermions in different excitation states.

In accordance to our theory of sub-elementary fermion/antifermion origination (section 4), the former set (a) describes their [C] phase with different diameters of opposite cones bases, characterizing symmetry shift between torus ($V^+$) and antitorus ($V^-$), correspondingly. The asymmetry of torus and antitorus is increasing with Bivacuum fermion excitation state, accompanied by *decreasing* of spatial separation between them. From formula (1.4) for this separation:

$$[\mathbf{d}_{\mathbf{V}^+ \updownarrow \mathbf{V}^-}]_n^i = \frac{h}{\mathbf{m}_0^i \mathbf{c}(1 + 2\mathbf{n})} \qquad 10.1$$

we can see, that the distance between torus and antitorus decreases with quantum number (**n**) increasing, indeed.

It was astounding to find out, that at maximum excitation and maximum asymmetry of Bivacuum dipole, corresponding to minimum diamond dimension (Fig.11b), the ratio of *down* diameter of cone/torus base to that of *upper* antitorus is $\mathbf{0.6}$, i.e. practically coincide with Golden mean ($\boldsymbol{\phi} = \mathbf{0.618}$). For the other hand, it follows from our Unified theory, that just this critical ratio of torus and antitorus diameters: $2\mathbf{L}_+ / 2\mathbf{L}_- = \boldsymbol{\phi}$ (see eq. 4.15) is a condition of the rest mass and charge origination, as a crucial stage of elementary fermions (electrons, protons, neutrons) fusion from sub-elementary ones (section 5).

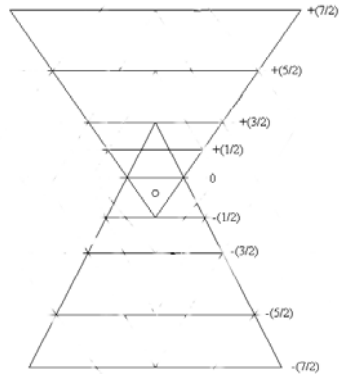

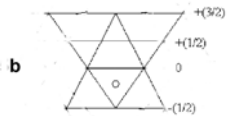

Fig. 11a. Part of Sri Yantra diagram, representing set of pairs of cones of opposite apex, corresponding to torus and antitorus of asymmetric Bivacuum fermions in [C] phase in different excitation states. The diameters of bases of pairs of cones corresponds to diameters of torus and antitorus of Bivacuum fermions.

Fig. 11b. Superposition of [C] and [W] phase of asymmetric Bivacuum fermion, corresponding to critical state of excitation and asymmetry, determined by Golden mean condition. This state is characterized by origination of the rest mass and charge, turning Bivacuum fermion to sub-elementary fermion. The next stage of matter organization from Bivacuum is fusion of triplets of elementary fermions from sub-elementary fermions.

The diamonds of increasing as respect to Fig.11b dimensions, incorporated in Sri Yantra diagram (Fig.10), reflects [W] phase of Bivacuum dipoles of different excitation states in form of Cumulative Virtual Clouds [**CVC**], emitted and absorbed in the process of quantum beats between asymmetric states of torus and antitorus.

The probability of coincidental correlation of quantitative and qualitative features of Sri Yantra diagram properties with key features of our theory of elementary particles is very low. It is a surprise, indeed, that only 10 millenniums after famous Sri Yantra diagram became known in mankind history, we became ready for understanding its encrypted information about principles of Universe construction.

## 11 The Link Between Maxwell's Formalism and Unified Theory

Using (7.18a), the quantization rule for photons can be expressed as:

$$\mathbf{n}\,\mathbf{E}_{el} = \mathbf{n}\,\hbar\boldsymbol{\omega}_{C\rightleftharpoons W} = \boldsymbol{\alpha}\,\mathbf{n}\hbar[\boldsymbol{\omega}_{\bar V}^{+} - \boldsymbol{\omega}_{\bar V}^{-}] = \boldsymbol{\alpha}\,\mathbf{n}(\mathbf{m}_{\bar V}^{+} - \mathbf{m}_{\bar V}^{-})\mathbf{c}^2 \qquad 11.1$$

where: $\mathbf{m}_{\bar V}^{+}\mathbf{c}^2 = \mathbf{n}\,\hbar\boldsymbol{\omega}_{\bar V}^{+}$ **and** $\mathbf{m}_{\bar V}^{-}\mathbf{c}^2 = \mathbf{n}\,\hbar\boldsymbol{\omega}_{\bar V}^{-}$ are the quantized energies of the actual vortex and complementary torus of sub-elementary particle.

From this formula one can see that the electromagnetic energy is a result of quantum beats with frequency $(\boldsymbol{\omega}_{C\rightleftharpoons W})$ between the actual and complementary corpuscular states of two uncompensated sub-elementary fermions with additive spins in composition of photons (*Fig.* 4).

The electromagnetic contribution to the total energy of wave B (11.1) is defined by the fine structure constant, as a factor:

$$\mathbf{E}_E = \boldsymbol{\alpha}\mathbf{E}_{C\rightleftharpoons W} = \boldsymbol{\alpha}\vec{\mathbf{n}}\,\hbar\boldsymbol{\omega}_B = \boldsymbol{\alpha}\vec{\mathbf{n}}\,\hbar(\boldsymbol{\omega}_{\bar V}^{+} - \boldsymbol{\omega}_{\bar V}^{-}) = \frac{\alpha}{2}\hbar[rot\,\vec{\mathbf{V}}^{+} - rot\,\vec{\mathbf{V}}^{-}] \qquad 11.2$$

where: $\vec{\mathbf{n}}$ is a unit-vector, common for both vortices; $\boldsymbol{\omega}_{CVC} = (\boldsymbol{\omega}_{\bar V}^{+} - \boldsymbol{\omega}_{\bar V}^{-})$ is a beats frequency between actual vortex and complementary toruses/vortices with angle velocities: $\vec{\mathbf{V}}^{+}$ and $\vec{\mathbf{V}}^{-}$, depending on radiuses of torus and antitorus.

It is assumed, that all of subquantum particles/antiparticles, forming actual and complementary vortices/toruses of [C] phase of sub-elementary fermions, have the same angle frequency: $\omega_{\bar V}^{+}$ and $\omega_{\bar V}^{-}$ and velocities, correspondingly.

We can express the divergency of Pointing vector: $\mathbf{P} = (c/4\pi)[\mathbf{EH}]$ via difference of contributions, related to actual and complementary toruses, using known relation of vector analysis:

$$div[\mathbf{EH}] = \frac{4\pi}{c}div\,\mathbf{P} = \mathbf{H}\,rot\,\mathbf{E} - \mathbf{E}\,rot\,\mathbf{H} \qquad 11.3$$

where $\mathbf{H}$ and $\mathbf{E}$ are the magnetic and electric energy contributions of subquantum particles, radiated and absorbed in a course of correlated $[C \rightleftharpoons W]$ pulsation of two uncompensated sub-elementary fermions of photon.

Two structures of photon, corresponding to its two polarization and spin $(S = \pm 1\hbar)$, can be presented as:

$$\langle 2[\mathbf{F}_{\uparrow}^{-} \bowtie \mathbf{F}_{\downarrow}^{+}] + [\mathbf{F}_{\downarrow}^{+} + \mathbf{F}_{\downarrow}^{-}]\rangle \qquad S = -1 \qquad 11.4$$

$$\langle 2[\mathbf{F}_{\downarrow}^{-} \bowtie \mathbf{F}_{\uparrow}^{+}] + [\mathbf{F}_{\uparrow}^{+} + \mathbf{F}_{\uparrow}^{-}]\rangle \qquad S = +1 \qquad 11.4a$$

The analogy between (11.2) and (11.3), illustrating the dynamics of [torus + antitorus] dipole, is evident, if we assume:

$$\hbar\omega_{\bar V}^{+} \sim \mathbf{H}\,rot\,\mathbf{E} \sim \frac{\alpha}{2}\hbar\,rot\,\vec{\mathbf{V}}^{+} \qquad 11.5$$

$$\hbar\omega_{\bar V}^{-} \sim \mathbf{E}\,rot\,\mathbf{H} \sim \frac{\alpha}{2}\hbar\,rot\,\vec{\mathbf{V}}^{-} \qquad 11.5a$$

Then, the divergence of Pointing vector will take a form:

$$\frac{4\pi}{c}div\,\mathbf{P} = \frac{\alpha}{2}\hbar\left[rot\,\vec{\mathbf{V}}^{+} - rot\,\vec{\mathbf{V}}^{-}\right] \sim \boldsymbol{\alpha}[\mathbf{m}_{\bar V}^{+} - \mathbf{m}_{\bar V}^{-}]c^2 \qquad 11.6$$

We can see from 11.5 and 11.5a, that the properties of both: magnetic and electric fields are implemented in each of our torus and antitorus of Bivacuum dipoles. The mechanism of this implementation was discussed in sections (8.6 - 8.8).

We may apply also the *Green theorems*, interrelating the volume and surface integrals, to our duality model. One of known Green theorems is:

$$\int\limits_{V} (\Psi\nabla^2\Phi - \Phi\nabla^2\Psi)\, dV = \int\limits_{S} dS \cdot (\Psi\nabla\Phi - \Phi\nabla\Psi)\, dV \qquad 11.7$$

If we define the scalar functions, as the instant energies of the actual and complementary states of [C] phase of sub-elementary particles as $\Phi = \mathbf{m}_{V}^{+}\mathbf{c}^2$ and $\Psi = \mathbf{m}_{V}^{-}\mathbf{c}^2$, then, taking into account that

$$\nabla^2\Phi = div\, grad\,\Phi = div\, grad\,(\mathbf{m}_{V}^{+}\mathbf{c}^2) \qquad 11.8$$

$$\nabla^2\Psi = div\, grad\,\Psi = div\, grad\,(\mathbf{m}_{V}^{-}\mathbf{c}^2) \qquad 11.8a$$

formula (11.7) can be presented in form:

$$\int\limits_{V} [(\mathbf{m}_{V}^{-}\mathbf{c}^2)\nabla^2(\mathbf{m}_{V}^{+}\mathbf{c}^2) - (\mathbf{m}_{V}^{+}\mathbf{c}^2)\nabla^2(\mathbf{m}_{V}^{-}\mathbf{c}^2)]\, dV \qquad 11.9$$

$$= \int\limits_{S} dS \cdot [(\mathbf{m}_{V}^{-}\mathbf{c}^2)\nabla(\mathbf{m}_{V}^{+}\mathbf{c}^2) - (\mathbf{m}_{V}^{+}\mathbf{c}^2)\nabla(\mathbf{m}_{V}^{-}\mathbf{c}^2)]\, dV \qquad 11.9a$$

The upper part (11.9) represents the energy of sub-elementary fermion in [C] phase and the lower part (11.9a) - the energy of cumulative virtual cloud (CVC), corresponding to [W] phase of the same particle.

## 12. The Principle of least action, the Second and Third laws of Thermodynamics. New Solution of Time Problem

### 12.1  The quantum roots of Principle of least action

Let us analyze the formula of *action* in Maupertuis-Lagrange form:

$$\mathbf{S}_{ext} = \int\limits_{t_0}^{t_1} 2\mathbf{T}_{k}^{ext}\, \mathbf{dt} \qquad 12.1$$

The action can be presented also using the Lagrange function, representing difference between the kinetic and potential energy: $L = \mathbf{T}_k - \mathbf{V}$. Using 6.8a, we can see, that $L = -\sqrt{1 - (\mathbf{v}/\mathbf{c})^2}\, \mathbf{m}_0\mathbf{c}^2$ and the action in Hamilton form can be expressed as:

$$\mathbf{S} = -\mathbf{m}_0\mathbf{c}^2 \int\limits_{t_0}^{t_1} \sqrt{1 - (\mathbf{v}/\mathbf{c})^2}\, \mathbf{dt} \qquad 12.1a$$

$$or : \mathbf{S} \simeq -\mathbf{m}_0\mathbf{c}^2 \sqrt{1 - (\mathbf{v}/\mathbf{c})^2} \cdot \mathbf{t} \qquad 12.1b$$

The *principle of Least action*, responsible for choosing one of number of possible particles trajectories from one configuration to another has a form:

$$\Delta\mathbf{S}_{ext} = 0 \qquad 12.2$$

This means, that the optimal trajectory of each particle corresponds to minimum variations of its external kinetic energy and time.

The time interval: $\mathbf{t} = \mathbf{t}_1 - \mathbf{t}_2 = \mathbf{n}\mathbf{t}_B$ we take as a quantized period of the de Broglie wave of particle ($\mathbf{t}_B = 1/\mathbf{v}_B$):

$$\mathbf{t} = \mathbf{t_1} - \mathbf{t_2} = \mathbf{nt}_B = \mathbf{n}/\mathbf{\nu}_B \qquad \text{12.3}$$

$$\mathbf{n} = \mathbf{1, 2, 3}\ldots.$$

Using eqs.(12.1 and 6.10a), we get for the dependence of action in Maupertuis-Lagrange form on introduced Bivacuum tuning energy (**TE**):

$$\mathbf{S}_{ext} = \mathbf{2T}_k^{ext} \cdot \mathbf{t} = \mathbf{m}_V^+ \mathbf{v}^2 \cdot \mathbf{t} = (\mathbf{1 + R})[\mathbf{m}_V^+ \mathbf{c}^2 - \mathbf{m_0 c}^2] \cdot \mathbf{t} \qquad \text{12.4}$$

$$or: \quad \mathbf{S}_{ext} = \mathbf{m}_V^+ \mathbf{v}^2 \cdot \mathbf{t} = (\mathbf{1 + R}) \, \mathbf{TE} \cdot \mathbf{t} \qquad \text{12.4a}$$

where relativistic factor: $\quad \mathbf{R} = \sqrt{1 - (\mathbf{v}/\mathbf{c})^2} \qquad \text{12.4b}$

$\mathbf{m}_V^+ \mathbf{v}^2 = \mathbf{2T}_k^{ext}$ is the doubled kinetic energy of particle.

We introduce here the new notion of *Bivacuum Tuning Energy* (**TE**), dependent on energy of Bivacuum virtual pressure waves ($\mathbf{E}_{\mathbf{VPW}_q}$) as:

$$\mathbf{TE} = \mathbf{E}_{tot} - \mathbf{E}_{\mathbf{VPW}_q} = \hbar\mathbf{\omega_{TE}} = \hbar[\mathbf{\omega_{C \rightleftharpoons W}} - \mathbf{q\omega_0}] \quad = \qquad \text{12.5}$$

$$= [\mathbf{m}_V^+ \mathbf{c}^2 - \mathbf{qm_0 c}^2] = \frac{\mathbf{m}_V^+ \mathbf{v}^2}{\mathbf{1 + R}} \qquad \text{12.6}$$

where: $\mathbf{q} = \mathbf{j} - \mathbf{k}$ ($\mathbf{q} = 1, 2, 3\ldots.$) is a quantum number, characterizing the excitation of Bivacuum Virtual Pressure Waves ($\mathbf{VPW}_q^\pm$), interacting with paired sub-elementary fermions of triplets $[\mathbf{F_\downarrow^-} \bowtie \mathbf{F_\uparrow^+}]$ in the process of $[\mathbf{C} \rightleftharpoons \mathbf{W}]$ pulsation:

$$< [\mathbf{F_\downarrow^-} \bowtie \mathbf{F_\uparrow^+}]_W + (\mathbf{F_\updownarrow^\pm})_C > \quad \rightleftharpoons \quad < [\mathbf{F_\downarrow^-} \bowtie \mathbf{F_\uparrow^+}]_C + (\mathbf{F_\updownarrow^\pm})_W > \qquad \text{12.7}$$

The frequency of beats ($\Delta\omega_{TE}$) equal to Bivacuum Tuning frequency is:

$$\Delta\omega_{TE} = (\mathbf{m}_V^+ - \mathbf{qm_0})\mathbf{c}^2/\hbar = [\mathbf{\omega_{C \rightleftharpoons W}} - q\mathbf{\omega_0}] \qquad \text{12.8}$$

where: $\mathbf{m}_V^+ \geq \mathbf{m_0}$ and $\mathbf{\omega_{C \rightleftharpoons W}} \geq \mathbf{\omega_0}.$

**Tending of TE and $\Delta\omega_{TE}$ to zero due to influence of basic $\mathbf{VPW}_{q=1}^\pm$ at q = 1 on triplets dynamics (forced resonance), minimizing their translational velocity and kinetic energy, provides realization of principle of Least action.**

At conditions, when $\mathbf{q = 1}$, the external *translational* velocity of particle is zero: $\mathbf{v}_{n=1} = \mathbf{0}$ without taking into account velocity of particle zero-point oscillations, induced by its $[\mathbf{C} \rightleftharpoons \mathbf{W}]$ pulsation.

### 12.2 The quantum roots of 2nd and 3d laws of thermodynamics

At the velocity of particles (**v**), corresponding to $\mathbf{q < 1.5}$, the interaction of these pulsing particles with basic ($\mathbf{q = 1}$) virtual pressure waves of Bivacuum ($\mathbf{VPW}_{q=1}^\pm$) due to forced resonance should slow down their velocity, driving translational mobility of particles to resonant conditions: $\mathbf{q = 1}$, $\mathbf{v \to 0}$.

*The second law of thermodynamics*, formulated as a spontaneous irreversible transferring of the heat energy from the warm body to the cooler body or surrounding medium, also means decreasing of kinetic energy of particles, composing this body. Consequently, the 2nd law of thermodynamics, as well as Principle of Least Action, can be a consequence of Tuning energy (**TE**) minimization, due to forced resonance of $\mathbf{VPW}_{q=1}$ with $\mathbf{C} \rightleftharpoons \mathbf{W}$ pulsation, slowing down particles thermal *translational* dynamics at pull-in range synchronization conditions at $(\mathbf{q < 1.5}) \overset{\mathbf{v \to 0}}{\to} (\mathbf{q = 1})$ :

$$\mathbf{TE} = \hbar(\omega_{\mathbf{C} \rightleftharpoons \mathbf{W}} \overset{\mathbf{v} \to \mathbf{0}}{\to} \omega_0))$$ 12.11

*The third law of thermodynamics* states, that the entropy of equilibrium system is tending to zero at the absolute temperature close to zero. Again, this may be a consequence of forced combinational resonance between basic $\mathbf{VPW}_{n=1}^{\pm}$ and particles $[\mathbf{C} \rightleftharpoons \mathbf{W}]$ pulsation, when *translational* velocity of particles $\mathbf{v} \to 0$ and $\mathbf{TE} = \hbar(\omega_{\mathbf{C} \rightleftharpoons \mathbf{W}} \to \omega_0) \overset{\mathbf{v} \to \mathbf{0}}{\to} \mathbf{0}$ at $(\mathbf{q} < \mathbf{1,5}) \overset{\mathbf{v} \to \mathbf{0}}{\to} (\mathbf{q} = \mathbf{1})$. At these conditions in accordance with Hierarchic theory of condensed matter (Kaivarainen, 1995; 2001; 2001a) the de Broglie wave length of atoms is tending to infinity and state of macroscopic Bose condensation of ultimate coherence and order, i.e. minimum entropy.

This result of our Unified theory could explain the energy conservation, notwithstanding of the Universe cooling. *Decreasing* of thermal kinetic energy of particles in the process of cooling is compensated by increasing of potential energy of particles interaction, accompanied the *increasing* of particles de Broglie wave length and their Bose condensation.

### 12.3 The new approach to problem of Time, as a "Time of Action"

Using formula (12.4a) at minimum and constant value of action in Maupertuis-Lagrange form:

$$\mathbf{S} = 2\mathbf{T}_k^{ext} \times \mathbf{t} = \mathbf{m}_V^+ \mathbf{v}^2 \times \mathbf{t} = \mathbf{min}$$

it is easy to show, that the *pace of time* ($\mathbf{dt/t}$) for any closed *conservative system* is determined by the pace of its kinetic energy change $(-\mathbf{dT}/\mathbf{T}_k)_{x,y,z}$, *anisotropic* in general case (Kaivarainen, 2004; 2005):

$$\left[ \frac{\mathbf{dt}}{\mathbf{t}} = \mathbf{d} \ln \mathbf{t} = -\frac{\mathbf{dT}_k}{\mathbf{T}_k} = -\mathbf{d} \ln \mathbf{T}_k \right]_{x,y,z}$$ 12.12

Similar relation can be obtained from principle of uncertainty for free particle with kinetic energy ($\mathbf{T}_k$) in coherent form: $\mathbf{T}_k \, \mathbf{t} = \hbar$. From formula (12.12) it is easy to derive a formula for *"Time of Action"* for conservative mechanical systems.

It is important to note, that in closed conservative mechanical or quantum system the total energy is permanent:

$$\mathbf{E}_{tot} = \mathbf{V} + \mathbf{T}_k = const$$

$$or : \Delta\mathbf{E}_{tot} = 0 \quad and \quad \Delta\mathbf{V} = -\Delta\mathbf{T}_k$$

and the *time of action* is always the *external one*.

By definition a *conservative system* is a system in which work done by a force is:
1. Independent of path;
2. Completely reversible.

Using relations (12.12) and relativistic expression for kinetic energy of *system or mechanical object*:

$$\mathbf{T}_k = \mathbf{m}_V^+ \mathbf{v}^2/2 = \tfrac{1}{2}\mathbf{m}_0 \mathbf{v}^2/\sqrt{1 - (\mathbf{v/c})^2}$$ 12.12a

the *pace of time* and *time of action* for closed system can be presented via *acceleration and velocity* of one or more parts, composing this system (Kaivarainen, 2004, 2005):

$$\left[ \left( \frac{\mathbf{dt}}{\mathbf{t}} = \mathbf{d}\ln\mathbf{t} \right) = -\frac{\mathbf{d\vec{v}}}{\mathbf{\vec{v}}} \frac{2-(\mathbf{v/c})^2}{1-(\mathbf{v/c})^2} \right]_{x,y,z} \qquad 12.13$$

We proceed from the fact, that the true *inertial frames* in our accelerating, rotating and gravitating Universe and in all of its lower levels formations and subsystems - *are nonexisting*.

The dynamics and accelerations in each *closed conservative system, where* $\mathbf{E}_{tot} = const$, are characterized by its dimensionless *pace of time* (12.13) and *time* itself:

$$\mathbf{t} = \left[ -\frac{\mathbf{\vec{v}}}{\mathbf{\vec{a}}} \frac{1-(\mathbf{v/c})^2}{2-(\mathbf{v/c})^2} \right]_{x,y,z} \qquad 12.14$$

where the acceleration in different kinds of motion can be expressed in different forms:

$$\mathbf{\vec{a}} = \mathbf{d\vec{v}/dt} = \frac{\mathbf{v}^2}{\mathbf{r}} = \omega^2\mathbf{r}$$

$$or : \quad \mathbf{\vec{a}} = \mathbf{G}\frac{\mathbf{M}}{r^2} \quad \text{- free fall acceleration}$$

The external reference frame for selected conservative system can be only the another inertialess system/frame, including the former one as a part and with other relativistic factor: $\mathbf{R}^2 = \mathbf{1} - (\mathbf{v/c})^2$. In such approach the *internal time* ($\mathbf{t}^{in}$) of smaller system can be analyzed as a part of external time of bigger conservative system ($\mathbf{t}^{ext}$) :

$$\mathbf{t}^{in} = \frac{\mathbf{t}^{ext}}{\sqrt{1-(\mathbf{v}^{ext}/\mathbf{c})^2}} = -\frac{\mathbf{\vec{v}}\sqrt{1-(\mathbf{v/c})^2}}{\mathbf{\vec{a}}\,[2-(\mathbf{v/c})^2]} \qquad 12.14a$$

The shape of this formula in conditions, when $\mathbf{t}^{ext} = const$ is close to to conventional formula of special relativity (12.15a) for time or clock, moving with velocity ($\mathbf{v} \lesssim \mathbf{c}$) relatively to the clock in rest ($\mathbf{v} \ll \mathbf{c}$).

From (12.14) we can see, that the time for selected object (microscopic or macroscopic) of conservative system is positive at velocity: $0 < \mathbf{v} < \mathbf{c}$, if its acceleration is negative ($\mathbf{d\vec{v}/dt} < \mathbf{0}$). On contrary, time is negative, if acceleration is positive ($\mathbf{d\vec{v}/dt} > \mathbf{0}$). For example, if temperature of conservative system and its kinetic energy are decreasing, the time and its pace are positive.

Thermal oscillations of atoms and molecules in condensed matter, like pendulums oscillation, are accompanied by alternation the sign of acceleration and, consequently, sign of time ($\pm\mathbf{t}^{ext}$ and $\pm\mathbf{t}^{in}$).

The **Corpuscle → Wave** transition of elementary particle, as it follows from Unified theory, is accompanied by decreasing of mass and kinetic energy of unpaired sub-elementary fermion and converting the kinetic energy of [C] phase to potential energy of $\mathbf{CVC}^{\pm}$ of [W] phase. Consequently, this semiperiod of pulsation is characterized by positive time ($\mathbf{t} > \mathbf{0}$)$_{C\to W}$. On contrary, the reverse [$\mathbf{W} \to \mathbf{C}$] transition corresponds to negative time ($\mathbf{t} < \mathbf{0}$)$_{W\to C}$.

In the absence of particles acceleration ($\mathbf{a} = \mathbf{d\vec{v}/dt} = \mathbf{0}$; $\mathbf{dT}_k/\mathbf{T}_k = 0$ and $\mathbf{c} > \mathbf{v} > \mathbf{0}$; ), the time of action ($\mathbf{t}$) is infinitive and its pace ($\mathbf{dt/t}$) is zero:

$$\mathbf{t} \to \infty \qquad \textbf{and} \qquad \left(\frac{\mathbf{dt}}{\mathbf{t}}\right) \to 0$$

$$at \quad \left(\vec{\mathbf{a}} = \mathbf{d}\vec{\mathbf{v}}/\mathbf{dt}\right) \to \mathbf{0} \quad \textbf{and} \quad \mathbf{v} = \textbf{const}$$

The infinitive life-time of the system means its absolute stability. The postulated by this author principle of conservation of internal kinetic energy of torus ($\mathbf{V}^+$) and antitorus ($\mathbf{V}^-$) of symmetric and asymmetric Bivacuum fermions/antifermions: $\left(\mathbf{BVF}_{as}^{\updownarrow}\right)^{\phi} \equiv \mathbf{F}_{\updownarrow}^{\pm}$ (eq.2.1), independently on their external velocity, in fact reflects the condition of infinitive life-time of Bivacuum dipoles in symmetric and asymmetric states. The latter means a stability of sub-elementary fermions and elementary particles, formed by them.

The permanent collective motion of the electrons in superconductors and atoms of $^4\mathbf{He}$ in superfluid liquids with constant velocity ($\mathbf{v} = \textbf{const}$) and $\left(\mathbf{d}\vec{\mathbf{v}}/\mathbf{dt}\right) = \mathbf{0}$ in the absence of collisions and accelerations are good examples, confirming validity of our formula (12.14), as far in these conditions $\mathbf{t} \to \infty$.

When the external translational velocity and external accelerations of Bivacuum dipoles ($\mathbf{BVF}$ and $\mathbf{BVB}^{\pm}$) are zero: $\mathbf{v} = 0$ and $\mathbf{d}\vec{\mathbf{v}}/\mathbf{dt} = \mathbf{0}$, like *in primordial Bivacuum*, the notion of time is uncertain: $\mathbf{t} = \mathbf{0/0}$.

Interesting, that similar uncertainty in time (12.14) corresponds to opposite limit condition, pertinent for photon or neutrino in primordial Bivacuum, when $\mathbf{v} = \mathbf{c} = \textbf{const}$ and $\mathbf{d}\vec{\mathbf{v}}/\mathbf{dt} = \mathbf{0}$. Just in such conditions when causality principle do not work the anomalous time effects are possible.

*In our approach, the velocity of light is the absolute value, determined by physical properties of Bivacuum, like sound velocity in any medium is determined by elastic properties of medium. The primordial Bivacuum superfluid matrix represents the Universal Reference Frame (URF) in contrast to conventional Relative Reference Frame (RRF). Consequently the Bivacuum has the Ether properties and Bivacuum dipoles - the properties of **ethons** - elements of the Ether.*

The positive acceleration of the Universe expansion ($\mathbf{d}\vec{\mathbf{v}}/\mathbf{dt} > 0$) at $\mathbf{c} > \mathbf{v} > 0$, in accordance to (12.13 and 12.14), means negative pace of external time and time itself for this highest Hierarchical level of Bivacuum organization. For the other hand, the process of cooling of each regular star system, like our Solar system, following gradual cooling of star, means slowing down the internal kinetic energy of thermal motion of atoms and molecules in such system, i.e. negative acceleration ($\mathbf{d}\vec{\mathbf{v}}/\mathbf{dt} < \mathbf{0}$) at $\mathbf{c} > \mathbf{v} > 0$. It corresponds to positive internal time and its pace in star systems. These opposite sign and the 'arrow' direction of *time of action* on different hierarchical levels of Universe organization, possibly is a consequence of tending of the Universe to keep its total energy permanent, following energy conservation law.

In accordance with Einstein relativistic theory (Landau and Lifshitz, 1988), the time of clock in the rest state ($\mathbf{t}^{ext}$), which can be considered, as the *external inertial frame* is interrelated with time ($\mathbf{t}^{in}$) in other inertial frame, moving relatively to former with velocity ($\mathbf{v}$) as:

$$\mathbf{t}^{ext} = \left(t_2' - t_1'\right)^{ext} = \mathbf{t}^{in}\sqrt{1 - (\mathbf{v/c})^2} \qquad\qquad 12.15$$

$$\mathbf{t}^{in} = \frac{\mathbf{t}^{ext}}{\sqrt{1 - (\mathbf{v/c})^2}} \qquad\qquad 12.15a$$

where: $\mathbf{t}^{ext} \equiv \left(t_2' - t_1'\right)^{ext}$ is the characteristic time of clock in the reference rest frame; $\mathbf{t}^{in} \equiv \left(t_2 - t_1\right)^{in}$ is the *internal proper time* of clock, moving with velocity: $\mathbf{v} \lesssim \mathbf{c}$, relatively to clock in the rest frame.

It is easy to see, that in relativistic conditions, when $\mathbf{v}^{in} \to \mathbf{c}$, the *proper time* of moving system/clock is tending to infinity: ($T \sim \mathbf{t}^{in} \to \infty$). This means that the moving clock is slower, than similar clock in state of rest relatively to moving one.

**If we consider the imaginary system**, containing only two clock in empty space, moving as respect to each other with permanent velocity, and use the 1st postulate of Special Relativity, i.e. similar laws of physics in any inertial system, we should get the similar time delay in both clocks, even if they move with different velocities in our Universal Reference Frame (URF) - Bivacuum. In other words, both clocks should display the *same time delay,* independently of difference of their velocities ratio to the light velocity $(\mathbf{v}/\mathbf{c})^2$. This result of special relativity is a consequence of assumption of the absence of Ether and absolute velocity. It sounds like a nonsense and has no experimental confirmation. It follows from our Unified theory, that the interpretation, given by Einstein to Michelson-Morley experiments, as the evidence of the Ether absence, was wrong in contrast to explanation, provided by the authors of this experiment themselves.

*Our formulas (12.14 and 12.14a), describing the properties of time (time of action) for conservative systems, are more advanced, than Einstein's (12.15a), as far they are not limited by inertialess frames and contain not only the relativistic factor, but also the velocity itself and acceleration. It will be demonstrated below, that our **time of action** concept better describe the dynamic processes on microscopic - quantum and macroscopic - cosmic scales.*

Different closed conservative systems of particles/objects, rotating around common center on stable orbits with radius ($r$), like in Cooper pairs of sub-elementary fermions, atoms, planetary systems, galactics, etc. are characterized by *centripetal* ($\mathbf{a}_{cp}$) and *centrifugal* ($\mathbf{a}_{cen}$) acceleration, equal by absolute value:

$$\mathbf{a}_{cp} = -\frac{\mathbf{d\vec{v}}}{\mathbf{dt}} = \frac{\vec{\mathbf{v}}^2}{\vec{\mathbf{r}}} = \omega^2 \vec{\mathbf{r}} = -\mathbf{a}_{cen} \qquad 12.16$$

where the *tangential* velocity of rotation is related to the radius $\vec{\mathbf{r}}$ and angular frequency of orbital rotation ($\omega$) as:

$$\left[\vec{\mathbf{v}} = 2\pi\vec{\mathbf{r}} \times \mathbf{v} = \omega\vec{\mathbf{r}}\right] \qquad 12.17$$

Consequently, we get for the ratio of tangential velocity of particle/object to its centripetal acceleration:

$$-\frac{\vec{\mathbf{v}}}{\mathbf{d\vec{v}/dt}} = \frac{1}{\omega} = \frac{\vec{\mathbf{r}}}{\vec{\mathbf{v}}} \qquad 12.17a$$

Putting (12.17a and 12.17) into (12.14), we get the dependence of *time of action* for Corpuscular phase of elementary particle, characterizing period of rotation of structure, like Fig.2 (electron) or Fig.4 (photon) around internal main axes with radius of rotation ($\mathbf{r}$) and angular frequency ($\omega = \vec{\mathbf{v}}/\vec{\mathbf{r}}$):

$$\mathbf{t} = \left[\frac{\vec{\mathbf{r}}}{\vec{\mathbf{v}}} \frac{1 - (\mathbf{v}/\mathbf{c})^2}{2 - (\mathbf{v}/\mathbf{c})^2}\right]_W = \left[\frac{1}{\omega} \frac{1 - (\vec{\mathbf{r}}\omega/\mathbf{L}_0\omega_0)^2}{2 - (\vec{\mathbf{r}}\omega/\mathbf{L}_0\omega_0)^2}\right]_C \qquad 12.18$$

The transition of elementary particles in [W] phase to [C] phase is accompanied by reversible of translational degrees of freedom to rotational ones.

For sub-elementary fermion in [C] phase, when the *translational* energy of elementary particle, pertinent for [W] phase, turns to *rotational* one, we have, using (12.16 and 12.17):

$$(\mathbf{v}/\mathbf{c})^2 = (\vec{\mathbf{r}}\boldsymbol{\omega}/\mathbf{L_0}\boldsymbol{\omega_0})^2 \qquad\qquad 12.19$$

$$where: \ \mathbf{L_0} = \hbar/\mathbf{m_0}\mathbf{c} \ \ and \ \ \boldsymbol{\omega_0} = \mathbf{m_0}\mathbf{c}^2/\hbar$$

From (12.18) we can see, that for *nonrelativistic* conditions of orbital rotation of system/object, when its tangential velocity $\mathbf{v} \ll \mathbf{c}$ and permanent angular frequency: $\boldsymbol{\omega} = \mathbf{v}/\mathbf{r} = \mathbf{const}$, we get from 12.18 the relation between characteristic time of this system and period of orbital rotation ($T$):

$$\mathbf{t}_{\mathbf{v}\ll\mathbf{c}}^{ext} \simeq \left| \frac{1}{2\boldsymbol{\omega}} \right| = \frac{1}{4\pi}T \qquad\qquad 12.20$$

For relativistic conditions of the same system, when $\mathbf{v} \simeq \mathbf{c}$ at angular velocity $(\boldsymbol{\omega} = \mathbf{v}/\mathbf{r}) = \mathbf{const}$, we get from (12.18), that characteristic time and period of orbiting elementary particle or macroscopic object is tending to zero, as far $\left[ 1 - (\mathbf{v}/\mathbf{c})^2 \right] \overset{\mathbf{v}\to\mathbf{c}}{\to} 0$ and $\left[ 2 - (\mathbf{v}/\mathbf{c})^2 \right] \overset{\mathbf{v}\to\mathbf{c}}{\to} 1$:

$$\mathbf{t}_{\mathbf{v}\leq\mathbf{c}} \to 0 \ \text{and the period} \ \left(T = \mathbf{1}/\mathbf{v}\right) \to 0 \ \ \text{at} \ \ \mathbf{v} \to \mathbf{c} \qquad\qquad 12.21$$

$$\text{and} \ \ \mathbf{r} \to \mathbf{r}_{max} \ \ \text{as far} \ \left(\boldsymbol{\omega} = \frac{\mathbf{v}}{\vec{\mathbf{r}}}\right) = \mathbf{const} \qquad\qquad 12.21a$$

For the case, under consideration, the increasing of radius of orbit ($\mathbf{r}$) proportional to increasing of velocity of orbiting particle/object at permanent angular frequency is a consequence of condition (12.21a).

For intermediate case, when $\mathbf{v} < \mathbf{c}$, using result (12.20), our formula for time (12.18) can be presented in a shape, symmetric to conventional relativistic formula for inherent time (12.15):

$$\mathbf{t} = 2\mathbf{t}_{\mathbf{v}\ll\mathbf{c}} \ \frac{1 - (\mathbf{v}/\mathbf{c})^2}{2 - (\mathbf{v}/\mathbf{c})^2} \qquad\qquad 12.22$$

where: $\mathbf{t} \sim \mathbf{t}_{\mathbf{v}>0}$ (12.15) and $2\mathbf{t}_{\mathbf{v}\ll\mathbf{c}} \sim \mathbf{t}_{\mathbf{v}=0}$ (12.15).

We may see, that for this intermediated case, the characteristic time in formula (12.15) of relativistic theory and our (12.22) is decreasing with velocity increasing in both description. However, in formula (12.22) the additional factor: $\left[ 2 - (\mathbf{v}/\mathbf{c})^2 \right]^{-1}$ makes the dependence of time of moving object (i.e. clock) on its velocity weaker than in (12.15).

Formula (12.14) determines, that at very low acceleration ($\mathbf{a} = \mathbf{d}\vec{\mathbf{v}}/\mathbf{dt}) \ll \mathbf{1}$, the ratio $[\mathbf{v}/\mathbf{a}]$ should dominate on ratio:

$$\frac{1 - (\mathbf{v}/\mathbf{c})^2}{2 - (\mathbf{v}/\mathbf{c})^2} \ll \left[ -\frac{\mathbf{v}}{\mathbf{a}} \right] \qquad\qquad 12.22a$$

Consequently, at condition (12.22a) the time of action should increase with velocity of rotating or pulsing object. The same qualitative result follows from special relativity (12.15a). *Consequently, at these condition the time delay in moving system, following from special relativity, is in accordance with our theory of time.*

The formula for time (12.14), determined by internal rotational degrees of freedom of stationary systems, like sub-elementary fermions in elementary particles, the electron orbiting in atom of hydrogen or any planet, rotating around the star, can be transformed to:

$$\mathbf{t} = \frac{1}{\boldsymbol{\omega}} \frac{\mathbf{m}_V^+ \mathbf{c}^2 \left[ 1 - (\mathbf{v}/\mathbf{c})^2 \right]}{\mathbf{m}_V^+ (2\mathbf{c}^2 - \mathbf{v}^2)} = \frac{1 - (\mathbf{v}/\mathbf{c})^2}{\boldsymbol{\omega}} \frac{\mathbf{E}_{tot}}{2\mathbf{V}} \qquad 12.23$$

where: $\mathbf{E}_{tot} = \mathbf{m}_V^+ \mathbf{c}^2 = const$ is a total energy of rotating with angular frequency $\boldsymbol{\omega}$ elementary particle with actual mass $\mathbf{m}_V^+$, as a conservative system;
$2\mathbf{V} = 2(\mathbf{E}_{tot} - \mathbf{T}_k) = \mathbf{m}_V^+ (2\mathbf{c}^2 - \mathbf{v}^2) = (\mathbf{m}_V^+ + \mathbf{m}_{\bar{V}})\mathbf{c}^2$ is a doubled potential energy of unpaired sub-elementary fermion of elementary particle with actual and complementary mass of torus and antitorus: $\mathbf{m}_V^+$ and $\mathbf{m}_{\bar{V}}$.

In the case of harmonic oscillation or standing wave, when $\mathbf{E}_{tot} = \mathbf{V} + \mathbf{T}_k = 2\mathbf{V}$ and $\mathbf{V} = \mathbf{T}_k$, the characteristic time of rotating with angular frequency ($\boldsymbol{\omega} = \mathbf{v}/\mathbf{r}$) particle is dependent only on the ratio of its absolute velocity to the light one $(\mathbf{v}/\mathbf{c})^2$.

### 12.4 The application of new time concept for explanation of Fermat principle

The Fermat principle states that light waves of a given frequency traverse the path between two points which takes the least time. Its modern form is "A light ray, in going between two points, must follow optical path length which is stationary with respect to variations of the path." In this formulation, the paths may be maxima, minima, or saddle points.

The most obvious example of this is the passage of light through a homogeneous medium in which the speed of light doesn't change with position. In this case shortest time is equivalent to the shortest distance between the points, which, as we all know, is a straight line. The examples are existing that time of light passage, including reflected beam, can be minimum or maximum like for light beams from source in the center of ellipsoid with mirror internal surface. There can be a number of trajectories of light beams with the same time of passion. For example, it is true for different beams from one focal point to another passing throw the lens on different distance from lens center. The most important condition for realization of Fermat principle is $\mathbf{t} = \mathbf{const}$. This principle explains the *law of reflection*, as the equality of angles of incidence and angle of reflection: $\boldsymbol{\theta}_I = \boldsymbol{\theta}_R$ and Snell's law of refraction: $\sin \boldsymbol{\theta}_I = \mathbf{n} \sin \boldsymbol{\theta}_R$.

However, it is not yet clear why the Fermat principle is working. Let us analyze the application of Fermat principle to light refraction, using our formula for time (12.14). In accordance to Fermat principle the variation of action time for photons at:
$\mathbf{E}_{tot} = \mathbf{V} + \mathbf{T}_k = \hbar\boldsymbol{\omega}_{ph} = const$ (condition of conservative system) should be zero: $\Delta \mathbf{t} = \mathbf{0}$.

The ratio of velocity of light in vacuum/bivacuum to its velocity ($\mathbf{v} \lesssim \mathbf{c}$) in gas, liquid or transparent solid determines the refraction index of corresponding medium: $(\mathbf{v}/\mathbf{c})^2 = 1/\mathbf{n}$. Taking this into account, the variation of (12.14) in [W] and [C] phase of photon can be presented as:

$$\Delta \mathbf{t} = \Delta \left[ -\frac{\vec{\mathbf{v}}}{\vec{\mathbf{a}}} \frac{1 - (1/\mathbf{n})}{2 - (1/\mathbf{n})} \right]_{W,C} = 0 \qquad 12.24$$

After differentiation (12.24), we get:

$$\frac{\Delta \mathbf{n}}{\mathbf{n} - 1} - \frac{2\Delta \mathbf{n}}{2\mathbf{n} - 1} = \frac{\Delta \mathbf{a}}{\mathbf{a}} - \frac{\Delta \mathbf{v}}{\mathbf{v}} \qquad 12.24a$$

At the conditions, when velocity of light in medium is close to this velocity in empty space: $\mathbf{n} = (\mathbf{c}/\mathbf{v})^2 \gtrsim 1$ we have $\frac{\Delta \mathbf{n}}{\mathbf{n}-1} >> \frac{2\Delta \mathbf{n}}{2\mathbf{n}-1}$ and (12.24a) turns to:

$$\Delta n \;\cong\; (n-1)\Big[\Big(-\frac{\mathbf{v}_2-\mathbf{v}_1}{\mathbf{v}_1}\Big)+\frac{\Delta\mathbf{a}}{\mathbf{a}}\,\Big]_W \qquad\qquad 12.24b$$

The relative change of acceleration $\Delta\mathbf{a}/\mathbf{a}$ describes the jump of light velocity on the interface between two different homogeneous medium.

It is easy to see from this formula, that if the light velocity in 2nd medium is lower, than in 1st and $(\mathbf{v}_2-\mathbf{v}_1) < 0$, the refraction index will increase: $\Delta\mathbf{n} > \mathbf{0}$. *This is in total accordance with empirical data and explains why the Fermat principle is working in geometrical optics.*

Formula (12.24b) describes the change of photon parameters it its Wave [W] phase.

The centripetal acceleration of photon in *Corpuscular [C] phase* can be expressed via tangential velocity and rotation radius of photon structure (Fig.4) as: $\mathbf{a}_{cp}=-\dfrac{\vec{\mathbf{v}}^2}{\vec{\mathbf{r}}}=-\boldsymbol{\omega}^2\mathbf{r}$ and

$$\frac{\Delta\mathbf{a}}{\mathbf{a}}\;=\;\Big(\frac{2\Delta\boldsymbol{\omega}}{\boldsymbol{\omega}}+\frac{\Delta\mathbf{r}}{\mathbf{r}}\Big)_C$$

The relative jump of tangential velocity of photon rotation in [C] phase $(\vec{\mathbf{v}}_{tn}=\boldsymbol{\omega}\vec{\mathbf{r}})$ on the interphase between two mediums is:

$$\frac{\Delta\mathbf{v}}{\mathbf{v}}\;=\;\Big(\frac{\Delta\boldsymbol{\omega}}{\boldsymbol{\omega}}+\frac{\Delta\mathbf{r}}{\mathbf{r}}\Big)_C$$

Consequently, the difference in relative increments for [C] phase of photon is:

$$\Big(\frac{\Delta\mathbf{a}}{\mathbf{a}}-\frac{\Delta\mathbf{v}}{\mathbf{v}}\Big)_C=\Big[\frac{\Delta\boldsymbol{\omega}}{\boldsymbol{\omega}}\Big]_C$$

Putting this expression to (12.24b), we get the increment of refraction index for photon in Corpuscular phase via relative jump of its angular frequency:

$$\Delta\mathbf{n}\;\cong\;(n-1)\Big[\frac{\Delta\boldsymbol{\omega}}{\boldsymbol{\omega}}\Big]_C \qquad\qquad 12.24c$$

This angular frequency of photon rotation coincides with frequency of its $[\mathbf{C}\rightleftharpoons\mathbf{W}]$ pulsation only in symmetric primordial Bivacuum. In the volume of liquids or solids the symmetry of Bivacuum dipoles and their dynamics are changed by elementary particles of medium. From 12.24c we get, that this should be accompanied by increasing of rotational frequency of photon in its [C] phase.

Our Unified theory, in contrast to relativistic one, considers the velocity as the *absolute* parameter, relative to translational velocity of symmetric Bivacuum dipoles equal to zero (see eq. 4.4). The light velocity ($\mathbf{c}$) is also absolute parameter, determined by properties of Bivacuum (ether) and independent on velocity of source of photons.

### 12.5  The quantitative evidence in proof of new theory of time and elementary particles formation from Bivacuum dipoles

Using eq. (12.14), it is possible to calculate the centrifugal acceleration in fast rotating Cooper pairs of sub-elementary fermions $[\mathbf{F}_{\downarrow}^{-}\bowtie\mathbf{F}_{\uparrow}^{+}]_C$ in triplets $<[\mathbf{F}_{\downarrow}^{-}\bowtie\mathbf{F}_{\uparrow}^{+}]_C+(\mathbf{F}_{\uparrow}^{\pm})_W>$, when paired sub-elementary fermions are rotating in corpuscular [C] phase and unpaired $(\mathbf{F}_{\uparrow}^{\pm})_W>$ is in the wave [W] phase. We analyze the condition of the rest state of the electron, when its *external translational* velocity is equal to zero and internal tangential velocity of sub-elementary fermion and antifermion rotation around common axis (Fig. 2), corresponds to Golden mean condition:

$$(\mathbf{v}/\mathbf{c})^2_\phi = \phi = 0.618$$

$$\mathbf{v}^\phi = \mathbf{c}\,(0.618)^{1/2} = 2.358 \times 10^7 \; m/s$$

In accordance to our theory of these conditions stand for the rest mass $(\mathbf{m}_0)$ and charge $(\mathbf{e}_0)$ origination (see chapter 5). The life-time $\mathbf{t_C}$ of Corpuscular phase of rotating $[\mathbf{F}^-_\downarrow \bowtie \mathbf{F}^+_\uparrow]_C$ of the electron is equal to semiperiod of $[\mathbf{C} \rightleftharpoons \mathbf{W}]$ pulsation of pair and triplet itself, determined by Compton angular frequency $\boldsymbol{\omega}^e_0 = \boldsymbol{\omega}^e_{\mathbf{C} \rightleftharpoons \mathbf{W}}$ :

$$\mathbf{t}^e_C \equiv \frac{1}{2}\mathbf{T}^e_{\mathbf{C} \rightleftharpoons \mathbf{W}} = \frac{1}{2\mathbf{v}^e_{\mathbf{C} \rightleftharpoons \mathbf{W}}} = \frac{\pi}{\boldsymbol{\omega}^e_0} = 4.02 \times 10^{-21} \; s \qquad 12.25$$

$$where: \quad \boldsymbol{\omega}^e_0 = \mathbf{m}^e_0 \mathbf{c}^2/\hbar \qquad 12.25a$$

Putting (12.24-12.25) in (12.14), we get for internal centrifugal acceleration of each of paired electronic sub-elementary fermions in [C] phase at Golden mean condition:

$$\left[ a^\phi_{cf} = (\mathbf{dv}/\mathbf{dt})^\phi \right]^e = \frac{\mathbf{v}^\phi}{\mathbf{t}^e_C}\frac{1-\phi}{2-\phi} = 1.62 \times 10^{28} \; m/s^2 \qquad 12.26$$

For comparisons, the free fall acceleration in gravitational field of the Earth is only: $g = 9.81$ m/s$^2$.

The corresponding centrifugal force is equal to product of acceleration (12.26) on the rest mass of rotating paired sub-elementary fermion:

$$\mathbf{F}^\phi_{cf} = \mathbf{m}_0 a^\phi = (9.1 \times 10^{-31})\,(0.162 \times 10^{29}) = 1.47 \times 10^{-2} \; kg \cdot m/s^2 \qquad 12.27$$

From conventional expression for centrifugal force in such a system and Golden mean conditions, we get:

$$\mathbf{F}^\phi_{cf} = \frac{2\mathbf{m}_0\,\phi\mathbf{c}^2}{\mathbf{L}_0} = \frac{2}{3.83 \times 10^{-13}} \times 9.1093897 \cdot 10^{-31} \times 5.56 \cdot 10^{14} = \qquad 12.27a$$
$$= 0.264 \times 10^{-2} \; kg \cdot m/s^2$$

This value is about 5.5 times less, than obtained using our expression for time and acceleration (12.26).

The condition of the electrons stability is that this centrifugal force is compensated by the opposite centripetal force in rotating pairs $[\mathbf{F}^-_\downarrow \bowtie \mathbf{F}^+_\uparrow]^\phi_C$. This compensation can be provided by Coulomb and in much less extent by gravitational attraction between torus and antitorus of paired sub-elementary fermion in triplets $< [\mathbf{F}^-_\downarrow \bowtie \mathbf{F}^+_\uparrow]_C + (\mathbf{F}^\pm_\updownarrow)_W >$:

$$\mathbf{F}^\phi_{Coul} = \frac{\mathbf{e}_+\mathbf{e}_-}{\boldsymbol{\varepsilon}^\phi(\mathbf{L}^\phi)^2} = \frac{\mathbf{e}^2_0}{\boldsymbol{\varepsilon}^\phi_0\mathbf{L}^2_0} = 1.98 \times 10^{-2} \; kg\;m/s^2 \qquad 12.28$$

$$\mathbf{F}^\phi_G = \mathbf{G}\,\frac{\mathbf{m}^2_0}{\mathbf{L}^2_0} = 6.67259 \times 10^{-11}\frac{(9.1093897 \times 10^{-31})^2}{(3.83 \times 10^{-13})^2} = 3.76 \times 10^{-46} \; kg\;m/s^2 \qquad 12.28a$$

where: $\mathbf{e}_-$ and $\mathbf{e}_+$ are the charges of $\mathbf{F}^-_\downarrow$ and $\mathbf{F}^+_\uparrow$ at Golden mean (GM) conditions (see paragraph 4.1 and eq. 4.18), equal to rest charge of the electron, in accordance to our model of elementary particles: $\mathbf{e}_0 = 1.602 \times 10^{-13}$ C.

The radius of rotation of this pair is equal to Compton radius at GM conditions (eq.5.4): $\mathbf{L}^\phi = \mathbf{L}_0 = \hbar/\mathbf{m}_0\mathbf{c} \simeq 3.83 \times 10^{-13}$ m. Assuming, that permittivity of Bivacuum between charges in pair $[\mathbf{F}^-_\downarrow \bowtie \mathbf{F}^+_\uparrow]_C$ is close to that of vacuum: $\boldsymbol{\varepsilon}^\phi \simeq \boldsymbol{\varepsilon}_0 = 8.85 \times 10^{-12}$ F m$^{-1}$, we get for Coulomb attraction force $\mathbf{F}_{Coul}^{\phi} = 1.98 \times 10^{-2}$ $kg\ m/s^2$.

The gravitational constant in (12.28a) $\mathbf{G} = 6.67259 \times 10^{-11}$ m$^3$ kg$^{-1}$ s$^{-2}$ and the rest mass of the electron squared: $\mathbf{m}_0^2 = \left(9.1093897 \times 10^{-31}\ kg\right)^2$. It is easy to see, that gravitational attraction is negligible small as respect to Coulomb one.

The calculated Coulomb force (12.28) is close to the opposite centrifugal force (12.27), providing stabilization of pairs $[\mathbf{F}_{\downarrow} \bowtie \mathbf{F}_{\uparrow}^{+}]_C^{\phi}$ in triplets of the electrons:

$$\frac{\mathbf{F}_{Coul}^{\phi}}{\mathbf{F}_{cf}^{\phi}} = \frac{1.98 \times 10^{-2}}{1.47 \times 10^{-2}} = 1.343 \qquad 12.29$$

A possible explanation of this small disbalance in Coulomb and centrifugal forces, can be a bigger permittivity of Bivacuum in the internal space of this pairs, as respect to empty Bivacuum/vacuum: $\boldsymbol{\varepsilon}^{\phi}/\boldsymbol{\varepsilon}_0 = 1.343$. The reason of bigger internal permittivity $\boldsymbol{\varepsilon}^{\phi} = 1/\boldsymbol{\mu}_0\mathbf{c}_{\phi}^2$ can be a bigger refraction index in space between two sub-elementary fermions in pairs $[\mathbf{F}_{\downarrow} \bowtie \mathbf{F}_{\uparrow}^{+}]_C^{\phi}$.

Like in the case of protons (see section 5.1), stabilization of electronic triplets in its [W] phase can be realized via electronic gluons, i.e. superposition of their Cumulative virtual clouds $[\mathbf{CVC}^{+} \bowtie \mathbf{CVC}^{-}]^e$ between paired sub-elementary fermions in [W] phase.

The close values of centrifugal and Coulomb interaction for the electrons and positrons, calculated on the base of parameters of paired sub-elementary fermions in their Corpuscular phase (angular frequency of $[\mathbf{C} \rightleftharpoons \mathbf{W}]$ pulsation and tangential velocity of their rotation), following from our model of elementary particles, is important fact, confirming our Unified theory of Bivacuum, the new model of stable elementary particles and time.

For much less stable triplet, like muon, the centrifugal force at Golden mean conditions (12.27a) exceeds many times the Coulomb attraction between its sub-elementary fermion and antifermion:

$$\mathbf{F}_{cf}^{\phi} = \frac{2\mathbf{m}_0\ \phi\mathbf{c}^2}{\mathbf{L}_0} = \frac{2}{\hbar}\mathbf{m}_0^2\ \phi\mathbf{c}^3 >> \frac{\mathbf{e}^2}{\boldsymbol{\varepsilon}_0\mathbf{L}_0^2} = \mathbf{F}_{Coul} \qquad 12.29a$$

This inequality is a result of the same charges of muon and electron at the mass of former exceeding the mass of latter about 200 times. It is a reason of muons much less stability and life-time, than that of electrons.

### 12.6 Shift of the period of elementary oscillations in gravitational field

The decreasing of the wavelength of photons (EM waves) and corresponding decreasing of their period in a gravitational field, predicted by general relativity theory (GRT), is dependent on mass ($\mathbf{M}$) and distance ($\mathbf{r}$) from center of mass to photons location and detection as:

$$\frac{\boldsymbol{\lambda}_G}{\boldsymbol{\lambda}_0} = \frac{\mathbf{T}_G}{\mathbf{T}_0} = \sqrt{1 - \frac{2GM}{c^2 r}} \qquad 12.30$$

$$or : \ \mathbf{T}_G \simeq \mathbf{T}_0\left(1 - \frac{GM}{c^2 r}\right) \ \ \text{at} \ \ \frac{2GM}{c^2 r} << 1 \qquad 12.30a$$

A heuristic Newtonian derivation gives similar result as (12.30a):

$$\frac{T_G}{T_0} = \frac{v_0}{v_G} = \frac{\lambda_G}{\lambda_0} = \frac{hc}{\lambda_0}\frac{\lambda_G}{hc} = \qquad\qquad 12.31$$

$$= \frac{E_0}{E_G} = \frac{m_G c^2 - \frac{GMm_G}{r}}{m_G c^2} = 1 - \frac{GM}{c^2 r} \qquad 12.31a$$

where: $T_G$, $v_G$ and $\lambda_G$ are the shifted by G - field period, frequency and wave length of elementary wave; $h$ is Planck's constant, $c$ is the speed of light, $E_0$ is the unperturbed energy, $E_G$ is the shifted energy; $m_G$ is the effective mass of photon in field.

In the absence of gravitational field, when $M = 0$ or $r = \infty$, the period of oscillation is maximum $\mathbf{T}_G \simeq \mathbf{T}_0$.

As far the Newtonian gravitational force can be expressed via gravitational acceleration $(a_G = G\frac{M}{r^2})$ as:

$$\mathbf{F}_G = G\frac{M\,m}{r^2} = a_G m \qquad\qquad 12.32$$

$$where: \quad a_G = G\frac{M}{r^2} = g \qquad\qquad 12.32a$$

Near surface of the Earth this acceleration is equal to free fall acceleration:
$a_G = g = 9.8$ m/s$^2$.

Using (12.31a), formula (12.30a) can be presented as:

$$\mathbf{T}_G \simeq \mathbf{T}_0\left(1 - \frac{a_G r}{c^2}\right) = \mathbf{T}_0\left(1 - \frac{GM}{c^2 r}\right) \qquad 12.32$$

In accordance to this formula, the period of oscillation ($\mathbf{T}_G$) of test system, like photon or electron [$\mathbf{C} \rightleftharpoons \mathbf{W}$] pulsation period, should increase with increasing of separation between the test system and center of gravitation body ($r$). The same result we get from our (12.14) in nonrelativistic conditions: $(\mathbf{v}/c)^2 \ll 1$.

For the other hand, from (12.30a) it follows that increasing of ($r$) at permanent $M$ should increase the period of pulsation ($\mathbf{T}_G$) and *decrease its frequency* - red Doppler shift.

The experiment for confirmation of described above consequences of General relativity theory (GR) was set up by Pound and Rebka (1959) in the Harvard tower, using Mössbauer effect. The Harvard tower is just 22.6 m, so the fractional gravitational red shift between the frequency $v^{bottom}$ of $\gamma$ −quantums *emitted at the bottom* of tower and frequency $v^{top}$ *absorbed at the top* of tower predicted by GRT, similar to simple classical approach (12.31), is given by the formula:

$$\frac{\Delta E}{E} = \frac{\mathbf{v}^{bottom} - \mathbf{v}^{top}}{\mathbf{v}^{top}} = \frac{T^{top} - T^{bottom}}{T^{top}} = \frac{Gl}{c^2} = 2.45 \times 10^{-15} \qquad 12.33$$

where: $\mathbf{G}$ is the gravitational constant; $l = r_2 - r_1 = 22.6\ m$ is the tower height and $c$ is the speed of light.

Pound and Rebka used the 14.4 keV gamma ray from the iron-57 isotope that has a high enough resolution to detect such a small difference in energy and frequency:
$\Delta E = h(\mathbf{v}^{bottom} - \mathbf{v}^{top})$. In other set of experiments the source of $\gamma$ −quantum was placed at the top of tower and detector at the bottom.

The predicted theoretically relative frequency shifts on the upward and downward paths where opposite by sign, but the same by absolute values. Their sum: $4.9 \times 10^{-15}$ appears to be very close to measured: $5.1 \times 10^{-15}$. Consequently, as it follows from our formula for period of elementary pulsations (12.14), it is smaller in locations, where gravitational or centrifugal accelerations are bigger.

The coincidence of quantitative experimental relative shifts values with theoretical ones, following from GTR and simple classical Newton's formalism (12.31a) is excellent.

However, it does not contain a strong evidence that GTR works better, than classical Newtonian approach.

### 12.7 The explanation of Hefele-Keating experiments

The additional confirmation of validity of our formula for time (12.14) is its ability to explain well known experiments of Hefele-Keating (1971) for verification of special and general theories of relativity (SR and GR).

They flew four *cesium atomic clocks* around the Earth in jets, first eastbound, then westbound. These experiments proved that atomic clocks period is dependent on the direction, velocity and altitude of jet airplanes. The direction and velocity of the airplanes where factors of the SR and the altitude was a factor of GR.

Compared to the time kept by control atomic clock fixed on the ground (USA), the *eastbound* clocks on the jets where slower (period of oscillation bigger) and *westbound clocks* - faster (period of oscillation shorter).

The velocity of *eastbound* clocks are the sum of tangential velocity of jet and tangential velocity of atmosphere at the altitude of jet flight: $\mathbf{v}_{res}^{east} = \mathbf{v}_{jet}{}' + \mathbf{v}_{at}$. For the other hand, the resulting velocity of *westbound* clock is a difference of these velocities: $\mathbf{v}^{west} = \mathbf{v}_{jet}{}' - \mathbf{v}_{at}$. The correct position of reference clock (non rotating) should be at the axes of the earth rotation (i.e. poles) of the earth. The velocity of the earth orbiting around the Sun and Sun system velocity in the universe was not taken into account.

Webster Kehr (2002) in his book "The detection of Ether" points out, that in original version of special relativity (1905) each of jets flying with permanent velocity should be considered as the *rest* reference frames.

However, even in such approximate approach, where the *local reference frames* instead Universal reference frame (URF) was used, Hafele and Keating found out, that the time effects, *calculated* using relativity theory, *coincide well* with experimental ones.

We will show below, that these experiments can be explained also on the base of our theory of time and simple Newtonian formula for gravitation and free fall acceleration, as a part of Unified theory.

The free fall acceleration following from Newton formula (12.32 and 12.32a) is:

$$a_G = (d\mathbf{v}/\mathbf{dt})_G = G\frac{M}{r^2} = g \qquad\qquad 12.34$$

Formula (12.14) can be presented in form, interrelating characteristic time of object with gravitational free fall acceleration ($a_G = g$), velocity of object and the increments of these parameters at permanent velocity:

$$\frac{\mathbf{T}^{ext}}{4\pi} \overset{\mathbf{v}<<\mathbf{c}}{\simeq} \mathbf{t}^{ext} = \frac{\vec{\mathbf{v}}\, r^2}{\mathbf{GM}} \frac{1 - (\mathbf{v}/\mathbf{c})^2}{2 - (\mathbf{v}/\mathbf{c})^2} \qquad\qquad 12.35$$

$$or: \left[\frac{1}{4\pi}\Delta\mathbf{T}^{ext}\right]_{\mathbf{v}=const} \overset{\mathbf{v}<<\mathbf{c}}{\simeq} \frac{1}{2\mathbf{GM}}(2\vec{\mathbf{v}}\, r\Delta r + r^2\Delta\vec{\mathbf{v}}) \qquad\qquad 12.35a$$

where: $T = 2\pi/\omega$ is the period of elementary oscillation in external reference frame (atomic clock in private case).

*Formula (12.35) interrelate our concept of time with gravitation, however, in different way, than general theory of relativity.*

At permanent tangential velocity of jets respectively to the Earth surface: $\mathbf{v} = const$, $\Delta\mathbf{v} = 0$ for nonrelativistic case: $\mathbf{v} << \mathbf{c}$ we get from (12.35a) the confirmation of (12.33), that the external period is increasing and frequency decreasing with distance from the earth

center:

$$[\Delta T = -\Delta\mathbf{v}]_{\mathbf{v}=\text{const}}^{ext} \overset{\mathbf{v}<<\mathbf{c}}{\simeq} 4\pi\frac{\mathbf{v}\,r\Delta r}{\mathbf{GM}} = 4\pi\frac{\mathbf{v}}{\mathbf{g}}\frac{\Delta r}{r} \qquad 12.36$$

where: $\Delta r = r_2 - r_1$ in private case corresponds to $l$ in eq.(12.33).

For the other case of permanent distance to the Earth center and surface: $r = const$; $\Delta r = 0$ and (12.35a) turns to:

$$[\Delta T = -\Delta\mathbf{v}]_{r=\text{const}} \overset{\mathbf{v}<<\mathbf{c}}{\simeq} 2\pi\frac{r^2\Delta\vec{\mathbf{v}}}{\mathbf{GM}} = 2\pi\frac{\Delta\vec{\mathbf{v}}}{\mathbf{g}} \qquad 12.37$$

where: $\mathbf{G} = 6.67259 \times 10^{-11}\,\mathrm{m^3\,kg^{-1}\,s^{-2}}$; $\mathbf{M} = 5.9742 \times 10^{24}$ kg is the earth mass; $r = 6.378164 \times 10^6$ m is the equatorial radius of the Earth; $\mathbf{g} = \mathbf{GM}/r^2 = 9.8$ m/s$^2$ free fall acceleration.

From this formula we can see, that as far velocity of *eastbound* clocks are the sum of tangential velocity of jet and tangential velocity of atmosphere at the altitude of jet flight: $\mathbf{v}_{res}^{east} = \mathbf{v}_{jet}{}' + \mathbf{v}_{at}$, the period of atomic clock should increase - time is slowing down. For the *westbound* clock the decreasing of actual velocity of clock: $\mathbf{v}^{west} = \mathbf{v}_{jet}{}' - \mathbf{v}_{at}$ should decrease the period of atomic clock and they show 'faster' time. These consequences are in total accordance with experiment of Hafele-Keating (1971).

### 12.8 Interrelation between period of the Earth rotation, its radius, free fall acceleration and tangential velocity

If we take the local reference frame, as a center of Earth, where the tangential velocity is zero ($\mathbf{v}_{tn} = 0$; $\Delta\mathbf{v}_{tn} = 0$), then the time and frequency increments should be also zero , as it follows from both formulas (12.36 and 12.37): $[\Delta T = -\Delta\mathbf{v}]_{\mathbf{v}=0;\,r=0} = 0$

The tangential velocity of the point on the Earth surface rotation is:

$$\mathbf{v}_{Earth}^{tn} = 2\pi r/T_{Earth} = \frac{6.28 \times 6.378164 \times 10^6\,\mathrm{m}}{24 \times 60 \times 60\,\mathrm{s}} = \frac{4.0 \times 10^7}{0.864 \times 10^5} = 4.63 \times 10^2\,\mathrm{m/s} \qquad 12.38$$

where: $T_{Earth} = 24\,h = 8.64 \times 10^4 s$ is the period of the Earth rotation.

We may assume, that the atmosphere of the Earth has the same tangential velocity, i.e. rotate with Earth.

The velocity of jet as respect to this rotating atmosphere is about $\mathbf{v}_{jet} = 700 km/h = 2 \times 10^2$ m/s.

Putting value (12.38) and others in (12.36) and assuming $(\Delta r/r) = 1$, we get for corresponding increment of period, corresponding to change of the radius of rotation from zero to the earth radius:

$$T_{Earth}^{cal} \sim [\Delta T]_{r=\text{const}} = 4\pi\frac{\mathbf{v}}{\mathbf{g}}\frac{\Delta r}{r} = 12.56\frac{4.63 \times 10^2}{9.8} = 5.93 \times 10^3\mathrm{s} \qquad 12.39$$

*This calculated value is about 15 times less*, than real period of the Earth rotation: $T_{Earth}/T_{Earth}^{cal} \simeq 15$. This discrepancy may be a result of following factors:

1) The opposite direction of rotation of the inner volumes of the earth, for example its nuclear, as respect to its surface core, keeping the resulting angular momentum equal to zero:

$$M_{ext}\mathbf{v}_{ext}\,\mathbf{r}_{ext} + M_{in}\mathbf{v}_{in}\,\mathbf{r}_{in} = 0 \qquad 12.40$$

where $M_{ext}$; $\mathbf{v}_{ext}$; and $\Delta\mathbf{r}_{ext}$ are the averaged mass, velocity and effective radius of corresponding regions of the earth, rotation in opposite direction.

This factor may strongly increase the effective tangential velocity of the earth surface (**v**) as respect to axis of its rotation in (12.39).

2) nonlinear dependence of (**g**) on the distance from center of the Earth in the internal region of planet, i.e. $\mathbf{g} = \mathbf{f}(\Delta r/r)$;

3) contribution to (**v**) in (12.39) the Earth velocity motion on the orbit around Sun $(30 \times 10^3$ m/s) and Solar system in the Universe $(370 \times 10^3$ m/s);

4) slowing down the frequency of the Earth rotation with time (billions of years) due to different kind of energy dissipation, like interaction with moon, etc.

Formula (12.36) points to qualitatively similar time effects, as general relativity and our formula (12.37) to the same effects, as special relativity when $\mathbf{v} << \mathbf{c}$.

Consequently, our Unified theory, including new approach to time problem and accepting simple Newtonian formula for gravitational force, can explain all most important experiments, which where used for confirmation of special and general relativity.

The time in our approach is a characteristic parameter of any closed system (classical and quantum) dynamics, involving not only velocity but also acceleration. In contrast to time definition, following from special relativity (12.15), the time in our Unified theory is infinitive and independent on velocity in any inertial system of particles, when $(\mathbf{dv/dt}) = \mathbf{0}$.

*However, at any nonzero acceleration (dv/dt) = const > 0 the time is dependent on velocity of these objects in more complex manner, than it follows from special relativity. In fact, there are no physical systems in our expanding with acceleration Universe, formed by rotating galactics and stabilized by gravitational field, which can be considered, as perfectly inertial, i.e. where the acceleration is absent totally. This means, that conventional relativistic formula for time (12.15) is not applicable for real physical systems in general case.*

### 13. The Virtual Replica (VR) of Material Objects and its Multiplication (VRM)

Theory of Virtual Replica (**VR**) of material objects in Bivacuum and **VR** Multiplication in space and time: **VRM(r,t)** is proposed. The *primary* $\mathbf{VR}_0$ in initial time represents a three-dimensional (3D) superposition of Bivacuum virtual standing waves $\mathbf{VPW}_m^{\pm}$ and $\mathbf{VirSW}_m^{\pm 1/2}$, modulated by $[\mathbf{C} \rightleftharpoons \mathbf{W}]$ pulsation of elementary particles and translational and librational de Broglie waves of molecules of macroscopic object (http://arxiv.org/abs/physics/0207027).

*For the end of energy, charge and spin conservation* in Bivacuum, we have to assume, that symmetry shifts of Bivacuum dipoles, involved in **VR** formation, should compensate each other. This condition is satisfied, if we assume, that the primary and secondary **VR** is formed by certain number (N) of virtual Cooper pairs of Bivacuum fermions and antifermions of opposite spins and symmetry shifts:

$$\mathbf{VR} = \sum_{n}^{N} [\mathbf{BVF}^{\uparrow} \bowtie \mathbf{BVF}^{\downarrow}]_n$$

The *isotropic* infinitive multiplication of primary $\mathbf{VR}_0$ in space and time in form of 3D packets of virtual standing waves, representing huge number (M) of *secondary* $\mathbf{VR}_m$, is a result of interference of all pervading external coherent basic *reference waves* - Bivacuum Virtual Pressure Waves ($\mathbf{VPW}_{q=1}^{\pm}$) and Virtual Spin Waves ($\mathbf{VirSW}_{q=1}^{\pm 1/2}$) with similar kinds of modulated standing waves, like that, forming the primary **VR** and changing of the object itself with time. The latter has a properties of the *object waves*. Consequently, the **VRM(r,t)**, as a result of mixing of the *object waves* with *reference waves* can be named

**Holoiteration** by analogy with hologram (in Greece *'holo'* means the 'whole' or 'total'). The **VRM(r,t)** can be considered as a result of linear superposition of primary **VR₀** of different states with corresponding amplitude of probability ($c_m$):

$$\mathbf{VRM(r,t)} = \sum_m^M c_m[\mathbf{VR}_m >_m$$

The frequencies of basic reference virtual pressure waves (**VPW**$_{q=1}^{\pm}$ ≡ **VPW**$_0^{\pm}$) and virtual spin waves (**VirSW**$_{q=1}^{\pm 1/2}$ ≡ **VirSW**$_0^{\pm 1/2}$) of Bivacuum are equal to Compton frequencies of three electron generation ($i = e, \mu, \tau$):

$$[\boldsymbol{\omega}_{VPW_0} = \boldsymbol{\omega}_{VirSW_0} = \boldsymbol{\omega}_0 = \boldsymbol{\omega}_{\mathbf{C \rightleftharpoons W}}^{\mathbf{v=0}} = \mathbf{m}_0\mathbf{c}^2/\hbar]^i \qquad 13.1$$

The *Bivacuum virtual pressure waves modulation (***VPW**$_\mathbf{m}^{\pm}$) can be realized by pairs of positive and negative cumulative virtual clouds (**CVC**$^+$ ⋈ **CVC**$^-$), emitted/absorbed in the process of [**C ⇌ W**] pulsation of *pairs:* [**F**$_\uparrow^+$ ⋈ **F**$_\downarrow^-$]$_C$ ⇌ [**F**$_\uparrow^+$ ⋈ **F**$_\downarrow^-$]$_W$ of elementary triplets (electrons, protons, neutrons) < [**F**$_\uparrow^+$ ⋈ **F**$_\downarrow^-$] + **F**$_\updownarrow^\pm$ >$^i$ of the object. These kinds of waves superposition are responsible for gravitational attraction or repulsion between two or more objects and do not depend on the charge of triplets (http://arxiv.org/abs/physics/0207027).

The *Bivacuum virtual spin waves modulation* (**VirSW**$^{\pm 1/2}$) can be a consequence of *recoil angular momentum oscillation*, accompanied the **CVC**$^\pm$ emission ⇌absorption in the process of [**C ⇌ W**] pulsation of *unpaired* sub-elementary fermion or antifermion **F**$_\updownarrow^\pm$ >$^i$ of triplets:

$$[(\mathbf{F}_\uparrow^+ \bowtie \mathbf{F}_\downarrow^-) + (\mathbf{F}_\updownarrow^\pm)_W] \underset{-\mathbf{CVC}^\pm + \mathbf{Antirecoil}}{\overset{+\mathbf{CVC}^\pm - \mathbf{Recoil}}{<\!\!=\!\!=\!\!=\!\!=\!\!=\!\!=>}} [(\mathbf{F}_\uparrow^+ \bowtie \mathbf{F}_\downarrow^-)_W + (\mathbf{F}_\updownarrow^\pm)_C] \qquad 13.1a$$

The recoil energy of the in-phase [**C ⇌ W**] pulsation of a sub-elementary fermion **F**$_\uparrow^+$ and antifermion **F**$_\downarrow^-$ of pair [**F**$_\downarrow^-$ ⋈ **F**$_\uparrow^+$] and the angular momenta of CVC$^+$ and CVC$^-$ of **F**$_\uparrow^+$ and **F**$_\downarrow^-$ in pairs compensate each other and the resulting recoil momentum and energy of [**F**$_\downarrow^-$ ⋈ **F**$_\uparrow^+$] is zero.

The stability of VR of object, as a *hierarchical system of quantized metastable torus-like and vortex filaments structures formed by* **VPW**$_\mathbf{m}^{\pm}$ *and by* **VirSW**$_\mathbf{m}^{\pm 1/2}$ *excited by paired and unpaired sub-elementary fermions, correspondingly,* in superfluid Bivacuum, could be responsible for so-called "**phantom effect**" of object after its destroymet or removing to remote place.

For free elementary particles the notion of secondary virtual replica, as one of multiplicated primary VR₀ coincide with notion of one of possible *'anchor sites'*, as a conjugated dynamic complex of three Cooper pair of asymmetric fermions.

### 13.1 Bivacuum perturbations, induced by dynamics of triplets and their paired sub-elementary fermions

In contrast to the situation with unpaired sub-elementary fermion (**F**$_\updownarrow^\pm$) in triplets, the recoil/antirecoil momenta and energy, accompanying the in-phase emission/absorption of **CVC**$_{\mathbf{S=+1/2}}^+$ and **CVC**$_{\mathbf{S=-1/2}}^-$ by **F**$_\uparrow^+$ and **F**$_\downarrow^-$ of pair [**F**$_\uparrow^+$ ⋈ **F**$_\downarrow^-$], compensate each other in the process of their [**C ⇌ W**] pulsation. Such pairs display the properties of neutral particles with zero spin and zero rest mass:

$$[\mathbf{F}_\uparrow^- \bowtie \mathbf{F}_\downarrow^+]_C \quad \xleftrightarrow{\frac{[\mathbf{E}_{CVC^+} + \mathbf{E}_{CVC^-}] + \Delta\mathbf{VP}^{\mathbf{F}_\uparrow^+ \bowtie \mathbf{F}_\downarrow^-}}{[\mathbf{E}_{CVC^+} + \mathbf{E}_{CVC^-}] - \Delta\mathbf{VP}^{\mathbf{F}_\uparrow^+ \bowtie \mathbf{F}_\downarrow^-}}} \quad [\mathbf{F}_\uparrow^- \bowtie \mathbf{F}_\downarrow^+]_W \qquad 13.2$$

The total energy increment of elementary particle, equal to that of each of sub-elementary fermions of triplet, generated in nonequilibrium processes, accompanied by entropy change, like melting, boiling, etc., can be presented in a few manners:

$$\Delta\mathbf{E}_{tot} = \Delta(\mathbf{m}_V^+ \mathbf{c}^2) = \Delta\left(\frac{\mathbf{m}_0 \mathbf{c}^2}{[1 - (\mathbf{v}/\mathbf{c})^2]^{1/2}}\right) = \qquad 13.3$$

$$= \frac{\mathbf{m}_0 \mathbf{v}}{\mathbf{R}^3}\Delta\mathbf{v} = \frac{\mathbf{p}}{\mathbf{R}^2}\Delta\mathbf{v} = \frac{\mathbf{h}}{\lambda_B \mathbf{R}^2}\Delta\mathbf{v}$$

$$or: \quad \Delta\mathbf{E}_{tot} = \Delta[(\mathbf{m}_V^+ - \mathbf{m}_V^-)\mathbf{c}^2(\mathbf{c}/\mathbf{v})^2] = \frac{2\mathbf{T}_k}{\mathbf{R}^2}\frac{\Delta\mathbf{v}}{\mathbf{v}} \qquad 13.4$$

$$or: \quad \Delta\mathbf{E}_{tot} = \frac{2\mathbf{T}_k}{\mathbf{R}^2}\frac{\Delta\mathbf{v}}{\mathbf{v}} = \Delta[\mathbf{R}(\mathbf{m}_0\mathbf{c}^2)_{rot}^{in}] + \Delta(\mathbf{m}_V^+ \mathbf{v}^2)_{tr}^{ext} \qquad 13.4a$$

where: $\mathbf{R} = \sqrt{1 - (\mathbf{v}/\mathbf{c})^2}$ is the relativistic factor; $\Delta\mathbf{v}$ is the increment of the external translational velocity of particle; the actual inertial mass of sub-elementary particle is: $\mathbf{m}_V^+ = \mathbf{m}_0/\mathbf{R}$; $\mathbf{p} = \mathbf{m}_V^+ \mathbf{v} = \mathbf{h}/\lambda_B$ is the external translational momentum of unpaired sub-elementary particle $\mathbf{F}_\uparrow^\pm >^i$, equal to that of whole triplet $< [\mathbf{F}_\uparrow^+ \bowtie \mathbf{F}_\downarrow^-] + \mathbf{F}_\uparrow^\pm >^i$; $\lambda_B = \mathbf{h}/\mathbf{p}$ is the de Broglie wave of particle; $2\mathbf{T}_k = \mathbf{m}_V^+ \mathbf{v}^2$ is a doubled kinetic energy; $\Delta\ln\mathbf{v} = \Delta\mathbf{v}/\mathbf{v}$.

The increments of *internal* rotational and *external* translational contributions to total energy of the de Broglie wave (see eq. 13.4a) are, correspondingly:

$$\Delta[\mathbf{R}(\mathbf{m}_0\mathbf{c}^2)_{rot}^{in}] = -2\mathbf{T}_k(\Delta\mathbf{v}/\mathbf{v}) \qquad 13.5$$

$$\Delta(\mathbf{m}_V^+ \mathbf{v}^2)_{tr}^{ext} = \Delta(2\mathbf{T}_k)_{tr}^{ext} = 2\mathbf{T}_k\frac{1+\mathbf{R}^2}{\mathbf{R}^2}\frac{\Delta\mathbf{v}}{\mathbf{v}} \qquad 13.5a$$

The time derivative of total energy of elementary de Broglie wave is:

$$\frac{d\mathbf{E}_{tot}}{dt} = \frac{2\mathbf{T}_k}{\mathbf{R}^2\mathbf{v}}\frac{d\mathbf{v}}{dt} = \frac{2\mathbf{T}_k}{\mathbf{R}^2}\frac{d\ln\mathbf{v}}{dt} \qquad 13.5b$$

Between the increments of energy of triplets, equal to that of unpaired $\Delta\mathbf{E}_{tot} = \Delta\mathbf{E}_{\mathbf{F}_\uparrow^\pm}$ and increments of modulated $\mathbf{CVC_m^+}$ and $\mathbf{CVC_m^-}$, emitted by pair $[\mathbf{F}_\uparrow^- \bowtie \mathbf{F}_\downarrow^+]$ in the process of $[\mathbf{C} \to \mathbf{W}]$ transition, the direct correlation is existing.

These cumulative virtual clouds modulated by particle's de Broglie wave ($\lambda_B = \mathbf{h}/\mathbf{m}_V^+ \mathbf{v}$): $\mathbf{CVC_m^+}$ and $\mathbf{CVC_m^-}$ of paired sub-elementary fermions, superimposed with basic virtual pressure waves ($\mathbf{VPW}_0^\pm$) of Bivacuum, turn them to the *object waves* ($\mathbf{VPW_m^\pm}$), necessary for virtual hologram of the object formation:

$$\Delta\mathbf{E}_{\mathbf{F}_\uparrow^+}^{\mathbf{F}_\uparrow^- \bowtie \mathbf{F}_\downarrow^+} = \frac{\mathbf{h}}{\lambda_B \mathbf{R}^2}\Delta\mathbf{v} = \frac{2\mathbf{T}_k}{\mathbf{R}^2}\Delta\ln\mathbf{v} \quad \xrightarrow{\mathbf{CVC_m^+}} \Delta(\mathbf{VPW_m^+}) \qquad 13.6$$

$$-\Delta\mathbf{E}_{\mathbf{F}_\downarrow^-}^{\mathbf{F}_\uparrow^- \bowtie \mathbf{F}_\downarrow^+} \quad \xrightarrow{\mathbf{CVC_m^-}} \Delta(\mathbf{VPW_m^-}) \qquad 13.6a$$

The virtual pressure waves represent oscillations of corresponding virtual pressure ($\mathbf{VirP}_m^\pm$).

The increment of total energy of fermion or antifermion, equal to increment of its unpaired sub-elementary fermion can be presented via increments of paired sub-elementary fermions (13.5 and 13.5a), like:

$$\Delta \mathbf{E}_{tot} = \Delta \mathbf{E}_{\mathbf{F}_{\updownarrow}^{\pm}} = \frac{1}{2}\left(\Delta \mathbf{E}_{\mathbf{F}_{\uparrow}^{+}}^{\mathbf{F}_{\uparrow}^{+} \bowtie \mathbf{F}_{\downarrow}^{-}} - \Delta \mathbf{E}_{\mathbf{F}_{\downarrow}^{-}}^{\mathbf{F}_{\uparrow}^{+} \bowtie \mathbf{F}_{\downarrow}^{-}}\right) + \frac{1}{2}\left(\Delta \mathbf{E}_{\mathbf{F}_{\uparrow}^{+}}^{\mathbf{F}_{\uparrow}^{+} \bowtie \mathbf{F}_{\downarrow}^{-}} + \Delta \mathbf{E}_{\mathbf{F}_{\downarrow}^{-}}^{\mathbf{F}_{\uparrow}^{+} \bowtie \mathbf{F}_{\downarrow}^{-}}\right) = \Delta \mathbf{T}_{k}^{+} + \Delta \mathbf{V}^{+} \qquad 13.7$$

where, the contributions of the kinetic and potential energy increments to the total energy increment, interrelated with increments of positive and negative virtual pressures ($\Delta \mathbf{VirP}^{\pm}$), are, correspondingly:

$$\Delta \mathbf{T}_{k} = \frac{1}{2}\left(\Delta \mathbf{E}_{\mathbf{F}_{\uparrow}^{+}}^{\mathbf{F}_{\uparrow}^{+} \bowtie \mathbf{F}_{\downarrow}^{-}} - \Delta \mathbf{E}_{\mathbf{F}_{\downarrow}^{-}}^{\mathbf{F}_{\uparrow}^{+} \bowtie \mathbf{F}_{\downarrow}^{-}}\right) \sim \frac{1}{2}(\Delta \mathbf{VirP}^{+} - \Delta \mathbf{VirP}^{-}) \sim \alpha \Delta (\mathbf{m}_{V}^{+} \mathbf{v}^{2})_{\mathbf{F}_{\updownarrow}^{\pm}} \qquad 13.8$$

$$\Delta \mathbf{V} = \frac{1}{2}\left(\Delta \mathbf{E}_{\mathbf{F}_{\uparrow}^{+}}^{\mathbf{F}_{\uparrow}^{+} \bowtie \mathbf{F}_{\downarrow}^{-}} + \Delta \mathbf{E}_{\mathbf{F}_{\downarrow}^{-}}^{\mathbf{F}_{\uparrow}^{+} \bowtie \mathbf{F}_{\downarrow}^{-}}\right) \sim \frac{1}{2}(\Delta \mathbf{VirP}^{+} + \Delta \mathbf{VirP}^{-}) \sim \beta \Delta \left(\mathbf{m}_{V}^{+} + \mathbf{m}_{V}^{-}\right)\mathbf{c}_{\mathbf{F}_{\updownarrow}^{\pm}}^{2} \qquad 13.8a$$

*The specific information of any object is imprinted in its Virtual Replica (VR)*, because cumulative virtual clouds ($\mathbf{CVC}_{\mathbf{m}}^{\pm}$) of the object's elementary particles and their superposition with Bivacuum pressure waves and Virtual spin waves: $\mathbf{VPW}_{\mathbf{m}}^{\pm}$ and $\mathbf{VirSW}_{\mathbf{m}}^{\pm 1/2}$ are modulated by frequency, phase and amplitude of de Broglie waves of molecules, composing this object. Comparing eqs. 8.10ab and 13.8a we may see, that the modulated gravitational virtual pressure waves form a part of VR.

### 13.2 Modulation of Virtual Pressure Waves ($VPW_q^{\pm}$) and Virtual Spin Waves ($VirSW_q^{\pm 1/2}$) of Bivacuum by molecular translations and librations

The external translational/librational kinetic energy of particle ($\mathbf{T}_{k})_{tr,lb}$ is directly related to its de Broglie wave length ($\boldsymbol{\lambda}_{B}$), the group ($\mathbf{v}$), phase velocity ($\mathbf{v}_{ph}$) and frequency ($\mathbf{v}_{B} = \boldsymbol{\omega}_{B}/2\pi$):

$$\left(\boldsymbol{\lambda}_{B} = \frac{\mathbf{h}}{\mathbf{m}_{V}^{+}\mathbf{v}} = \frac{\mathbf{h}}{2\mathbf{m}_{V}^{+}\mathbf{T}_{k}} = \frac{\mathbf{v}_{ph}}{\mathbf{v}_{B}} = 2\pi\frac{\mathbf{v}_{ph}}{\boldsymbol{\omega}_{B}}\right)_{tr,lb} \qquad 13.9$$

where the de Broglie wave frequency is related to its length and kinetic energy of particle as:

$$\left[\mathbf{v}_{B} = \frac{\boldsymbol{\omega}_{B}}{2\pi} = \frac{h}{2\mathbf{m}_{V}^{+}\boldsymbol{\lambda}_{B}^{2}} = \frac{\mathbf{m}_{V}^{+}\mathbf{v}^{2}}{2h}\right]_{tr,lb} \qquad 13.10$$

The total energy/frequency of de Broglie wave and *resulting* frequency of pulsation ($\boldsymbol{\omega}_{\mathbf{C} \rightleftharpoons \mathbf{W}})_{tr,lb}$ (see eq. 7.4) is a result of modulation/superposition of the internal frequency, related to the rest mass of particle, by the external most probable frequency of de Broglie wave of the whole particle ($\boldsymbol{\omega}_{B})_{tr,lb}$, determined by its most probable external momentum: ($\mathbf{p} = \mathbf{m}_{V}^{+}\mathbf{v})_{tr,lb}$, related to translations or librations:

$$[\mathbf{E}_{tot} = \mathbf{m}_{V}^{+}\mathbf{c}^{2} = \hbar\boldsymbol{\omega}_{\mathbf{C} \rightleftharpoons \mathbf{W}}]_{tr,lb} = \mathbf{R}(\hbar\boldsymbol{\omega}_{0})_{rot}^{in} + (\hbar\boldsymbol{\omega}_{B}^{ext})_{tr,lb} = \mathbf{R}(\mathbf{m}_{0}\boldsymbol{\omega}_{0}^{2}\mathbf{L}_{0}^{2})_{rot}^{in} + \left(\frac{h^{2}}{\mathbf{m}_{V}^{+}\boldsymbol{\lambda}_{B}^{2}}\right)_{tr,lb}^{ext} \qquad 13.10a$$

where relativistic factor: $\mathbf{R} = \sqrt{1 - (\mathbf{v}/\mathbf{c})^{2}}$ is tending to zero at $\mathbf{v} \rightarrow \mathbf{c}$.

In composition of condensed matter the value of ($\boldsymbol{\lambda}_{B})_{tr,lb}$ is bigger for librations than for translation of molecules. The corresponding most probable modulation frequencies of translational and librational de Broglie waves is possible to calculate, using our Hierarchic theory of condensed matter and based on this theory computer program (Kaivarainen, 2001; 2003; 2004; 2005).

The *frequencies* of Bivacuum virtual pressure waves ($\mathbf{VPW_m^\pm}$) and virtual spin waves ($\mathbf{VirSW_m^{\pm1/2}}$) are modulated by the *resulting* frequencies of de Broglie waves of the object molecules, related to librations ($\boldsymbol{\omega}_{lb}$) and translations ($\boldsymbol{\omega}_{tr}$), correspondingly.

The combinational resonance between the basic Bivacuum virtual waves ($q = 1$) and resulting frequency of $[\mathbf{C} \rightleftharpoons \mathbf{W}]$ pulsation of electrons, protons and neutrons, composing atoms and molecules of the object, is possible at conditions:

$$\boldsymbol{\omega}_{\mathbf{VPW_{q=1}^\pm}}^i = \mathbf{z}\,\mathbf{R}\,\boldsymbol{\omega}_0^i + \mathbf{g}\,\boldsymbol{\omega}_B^{tr} + \mathbf{r}\,\boldsymbol{\omega}_B^{lb} \cong \mathbf{z}\,\mathbf{R}\,\boldsymbol{\omega}_0^i + \mathbf{g}\,\boldsymbol{\omega}_B^{tr} \qquad 13.11$$

$$\boldsymbol{\omega}_{\mathbf{VirSW_{q=1}^{\pm1/2}}}^i = \mathbf{z}\,\mathbf{R}\,\boldsymbol{\omega}_0^i + \mathbf{g}\,\boldsymbol{\omega}_B^{tr} + \mathbf{r}\,\boldsymbol{\omega}_B^{lb} \cong \mathbf{z}\,\mathbf{R}\,\boldsymbol{\omega}_0^i + \mathbf{r}\,\boldsymbol{\omega}_B^{lb} \qquad 13.11a$$

$$\mathbf{R} = \sqrt{1 - (\mathbf{v}/\mathbf{c})^2}\,; \quad \mathbf{z}, \mathbf{g}, \mathbf{r} = 1, 2, 3 \dots \text{(integer numbers)}$$

Each of 24 collective excitations of condensed matter, introduced in our Hierarchic theory (Kaivarainen, 1995; 2001, 2004), has the own characteristic frequency and can contribute to Virtual Replica of the object.

In contrast to regular hologram, VR contains information not only about surface and shape properties of the object, but also about its internal properties.

Three kind of modulations: *frequency, amplitude and phase* of Bivacuum virtual waves ($\mathbf{VPW_m^\pm}$) and ($\mathbf{VirSW_m^{\pm1/2}}$) by $[\mathbf{C} \rightleftharpoons \mathbf{W}]$ pulsation of elementary particles of molecules and their de Broglie waves may be described by known relations (Prochorov, 1999):

1. *The frequencies* of virtual pressure waves ($\omega_{VPW^\pm}^M$) and spin waves ($\omega_{VirSW^\pm}^M$), *modulated* by translational and librational de Broglie waves of the object's molecules, can be presented as:

$$\boldsymbol{\omega}_{VPW_m^\pm}^M = \mathbf{R}\boldsymbol{\omega}_0^i + \Delta\boldsymbol{\omega}_B^{tr}\cos\boldsymbol{\omega}_B^{tr}\,t \qquad 13.12$$

$$\boldsymbol{\omega}_{VirSW_m^{\pm1/2}}^M = \mathbf{R}\boldsymbol{\omega}_0^i + \Delta\boldsymbol{\omega}_B^{lb}\cos\boldsymbol{\omega}_B^{lb}\,t \qquad 13.12a$$

The Compton pulsation frequency of elementary particles (section 1.4; 1.5) is equal to basic frequency of Bivacuum virtual waves at $\mathbf{q} = \mathbf{j} - \mathbf{k} = \mathbf{1}$:

$$\boldsymbol{\omega}_0^i = \mathbf{m}_0^i\mathbf{c}^2/\hbar = \boldsymbol{\omega}_{VPW_{q=1}^\pm,ViSW_{q=1}}^i \qquad 13.12b$$

Such kind of modulation is accompanied by two satellites with frequencies: $(\boldsymbol{\omega}_0^i + \boldsymbol{\omega}_B^{tr,lb})$ and $(\boldsymbol{\omega}_0^i - \boldsymbol{\omega}_B^{tr,lb}) = \Delta\boldsymbol{\omega}_{tr,lb}^i$. The latter is named frequency deviation. In our case: $\boldsymbol{\omega}_0^e\,(\sim 10^{21}s^{-1}) >> \boldsymbol{\omega}_B^{tr,lb}(\sim 10^{12}s^{-1})$ and $\Delta\boldsymbol{\omega}_{tr,lb} >> \boldsymbol{\omega}_B^{tr,lb}$.

The temperature of condensed matter and phase transitions may influence the modulation frequencies of de Broglie waves of its molecules.

2. *The amplitudes of virtual pressure waves (VPW$_m^\pm$) and virtual spin waves VirSW$_m^{\pm1/2}$ (informational waves) modulated by the object are dependent on translational and librational de Broglie waves frequencies as:*

$$\mathbf{A}_{VPW_m^\pm} \approx \mathbf{A}_0(\sin\mathbf{R}\boldsymbol{\omega}_0^i\mathbf{t} + \gamma\,\boldsymbol{\omega}_B^{tr}\sin\mathbf{t}\cdot\cos\boldsymbol{\omega}_B^{tr}t) \qquad 13.13$$

$$\mathbf{I}_{VirSW_m^\pm} \approx \mathbf{I}_0(\sin\mathbf{R}\boldsymbol{\omega}_0^i\mathbf{t} + \gamma\,\boldsymbol{\omega}_B^{lb}\sin\mathbf{t}\cdot\cos\boldsymbol{\omega}_B^{lb}\mathbf{t}) \qquad 13.13a$$

where: the informational/spin field amplitude is determined by the amplitude of Bivacuum fermions $[BVF^\uparrow \rightleftharpoons BVF^\downarrow]$ equilibrium constant oscillation: $\mathbf{I}_S \equiv \mathbf{I}_{\mathbf{VirSW^{\pm1/2}}} \sim \mathbf{K}_{BVF^\uparrow \rightleftharpoons BVF^\downarrow}(\mathbf{t})$

The index of frequency modulation is defined as: $\gamma = (\Delta\boldsymbol{\omega}_{tr,lb}/\boldsymbol{\omega}_B^{tr,lb})$. The carrying zero-point pulsation frequency of particles is equal to the basic frequency of Bivacuum virtual waves: $\omega_{VPW_0^\pm, ViSW_0}^i = \omega_0^i$. Such kind of modulation is accompanied by two satellites with frequencies: $(\omega_0^i + \omega_B^{tr,lb})$ and $(\omega_0^i - \omega_B^{tr,lb}) = \Delta\omega_{tr,lb}$. In our case: $\omega_0^e(\sim 10^{21}s^{-1}) \gg \omega_B^{tr,lb}(\sim 10^{12}s^{-1})$ and $\gamma \gg 1$.

The fraction of molecules in state of mesoscopic molecular Bose condensation (mBC), representing, coherent clusters (Kaivarainen, 2001a,b; 2004) is a factor, influencing the amplitude ($A_0$) and such kind of modulation of Virtual replica of the object.

3. *The phase modulated* $VPW_m^\pm$ *and* $VirSW_m^{\pm 1/2}$ by de Broglie waves of molecules, related to their translations and librations, can be described like:

$$\mathbf{A}_{VPW_m^\pm}^M = \mathbf{A}_0 \sin\left(\mathbf{R}\omega_0 \mathbf{t} + \Delta\varphi_{tr} \sin\omega_B^{tr}\mathbf{t}\right) \qquad 13.14$$

$$\mathbf{I}_{VirSW_m^{\pm 1/2}}^M = \mathbf{I}_0 \sin\left(\mathbf{R}\omega_0 \mathbf{t} + \Delta\varphi_{lb} \sin\omega_B^{lb}\mathbf{t}\right) \qquad 13.14a$$

The value of phase increment $\Delta\varphi_{tr,lb}$ of modulated virtual waves of Bivacuum ($\mathbf{VPW}_m^\pm$ and $\mathbf{VirSW}_m^{\pm 1/2}$), contains the information about geometrical properties of the object. The phase modulation takes place, if the phase increment $\Delta\varphi_{tr,lb}$ is independent on the modulation frequency $\omega_B^{tr,lb}$.

### 13.3 The superposition of internal and surface Virtual Replicas of the object, as the "Ether Body"

The superposition of individual microscopic $\mathbf{VR}_{mic}$ of the electrons, protons, neutrons and atoms/molecules of the object (internal and surface ones), formed by interference of de Broglie waves of these particles with basic virtual waves of Bivacuum ($\mathbf{VPW}_{q=1}^\pm$ and $\mathbf{VirSW}_{q=1}^{\pm 1/2}$), stands for *internal macroscopic virtual replica* of the object ($\mathbf{VR}^{in}$), describing its *internal* bulk properties. The overall shape of ($\mathbf{VR}^{in}$) should be close to shape of the object itself, for example, such as the human's body and it organs shape.

*Spatial stability of condensed systems* means that the macroscopic internal virtual replica: $\mathbf{VR}^{in} = \sum \mathbf{VR}_{mic}^{in}$, as a result of 3D standing waves superposition of microscopic $\mathbf{VR}_{mic}^{in}$ in superfluid Bivacuum, should have location of nodes, coinciding with the most probable positions of the atoms and molecules in condensed matter.

The superposition of coherent de Broglie waves of atoms and molecules in clusters, forming 3D standing waves B, determined by their librations and translations, represents the *mesoscopic Bose condensate:* [$\mathbf{mBC}$] (Kaivarainen, 2001 b,c). In accordance to our theory, this means also the coherent [$\mathbf{C} \rightleftharpoons \mathbf{W}$] pulsations of elementary particles of these molecules and atoms. The violation of this coherency is accompanied by density fluctuation and defects origination or cavitational fluctuations in solids and liquids.

The *surface macroscopic virtual replica of the object:* $\mathbf{VR}^{sur} = \sum \mathbf{VR}_{mic}^{sur}$ is a part of the **Ether body**. The mechanism of its origination is similar to *internal macroscopic virtual replica* of the object $\mathbf{VR}^{in} = \sum \mathbf{VR}_{mic}^{in}$. It is a result of modulation of Bivacuum virtual waves by de Broglie waves of elementary particles of the atoms and molecules on the surface of the object. Its dimension can exceed the dimensions of the object.

The superposition of the internal and surface virtual replicas corresponds to notion of the "*ether body*" in Eastern philosophy:

$$\mathbf{Ether\ Body} \equiv \mathbf{VR} = \mathbf{VR}^{in} + \mathbf{VR}^{sur} = \sum(\mathbf{VR}_{mic}^{in} + \mathbf{VR}_{mic}^{sur}) \qquad 13.15$$

Stability of hierarchic system of whirls, forming Ether Body, as a hierarchical system of virtual standing waves and *curls* in superfluid Bivacuum (like permanent circular currents and whirls in superfluid $^4\mathbf{He}$), could be responsible for so-called *"phantom effect"* of this object.

### 13.4  The infinitive spatial Virtual Replica Multiplication VPM(r).
### The "Astral" and "Mental" bodies, as a distant and nonlocal components of VRM(r)

The mechanism of primary *Virtual Replica Multiplication* (**VPM***)* have general features with hologram origination, however without photomaterials or screens, fixing **VR**. The role of coherent *reference waves* play unperturbed by the object basic Bivacuum virtual waves (**VPW**$_{q=1}^{\pm}$ and nonlocal **VirSW**$_{q=1}^{\pm 1/2}$). The role of *subject waves* is represented by the primary Virtual Replica of the object, containing information not only on shape/surface, but also about internal properties of the subject.

The VRM is a spatially *isotropic* process, like excitation of spherical waves. It can be subdivided on two components - *distant (translational)* and *nonlocal (rotational or librational)*:

1) the *distant component* of **VRM(r)**$^{dis}$ is a result of replication of the translational component of primary **VR** outside the volume of the object, by means of Virtual Pressure Waves (**VPW**$_{q=1}^{\pm}$). The front of 3D VRM$^{dis}$ in form of huge number of *secondary* VR isotropicaly expand in space like gravitation waves with light velocity (http://arxiv.org/abs/physics/0207027).

The virtual replica of the object can be reproducible in form of distant **VRM**$^{dis}$ (like the hologram) in any remote space regions, where the interference pattern of the *reference wave* **VPW**$_{q=1}^{\pm}$ with *primary* **VR** as the object wave, is existing. The volume of space, occupied by distant **VRM**$^{dis}$, is expanding with light velocity (**c**) during the life-time of primary **VR** and atoms, composing the object.

The expanding with light velocity population of **VR(ct)**, spatially separated from the body/object, may correspond to Eastern ancient notion of the *"astral body"*:

$$\textbf{Astral Body} = \sum^{t} \textbf{VR}_{tr}(\textbf{ct}) = \textbf{VRM}^{dis} \qquad\qquad 13.16$$

As far each individual secondary **VR** in population $\sum^{t} \textbf{VR(ct)}$ in the absence of dissipation in superfluid Bivacuum is the exact copy of the primary **VR**, they should be spatially also indistinguishable, like particles in state of Bose condensate. The detected by psychic or by special device secondary replica displays its properties

The dielectric permittivity ($\varepsilon_0$) and permeability ($\mu_0$) in the volume of the Astral bodies may differ from their averaged values in Bivacuum because of small charge symmetry shift in Bivacuum fermions (**BVF**$^{\updownarrow}$): $\Delta e = |e_+ - e_-| > 0$, induced by *recoil ⇌ antirecoil* effects, accompanied [**C ⇌ W**] pulsation of elementary particles. Consequently, the probability of atoms and molecules excitation and ionization (dependent on Coulomb interaction between electrons and nuclears), as a result of their thermal collisions with excessive kinetic energy, may be higher in volumes of the Astral bodies, than outside of them. This may explain their special optical effects - a shining of some objects phantoms (ghosts) in darkness, or their specific spectrogram, representing astral bodies. Their spatial instability of phantoms can be explained by spatial similarity of astral bodies, composing **VRM**$^{dis}$. The possibility of phenomena like *remote vision and remote healing* also follow from our holomovement like mechanism of **VRM(r)**$_S$ ⋈ **VRM(r)**$_R$ superposition of Sender and Receiver and their 'tuning'.

The sensitivity of Kirlian effect or Gas Discharge Visualization (GDV) to internal process of macroscopic object, like human body, also can be explained by specific properties of the surface Ether and Astral bodies, changing the probability of the air molecules excitation/ionization after thermal collision;

2) the *nonlocal component* of **VRM**$^{nl}$ is a result of 3D replication of the

rotational/librational component of *primary* virtual replica ($\mathbf{VR}_{lb}$) outside the volume of the object, by means of nonlocal (informational) Virtual Spin Waves ($\mathbf{VirSW}_{q=1}^{\pm}$), propagating in symmetric Bivacuum instantly, i.e. without light velocity limitation.

The *nonlocal macroscopic virtual replica multiplication* ($\mathbf{VRM}^{nl}$) or $\mathbf{VR}$ *iteration*, is a result of interference of modulated by librational de Broglie waves the *recoil $\rightleftharpoons$ antirecoil* effects the Bivacuum virtual spin waves: $\mathbf{VirSW}_{\mathbf{m}}^{\pm 1/2} - object\ spin\ waves$ with corresponding *reference spin waves* of Bivacuum ($\mathbf{VirSW}_{\mathbf{q}}^{\pm 1/2}$).

The Eastern notion of *mental body* may correspond to $\mathbf{VRM}^{nl}$, as a multiplication (holoiteration or holomovement after Bohm) of informational Virtual Replicas [$\mathbf{VR}_{lb}$]:

$$\mathbf{Mental\,Body} \;=\; \sum^{t} \mathbf{VR}_{lb} = \mathbf{VRM}^{nl} \qquad\qquad 13.17$$

Hierarchical superposition of huge number of Astral and Mental Bodies of all human population on the Earth can be responsible for Global Informational Field origination, like Noosphere, proposed by Russian scientist Vernadsky in the beginning of 20th century. The Astral and Mental bodies may partly be overlapped with Ether body. This provide the possibility of dynamic exchange interaction and feedback reaction between all three virtual bodies of the object: Ether, Astral and Mental.

*One important conjecture,* following from our approach to distant $\mathbf{VRM}^{dis}$ can be discussed. We proceed from the consequence of our theory, that the volume of space, occupied by distant $\mathbf{VRM}^{dis}$ is expanding isotropicaly with light velocity (**c**) in 3D space during the life-time of $\mathbf{VR}$ and atoms, composing the object.

The life span of the individual stable atoms, including hydrogen, carbons, oxygen, composing biological objects is comparable with life-time of the Universe, i.e. over ten billions of years. This means, that not only nonlocal $\mathbf{VRM}^{nl}$, but as well the distant $\mathbf{VRM}^{dis}$ of these atoms may involve all the Universe. It is a conditions of Virtual Guides of spin, momentum and energy ($\mathbf{VirG}_{SME}$) 3D net formation in the Universe, connecting virtually all similar and coherent elementary particles and atoms of the equal de Broglie wave length. We suppose, however, that only in composition of biosystems these atoms may become enough coherent and orchestrated to provide a strong enough *cumulative interaction* between Sender and Receiver, for example, between *psychic* and very remote objects (inorganic or biological) via 3D net of $\mathbf{VirG}_{SME}$ and $\mathbf{VRM}$ as a factors of Bivacuum mediated interaction (BMI). The construction of $\mathbf{VirG}_{SME}$ and mechanism of their action will be described in the next section.

*A complex Hierarchical system* $\sum \mathbf{VRM(r,t)}$ *of Solar system, galactics, including Noosphere, may be considered as Hierarchical quantum supercomputer or Superconsciousness, able to simulate all probable situations of virtual future and past. It is possible in conditions of time uncertainty: $t = 0/0$ when the translational velocity $\mathbf{v} = \mathbf{0}$ and accelerations* ($\mathbf{dv/dt}$) = 0 *in the volume of* $\sum \mathbf{VRM(r,t)}$ *are zero (Kaivarainen, 2005: http://arXiv.org/abs/physics/0103031 ).*

Our theory admit a possibility of feedback reaction between the iterated VR and primary one and between primary VR and the object physical properties. Consequently, the phenomena of most probable event anticipation by enough sensitive physical detectors and human beings (psychics) is possible in principle. This may explain the reproducible results of unconsciousness response (by changes of human skin conductance) of future events (presponse), obtained by Dick Bierman and Dean Radin (2002). However, in these experiments the possibility of influence of intention of participant on random events generator (REG), choosing next photo (calm or emotional), like in Bierman's experiments, also should be taken into account. Such kind of weak influence of humans intention on REG was demonstrated in long term studies of Danne and Jahn (2003).

*In contrast to virtual time, the reversibility of real time looks impossible,* as far it needs the reversibility of all dynamic process in Universe due to interrelations between closed systems of different levels of hierarchy. It is evident that such 'play back' of the Universe history needs the immense amount of energy redistribution in the Universe.

*All three described Virtual Replicas: Ether, Astral and Mental bodies are interrelated with each other and physical body.* The experimental evidences are existing, that between properties of the *Ether* bodies and corresponding physical bodies of living organisms or inorganic matter, the correlation takes a place. It is confirmed by the Kirlian effect, reflecting the ionization/excitation threshold of the air molecules in volume of Ether and astral bodies.

The perturbation of the Ether body of one object (Receptor) by the astral or mental body of the other object (Sender) can be imprinted in properties of physical body (condensed matter) of Receptor for a long time in form of subtle, but stable structural perturbations. The stability of such kind of informational 'taping' is determined by specific properties of material, as a matrix for imprinting. For example, water and aqueous systems, like biological ones, are very good for stable imprinting of virtual information and energy via introduced VRM and Virtual Guides (see next chapter). However, some 'sensitive' stones or other rigid materials have a much longer life span.

*The Ghost phenomena can be explained by reproducing of such imprinted in walls, cells and floor information, mediated by distant virtual replica multiplication (VRM$^{dis}$). The reproduction of VR from imprinted in condensed matter VRM$^{dis}$ is a process, similar to treatment of regular hologram by the reference waves. In the case of 'Ghost' the reference waves can be presented by basic VPW$_m^\pm$ and VirSW$_m^{\pm 1/2}$, modulated by special superposition of Virtual replicas of other objects, for example, Earth, Moon and Sun.*

The *nonlocal* Mental - Informational body formation in living organisms and humans, in accordance to our theory (Kaivarainen, 2001; 2003), is related to equilibrium shift of dynamic equilibrium of [assembly ⇌ disassembly] of coherent water clusters in microtubules of the neurons (librational effectons), accompanied series of elementary acts of consciousness in *nonequilibrium processes* of meditation, intention and braining. Corresponding variations of kinetic energy and momentum of water molecules can be transmitted  from Sender to remote Receiver via nonlocal virtual spin-momentum-energy guides **VirG**$_{SME}$ (**S <=> R**), described in next chapter.

In complex process of Psi phenomena, the first stage is a 'target searching' by nonlocal [mental body] of psychic, then formation of **VirG**$_{SME}$ (**S <=> R**)$^i$, then activation of psychic's [astral body] by its [ether body]. The latter can be interrelated with specific processes of physical body of psychic, like dynamics of water in microtubules of neurons ensembles, realizing elementary acts of perception and consciousness, in accordance to our model (Kaivarainen, 2000; 2005).

The possible mechanism of entanglement between microscopic and macroscopic objects, based on our Unified theory, will be described in Chapters 14 and 15.

### 13.5  Contributions of different kind of internal dynamics of matter to Virtual Replica of the object

For each of 24 selected collective excitation of condensed matter, considered in our Hierarchic theory of matter (Kaivarainen, 2000a), the averaged thermal vibrations contribution to VR of the object can be evaluated, using special computer program, named Comprehensive Analyzer of Matter Properties - CAMP.

The most effective source of coherent Virtual pressure waves (VPW$^\pm$) amplitude oscillations are the [disassembly ⇌ assembly] of coherent clusters, existing in liquids *(librational primary effectons)* and solids *(librational and translational primary effectons)*.

Such clusters are the result of the ambient temperature mesoscopic Bose condensation (*mBC*) and may contain from tens (in liquids) to thousands (in solids) of coherent molecules. *Primary convertons* - transition states between primary librational and translational effectons in liquids represents assembly - disassembly of clusters. These processes are accompanied by oscillation of molecular de Broglie waves length and frequency, modulating the carrying frequency of Bivacuum virtual pressure waves ($\mathbf{VPW}_{q=1}^{\pm}$). In accordance to described in section (15.1) mechanism, such kind of modulation follows by formation of hologram-like Virtual Replica of the object. Other kinds of collective excitations in condensed matter are not so coherent (Kaivarainen, 2001; 2003) and corresponding VR components are not stable. This means that variation of *mBC* fraction in the object influence on the life-time of its virtual replica.

The internal kinetic energy of collective excitations: primary effectons ($T_{kin}^{eff}$), transitons ($T_{kin}^{t}$) and convertons ($T_{kin}^{con}$) vary, as a result of temperature change or more strongly as a result of nonequilibrium cooperative process, like melting. The values of these contributions and their changes may be calculated using Hierarchic theory of condensed matter, based on CAMP computer program (Kaivarainen, 2000a). The translational dynamics dynamics turns the basic virtual Pressure Waves ($\mathbf{VPW}_{q=1}^{\pm}$) to modulated ones and librational dynamics modulate the basic virtual spin waves ($\mathbf{VirSW}_{q=1}^{\pm 1/2}$), as was demonstrated in previous section:

$$2\Delta(T_{kin}^{tot}) = 2\Delta\left[\left(T_{kin}^{eff} + T_{kin}^{t}\right)_{tr,lb} + T_{kin}^{con}\right] = \qquad 13.18$$

$$= \Delta\left\{V_0 \frac{2}{Z} \sum_{tr,lb}\left[n_{ef}\frac{\sum(E^a)_{1,2,3}^2}{2M_{ef}(\mathbf{v}_{ph}^a)^2}(P_{ef}^a + P_{ef}^b)\right]^{eff} + \left[n_t\frac{\sum(E_t)_{1,2,3}^2}{2M_t(\mathbf{v}_s^{res})^2}P_d\right]^t\right\} \quad 13.18a$$

$$+ \Delta\left\{V_0\frac{n_{con}}{Z}\frac{\left(E_{ac}\right)^2}{6M_c(\mathbf{v}_s^{res})^2}P_{ac} + \frac{\left(E_{bc}\right)^2}{6M_c(\mathbf{v}_s^{res})^2}P_{bc} + \frac{\left(E_{cMd}\right)^2}{6M_c(\mathbf{v}_s^{res})^2}\right\}^{con} \qquad 13.18b$$

$$\sim[\mathbf{VPW}_m^+ + \mathbf{VPW}_m^-]_{tr} + [\mathbf{VirSW}_m^{+1/2} + \mathbf{VirSW}_m^{-1/2}]_{lb} \sim [\mathbf{VirP}^+ + \mathbf{VirP}^-]_{tr,lb} \quad 13.18c$$

where: $V_0$ molar volume of water; $Z$ partition function; $n_{ef}$ concentration of primary effectons; $E^a$ energy of the (a) state of the effectons; $P_{ef}^a$ and $P_{ef}^b$ probabilities of (a) and (b) states of the effectons; $M_{ef}$, $M_t$ and $M_c$ are the masses of primary effectons, transitons and convertons; $\mathbf{v}_{ph}^a$ is phase velocity of the effecton in *a*- state; $n_t$ and $E_t$ are concentration and energy of transitons; $n_{con}$ is concentration of convertons; $E_{ac}$ and $E_{bc}$ and $E_{cMd}$ are the energies of (a), (b) [lb/tr] convertons and macroconvertons, correspondingly.

For more detailed description of these parameters see paper: Hierarchic Theory of Condensed Matter and its Computerized Application to Water and Ice, available on-line: http://arXiv.org/abs/physics/0102086.

## 14 Possible Mechanism of entanglement between remote elementary particles via Virtual Guides of spin, momentum and energy ($\mathbf{VirG}_{S,M,E}^i$)

The instant entanglement between two or more remote *similar* elementary particles (electrons, protons, neutrons, photons), named [Sender (S)] and [Receiver (R)], revealed in a lot of experiments, started by Aspect and Grangier (1983). In accordance to our theory, the entanglement involves a few stages:

**1**. Tuning of the frequency and phase of [$\mathbf{C} \rightleftharpoons \mathbf{W}$] pulsation of remote elementary particles, like photons electrons, protons, neutrons - free or in composition of atoms and molecules, under the action of basic Bivacuum virtual pressure waves: $\mathbf{VPW_{q=1}^+}$ and $\mathbf{VPW_{q=1}^-}$ and virtual spin waves: $\mathbf{VirSW_{q=1}^{\pm 1/2}}$ and $\mathbf{VPW_{q=1}^{\pm}}$;

**2**. A superposition of two virtual spin waves, excited by similar elementary particles (electrons or protons) of Sender $(\mathbf{VirSW^{S=+1/2}})_S$ and Receiver $(\mathbf{VirSW_m^{S=-1/2}})_R$ of the same pulsation frequency and opposite spins, i.e. opposite phase of $[\mathbf{C} \rightleftharpoons \mathbf{W}]$ pulsation, as the 1st stage of *Virtual Guide* of spin, momentum and energy $\mathbf{VirG}_{SME}\,(\mathbf{S} <=> \mathbf{R})^i$ (Fig.12) formation.

**3**. This stage stimulate the 2nd stage of $\mathbf{VirG}_{SME}\,(\mathbf{S} <=> \mathbf{R})$ formation - the assembly of Cooper pairs of Bivacuum fermions ($\mathbf{BVF}^- \bowtie \mathbf{BVF}^+$) or single Bivacuum bosons ($\mathbf{BVB}^+$) in quasi 1-dimensional virtual Bose condensate

$$\left[ < [\mathbf{F}_\downarrow^+ \bowtie \mathbf{F}_\uparrow^-]_C + (\mathbf{F}_\downarrow^-)_W >_S \xrightarrow{(\mathbf{VirSW^{S=+1/2}})_S} \overset{\mathbf{VirG}_{SME}(\mathbf{S}<=>\mathbf{R})}{\boxed{\begin{array}{c} \mathbf{BVB}^+ \\ \Longrightarrow\Longleftarrow \\ \mathbf{BVF}^-\bowtie\mathbf{BVF}^+ \end{array}}} \xleftarrow{(\mathbf{VirSW_m^{S=-1/2}})_R} < (\mathbf{F}_\uparrow^-)_C + [\mathbf{F}_\downarrow^- \bowtie \mathbf{F}_\uparrow^+]_W >_R \right]$$

The radius of virtual microtubules of $\mathbf{VirG}_{SME}^i$ is determined by Compton radius of three generation of torus and antitorus ($i = e, \mu, \tau$), forming them:

$$\mathbf{L}_V^e = \hbar/\mathbf{m}_0^e \mathbf{c} >> \mathbf{L}_V^\mu = \hbar/\mathbf{m}_0^\mu \mathbf{c} > \mathbf{L}_V^\tau = \hbar/\mathbf{m}_0^\mu \mathbf{c} \qquad \text{14.1a}$$

The radius of $\mathbf{VirG}_{SME}^e\,(\mathbf{S} <=> \mathbf{R})$, connecting two remote electrons, is the biggest one ($\mathbf{L}^e$). The radius of $\mathbf{VirG}_{SME}^\tau$, connecting two protons or neutrons ($\mathbf{L}^\tau$) is about $3.5 \times 10^3$ times smaller. The entanglement between similar and tuned by virtual waves atoms in pairs, like hydrogen, oxygen, carbon or nitrogen can be realized via complex system of virtual guides of atomic $\mathbf{VirG}_{SME}^{at}\,(\mathbf{S} <=> \mathbf{R})$, representing *multishell constructions*.

The formation of two spatial configurations of Virtual Guides, representing quasi one-dimensional virtual Bose condensate (vBC), is possible:

a) single *nonlocal virtual guides* $\mathbf{VirG}_{SME}^{(\mathbf{BVB}^\pm)^i}\,(\mathbf{S} <=> \mathbf{R})$ - from big number of Bivacuum bosons $\mathbf{N} \times (\mathbf{BVB}^\pm)^i$. In this case the $\mathbf{VirG}_{SME}^{(\mathbf{BVB}^\pm)^i}\,(\mathbf{S} <=> \mathbf{R})$ is not rotating as a whole around its main axis and the resulting *angular* momentum (spin) is zero. The longitudinal momentum of $(\mathbf{BVB}^\pm)^i = [\mathbf{V}^+ \Updownarrow \mathbf{V}^-]$ is zero also, providing conditions for 1D virtual BC;

b) double *nonlocal virtual guides* $\mathbf{VirG}_{SME}^{[\mathbf{BVF}^\uparrow \bowtie \mathbf{BVF}^\downarrow]^i}\,(\mathbf{S} <=> \mathbf{R})$, assembled by 'head-to-tail' principle from Cooper pairs of Bivacuum fermions $\mathbf{N} \times [\mathbf{BVF}^\uparrow \bowtie \mathbf{BVF}^\downarrow]^i$. In this case each of two adjacent microtubules from $\mathbf{BVF}^\uparrow$ or $\mathbf{BVF}^\downarrow$ may rotate as respect to each other and around their own axes in opposite directions.

Two remote coherent triplets - elementary particles, like: electron - electron, proton - proton or neutron-neutron with similar frequency of $[\mathbf{C} \rightleftharpoons \mathbf{W}]_{e,p}$ pulsation and opposite spins (phase) can be connected by corresponding Virtual guides: $\mathbf{VirG}_{SME}^{e,p,n}(\mathbf{S} <=> \mathbf{R})$ of spin (S), momentum (M) and energy (E) from Sender to Receiver. The spin - information (qubits), momentum and kinetic energy instant transmission via such $\mathbf{VirG}_{SME}^i(\mathbf{S} <=> \mathbf{R})$ from [S] and [R] is possible. The same mechanism is valid for two synchronized photons (bosons) of opposite spins. Such information transmission can be instant, accompanied by 'flip-flop' spin exchange between $\mathbf{BVF}^\uparrow$ and $\mathbf{BVF}^\downarrow$ in Cooper pairs $[\mathbf{BVF}^\uparrow \bowtie \mathbf{BVF}^\downarrow]$ or between torus and antitorus: $\mathbf{V}^+ \uparrow$ and $\mathbf{V}^- \downarrow$ of Bivacuum bosons $(\mathbf{BVB}^\pm)^i = [\mathbf{V}^+ \uparrow\downarrow \mathbf{V}^-]$.

The double $\mathbf{VirG}_{SME}^{[\mathbf{BVF}^\uparrow \bowtie \mathbf{BVF}^\downarrow]^i}\,(\mathbf{S} <=> \mathbf{R})$ can be transformed to single $\mathbf{VirG}_{SME}^{(\mathbf{BVB}^\pm)^i}\,(\mathbf{S} <=> \mathbf{R})$ by conversion of opposite Bivacuum fermions: $\mathbf{BVF}^\uparrow = [\mathbf{V}^+ \uparrow\uparrow \mathbf{V}^-]$ and $\mathbf{BVF}^\downarrow = [\mathbf{V}^+ \downarrow\downarrow \mathbf{V}^-]$ to the pair of Bivacuum bosons of two possible polarization $\mathbf{BVB}^+$

and $\mathbf{BVB}^-$:

$$\mathbf{VirG_{BVB^+}}\,(\mathbf{S}<=>\mathbf{R}) = [\,\mathbf{n_+BVB^+}(\mathbf{V^+\uparrow\downarrow V^-})\,]^i \qquad 14.2$$

$$\mathbf{VirG_{BVB^-}}\,(\mathbf{S}<=>\mathbf{R}) = [\,\mathbf{n_-BVB^-}(\mathbf{V^+\downarrow\uparrow V^-})\,]^i \qquad 14.2a$$

### 14.1. The mechanism of momentum and energy transmission between similar elementary particles of Sender and Receiver via $VirG_{\mathbf{SME}}(\mathbf{S} <=> \mathbf{R})^i$

The increments or decrements of momentum $\pm\Delta\mathbf{p} = \Delta(\mathbf{m}_V^+\mathbf{v})_{tr,lb}$ and kinetic $(\pm\Delta\mathbf{T}_k)_{tr,lb}$ energy transmission from [S] to [R] of *selected generation of elementary particles* is determined by the translational and librational velocity variation $(\Delta\mathbf{v})$ of nuclei of (S). This means, that energy/momentum transition from [S] to [R] is possible, if they are in nonequilibrium state.

The variation of kinetic energy of atomic nuclei under external force application, induces nonequilibrium in a system $(\mathbf{S} + \mathbf{R})$ and decoherence of $[\mathbf{C} \rightleftharpoons \mathbf{W}]$ pulsation of protons and neutrons of [S] and [R]. The nonlocal energy transmission from [S] to [R] is possible, if the decoherence is not big enough for disassembly of the virtual microtubules and their systems in the case of atoms. The electronic $\mathbf{VirG}_{SME}^e$, as more coherent (not so dependent on thermal vibrations), can be responsible for stabilization of the complex atomic Virtual Guides $\sum \mathbf{VirG}_{SME}^{e,p,n}(\mathbf{S} <=> \mathbf{R})$.

The values of the energy and velocity increments or decrements of free elementary particles are interrelated by (13.3).

The instantaneous energy flux via $(\mathbf{VirG}_{SME})^i$, is mediated by pulsation of energy and radii of torus $(\mathbf{V^+})$ and antitorus $(\mathbf{V^-})$ of Bivacuum bosons: $\mathbf{BVB^+}= [\mathbf{V^+\uparrow\downarrow V^-}]$. Corresponding energy increments of the actual torus and complementary antitorus of $\mathbf{BVB}^\pm$, forming $(\mathbf{VirG}_{SME})^i$, are directly related to increments of Sender particle external velocity $(\Delta\mathbf{v})$:

$$\Delta\mathbf{E}_{V^+} = +\Delta\mathbf{m}_V^+c^2 = \left(+\frac{\mathbf{p}^+}{\mathbf{R}^2}(\Delta\mathbf{v})_{\mathbf{F}_\uparrow^+}^{[\mathbf{F}_\uparrow^+\bowtie\mathbf{F}_\downarrow^-]} = \mathbf{m}_V^+c^2\,\frac{\Delta\mathbf{L}_{V^+}}{\mathbf{L}_{V^+}}\right)_{N,S} \quad \text{actual} \qquad 14.3$$

$$\Delta\mathbf{E}_{V^-} = -\Delta\mathbf{m}_V^-c^2 = \left(-\frac{\mathbf{p}^-}{\mathbf{R}^2}(\Delta\mathbf{v})_{\mathbf{F}_\uparrow^+}^{[\mathbf{F}_\uparrow^+\bowtie\mathbf{F}_\downarrow^-]} = -\mathbf{m}_V^-c^2\,\frac{\Delta\mathbf{L}_{V^-}}{\mathbf{L}_{V^-}}\right)_{N,S} \quad \text{complementary} \qquad 14.4$$

where: $\mathbf{p}^+= \mathbf{m}_V^+\mathbf{v}$; $\mathbf{p}^-= \mathbf{m}_V^-\mathbf{v}$ are the actual and complementary momenta; $\mathbf{L}_{V^+} = \hbar/\mathbf{m}_V^+\mathbf{c}$ and $\mathbf{L}_{V^-} = \hbar/\mathbf{m}_V^-\mathbf{c}$ are the radii of torus and antitorus of $\mathbf{BVB}^\pm = [\mathbf{V^+\Updownarrow V^-}]$, forming $\mathbf{VirG}_{SME}(\mathbf{S} <=> \mathbf{R})^i$.

The nonlocal energy exchange between [S] and [R] is accompanied by the *instant pulsation of radii* of tori $(\mathbf{V^+})$ and antitori $(\mathbf{V^-})$ of $\mathbf{BVF}^\updownarrow$ and $\mathbf{BVB}^\pm$, accompanied by corresponding pulsation $|\Delta\mathbf{L}_{V^\pm}/\mathbf{L}_{V^\pm}|$ of the whole virtual microtubule $\mathbf{VirG}_{SME}$ (Fig.12).

*The nonequilibrium state* of elementary particles of [S] and [R], connected by $\mathbf{VirG}_{S,M,E}$, means difference in their kinetic and total energies and frequency of de Broglie waves and that of $[\mathbf{C} \rightleftharpoons \mathbf{W}]$ pulsation. The consequence of this difference are beats between states of [S] and [R], equal to frequency of $\mathbf{VirG}_{SME}$ radius pulsation. Using eqs. 7.4 and 7.4a, we get:

$$\Delta \mathbf{v}_{\mathbf{VirG}}^{S,R} = \mathbf{v}_{\mathbf{C} \rightleftharpoons \mathbf{W}}^{S} - \mathbf{v}_{\mathbf{C} \rightleftharpoons \mathbf{W}}^{R} = \frac{\mathbf{c}^2}{h}\left[(\mathbf{m}_V^+)^S - (\mathbf{m}_V^+)^R\right] = \quad\quad\quad \text{14.4a}$$

$$= \frac{1}{h}\left[\mathbf{m}_0\mathbf{c}^2(\mathbf{R}^S - \mathbf{R}^R) + \left(\frac{h^2}{(\mathbf{m}_V^+\boldsymbol{\lambda}_B^2)^S} - \frac{h^2}{(\mathbf{m}_V^+\boldsymbol{\lambda}_B^2)^R}\right)\right]$$

The beats between the total frequencies of [S] and [R] states (electrons, protons or neutrons), connected by $\mathbf{VirG}_{S,M,E}$ and different excitation states $(j-k)$ of $[\mathbf{BVF}^\uparrow \bowtie \mathbf{BVF}^\downarrow]_{j-k}$ are accompanied by *emission* $\rightleftharpoons$ *absorption* of positive and negative virtual pressure waves: $\mathbf{VPW}^+$ and $\mathbf{VPW}^-$, generating positive and negative virtual pressure: $\mathbf{VirP}^+$ and $\mathbf{VirP}^-$.

The difference between total energies of elementary particles of Sender and Receiver can be expressed via these virtual pressures, using eq.7.4c and 14.4a, as:

$$\mathbf{E}_{tot}^S - \mathbf{E}_{tot}^S = h\Delta\mathbf{v}_{\mathbf{VirG}}^{S,R} = \Delta(\mathbf{m}_V^+\mathbf{c}^2)^{S,R} = \Delta\mathbf{V} + \Delta\mathbf{T_k} = \quad\quad \text{14.4b}$$

$$= \frac{1}{2}\Delta(\mathbf{m}_V^+ + \mathbf{m}_V^-)^{S,R}\mathbf{c}^2 + \frac{1}{2}\Delta(\mathbf{m}_V^+ - \mathbf{m}_V^-)^{S,R}\mathbf{c}^2 \sim$$

$$\sim \Delta(\mathbf{VirP}^+ + \mathbf{VirP}^-)^{S,R} + \Delta(\mathbf{VirP}^+ - \mathbf{VirP}^-)^{S,R} \quad\quad \text{14.4c}$$

If the temperature or kinetic energy of [S] is higher, than that of [R]: $\mathbf{T}_S > \mathbf{T}_R$, then $\Delta\mathbf{v}_{\mathbf{VirG}}^{S,R} > 0$ and the *direction* of momentum and energy flux, mediated by positive and negative virtual pressure of subquantum particles and antiparticles: $\Delta\mathbf{VirP}^+$ and $\Delta\mathbf{VirP}^-$, is from [S] *to* [R]. The opposite nonequilibrium state of system, i.e. $\mathbf{T}_S < \mathbf{T}_R$ provides the opposite direction of energy/momentum flux - from [R] to [S].

The proposed mechanism of Pauli repulsion between fermions of the same spin state (section 9) also may realize the repulsion between Sender and Receiver.

The length of $\mathbf{VirG}_{SME}(\mathbf{S} <=> \mathbf{R})$, connecting tuned elementary particles, also can vary in the process of [S] *and* [R] interaction because of immediate self-assembly of Bivacuum dipoles into virtual guides.

### 14.2 The mechanism of spin/information exchange between tuned particles of Sender and Receiver via VirG$_{\mathbf{SME}}$

Most effectively the proposed mechanism of spin (information), momentum and energy exchange can work between Sender and Receiver, containing coherent molecular clusters with dimensions of 3D standing de Broglie waves of molecules in state of mesoscopic Bose condensate (mBC) (Kaivarainen, 2001, 2005).

The nonlocal spin/qubit exchange between [S] and [R] via single or double $\mathbf{VirG}_{SME}^i(\mathbf{S} <=> \mathbf{R})^i$ does not need the radius pulsation, but only the instantaneous polarization change of Bivacuum bosons $(\mathbf{BVB}^+ \rightleftharpoons \mathbf{BVB}^-)^i$ or instant spin state exchange of two Bivacuum fermions, forming virtual Cooper pairs in the double virtual guide:

$$[\mathbf{BVF}^\uparrow \bowtie \mathbf{BVF}^\downarrow]^i \quad \overset{(S=+1/2)\to(S=-1/2)}{\rightleftharpoons} \quad [\mathbf{BVF}^\downarrow \bowtie \mathbf{BVF}^\uparrow]^i$$

The instantaneous spin state/information exchange frequency is determined by frequency of spin change of fermion of Sender, accompanied by counterphase spin state change of fermion of Receiver.

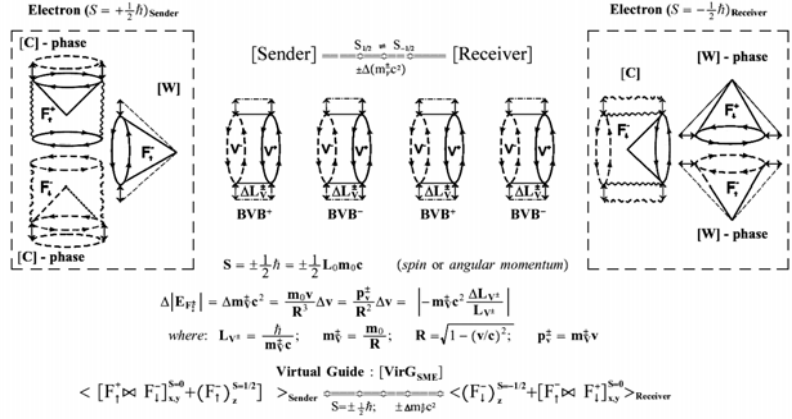

**Fig. 12**. The mechanism of nonlocal Bivacuum mediated interaction (entanglement) between two distant unpaired sub-elementary fermions of 'tuned' elementary triplets (particles) of the opposite spins $< [\mathbf{F}_\uparrow^+ \bowtie \mathbf{F}_\downarrow^-] + \mathbf{F}_\uparrow^- >_{\mathbf{Sender}}^i$ and $< [\mathbf{F}_\uparrow^+ \bowtie \mathbf{F}_\uparrow^-] + \mathbf{F}_\downarrow^- >_{\mathbf{Receiver}}^i$, with close frequency of $[\mathbf{C} \rightleftharpoons \mathbf{W}]$ pulsation and close de Broglie wave length ($\lambda_\mathbf{B} = \mathbf{h}/\mathbf{m}_\uparrow^+ \mathbf{v}$) of particles. The tunnelling of momentum and energy increments: $\Delta|\mathbf{m}_V^\pm \mathbf{c}^2|$ $\sim \Delta|\mathbf{VirP}^+| \pm \Delta|\mathbf{VirP}^-|$ from Sender to Receiver and vice-verse via Virtual spin-momentum-energy Guide $[\mathbf{VirG}_{\mathbf{SME}}^i]$ is accompanied by instantaneous pulsation of diameter $(\mathbf{2}\Delta\mathbf{L}_V^\pm)$ of this virtual guide, formed by Bivacuum bosons $\mathbf{BVB}^\pm$ or double microtubule, formed by Cooper pairs of Bivacuum fermions $[\mathbf{BVF}^\uparrow \bowtie \mathbf{BVF}^\downarrow]$. The nonlocal spin state exchange between [S] and [R] can be induced by the change of polarization of Cooper pairs: $[\mathbf{BVF}^\uparrow \bowtie \mathbf{BVF}^\downarrow] \rightleftharpoons [\mathbf{BVF}^\downarrow \bowtie \mathbf{BVF}^\uparrow]$ and Bivacuum bosons: $\mathbf{BVB}^+ \rightleftharpoons \mathbf{BVB}^-$,. composing the double or single $\mathbf{VirG}_{\mathbf{SME}}(\mathbf{S} <=> \mathbf{R})^i$, correspondingly.

The described above mechanisms of nonlocal/instant transmission of spin/information, momentum and energy between coherent clusters of elementary particles and atoms of Sender and Receiver, connected by Virtual Guides, may describe a lot of unconventional experimental results, like Kozyrev, Tiller ones (section 18) and lot of Psi phenomena.

In virtual microtubules $\mathbf{VirG}_{SME}^i(\mathbf{S} <=> \mathbf{R})^i$ the time and its 'pace' are uncertain: $\mathbf{t} = \mathbf{0}/\mathbf{0}$, if the external translational or tangential velocities $(\mathbf{v})$ and accelerations $(\mathbf{dv}/\mathbf{dt})$ of Bivacuum dipoles, composing them, are zero (see eqs. 12.13 and 12.14).

### 14.3  The role of tuning force $(\mathbf{F}_{\mathbf{VPW}^\pm})$ of virtual pressure waves $\mathbf{VPW}_q^\pm$ of Bivacuum in entanglement

The tuning between **two similar elementary** particles: 'sender (S)' and 'receiver (R)' via $\mathbf{VirG}_{SME}(\mathbf{S} <=> \mathbf{R})^i$ may be qualitatively described, using well known model of *damped harmonic oscillators,* interacting with all-pervading virtual pressure waves $(\mathbf{VPW}_{q=1}^\pm)$ of Bivacuum with fundamental frequency $\boldsymbol{\omega}_0 = \mathbf{m}_0 \mathbf{c}^2/\hbar$. The criteria of tuning - synchronization of [S] and [R] is the equality of the amplitude probability of resonant energy exchange of Sender and Receiver with virtual pressure waves $(\mathbf{VPW}_{q=1}^\pm)$: $\mathbf{A}_{C \rightleftharpoons W}^S = \mathbf{A}_{C \rightleftharpoons W}^R$, resulting from minimization of frequency difference $(\boldsymbol{\omega}_S - \boldsymbol{\omega}_0) \to 0$ and $(\boldsymbol{\omega}_R - \boldsymbol{\omega}_0) \to 0$:

$$\mathbf{A}_{C \rightleftharpoons W}^{S} \sim \left[ \frac{1}{(\mathbf{m}_V^+)_S} \frac{\mathbf{F_{VPW^\pm}}}{(\boldsymbol{\omega}_S^2 - \boldsymbol{\omega}_0^2) + \text{Im } \boldsymbol{\gamma \omega}_S} \right] \qquad 14.5$$

$$[\mathbf{A}_{C \rightleftharpoons W}^{R}]_{x,y,z} \sim \left[ \frac{1}{(\mathbf{m}_V^+)_R} \frac{\mathbf{F_{VPW^\pm}}}{(\boldsymbol{\omega}_R^2 - \boldsymbol{\omega}_0^2) + \text{Im } \boldsymbol{\gamma \omega}_R} \right] \qquad 14.5a$$

where the frequencies of $\mathbf{C} \rightleftharpoons \mathbf{W}$ pulsation of particles of Sender ($\boldsymbol{\omega}_S$) and Receiver ($\boldsymbol{\omega}_R$) are:

$$\boldsymbol{\omega}_R = \boldsymbol{\omega}_{\mathbf{C} \rightleftharpoons \mathbf{W}} = \mathbf{R}\boldsymbol{\omega}_0^{in} + (\boldsymbol{\omega}_B^{ext})_R \qquad 14.6$$

$$\boldsymbol{\omega}_S = \boldsymbol{\omega}_{\mathbf{C} \rightleftharpoons \mathbf{W}} = \mathbf{R}\boldsymbol{\omega}_0^{in} + (\boldsymbol{\omega}_B^{ext})_S \qquad 14.6a$$

$\boldsymbol{\gamma}$ is a damping coefficient due to *decoherence effects*, generated by local fluctuations of Bivacuum deteriorating the phase/spin transmission via $\mathbf{VirG}_{SME}$; $(\mathbf{m}_V^+)_{S,R}$ are the actual mass of (S) and (R); $[\mathbf{F_{VPW}}]$ is a *tuning force of virtual pressure waves* $\mathbf{VPW^\pm}$ *of Bivacuum with tuning energy* $\mathbf{E}_{VPW} = \mathbf{q m}_0 \mathbf{c}^2$ *and wave length* $\mathbf{L}_{VPW} = \hbar/\mathbf{m}_0 \mathbf{c}$

$$\mathbf{F_{VPW_q^\pm}} = \frac{\mathbf{E}_{VPW_q}}{\mathbf{L}_{VPW_q}} = \frac{\mathbf{q}}{\hbar} \mathbf{m}_0^2 \mathbf{c}^3 \qquad 14.7$$

The most probable Tuning force has a minimum, corresponding to $\mathbf{q} = \mathbf{j} - \mathbf{k} = \mathbf{1}$.

The influence of *virtual pressure force* ($\mathbf{F_{VPW_q}}$) stimulates the synchronization of [S] and [R] pulsations, i.e. $\omega_R \rightarrow \omega_S \rightarrow \omega_0$. This fundamental frequency $\boldsymbol{\omega}_0 = \mathbf{m}_0 \mathbf{c}^2/\hbar$ is the same in any space volume, including those of Sender and Receiver.

The $\mathbf{VirG}_{SME}$ represent quasi $\mathbf{1D}$ macroscopic virtual Bose condensate with a configuration of single microtubules, formed by Bivacuum bosons ($\mathbf{BVB^\pm}$) or with configuration of double microtubules, composed from Cooper pairs as described in previous section.

The effectiveness of entanglement between number of similar elementary particles of Sender and Receiver - free or in composition of atoms and molecules via highly anisotropic nonlocal virtual guide bundles

$$\left[ \mathbf{N(t,r)} \times \sum^{\mathbf{n}} \mathbf{VirG}_{SME} (\mathbf{S} <=> \mathbf{R}) \right]_{x,y,z}^{i} \qquad 14.7a$$

is dependent on synchronization of $[\mathbf{C} \rightleftharpoons \mathbf{W}]$ pulsation frequency of these particles.

In this expression ($\mathbf{n}$) is a number of pairs of similar tuned elementary particles (protons, neutrons and electrons) in atoms/molecules of $\mathbf{S}$ and $\mathbf{R}$; $\mathbf{N(t,r)}$ is a number of coherent atoms/molecules in the coherent molecular clusters - mesoscopic BC (Kaivarainen, 2001; 2004).

The 'tuning' of particles phase and frequency pulsation occur under the forced resonance exchange interaction between virtual pressure waves $\mathbf{VPW}_q^+$; $\mathbf{VPW}_q^-$ and pulsing particles.

The mechanism proposed may explain the experimentally confirmed nonlocal interaction between coherent elementary particles (Aspect and Gragier, 1983), atoms and their remote coherent clusters.

Our theory predicts that the same mechanism, involving nonlocal bundles $[\mathbf{N(t,r)} \times \sum \mathbf{VirG}_{SME} (\mathbf{S} <=> \mathbf{R})]_{x,y,z}^{i}$, may provide the entanglement between macroscopic systems, including biological ones.

*14.4 The vortical filaments in superfluids, as the analogs of virtual guides of Bivacuum*

When the rotation velocity of a cylindrical vessel containing **He II** becomes high enough, then the emergency of so-called vortex filaments becomes thermodynamically favorable. The filament is formed by the superfluid component of **He II** in such a way that their momentum of movement decreases the total energy of **He II** in a rotating vessel. The shape of filaments in this case is like a straight rod and their *thickness* is of the order of atom's dimensions, increasing with lowering the temperature at $T < T_\lambda$.

Vortex filaments are continuous. They may be closed or limited within the boundaries of vessel.

The hydrodynamics of normal and superfluid components of He II in container of radius (**r**), rotating with angular frequency $\Omega$ are characterized by two velocities, correspondingly

$$\mathbf{v}_n = \Omega \, \mathbf{r} \qquad\qquad 14.8$$

$$\mathbf{v}_{sf} = \frac{\hbar}{\mathbf{m}} \nabla \phi = N \frac{\hbar}{\mathbf{m} \, \mathbf{r}} \qquad\qquad 14.8a$$

where $\nabla \phi \sim k_{sf} = 1/\mathbf{L}_{sf}$ is a phase of Bose condensate wave function: $\mathbf{\Psi} = \mathbf{\rho}_s^{1/2} \times \mathbf{e}^{i\phi}$ ($\mathbf{\rho}_s$ is a density of superfluid component); $N$ is a number of rectilinear vortex lines.

The motion of superfluid component is potential, as far its velocity ($\mathbf{v}_{sf}$) is determined by eq. 14.8a and:

$$rot \, \mathbf{v}_{sf} = 0 \qquad\qquad 14.8b$$

The values of velocity of circulation of filaments are determined (Landau, 1941) as follows:

$$\oint \mathbf{v}_{sf} dl = 2\pi r \, \mathbf{v}_{sf} = 2\pi\kappa = \frac{\hbar}{m} \Delta\Phi \qquad\qquad 14.9$$

where: $\mathbf{\Delta\Phi} = \mathbf{n} \, \mathbf{2\pi}$ is a phase change as a result of circulation, $n = 1, 2, 3\ldots$ is the integer number.

and

$$\mathbf{v}_{sf} = \kappa/r \qquad\qquad 14.9a$$

Increasing the radius of circulation (r) leads to decreased circulation velocity ($\mathbf{v}_{sf}$).

Comparing (14.9a) and (14.9) gives:

$$\kappa = n \frac{\hbar}{m} \qquad\qquad 14.10$$

It has been shown that only vortical structures with $n = 1$ are thermodynamically stable. Taking this into account, we have from (14.9a) and (14.10):

$$r = n \frac{\hbar}{\mathbf{m} \mathbf{v}_{sf}} \qquad\qquad 14.11$$

An increase in the angle frequency of rotation of the cylinder containing **HeII** results in the increased density distribution of vortex filaments on the cross-section of the cylinder.

As a result of interaction between the filament and the normal component of **HeII**, the filaments move in the rotating cylinder with normal liquid.

The flow of **He II** through the capillaries also can be accompanied by appearance of vortex filaments.

In ring-shaped vessels the circulation of closed vortex filaments is stable. Stability is related to the quantum pattern of circulation change (eqs. 14.9 and 14.10).

Let us consider now the phenomena of superfluidity in **He II** in the framework of our hierarchic concept (Kaivarainen, 2001).

### 14.4 Theory of superfluidity, based on hierarchic model of condensed matter

It will be shown below how our hierarchic model (Table 1 in http://arXiv.org/abs/physics/0102086) can be used to explain **He II** properties, its excitation spectrum (Fig. 13), increased heat capacity at $\lambda$-point and the vortex filaments formation.

We assume here, that the formulae obtained earlier (Kaivarainen, 2001) for internal energy, viscosity, thermal conductivity and vapor pressure remain valid for both components of He II.

The theory proposed by Landau (Lifshits, Pitaevsky, 1978) qualitatively explains only the lower branch (a) in experimental spectrum (Fig. 13), as a result of phonons and rotons excitation.

But the upper branch (b) points that the real process is more complicated and needs introduction of other quasiparticles and excited states for its explanation.

Our hierarchic model of superfluidity (Kaivarainen, 2006) interrelates the lower branch with the ground acoustic (a) state of primary effectons in liquid $^4$He and the upper branch with their excited optical (b) state. In accordance with our model, the dissipation and viscosity friction (see section 11.6 in ) arise in the normal component of He II due to thermal phonons radiated and absorbed in the course of the $\bar{b} \to \bar{a}$ and $\bar{a} \to \bar{b}$ transitions of secondary effectons, correspondingly.

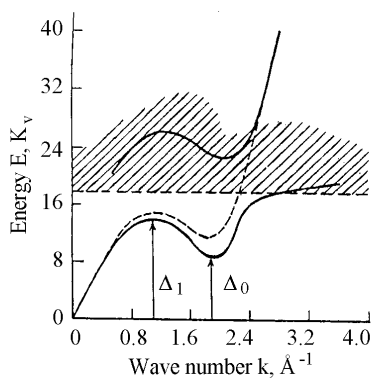

**Fig. 13**. Excitation spectrum of liquid $^4$**He** from neutron scattering measurements (March and Parrinello, 1982). Spectrum is characterized by two branches, corresponding to (a-acoustic) and (b-optical) states of the primary librational effectons according to the hierarchic model (Kaivarainen, 2001).

Landau described the minimum in the region of $\lambda$-point using the expression:

$$E = \Delta_0 + \frac{(P - P_0)^2}{2m^*},$$

14.12

where $\Delta_0$ and $P_0$ are the energy and momentum of liquid $^4$**He** at $\lambda$-point (Fig. 1) and $m^* = 0.16m$ is the effective mass of the $^4$**He** atom $(m_{He} = 4 \times 1.44 \cdot 10^{-24}g = 5.76 \cdot 10^{-24}g)$. The effective mass $m^*$ is determined experimentally.

Feynman (1953) explained the same part of the excitation spectra by the non-monotonic behavior of the structure factor $S(k)$ and the formula:

$$E = \hbar\omega = \frac{\hbar^2 k^2}{2mS} = \frac{\hbar^2}{2mL^2 S} \qquad 14.13$$

where:

$$k = 1/L = 2\pi/\lambda \qquad 14.14$$

is the wave number of neutron interacting with liquid $^4He$.

*Our hierarchic theory of condensed matter allows to unify Landau's and Feynman's approaches.* The total energy of de Broglie wave either free or as part of condensed matter can be expressed through its amplitude squared ($A^2$), length squared ($L^2$) and effective mass ($m^*$) in the following manner (Kaivarainen, 2001):

$$E_{tot} = T_k + V = \frac{\hbar^2}{2mA^2} = \frac{\hbar^2}{2m^*L^2} \qquad 14.15$$

In accordance with our Hierarchic theory (Kaivarainen, 2001), the structural factor S(k) is equal to the kinetic ($T_k$) to total ($E_{tot}$) energy ratio of wave B:

$$S = T_k/E_{tot} = A^2/L^2 = m^*/m \qquad 14.16$$

where:

$$T_k = P^2/2m = \frac{\hbar}{2mL^2} \qquad 14.17$$

Combining (14.15), (14.16) and (14.17), we get the following set of equation for the energy of $^4$He at transition $\lambda$-point:

$$\left.\begin{array}{l} \Delta_0 = E_0 = \dfrac{\hbar^2}{2mA_0^2} = \dfrac{\hbar^2}{2m^*L_0^2} \\[2mm] \Delta_0 = \dfrac{\hbar^2}{2mL_0^2 S} = \dfrac{T_k^0}{S} \end{array}\right\} \qquad 14.18$$

These approximate formulae for the total energy of liquid $^4$He made it possible to estimate the most probable wave B length, forming the primary librational (or rotational effectons) at $\lambda$-point:

$$\lambda_0 = \frac{h}{mv_{gr}^0} = 2\pi L_0 = 2\pi A_0 \left( m/m^* \right)^{1/2}, \qquad 14.19$$

where the critical amplitude of wave B:

$$A_0 = \hbar \left( \frac{1}{2mE_0} \right)^{1/2} \qquad 14.20$$

can be calculated from the experimental $E_0$ values (Fig...............). Putting in (14.20) and (14.19) the available data:

$$\Delta_0 = E_0 = k_B \cdot 8.7K = 1.2 \cdot 10^{-15} \text{ erg};$$

the mass of atom: $m(^4\text{He}) = 5.76 \cdot 10^{-24}g$ and $(m^*/m) = 0.16$, we obtain:

$$\lambda_0 \cong 14 \cdot 10^{-8} cm = 14 Å \qquad 14.21$$

the corresponding most probable group velocity of $^4\textbf{He}$ atoms is: $\textbf{v}_{gr}^0 = 8.16 \cdot 10^3 cm/s$.

It is known from the experiment that the volume occupied by *one atom of liquid* $^4\textbf{He}$ is equal: $\textbf{V}_{^4He} = 46 \mathring{A}^3$/atom. The edge length of the corresponding cubic volume is:

$$\textbf{l} = \left( \textbf{V}_{^4He} \right)^{1/3} = 3.58 \mathring{A} \qquad\qquad 14.22$$

From (14.21) and (14.22) we can calculate the number of $^4\textbf{He}$ atoms in the volume of primary librational (rotational) effecton at $\lambda$-point:

$$n_V^0 = \frac{V_{ef}}{\textbf{V}_{^4He}} = \frac{(9/4\pi)\lambda_0}{l^3} = 43 \text{ atoms} \qquad\qquad 14.23$$

One edge of such an effecton of cube shape contains: $q = (43)^{1/3} \cong 3.5$ atoms of liquid $^4\textbf{He}$.

We must take into account, that these parameters can be *lower than the real* ones, as far in above simple calculations we did not consider the contributions of secondary effectons, transitons and deformons to total internal energy (Kaivarainen, 2001).

On the other hand, in accordance with Hierarchic model, the conditions of the maximum stability of primary effectons correspond to the *integer* number of particles in the edge of these effectons (Kaivarainen, 2001).

Consequently, we have to assume that the true number of $^4He$ atoms forming a primary effecton at $\lambda$-point is equal to $n_V^0 = 64$. It means that the edge of cube as the effecton shape approximation contains $q^0 = 4$ atoms of $^4He$:

$$q^0 = (n_V^0)^{1/3} = 64^{1/3} = 4 \qquad\qquad 14.24$$

The *primary librational effectons* of such a type may correspond to *rotons* introduced by Landau to explain the high heat capacity of **HeII**.

The thermal momentums of $^4$He atoms in these coherent clusters can totally compensate each other and the resulting momentum of primary effectons is equal to zero. Further decline in temperature gives rise to dimensions of primary effectons, representing *mesoscopic Bose condensate* (mBC). The most stable of them contain in their ribs the integer number of helium atoms:

$$q = q^0 + n \qquad\qquad 14.25$$

where: $q_0 = 4$ and $n = 1, 2, 3 \dots$

$\lambda_0$, $n_V^0$ and $n_e^0$ can be calculated more accurately, using our computer program, based on Hierarchic theory, if the required experimental data on IR spectroscopy and sound velocimetry are available.

Let us consider now the consequence of the phenomena observed in $^4\textbf{He}$ in the course of temperature decline to explain Fig. 13 in the framework of hierarchic model:

**1**. In accordance to our model, the lowering of the temperature till the 4.2 K and gas-liquid first order phase transition occurs under condition, when the most probable wave B length of atoms related to their librations/rotations starts to exceed the average distance between $^4\textbf{He}$ atoms in a liquid phase and mesoscopic Bose condensation (**mBC**) in form of coherent atomic clusters becomes possible:

$$\lambda = h/mv_{gr} \geq 3.58 \mathring{A} \qquad\qquad 14.26$$

The corresponding value of the most probable group velocity is

$$\mathbf{v}_{gr} \leq 3.2 \cdot 10^4 cm/\text{s}.$$

The translational thermal momentums of particles are usually bigger and waves B length smaller than those related to librations. In accordance with our model of first order phase transitions (Kaivarainen, 2001, section 6.2), this fact determines the difference in the temperatures of [gas → liquid] and [liquid → solid] transitions.

The freezing of liquid $^4$He occurs at a sufficiently high pressure of ~ 25 atm. only and means the emergency of primary translational effectons in accordance to our theory of 1st order phase transitions (Kaivarainen, 2001). The pressure increasing, as well as temperature decreasing, decline the translational thermal momentum of particles and stimulates Bose condensation, responsible for coherent clusters formation of corresponding type.

In normal component of liquid $^4$**HeII**, like in a usual liquid at $T > 0$ $K$, the existence of primary and secondary effectons, convertons, transitons and deformons is possible. The contributions of each of these quasiparticles determine the total internal energy, kinetic and potential energies, viscosity, thermal conductivity, vapor pressure and many other parameters (Kaivarainen, 2001).

We assume that the lower branch in the excitation spectrum of Fig. 13 reflects the acoustic (a) state and the upper branch the optic (b) state of primary (lb and tr) effectons.

**2**. Decreasing the temperature to $\lambda$-point: $T_\lambda = 2.17K$ is accompanied by condition (14.24), which stimulates Bose-condensation of atoms, increasing the dimensions of primary effectons. This leads to emergency of primary polyeffectons as superfluid subsystem due to distant Van der Waals interactions and Josephson junctions between neighboring primary effectons. *This second order phase transition* is accompanied by (a)-states probability increasing ($P_a \to 1$) and that of (b)-states decreasing ($P_b \to 0$). The probability of primary and secondary deformons ($P_d = P_a \cdot P_b$; $\bar{P}_d = \bar{P}_a \cdot \bar{P}_b$) decreases correspondingly. In the excitation spectrum (Fig.1) these processes are displayed as a tending of (b)-branch closer to (a)-branch, as a consequence of degeneration of b-branch at very law temperature.

Like in the theory of 2nd order phase transitions proposed by Landau (Landau and Lifshits, 1976), we can introduce here the *order parameter* as:

$$\eta = 1 - \kappa = 1 - \frac{P_a - P_b}{P_a + P_b} \qquad\qquad 14.27$$

where: $\kappa = \frac{P_a - P_b}{P_a + P_b}$ is an equilibrium parameter.

One can see that at $P_a = P_b$, the equilibrium parameter $\kappa = 0$ and $\eta = 1$ (the system is far from 2nd order phase transition). On the other hand, at conditions of phase transition: $T \to T_\lambda$ when $P_b \to 0$, $\kappa \to 1$ and parameter of order tends to zero ($\eta \to 0$).

Similar to Landau's theory, the equality of specific parameter of order to zero, is a criteria of 2nd order phase transition. As usual, this transition is followed by a decrease in structural symmetry with a decline in temperature.

The important point of our scenario of superfluidity is a statement that the leftward shift of ($a \Leftrightarrow b$) equilibrium of the primary effectons (tr and lb) becomes stable starting from $T_\lambda$ due to their polymerization "side by side". This process of *macroscopic* Bose-condensation in real quantum liquids, including conversion of secondary effectons to primary ones, differs from condensation of an ideal Bose-gas. Such kind of Bose-condensation means the enhancement of the concentration of primary effectons in (a) state with lower energy, accompanied by degeneration of the all other kind of collective excitations. The polymerization of primary effectons in He II gives rise to macroscopically long filament-like polyeffectons.

*Such process can be considered as self-organization on macroscopic scale. These filament-like polyeffectons, standing for superfluid component in quantum, can form closed circles or three-dimensional (3D) isotropic networks in a vessel with He II. The remnant fraction of liquid represent normal fraction of He II.*

*14.5 The vortical filaments in superfluids as the analogs of virtual guides of Bivacuum*

Polyeffectons are characterized by the dynamic equilibrium: $[assembly \Leftrightarrow disassembly]$. Temperature decreasing and pressure increasing shift this equilibrium to the left, increasing the surface of the primary effectons side-by-side interaction and number of Josephson junctions. The probability of tunneling between coherent clusters increases also correspondingly.

The relative movement (sliding) of flexible "snake-like" polyeffectons occurs without phonons excitation in the volumes of IR deformons, equal to that of macrodeformons. Just macrodeformons excitation is responsible for dissipation and viscosity in normal liquids (Kaivarainen, 2001; 2006). The absence of macrodeformons excitation, making it possible the polyeffectons emergency (macroscopic Bose condensation), explains the absence of dissipation and superfluidity phenomenon according to our model.

Breaking of symmetry in a three-dimensional polyeffectons network and its violation can be induced by external fields, like the gravitational gradient, mechanical perturbation and surface effects. It is possible because coherent polyeffecton system is highly cooperative and unstable.

In rotating cylindrical vessel, the colinear filament-like polyeffectons originate from 3D isotropic net of polyeffectons and they tend to be oriented along the cylinder axis with their own rotation round their own axis in the direction opposite to that of cylinder rotation, as a consequence of angular momentum conservation. In accordance with our model, this phenomenon represents the vortex filaments in He II, discussed above. The radius of the filaments (42) is determined by the group velocity of the coherent $^4$He atoms, which form part of the primary effectons($\mathbf{v}_{gr} = \mathbf{v}_{sf}$). The numerical value of $\mathbf{v}_{gr}$ must be equal to or less than $6 \cdot 10^3 cm/s$, this corresponding to conditions (14.23 and 14.24). At $T \rightarrow 0$, $\mathbf{v}_{gr}$ decreases, providing the filament radius (14.11) increasing. Finally most probable velocity declines to the values corresponding to $\mathbf{v}_{gr}^{\min} = \mathbf{v}^0$ determined by zero-point oscillations of $^4\mathbf{He}$ atoms. Under these conditions the aggregation or polymerization of translational primary effectons in (a)-state can occur, following by liquid-solid phase transition in $^4$He.

The self-organization of highly cooperative coherent polyeffectons in $\lambda$-point and strong ($\mathbf{a} \rightleftharpoons \mathbf{b}$) equilibrium leftward shift should be accompanied by a heat capacity jump.

The mechanism, leading to stabilization of (a)- state of primary effectons as the first stage of their polymerization, is a formation of *coherent superclusters* from primary effectons. Stabilization of (a) states in *superclusters or bundles of vortical superfluid filaments* could be resulted from macroscopic self-organization of matter, turning mesoscopic Bose condensation to macroscopic one. Corresponding process stabilize the acoustic (a) state of primary effectons and destabilize the optic (b) state.

*The successive mechanisms of super-clusterization of primary effectons and polymerization of these superclusters could be responsible for second order phase transitions, leading to emergency of superfluidity and superconductivity.*

*The second sound* in such a model can be attributed to phase velocity in a system of polyeffectons or superclusters. The propagation of the second sound through chain polyeffectons or superclusters should be accompanied by their elastic deformation and [assembly ⇔ disassembly] equilibrium oscillations.

*The third sound* can be also related to the elastic deformation of polyeffectons and equilibrium constant oscillations of superclusters, *however only in the surface layers* with properties different from those in bulk volume. In accordance with hierarchic theory of surface tension for regular liquids (Kaivarainen, 2001), such a difference between surface and volume parameters is responsible for surface tension ($\sigma$) in quantum liquid, like **HeII**, and its increasing at $\lambda$-point. Such enhancement of $\sigma$ explains disappearance of cavitational bubbles at $T < T_\lambda$.

*The fourth sound* is a consequence of primary effectons volume increasing and the change in their phase velocity as a result of He II interaction with narrow capillary's walls and their thermal movement stabilization.

*The normal component of He II* is related to the fraction of He II atoms, not involved in polyeffectons formation. This fraction composes individual primary and secondary effectons, maintaining the ability for $(a \Leftrightarrow b)$ and $(\bar{a} \Leftrightarrow \bar{b})$ transitions. In accordance with our hierarchic model, these transitions in composition of macroeffectons and macrodeformons are accompanied by the emission and absorption of heat phonons.

The manifestation of viscous properties in normal liquid and normal component of He II is related to fluctuations of macrodeformons ($V_D^M$), accompanied by dissipation (Kaivarainen, 2001).

On the other hand, macro- and superdeformons are absent in the superfluid component, as far in primary polyeffectons at $T < T_\lambda$: the probability of B-state of macroeffectons: $P_B = P_b \cdot \bar{P}_b \to 0$; the probability of A-state of the macroeffectons: $P_A = P_a \cdot \bar{P}_a \to 1$ and, consequently, the probability of macrodeformons tends to zero: $P_D^M = P_B \cdot P_A \to 0$. Decreasing the probability of superdeformons $P_D^S = (P_D^M)_{tr} \cdot (P_D^M)_{lb} \to 0$ means the decreased concentration of cavitational bubbles and vapor pressure.

**3**. We can explain the decrease in E(k) in Fig. 13 around $T = T_\lambda$ by reducing the contributions of (b) state of the primary effectons, due to their Bose-condensation, decreasing the fraction of secondary effectons and concomitant elimination of the contribution of secondary acoustic deformons (i.e. phonons) to the total energy of liquid [4]He. One can see from our theory of viscosity (Kaivarainen, 1995; 2001), that in the absence of secondary effectons and macroeffectons excitations, providing dissipation in liquids, the viscosity of liquid tends to zero: $\eta \to 0$. In accordance with hierarchic theory of thermal conductivity (Kaivarainen, 1995; 2001), the elimination of secondary acoustic deformons at $T \leq T_\lambda$ must lead also to enhanced thermal conductivity. *This effect was registered experimentally in superfluids, indeed.*

**4**. The increase in $E(k)$ in Fig. 1 at $T < T_\lambda$ can be induced by the enhanced contribution of primary polyeffectons to the total energy of He II and the factor: $U_{tot}/T_k = S^{-1}$ in new state equation, derived in Hierarchic theory. The activity of the normal component of He II, as a solvent for polyeffectons, reduces and tends to zero at $T \to 0$. Under such condition ($T = 0$) super-polymerization and total Bose-condensation occur in the volume of [4]**He**.

The maximum in Fig. 13 at $0 < T < T_\lambda$ is a result of competition of two opposite factors: rise in the total energy of **He II** due to progress of primary effectons polymerization and its reduction due to the decline of the most probable group velocity ($\mathbf{v}_{gr}$), accompanied by secondary effectons and deformons degeneration. The latter process predominates at $T \to 0$. The development of a polyeffectons superfluid subsystem is accompanied by corresponding diminution of the normal component in He II ($\rho_S \to 1$ and $\rho \to 0$). The normal component has a bigger internal energy than superfluid one.

The own dimensions of primary translational and librational effectons in composition of polyeffectons increases at $T \to 0$.

*Inaccessibility of b-state of primary effectons at $T \leq T_\lambda$*

Let us analyze our formula (Kaivarainen, 2001) for phase velocity of primary effectons in the acoustic (a)-state at condition $T \leq T_\lambda$, when filament - like polyeffectons from molecules of liquid originate:

$$\mathbf{v}_{ph}^a = \frac{\mathbf{v}_S \frac{1-f_d}{f_a}}{1 + \frac{P_b}{P_a}\left(\frac{v_{\text{res}}^b}{v_{\text{res}}^v}\right)}$$                14.28

where: $\mathbf{v}_S$ is the sound velocity; $P_b$ and $P_a$ are the thermal accessibilities of the (b) and (a) states of primary effectons; $f_d$ and $f_a$ are the probabilities of primary deformons and primary effectons in (a) state excitations.

One can see from (14.28), that if:

$$P_b \to 0, \;\; then \;\; P_d = P_b P_a \to 0 \;\; and \;\; f_d \to 0 \;\; at \;\; T \leq T_\lambda$$

then phase velocity of the effecton in (a) state tends to sound velocity:

$$\mathbf{v}_{ph}^a \to \mathbf{v}_S$$                14.29

For these $\lambda -$ point conditions, the total energy of $^4\mathbf{He}$ atoms, forming polyeffectons due to Bose-condensation of secondary effectons can be presented as:

$$E_{\text{tot}} \sim E_a = m\mathbf{v}_{gr}\mathbf{v}_{ph}^a \to m\mathbf{v}_{gr}\mathbf{v}_S$$                14.30

where the empirical sound velocity in He II is $\mathbf{v}_S = 2.4 \cdot 10^4 cm/s$.

The kinetic energy of wave B at the same conditions is $T_k = m\mathbf{v}_{gr}^2/2$. Dividing $E_{\text{tot}}$ by $T_k$ we have, using (14.16):

$$\frac{\mathbf{v}_S}{\mathbf{v}_{gr}} = \frac{E_{\text{tot}}}{2T_k} = \frac{1}{2S} = \frac{1}{2(m^*/m)}$$                14.31

and

$$\mathbf{v}_{gr}^0 = \mathbf{v}_s \cdot 2S^0 = 2.4 \cdot 10^4 \times 0.32 = 7.6 \cdot 10^3 cm/s.$$                14.32

$m^* = 0.16m$ is the semiempirical effective mass at $T = T_\lambda$.

The most probable wave B length corresponding to (14.32) at $\lambda$-point:

$$\lambda^0 = h/m\mathbf{v}_{gr}^0 = 15.1 \, \mathring{A}$$                64

The number of $^4\mathbf{He}$ atoms in the volume of such effecton, calculated in accordance with (14.23) is equal to: $q^0 = (n_v^0)^{1/3} = 3.8$.

This result is even closer to one predicted by the hierarchic model (eq. 14.24) than (14.22). It confirms that at $T \leq T_\lambda$ the probability of b-state $P_b \to 0$ and conditions (14.29) and (14.30), following from our model, take a place indeed. In such a way our hierarchic model of superfluidity explains the available experimental data on liquid $^4\mathbf{He}$ in a non contradiction manner, as a limit case of our hierarchic viscosity theory for normal liquids.

### *Superfluidity in $^3He$*

The scenario of superfluity, described above for Bose-liquid of $^4\mathbf{He}$ ($S = 0$) in principle is valid for Fermi-liquid of $^3\mathbf{He}$ ($S = \pm1/2$) as well. A basic difference is determined by an additional preliminary stage related to the formation of Cooper pairs of $^3\mathbf{He}$ atoms with total spins, equal to $S = 1\hbar$, i.e. with boson's properties. The bosons only can form primary effectons, as a coherent clusters containing particles with *equal kinetic* energies.

We assume in our model that Cooper's pairs $[^3\mathbf{He}^\uparrow \Leftrightarrow \;^3\mathbf{He}^\uparrow]_{S=1}$ can be formed between neighboring $^3$**He** atoms of opposite spins by head-to principle, when their spins are the additive values. It means that the minimum number of $^3$**He** atoms forming part of the primary effecton's edge at $\lambda$-point must be 8, i.e. two times more than that in $^4$**He** (condition 14.24). Correspondingly, the number of $^3$**He** atoms in the volume of an effecton is $(n_V^0)_{3_{He}} = 8^3 = 312$. These conditions explains the fact that superfluidity in $^3$**He** arises at temperature $T = 2.6 \cdot 10^{-3} K$, i.e. much lower than that in $^4$**He**. For the other hand, the length of coherence in superfluid $^3$**He** is much bigger that in $^4$**He**.

The formation of flexible filament-like polyeffectons, representing macroscopic Bose-condensate in liquid $^3$**He** responsible for superfluidity, is a process, similar to that in $^4$**He** described above. Good review of vortex formation in superfluid $^3$He and analogies in in quantum field theory is presented by Eltsov, Krusius and Volovik (2004).

In contrast to $^4$**He II** there are two major phases of superfluid $^3$**He**, the A and B phases. The important for us neutron - induced vortical filaments formation have been performed in the quasi-isotropic $^3$He-B (Ruutu et al. 1966). In the present context the vortices in $^3$**He-B** are similar to those in superfluid $^4$He-II.

### 14.6 Stimulation of vortex bundles formation in $^3$He-B by spinning elementary particles

A cylindrical sample container with superfluid $^3$He $-$ **B** was rotated at constant angular velocity and temperature T, under NMR absorption monitoring. When the sample is irradiated with neutrons, vortex lines are observed to form. The neutron source was located at a proper distance (few tens of cm) from the cryostat so that vortex lines are observed to form in well resolved individual events. The experimental signal for the appearance of a new vortex line is an abrupt jump in NMR absorption.

Liquid $^3$He-B can be locally heated with the absorption reaction of a thermal neutron:

$$\mathbf{n} + {}_2^3\mathbf{He} \rightarrow \mathbf{p} + {}_1^3\mathbf{H} + E_0 \quad (E_0 = 764 \text{ keV})$$

The reaction products, a proton and a triton (${}_1^3$**H**) produce two collinear ionization tracks (Meyer and Sloan 1997). The ionized particles, electrons and $^3$He ions, diffuse in the liquid and recombine. About 80 % or more of $E_0$ is spent to heat a small volume with a radius about 50 $\mu$m, turning its superfluid state into the normal one.

Subsequently, the heated volume of normal liquid cools back through $T_c$ in microseconds. The measurements demonstrate that vortex lines are stimulated by neutron absorption indeed. In the rotating experiments in Helsinki these rectilinear vortex lines are counted with NMR methods (Ruutu et al. 1996a).

In other series of $^3$He experiments, performed in Grenoble (Bauerle et al. 1996, 1998a), the vortices formed in a neutron absorption event are detected calorimetrically. In zero temperature limit the mutual friction becomes vanishingly small and the life time of the vorticity very long.

Yarmchuk and Packard (1982) obtained images of a vortex in superfluid by imaging of electrons, initially trapped by the vortex cores.

We consider stimulation of vortex bundles formation in superfluids by elementary particles, as a confirmation of our model of fermions as a triplets of sub-elementary fermions, rotation around joint axis (Fig.2). Corresponding superfluid vortical filaments are a structures, analogues to introduced Virtual Guides of spin, momentum and energy, formed by Bivacuum dipoles, connecting coherent elementary particles (see Fig.12 and corresponding comments).

The ability of quantum objects rotation to induce the vortical structures in quantum liquid was obtained in work of Madison et al. (2000). They stir with a focused laser beam a Bose-Einstein condensate of 87Rb atoms confined in a magnetic trap. The formation from single to eleven vortices, increasing with frequency of beam rotation was observed. The measurements of the decay of a vortex array once the stirring laser beam is removed was performed.

This author propose, that the orbits of planets around rotating stars and star systems around rotating center of galactic (supermassive black hole) may correspond to vortical filaments of superfluid fraction of Bivacuum, induced by central object rotation. In accordance to presented theory, these filaments are formed by closed bundles of virtual guides $[\mathbf{N(t,r)} \times \sum \mathbf{VirG}_{SME} (\mathbf{S} <==> \mathbf{R})]^i_{x,y,z}$. These orbits quantization may follow the rules of angular momentum quantization, induced by rotating objects in superfluids. The evidence supporting such idea is existing (Dinicastro, 2005).

## 15  New kind of Bivacuum mediated nonlocal interaction between macroscopic objects

### 15.1  The stages of Bivacuum mediated interaction (BMI) activation between Sender and Receiver

Theories of Virtual Replica (**VR**) of material objects in Bivacuum and primary **VR** Multiplication (**VRM**), described in chapter 13, in combination with theory of Virtual Guides (**VirG**$_{SME}$) (see chapter 14), are the background for explanation of different kind of paranormal phenomena, including parapsychology. The primary **VR** represents a three-dimensional (3D) superposition of Bivacuum virtual standing waves $\mathbf{VPW}^{\pm}_m$ and $\mathbf{VirSW}^{\pm 1/2}_m$, modulated by $[\mathbf{C} \rightleftharpoons \mathbf{W}]$ pulsation of elementary particles and translational and librational de Broglie waves of molecules of macroscopic object.

The infinite multiplication of **VR** in space and time: **VRM(r,t)** in form of 3D packets of virtual standing waves, representing *iterated* primary **VR**, is a result of interference of all pervading external coherent basic *reference waves* - Bivacuum Virtual Pressure Waves ($\mathbf{VPW}^{\pm}_{q=1}$) and Virtual Spin Waves ($\mathbf{VirSW}^{\pm 1/2}_{q=1}$) with similar kinds of modulated standing waves ($\mathbf{VPW}^{\pm}_m$ and $\mathbf{VirSW}^{\pm 1/2}_m$), forming  primary **VR**. The latter can be considered as the *object waves,* making it possible to name the **VRM**, as **Holoiteration** by analogy with regular hologram (see chapter 13).

We put forward a conjecture, that the dependence of complex **VRM(t)** on time is a consequence of its ability to self-organization in both directions - positive (evolution) and negative (devolution) in nonequilibrium conditions. Virtuality of such systems, means by definition, that relativistic mechanics and causality principle do not work for them.

Depending on the type modulation (section 13.2) **VR** and **VRM(r,t)** are subdivided on the:

a) frequency modulated;
b) amplitude modulated;
c) phase modulated;
d) polarization modulated.

Only their superposition contains all the information about positions and dynamics of atoms/molecules, composing object, possibilities of their evolution and devolution.

 The nonlocal Virtual Guides (**VirG**$_{SME}$) of spin, momentum and energy (chapter 14), represent virtual microtubules with properties of one-dimensional virtual Bose condensate, constructed from 'head-to-tail' polymerized Bivacuum bosons ($\mathbf{BVB}^{\pm}$) or Cooper pairs of Bivacuum fermions ($\mathbf{BVF}^{\uparrow} \bowtie \mathbf{BVF}^{\downarrow}$). The bundles of **VirG**$_{SME}$($\mathbf{S}$ <==> $\mathbf{R}$), connecting coherent atoms of Sender (S) and Receiver (S) are responsible for nonlocal Bivacuum mediated interaction between them. The introduced in our theory *Bivacuum Mediated Interaction* (**BMI**) is a new fundamental interaction due to superposition of Virtual replicas of Sender and Receiver and connection of their coherent atoms via **VirG**$_{\mathbf{SME}}$($\mathbf{S}$ <==> $\mathbf{R}$) bundles (eq.14.7a):

$$\left[ \mathbf{N(t,r)} \times \sum^{\mathbf{n}} \mathbf{VirG}_{SME}\,(\mathbf{S} <==> \mathbf{R}) \right]^{i}_{x,y,z}$$

where: (**n**) is a number of pairs of similar tuned elementary particles (protons, neutrons and electrons) in atoms and molecules of **S** and **R**; $\mathbf{N(t,r)}$ is a number of coherent atoms/molecules in the coherent molecular clusters - mesoscopic BC (Kaivarainen, 2001; 2004).

Just $\mathbf{BMI(r,t)}$ is responsible for remote ultraweak nonlocal interaction and different psi-phenomena. For activation of psi-channels the system: [S + R] should be in nonequilibrium state.

**After our Unified Model, the informational (spin), momentum and energy exchange interaction between Sender [S] and Receiver [R], representing *Virtual tunnel* formation, involves following three stages**:

**1**. Superposition of nonlocal (informational/spin) components of [S] and [R] Virtual Replicas Multiplication:

$$\mathbf{VRM}^{nl}_{S} \bowtie \mathbf{VRM}^{nl}_{R} \qquad\qquad 15.1$$

formed by modulated by the objects de Broglie waves virtual spin waves of Sender and Receiver: $\mathbf{VirSW}^{S=\pm 1/2}_{S}$ and $\mathbf{VirSW}^{S=\pm 1/2}_{R}$;

**2**. Formation of bundles of nonlocal Virtual guides of spin, momentum and energy, connecting coherent nucleons and electrons of [S] and [R]:

$$\left[ \mathbf{N(t,r)} \times \sum^{\mathbf{n}} \mathbf{VirG}_{SME}\,(\mathbf{S} <=> \mathbf{R}) \right]^{i}_{x,y,z} = \mathbf{N(t,r)} \times \sum^{\mathbf{n}} \left[ \mathbf{VirSW}^{S=+1/2}_{S} \; {\overset{\mathbf{BVB^{\pm}}}{\underset{\mathbf{BVF^{\uparrow}\bowtie BVF^{\downarrow}}}{=\diamond=}}} \; \mathbf{VirSW}^{S=-1/2}_{R} \right]^{i}_{x,y,z}$$

$\mathbf{VirG}_{SME}\,(\mathbf{S} <=> \mathbf{R})$ is quasi-1D virtual microtubule (quasi one-dimensional virtual Bose condensate), formed primarily by standing $\mathbf{VirSW}^{S=+1/2}_{S} \;{\overset{\mathbf{BVB^{\pm}}}{\underset{\mathbf{BVF^{\uparrow}\bowtie BVF^{\downarrow}}}{<=\diamond=>}}}\; \mathbf{VirSW}^{S=-1/2}_{R}$ of opposite spins, following by self-assembly of Cooper pairs of $[\mathbf{BVF^{\uparrow}} \bowtie \mathbf{BVF^{\downarrow}}]^{i}$ or Bivacuum bosons $(\mathbf{BVB^{\pm}})^{i}$;

**3**. Superposition of distant components of Virtual Replicas Multiplication of [S] and [R], formed by standing virtual pressure waves $[\mathbf{VPW}^{+}_{m} \bowtie \mathbf{VPW}^{-}_{m}]^{i}_{S} =\diamond= [\mathbf{VPW}^{+}_{m} \bowtie \mathbf{VPW}^{-}_{m}]^{i}_{R}$, modulated by [S] and [R]:

$$\mathbf{VRM}^{dis}_{S} \bowtie \mathbf{VRM}^{dis}_{R} = \sum \{ [\mathbf{VPW}^{+}_{m} \bowtie \mathbf{VPW}^{-}_{m}]^{i}_{S} =\diamond= [\mathbf{VPW}^{+}_{m} \bowtie \mathbf{VPW}^{-}_{m}]^{i}_{R} \} \qquad 15.3$$

The described above three stages of [S] and [R] Bivacuum mediated interaction (**BMI**) involves formation of *Virtual tunnel*. For activation of this channel, the whole system: ([S] + [R]) should be in nonequilibrium state.

**We put forward a conjecture, that even teleportation or spatial exchange of macroscopic number of coherent atoms between very remote regions of the Universe (teleportation) is possible via coherent *Virtual tunnels*. If this consequence of this theory will be confirmed, we get a new crucial method of the instant inter-stars propulsion**.

For special case if Sender [S] or Receiver [R] is psychic, the double conducting membranes of the coherent nerve cells (like in axons) may provide the cumulative Casimir effect, contributing Virtual Replica of [S] and [R].

The quantum neurodynamics processes in Sender (Healer) may be accompanied by radiation of electromagnetic waves or magnetic impulses, propagating in Bivacuum via virtual guides: $\mathbf{VirG}_{SME}(\mathbf{S} <=> \mathbf{R})$. Such kind of radiation from different regions of Sender/Healer has been revealed experimentally.

The important role in Bivacuum mediated Mind-Matter and Mind-Mind interaction, plays the coherent fraction of water in **microtubules** of neurons in state of *mesoscopic molecular Bose condensate (mBC)* (Kaivarainen: http://arxiv.org/abs/physics/0102086). This fraction of **mBC** is a variable parameter, dependent on structural state of microtubules and number of simultaneous elementary acts of consciousness (Kaivarainen: http://arxiv.org/abs/physics/0003045). It can be modulated not only by excitation of nerve cells, but also by *specific interaction with virtual replica of one or more chromosomes* $(VR^{DNA})$ *of the same or other cells.*

*The change of frequency of selected kind of thermal fluctuations, like cavitational ones, in the volume of receiver [R], including cytoplasm water of nerve cells, is accompanied by reversible disassembly of microtubules and actin' filaments, i.e.* [**gel** $\rightleftharpoons$ **sol**] *transitions. These reactions, responsible for elementary act of consciousness,* are dependent on the changes of corresponding activation barriers.

*The mechanisms of macroscopic quantum entanglement, proposed in our work, is* responsible for change of intermolecular Van der Waals interaction in the volume of [R] and probability of selected thermal fluctuations (i.e. cavitational fluctuations), induced by [S]. In this case, realization of certain series of elementary acts of consciousness of [S] will induce similar series in nerve system of [R]. This means informational exchange between $VR^R$ and $VR^S$ of two psychics via Virtual Guides: $\mathbf{VirG}_{SME}^i(\mathbf{S} <=> \mathbf{R}),$ and *their bundles, forming Virtual tunnels, which may be named Psi-channels*:

$$\left[ \mathbf{N(t,r)} \times \sum^{\mathbf{n}} \mathbf{VirG}_{SME}\,(\mathbf{S} <=> \mathbf{R}) \right]_{x,y,z}^{i}$$

The *specific character* of telepathic signal transmission from [S] to [R] may be provided by modulation of $\mathbf{VRM}_{MT}^{S}$ of microtubules by $\mathbf{VRM}_{DNA}^{S}$ of DNA of sender's chromosomes in neuron ensembles, responsible for subconsciousness, imagination and consciousness. The resonance - the most effective remote informational/energy exchange between two psychics is dependent on corresponding 'tuning' of their nerve systems. As a background of this tuning can be described Bivacuum mediated interaction (BMI) between the crucial neurons components of [S] to [R]:

$$\sum \left[ 2\ \mathbf{centrioles + chromosomes} \right]_{\mathbf{S}} \overset{\mathbf{BMI}}{<===>} \sum \left[ 2\ \mathbf{centrioles + chromosomes} \right]_{\mathbf{R}} \qquad 15.4$$

In accordance to our theory of elementary act of consciousness and *three stages of BMI mediated Psi channel formation*, described above, the modulation of dynamics of [assembly $\rightleftharpoons$ disassembly] of microtubules by influence on probability of cavitational fluctuations in the nerve cells and corresponding [*gel* $\rightleftharpoons$ *sol*] transitions by directed mental activity of [Sender] can provide **telepathic contact and remote viewing** between [Sender] and [Receiver].

The mechanism of **remote healing** could be the same, but the local targets in the body of patient [R] are not necessarily the MTs and chromosomes of the nerve cells, but **centrioles** + **chromosomes** of the ill organs (heart, liver, etc.).

The **telekinesis**, as example of mind-matter interaction, should be accompanied by significant nonequilibrium process in the nerve system of Sender, related to increasing of kinetic energy of coherent molecules in neurons of Sender, like cumulative momentum of water clusters, coherently melting in microtubules of centrioles and inducing their disassembly. Corresponding momentum and kinetic energy are transmitted to 'receiver - target' via multiple correlated bundles of $\mathbf{VirG}_{SME}$ in superimposed $\mathbf{VRM}_{S,R}$ (Psi-channels).

The specific magnetic potential exchange between [S] and [R] via *Virtual tunnel* can be generated by the nerve impulse regular propagation along the axons and depolarization of nerve cells membranes (i.e. electric current) in the 'tuned' ensemble of neuron cells of psychic - [Sender], accompanied by magnetic flux. These processes are accompanied by $\mathbf{BVF}^{\uparrow} \rightleftharpoons \mathbf{BVB}^{\pm} \rightleftharpoons \mathbf{BVF}^{\downarrow}$ equilibrium shift to the right or left, representing magnetic field excitation.

The evidence are existing, that *Virtual tunnel* between [S] and [R] works better, if the frequencies of geomagnetic Schumann waves - around 8 Hz (close to brain waves frequency) are the same in location of [S] and [R]. However, the main coherence factor in accordance to our theory, are all-pervading Bivacuum virtual pressure waves ($\mathbf{VPW}_{q=1}^{\pm}$), with basic Compton frequency $[\omega_0 = \mathbf{m}_0\mathbf{c}^2/\hbar]^i$, equal to carrying frequency of [Corpuscle $\rightleftharpoons$ Wave] pulsations of the electrons, protons, neurons, composing real matter and providing entanglement. The macroscopic Bivacuum flicker fluctuation, activated by nonregular changes/jumps in properties of complex Hierarchical Virtual replica of Solar system and even galactic, related to *sideral time*, also may influence on quality of Psi-chanells between Sender and Receiver.

Formation of the different kinds of virtual standing waves, representing nonlocal and distant fractions of Virtual Replicas $(VR)_{S,R}$ of Sender [S] and Receiver [R], necessary

*Virtual tunnel:* $\left[ \mathbf{N(t,r)} \times \sum^{\mathbf{n}} \mathbf{VirG}_{SME}\left(\mathbf{S} <=> \mathbf{R}\right) \right]_{x,y,z}^{i}$ formation, are presented in Table 1

**TABLE 1**

# The role of paired and unpaired sub-elementary particles
## of the electron's [Corpuscle ⇌ Wave] pulsation and rotation:

$$\langle [\mathbf{F}_\uparrow^+ \bowtie \mathbf{F}_\downarrow^-]_W + (\mathbf{F}_\uparrow^-)_C \rangle \rightleftharpoons \langle [\mathbf{F}_\uparrow^+ \bowtie \mathbf{F}_\downarrow^-]_C + (\mathbf{F}_\downarrow^-)_W \rangle$$

## in Bivacuum mediated interaction between sender [S] and receiver [R]

**Pair of sub-elementary particle**

**and antiparticle pulsation and rotation:**

$$[\mathbf{F}_\uparrow^+ \bowtie \mathbf{F}_\downarrow^-]_W \rightleftharpoons \langle [\mathbf{F}_\uparrow^+ \bowtie \mathbf{F}_\downarrow^-]_C$$

1. Virtual Pressure Waves: $\left[ \mathbf{VPW}^+ \bowtie \mathbf{VPW}^- \right]$

2. Total Virtual Pressure energy increment,

equal to that of total and unpaired $(\Delta \mathbf{E}_{\mathbf{F}_\uparrow^+})$:

$$\Delta \mathbf{E}_{\mathbf{F}_\uparrow^+} \sim \Delta \mathbf{VirP}_{\mathbf{F}_\uparrow^+}^\pm = \frac{1}{2} \left| \mathbf{VirP}_{\mathbf{F}_\uparrow^+}^+ - \mathbf{VirP}_{\mathbf{F}_\downarrow^-}^- \right|^{[\mathbf{F}_\uparrow^+ \bowtie \mathbf{F}_\downarrow^-]} +$$

$$+ \frac{1}{2} \left| \mathbf{VirP}_{\mathbf{F}_\uparrow^+}^+ + \mathbf{VirP}_{\mathbf{F}_\downarrow^-}^- \right|^{[\mathbf{F}_\uparrow^+ \bowtie \mathbf{F}_\downarrow^-]}$$

where the kinetic and potential energy increments:

$$\Delta \mathbf{T}_k = \frac{1}{2} \left| \mathbf{VirP}_{\mathbf{F}_\uparrow^+}^+ - \mathbf{VirP}_{\mathbf{F}_\downarrow^-}^- \right|^{[\mathbf{F}_\uparrow^+ \bowtie \mathbf{F}_\downarrow^-]}$$

$$\Delta \mathbf{V} = \frac{1}{2} \left| \mathbf{VirP}_{\mathbf{F}_\uparrow^+}^+ + \mathbf{VirP}_{\mathbf{F}_\downarrow^-}^- \right|^{[\mathbf{F}_\uparrow^+ \bowtie \mathbf{F}_\downarrow^-]}$$

3. Virtual Replica of the Object (VR = VR$^{in}$ + VR$^{sur}$)

4. Virtual Replicas of [S] and [R] Multiplication:

$$\mathbf{VRM}_S = \sum \mathbf{VR}_S \iff \sum \mathbf{VR}_R = \mathbf{VRM}_R$$

**Unpaired sub-elementary**

**fermion pulsation and rotation:**

$$(\mathbf{F}_{\uparrow\downarrow}^\pm)_C \overset{\mathbf{BvSO}}{\rightleftharpoons} (\mathbf{F}_{\uparrow\downarrow}^\pm)_W \overset{\mathbf{CVC}^{\circlearrowleft\circlearrowright}}{\Longleftrightarrow} \mathbf{VirSW}^{\circlearrowleft\circlearrowright}$$

1. Electromagnetic potential:

$$\mathbf{E}_{EM} = \alpha \mathbf{m}_V^+ \mathbf{c}^2 \sim$$

$$\sim \frac{1}{2} \left| \mathbf{VirP}_{\mathbf{F}_\uparrow^+}^+ - \mathbf{VirP}_{\mathbf{F}_\downarrow^-}^- \right|^{[\mathbf{F}_\uparrow^+ \bowtie \mathbf{F}_\downarrow^-]}$$

2. Gravitational potential:

$$\mathbf{E}_G = \beta [\mathbf{m}_V^+ + |\mathbf{m}_V^-|] \mathbf{c}^2 \sim$$

$$\sim \frac{1}{2} \left| \mathbf{VirP}_{\mathbf{F}_\uparrow^+}^+ + \mathbf{VirP}_{\mathbf{F}_\downarrow^-}^- \right|^{[\mathbf{F}_\uparrow^+ \bowtie \mathbf{F}_\downarrow^-]}$$

3. Virtual Spin Waves (VirSW):

$$\mathbf{I}_S \equiv \mathbf{I}_{\mathbf{VirSW}^{\pm 1/2}} \sim \mathbf{K}_{BVF^\uparrow \Leftrightarrow BVF^\downarrow}(\mathbf{t}) =$$

$$(\mathbf{K}_{BVF^\uparrow \Leftrightarrow BVF^\downarrow})_0 [\sin(\omega_0^i t) + \gamma \omega_B^{lb} \sin(\omega_B^{lb} t)]$$

4. The bundles of Virtual Guides:

$$\left[ \mathbf{N}(\mathbf{t,r}) \times \sum_{}^{n} \mathbf{VirG}_{SME}(\mathbf{S} \Longleftrightarrow \mathbf{R}) \right]_{x,y,z}^{i}$$

formation between remote [S] and [R]:

$$\mathbf{VirSW}_S^{S=+1/2} \overset{\mathbf{BVB}^\pm}{\underset{BVF^\uparrow \bowtie BVF^\downarrow}{\Longleftrightarrow}} \mathbf{VirSW}_R^{S=-1/2}$$

Pauli attraction (Cooper pairs formation) or repulsion between $\mathbf{BVF}^{\updownarrow}$

of the opposite or similar spins

---

*One of the result of Virtual tunnel formation, as a superposition of VRM$_{S,R}$ and bundles of VirG$_{SME}^{ext}$,*

*is a change of permittivity $\varepsilon_0$ and permeability $\mu_0$ of Bivacuum $[\varepsilon_0 = n_0^2 = 1/(\mu_0 c^2)]$.*

*In turn, $(\pm \Delta \varepsilon_0)$ influence Van-der-Waals interactions in condensed matter,*

*changing the probability of defects origination in solids and cavitational fluctuations in liquids.*

*Bidirectional change of pH of water via Virtual tunnel can be a consequence*

*of $\pm \Delta VP^\pm$ and $\pm \Delta \varepsilon_0$ influence on cavitational fluctuations, accompanied by shift of dynamic equilibrium:*

$H_2O \rightleftharpoons HO^- + H^+$ *and assembly* $\rightleftharpoons$ disassembly of microtubules in nerve cells.

The coherency of all components of Virtual wave guide between [S] and [R], formed by nonlocal virtual spin waves (**VirSW**$^\circlearrowright$ and **VirSW**$^\circlearrowleft$) of two opposite angular momentums and virtual pressure waves (**VPW**$_q^+$ and **VPW**$_q^-$) of two opposite energies, corresponds to finest tuning of mind-matter and mind-mind interaction. The coherency between signals of [S] and [R] can be provided by *Tuning Force (TF) of Bivacuum* and modulation of nonlocal Virtual Guides (**VirG**$_{SME}$) by cosmic and geophysical magnetic flicker noise.

The [*dissociation* $\rightleftharpoons$ *association*] equilibrium oscillation of coherent water clusters in state of molecular Bose condensate (mBC) in microtubules of nerve cells, modulating (**VirSW**$^{\circlearrowright,\circlearrowleft}$) and **VPW**$^\pm$, is a crucial factor for realization of quantum Psi phenomena. The virtual replica (VR) of microtubules and its multiplication (VRM) can be modulated also by secondary virtual replicas of DNA.

### 15.2 The examples of Bivacuum mediated interaction (BMI) between macroscopic objects

In accordance to our approach, the remote interaction between macroscopic Sender [S] and Receiver [R] can be realized, as a result of *Bivacuum mediated interaction (BMI)*, like superposition of distant and nonlocal components of their Virtual Replicas Multiplication (**VRM**$_S \leftrightarrow$ **VRM**$_R$), described in previous sections.

Nonequilibrium processes in [Sender], accompanied by acceleration of particles, like evaporation, heating, cooling, melting, boiling etc. may stimulate the *nonelastic effects* in the volume of [Receiver] and increments of modulated virtual pressure and spin waves (**ΔVPW**$_\mathbf{m}^\pm$ and **ΔVirSW**$_\mathbf{m}^{\pm 1/2}$), accompanied [**C** $\rightleftharpoons$ **W**] pulsation of triplets $[\mathbf{F}_\uparrow^+ \bowtie \mathbf{F}_\downarrow^-] + \mathbf{F}_\updownarrow^\pm >^i$, formed by sub-elementary fermions of different generation, representing electrons, protons and neutrons.

The following unconventional kinds of effects of nonelectromagnetic and non-gravitational nature can be anticipated in the remote interaction between **macroscopic** nonequilibrium [Sender] and sensitive detector [Receiver] via multiple Virtual spin and energy guides **VirG**$_{SME}$ (Fig.4), if our theory of nonlocal spin, momentum and energy exchange between [S] and [R], described above is correct:

**I**. Weak repulsion and attraction between 'tuned' [S] and [R] and rotational momentum in [R] induced by [S], as a result of transmission of momentum/kinetic energy and angular momentum (spin) between elementary particles of [S] and [R]. The probability of such 'tuned' interaction between [S] and [R] is dependent on dimensions of coherent clusters of atoms and molecules of condensed matter in state of mesoscopic Bose condensation (**mBC**) (Kaivarainen, 1995; 2001; 2003; 2004). The number of atoms in such clusters $\mathbf{N(t,r)}$ is related to number of **VirG**$_{SME}$ in the bundles $\left[ \mathbf{N(t,r)} \times \sum_{}^{\mathbf{n}} \mathbf{VirG}_{SME} (\mathbf{S} \Longleftrightarrow \mathbf{R}) \right]_{x,y,z}^i$, connecting tuned **mBC** in [S] and [R]. The $\mathbf{N(t,r)}$ may be regulated by temperature, ultrasound, etc. The kinetic energy distant transmission from atoms of [S] to atoms of [R] may be accompanied by the temperature and local pressure/sound effects in [R];

**II**. Increasing the probability of thermal fluctuations in the volume of [R] due to decreasing of Van der Waals interactions, because of charges screening effects, induced by overlapping of distant virtual replicas of [S] and [R] and increasing of dielectric permittivity of Bivacuum. In water the variation of probability of cavitational fluctuations should by accompanied by the in-phase variation of pH and electric conductivity due to shifting the equilibrium: $H_2O \rightleftharpoons H^+ + HO^-$ to the right or left;

**III**. Small changing of mass of [R] in conditions, changing the probability of the inelastic recoil effects in the volume of [R] under influence of [S];

**IV**. Registration of metastable virtual particles (photons, electrons, positrons), as a result of Bivacuum symmetry perturbations.

*The first kind* (I) of new class of interactions between coherent fermions of [S] and [R] is a result of huge number (bundles) of correlated virtual spin-momentum-energy guides $\mathbf{VirG}_{SME} \equiv [\mathbf{VirSW}_S^\cup \Rightarrow\Leftrightarrow\Leftarrow \mathbf{VirSW}_R^\cup]$ formation by standing spin waves ($\mathbf{VirSW}_{S,R}$).

These guides can be responsible for:

a) virtual signals (phase/spin), momentum and kinetic energy instant transmission between [S] and [R], meaning the nonlocal information and energy exchange;

b) the regulation of Pauli repulsion effects between fermions of [S] and [R] with parallel spins;

c) the transmission of macroscopic rotational momentum from [S] of [R]. This process

provided by $\left[ \mathbf{N(t,r)} \times \sum\limits^n \mathbf{VirG}_{SME} \, (\mathbf{S} \mathrel{<=\!>} \mathbf{R}) \right]^i_{x,y,z}$ , is dependent on the difference between the *external* angular momentums of elementary fermions of [S] and [R].

*The second kind* (II) of phenomena: influence of [S] on probability of thermal fluctuations in [R], - is a consequence of the additional symmetry shift in Bivacuum fermions ($\mathbf{BVF}^{\ddagger}$), induced by superposition of distant and nonlocal multiplicated Virtual Replicas of [S] and [R]: $\mathbf{VRM}^S \bowtie \mathbf{VRM}^R$, which is accompanied by increasing of Bivacuum fermions ($\mathbf{BVF}^{\ddagger} = [\mathbf{V}^+ \Updownarrow \mathbf{V}^-]$) virtual charge: $\Delta\mathbf{e} = (\mathbf{e}_{V^+} - \mathbf{e}_{V^-}) \ll \mathbf{e}_0$ in the volume of [R]. Corresponding increasing of Bivacuum permittivity ($\boldsymbol{\varepsilon}_0$) and decreasing magnetic permeability ($\boldsymbol{\mu}_0$) : $\boldsymbol{\varepsilon}_0 = 1/(\boldsymbol{\mu}_0\mathbf{c}^2)$ is responsible for the charges screening effects in volume of [R], induced by [S]. This weakens the electromagnetic Van der Waals interaction between molecules of [R] and increases the probability of defects origination and cavitational fluctuations in solid or liquid phase of Receiver.

*The third kind of phenomena (III)*: reversible decreasing of mass of rigid [R] can be a result of reversible lost of energy of Corpuscular phase of particles, as a consequence of inelastic recoil effects, following the in-phase [$\mathbf{C} \rightarrow \mathbf{W}$] transition of $\mathbf{N}_{coh}$ coherent nucleons in the volume of [R].

The probability of recoil effects can be enhanced by heating the rigid object or by striking it by another hard object. This effect can be registered directly - by the object mass decreasing. In conditions, close to equilibrium, the Matter - Bivacuum energy exchange relaxation time, following the process of coherent [$\mathbf{C} \rightleftharpoons \mathbf{W}$] pulsation of macroscopic fraction of atoms is very short and corresponding mass defect effect is undetectable. *Such collective recoil effect of coherent particles* could be big in superconducting or superfluid systems of macroscopic Bose condensation or in good crystals, with big domains of atoms in state of Bose condensation.

*The fourth kind of the above listed phenomena* - increasing the probability of virtual particles and antiparticles origination in asymmetric Bivacuum in condition of forced resonance with exciting Bivacuum virtual waves was discussed earlier (Kaivarainen http://arxiv.org/abs/physics/0103031).

It was demonstrated (Kaivarainen: http://arxiv.org/abs/physics/0207027), that the listed nontrivial consequences of our Unified theory (I - IV) are consistent with unusual data, obtained by groups of Kozyrev (1984; 1991) and Korotaev (1999; 2000). It is important to note, that these experiments are incompatible with existing today paradigm. It means that the current paradigm is timed out and should be replaced by the new one.

### 15.3 The idea of nonlocal signals transmitter and detector construction and testing

The simple constructions of artificial physical devices with functions of [Sender] and one or more [Receiver] for verification of nonlocal mechanism of communication via Virtual Guides of spin/information, momentum and energy, following from our Unified theory, were suggested (Kaivarainen, 2004a; 2004b). They can represent two or more identical and 'tuned' to each other superconducting or superfluid multi-rings or torus/donuts systems.

The pair: [S] and [R] can be presented by two identical systems, composed from the same number (7 or more) of superconducting or superfluid rings of decreasing radius - from meters to centimeters, following Fibonacci series, because of fundamental role of Golden mean in Nature, enclosed in each other. The "tuning" of Virtual Replicas of [S] and [R] constructions in state of macroscopic Bose condensation (superconducting or superfluid) can be realized by keeping them nearby with parallel orientation of two set of rings during few hours for equalizing of their physical parameters, i.e. currents. After such tuning, they can be removed from each other, keeping their superconducting or superfluid state on at the same temperature, pressure and other conditions. The separation can be increased from hundreds of meters to hundreds of kilometers and tested for signals transmission in each equipped for such experiments laboratory.

The experiments for registration of nonlocal interactions could be performed, as follows. At the precisely *fixed time moment*, the superfluid or superconducting properties of one of rings of Sender [S], should be *switched off* by heating, ultrasound or magnetic field (Meissner effect). At the same moment of time the superconducting or superfluid parameters of all rings of Receiver [R] should be registered. *If the biggest changes will occur in the ring of [R]-system with the same radius, as that in [S]-system and faster, than light velocity, it will be a confirmation of possibility of nonlocal Bivacuum mediated information and momentum exchange (entanglement), following from our theory and based on resonant principles. The corresponding remote signals exchange via proposed in our work Virtual Guides (VirG$_{SME}$), should not be shielded by any screen.*

There are a number of laboratories over the World, capable to perform the proposed experimental project. In the case of success, such Nonlocal Signals Detector/Transmitter (NSD/T) with variable parameters would be the invaluable tool for extraterrestrial civilizations search in projects, like SETI and for distant cosmos exploration (NASA). On the Earth, the Internet, radio and TV - nets also will get a strong challenge.

### 15.4  GeoNet of CAMP based - Detectors of Water Properties, as a Supersensor of Terrestrial and Extraterrestrial Coherent Signals

The idea of GeoNet of equidistantly distributed over the surface of the Earth hundreds of water detectors, serving as a Supersensor is based on unique informational possibilities of new optoacoustic device: Comprehensive Analyzer of Matter Properties (CAMP). The CAMP is one of applications of new Hierarchic theory of condensed matter, general for liquids and solids (http://arxiv.org/abs/physics/0207114). Using theory based computer program (copyright, 1997, USA, Kaivarainen) and four input experimental parameters, measured at the same temperature and pressure:

  1) sound velocity;
  2) density
  3) refraction index and
  4) positions of translational and librational bands in IR or Raman spectra -

it is possible, using PC in less than second, to calculate more than 300 physical parameters of water, ice or other condensed matter. These parameters include internal energy, heat capacity, viscosity, self-diffusion, thermal conductivity, surface tension, dimensions and life-times of 24 quantum excitations, describing condensed matter dynamic structure.

Water is a sensitive detector for any kind of fields, including gravitational one via bundles of nonlocal Virtual Guides of spin, momentum and energy

$$\left[ \mathbf{N(t, r)} \times \sum_{x,y,z}^{\mathbf{n}} \mathbf{VirG}_{SME} (\mathbf{S} <=> \mathbf{R}) \right]_{x,y,z}^{i}$$ . The Sun, Moon and perhaps, the black hole in center of our galactic are the strongest sources of coherent oscillations of gravitational field (GF), existing in accordance to our theory, in form of modulated virtual pressure waves of positive and negative energy ($\mathbf{VPW^+}$ and $\mathbf{VPW^-}$), interacting with protons and electrons of water molecules.

The induced by GF *coherent* changes of water physical properties on the remote points of the Earth surface, registered by CAMP devices, can be analyzed by the global CAMP - GeoNet system via Internet.

The corresponding coherent variations of physical properties of standard aqueous solutions in EM screened vessels by Faraday cages at constant temperature and pressure could be monitored by CAMP. Such [water samples/detectors + CAMP], will be distributed over the surface of the Earth, forming a nodes of GeoNet.

I propose to use such GeoNet on the Earth surface, like giant Supersensor for terrestrial and extraterrestrial coherent signals registration. For this end a hundreds of standard water-filled cells, unified with CAMP, over the planet surface should be under permanent centralized control, using satellites and the Internet. The Fourier analysis of the input signals, inducing water perturbations, registered by CAMP, makes it possible to select only coherent patterns of dynamic changes of water properties in big number of water-filled cells over the Earth. These patterns will be analyzed for getting the detailed information about the amplitude and frequency of coherent signals.

Sensitivity of proposed global sensor system - GeoNet is much higher than existing currently technics due to its global scale and the CAMP huge informational possibilities. The localization and forecast of the Earthquakes are a minimum results of such global project realization. This forecasting compensate quickly all related to project of GeoNet expenses.

The valuable knowledge about the influence of gravitational dynamics of Sun, Moon and planets of Solar system on the geophysical process on the Earth could be obtained via proposed GeoNet of CAMP systems.

## Main Conclusions

1. A new Bivacuum model is developed, as the infinite dynamic superfluid matrix of virtual dipoles, named Bivacuum fermions ($\mathbf{BVF}^{\updownarrow})^i$ and Bivacuum bosons ($\mathbf{BVB}^{\pm})^i$, formed by correlated torus ($\mathbf{V^+}$) and antitorus ($\mathbf{V^-}$), as a collective excitations of subquantum particles and antiparticles of opposite energy, charge and magnetic moments, separated by energy gap. In primordial symmetric Bivacuum, i.e. in the absence of matter and fields, these parameters of torus and antitorus totally compensate each other. Their spatial and energetic properties correspond to three generations of electrons, muons and tauons ($i = e, \mu, \tau$). The symmetric primordial Bivacuum can be considered as the *Universal Reference Frame* (**URF**), i.e. *Ether*, in contrast to *Relative Reference Frame* (**RRF**), used in special relativity (SR) theory. The elements of *Ether - ethons* correspond to our Bivacuum dipoles. It is shown in our work, that the result of Michelson - Morley experiment can be a consequence of *ether drug* by the Earth or Virtual Replica of the Earth in terms of our theory. The positive and negative Virtual Pressure Waves ($\mathbf{VPW}_q^{\pm}$) and Virtual Spin Waves ($\mathbf{VirSW}_q^{S=\pm 1/2}$) are the result of emission and absorption of positive and negative energy Virtual Clouds ($\mathbf{VC}_q^{\pm}$), resulting from transitions of torus $\mathbf{V^+}$ and antitorus $\mathbf{V^-}$ between different states of excitation, symmetrical in realms of positive and negative energy: $j - k = q$;

2. The symmetry shift between $\mathbf{V}^+$ and $\mathbf{V}^-$ actual and complementary mass and charge to the left or right, opposite for Bivacuum fermions $\mathbf{BVF}^\uparrow$ and antifermions $\mathbf{BVF}^\downarrow$, has the relativistic and reverse to that dependence on these dipoles *external tangential* or pure translational velocity. It is shown, that the value of Bivacuum dipoles symmetry shift is a criteria of their external *absolute velocity*, characterizing properties of secondary Bivacuum. This shift is accompanied by sub-elementary fermion and antifermion formation. The formation of sub-elementary fermions/antifermions and their fusion to stable triplets of elementary fermions, like electrons or protons $\langle [\mathbf{F}^-_\uparrow \bowtie \mathbf{F}^+_\downarrow] + \mathbf{F}^{\pm}_\updownarrow \rangle^{e,p}$, following by the *rest mass and charge* origination, become possible at the certain rotation velocity ($\mathbf{v}$) of Cooper pairs of $[\mathbf{BVF}^{\uparrow} \bowtie \mathbf{BVF}^{\downarrow}]$ around their common axis. It is shown, that this rotational-translational velocity value is determined by Golden Mean condition: $(\mathbf{v}/\mathbf{c})^2 = \phi = 0.618$. The close values of centripetal and Coulomb interaction, calculated on the base of most important parameters of paired sub-elementary fermions in their Corpuscular phase, following from our model of elementary particles and time theory, is very important fact. *It is a strong evidence in proof of our Unified theory of Bivacuum, elementary particles, mass and charge origination at Golden mean conditions and theory of time;*

3. The fundamental physical roots of Golden Mean condition: $(\mathbf{v}/\mathbf{c})^2 = \mathbf{v}^{ext}_{gr}/\mathbf{v}^{ext}_{ph} = \phi$ are revealed, as the equality of internal and external group and phase velocities of torus and antitorus of sub-elementary fermions, correspondingly: $\mathbf{v}^{in}_{gr} = \mathbf{v}^{ext}_{gr}$; $\mathbf{v}^{in}_{ph} = \mathbf{v}^{ext}_{ph}$. These equalities are named 'Hidden Harmony Conditions';

4. The new expressions for total, potential ($\mathbf{V}$) and kinetic ($\mathbf{T}_k$) energies of de Broglie waves of elementary particles were obtained. One of the expressions represents the extended basic Einstein - de Broglie formula $\mathbf{E}_{tot} = \mathbf{m}_0\mathbf{c}^2 = \hbar\omega_0$ for free particle:

$$\mathbf{E}_{tisot} = \mathbf{V} + \mathbf{T}_k = \frac{1}{2}(\mathbf{m}^+_V + \mathbf{m}^-_V)\mathbf{c}^2 + \frac{1}{2}(\mathbf{m}^+_V - \mathbf{m}^-_V)\mathbf{c}^2$$

$$or : \ \mathbf{E}_{tot} = \mathbf{m}^+_V\mathbf{c}^2 = \hbar\omega_{\mathbf{C}\rightleftharpoons\mathbf{W}} = \sqrt{1 - (\mathbf{v}/\mathbf{c})^2}\,\mathbf{m}_0\mathbf{c}^2 + \mathbf{h}^2/\mathbf{m}^+_V\lambda^2_B$$

$$or : \ \mathbf{E}_{tot} = \hbar\omega_{\mathbf{C}\rightleftharpoons\mathbf{W}} = \sqrt{1 - (\mathbf{v}/\mathbf{c})^2}\,\hbar\omega_0 + \mathbf{h}\mathbf{v}_B$$

where: $\mathbf{V} = \frac{1}{2}(\mathbf{m}^+_V + \mathbf{m}^-_V)\mathbf{c}^2$; $\mathbf{T}_k = \frac{1}{2}(\mathbf{m}^+_V - \mathbf{m}^-_V)\mathbf{c}^2$; $\mathbf{m}^+_V = \mathbf{m}_0/\sqrt{1 - (\mathbf{v}/\mathbf{c})^2}$ and $\mathbf{m}^-_V = \mathbf{m}_0\sqrt{1 - (\mathbf{v}/\mathbf{c})^2}$ are the actual (inertial) and complementary (inertialess) mass of torus and antitorus of sub-elementary fermion, correspondingly; $\omega_{\mathbf{C}\rightleftharpoons\mathbf{W}} = \mathbf{m}^+_V\mathbf{c}^2/\hbar$ is the resulting frequency of $[\mathbf{C} \rightleftharpoons \mathbf{W}]$ pulsation of sub-elementary fermion; $\omega_0 = \mathbf{m}_0\mathbf{c}^2/h$ is the Compton frequency of internal $[\mathbf{C} \rightleftharpoons \mathbf{W}]$ pulsation.

The new formulas take into account the contributions of the actual mass/energy of torus ($\mathbf{V}^+$) and those of complementary antitorus ($\mathbf{V}^-$), correspondingly, of asymmetric sub-elementary fermions to the total ones. The shift of symmetry between the inertial and inertialess mass and other parameters of torus and antitorus of sub-elementary fermions are dependent on their *internal* rotational-translational dynamics in composition of triplets and the *external* translational velocity of the whole triplets. The latter determines the external translational momentum and the empirical de Broglie wave frequency: $\mathbf{v}_B = \mathbf{m}^+_V\mathbf{v}^2/\mathbf{h}$ and length: $\lambda_B = \mathbf{h}/\mathbf{m}^+_V\mathbf{v}$;

5. A dynamic mechanism of [corpuscle ($\mathbf{C}$) $\rightleftharpoons$ wave ($\mathbf{W}$)] duality is proposed. It involves the modulation of the internal (hidden) quantum beats frequency between the asymmetric 'actual' (torus) and 'complementary' (antitorus) states of sub-elementary fermions or antifermions by the external - empirical de Broglie wave frequency of the whole particles (triplets), equal to beats of similar states of the 'anchor' Bivacuum fermion.

In nonrelativistic conditions such modulation stands for the wave packets origination. The process of transition of corpuscular phase to the wave phase is accompanied by reversible change of translational degrees of freedom to rotational ones;

6. The high-frequency photon is a result of fusion (annihilation) of two triplets of particle and antiparticle. It represents a rotating sextet of sub-elementary fermions and antifermions with axial structural symmetry and minimum energy $2\mathbf{m}_0^e\mathbf{c}^2$. The regular photons are the result of excitation of *secondary anchor sites* of the electrons or protons excitation. The *secondary anchor site* represents three correlated Cooper pairs $3[\mathbf{BVF}^\uparrow \bowtie \mathbf{BVF}^\downarrow]^i_{as}$. Its excitation can be a result of charge acceleration, like in ondulators or that, accompanied the transitions between excited and ground states of atoms and molecules. The electromagnetic field is a result of Corpuscle - Wave pulsation of photons, exciting $[\mathbf{VPW}^+ \bowtie \mathbf{VPW}^-]$ and their fast rotation with angle velocity ($\omega_{rot}$), equal to $[\mathbf{C} \leftrightarrows \mathbf{W}]$ pulsation frequency. The clockwise or anticlockwise direction of photon rotation, as respect to direction of its propagation, corresponds to its spin sign: $s = \pm\hbar$;

7. It is shown, that the information, encrypted in ancient *Sri-Yantra diagram,* can be interpreted as a confirmation of proposed mechanisms of origination of the rest mass and charge of elementary particles just at *Golden Mean* conditions and their corpuscle - wave duality;

8. The electrostatic and magnetic fields origination is a consequence of reversible $[recoil \leftrightarrows antirecoil]$ effects in Bivacuum matrix, generated by correlated $[Corpuscle \leftrightarrows Wave]$ pulsation of sub-elementary fermions/antifermions of triplets and their fast rotation, accompanied by Bivacuum dipoles symmetry shift and shift of equilibrium of Bivacuum fermions $[\mathbf{BVF}^\uparrow \rightleftharpoons \mathbf{BVF}^\downarrow]$ to the left (North pole) or to the right (South pole). The linear and axial alignment of Bivacuum dipoles and their dynamics are responsible for electrostatic and magnetic fields 'force lines' origination, correspondingly. The zero-point vibrations of particles and evaluated zero-point velocity of these vibrations are also the result of $[recoil \leftrightarrows antirecoil]$ effects, accompanied by $[\mathbf{C} \leftrightarrows \mathbf{W}]$ pulsation of triplets in state of rest, when their external translational velocity is zero;

9. The gravitational waves and field are the result of positive and negative energy virtual pressure waves excitation ($\mathbf{VPW}_q^+$ and $\mathbf{VPW}_q^-$) by the in-phase $[\mathbf{C} \rightleftharpoons \mathbf{W}]$ pulsation of pairs $[\mathbf{F}_\uparrow^- \bowtie \mathbf{F}_\downarrow^+]$ of triplets $\langle[\mathbf{F}_\uparrow^- \bowtie \mathbf{F}_\downarrow^+] + \mathbf{F}_\updownarrow^\pm\rangle$, counterphase to that of unpaired $\mathbf{F}_\updownarrow^\pm\rangle$. Such virtual waves provide the attraction or repulsion between pulsing remote particles, depending on phase shift of pulsations, as in the case of hydrodynamic Bjerknes force interaction.

The potential gravitational energy of huge number of Bivacuum dipoles in space between gravitating objects is equal to sum of the absolute values of energies of torus and antitorus of these dipoles:

$$E_G^0 = \sum^{N\to\infty} \beta(\mathbf{m}_V^+ + \mathbf{m}_V^-)\mathbf{c}^2 = \sum^{N\to\infty} \beta\mathbf{m}_0\mathbf{c}^2(2n+1)$$

When the in-phase pulsations of pairs of remote triplets turns to counterphase, depending on distance between objects or under magnetic field action, changing spin state of these fermions, the gravitation turns to antigravitation. The *antigravitation* is responsible for so-called *negative pressure or dark energy*.

This attraction gravitational energy of *'empty'* Bivacuum, when $\mathbf{m}_V^+ = \mathbf{m}_V^- = \mathbf{m}_0$ is generated by $\mathbf{VPW}_q^\pm$, radiated and absorbed in the process of symmetric transitions of torus and antitorus between excited and ground states: $\mathbf{E}_{\mathbf{VPW}_q^\pm} = \pm\mathbf{q}\,\hbar\boldsymbol{\omega}_0$, compensating each other: $+\mathbf{q}\,\hbar\boldsymbol{\omega}_0 = -\mathbf{q}\,\hbar\boldsymbol{\omega}_0$. Such mechanism of huge volumes of 'empty' Bivacuum determines the *cold dark matter effect;*

10. Maxwell's displacement current and the additional instant currents are the consequences of Bivacuum dipoles ($\mathbf{BVF}^{\updownarrow}$ and $\mathbf{BVB}^{\pm}$) in empty space symmetric excitations and vibrations, correspondingly. Their vibrations, corresponding to properties of *secondary* Bivacuum, represent reversible elastic deformations of Bivacuum matrix, induced by presence of fields and remote matter. The increasing of the excluded for photons volume of toruses and antitoruses due to their rotations and vibrations, enhance the refraction index of Bivacuum and decrease the light velocity near gravitating and charged objects. The nonzero contribution of the rest mass energy to photons and neutrino energy is a consequence of the enhanced refraction index of secondary Bivacuum and corresponding decreasing of the effective light velocity (for details see section 8.11). The latter can be revealed by small shift of Doppler effect in EM radiation of the probe in gravitational field. The *'Pioneer anomaly'* is a good example of such phenomena;

11. It is shown that the Principle of least action and realization of 2nd and 3d laws of thermodynamics for closed systems - can be a result of slowing down the dynamics of particles and their kinetic energy decreasing, under the influence of basic - lower frequency Virtual Pressure Waves ($\mathbf{VPW}^{\pm}_{q=1}$) with minimum quantum number $q = j - k = 1$. This is a consequence of forced combinational resonance between [$\mathbf{C} \leftrightarrow \mathbf{W}$] pulsation of particles and basic $\mathbf{VPW}^{\pm}_{q=1}$ of Bivacuum;

12. The dimensionless 'pace of time' ($\mathbf{dt/t} = -\mathbf{dT}_k/\mathbf{T}_k$) and *time of action* ($\mathbf{t}$) itself for each closed conservative system are determined by the change of this system kinetic energy. The time is positive, if dynamics of particles is slowing down and negative in the opposite case. This new concept of time is more advanced, than that of Einstein relativistic theory. For example, our formula for time includes not only velocity, but also acceleration of the object and frequency of its orbital rotation:

$$\mathbf{t} = \left[ -\frac{\vec{\mathbf{v}}}{\mathbf{d}\vec{\mathbf{v}}/\mathbf{dt}} \frac{1 - (\mathbf{v/c})^2}{2 - (\mathbf{v/c})^2} \right]_W = \left[ \frac{1}{\boldsymbol{\omega}} \frac{1 - (\vec{\mathbf{r}}\boldsymbol{\omega}/\mathbf{c})^2}{2 - (\vec{\mathbf{r}}\boldsymbol{\omega}/\mathbf{c})^2} \right]_C$$

where: $\boldsymbol{\omega} = \left( -\frac{\vec{\mathbf{v}}}{\mathbf{d}\vec{\mathbf{v}}/\mathbf{dt}} \right)^{-1} = \vec{\mathbf{v}}/\vec{\mathbf{r}} = 2\pi\boldsymbol{\nu}$ is the angular frequency of the object rotation with radius of orbit $\vec{\mathbf{r}}$.

In contrast to time definition, following from special relativity, the time of action is infinitive and independent on velocity in any *inertial* system of particles, when acceleration is zero. However, at any nonzero acceleration: $\mathbf{a} = \mathbf{dv/dt} = \boldsymbol{\omega}^2 r = \mathbf{G}\frac{M}{r^2} = \mathbf{const} > \mathbf{0}$, including case of orbital rotation, the time is dependent on velocity of these objects in more complex manner, than it follows from special relativity. In fact, there are no physical systems in our expanding with acceleration Universe which can be considered, as perfectly inertial. This means, that relativistic formula for time (12.15) is not valid in general case. It is demonstrated, that proposed 'time of action' theory confirms our model of elementary particles from sub-elementary fermions, including mass and charge origination, explains the Fermat principle and all experiments, which where considered, as a confirmation of special and general relativity;

13. Theory of Virtual Replica (**VR**) of material objects in Bivacuum and **VR** Multiplication: **VRM** ($\mathbf{r,t}$). The primary **VR** represents a three-dimensional (3D) superposition of Bivacuum virtual standing waves $\mathbf{VPW}^{\pm}_m$ and $\mathbf{VirSW}^{\pm 1/2}_m$, modulated by [$\mathbf{C} \rightleftharpoons \mathbf{W}$] pulsation of elementary particles and translational and librational de Broglie waves of molecules of macroscopic object (http://arxiv.org/abs/physics/0207027). The infinitive multiplication of primary **VR** in space in form of 3D packets of virtual standing waves, representing set of *secondary* **VR**: **VRM(r)**, is a result of interference of all pervading external coherent basic *reference waves* - Bivacuum Virtual Pressure Waves ($\mathbf{VPW}^{\pm}_{q=1}$) and Virtual Spin Waves ($\mathbf{VirSW}^{\pm 1/2}_{q=1}$) with similar kinds of modulated standing waves, forming $\mathbf{VR}$. The $\mathbf{VR}$ plays the role of the *object waves*. This phenomena may stand for *remote vision* of psychic. The ability of enough complex system of $\mathbf{VRM(r,t)}$ to self-organization in nonequilibrium conditions, make it possible multiplication of primary $\mathbf{VR}$ not only in space but as well, in time in both time direction - positive (evolution) and negative (devolution). The feedback reaction between most probable/stable $\mathbf{VRM(r,t)}$ and nerve system of psychic, including visual centers of brain, can by responsible for *clairvoyance*. The $\mathbf{VRM}$ of elementary particles coincides with notion of their *secondary anchor sites*, representing three conjugated Cooper pairs $3[\mathbf{BVF}^{\uparrow} \bowtie \mathbf{BVF}^{\downarrow}]^{i}_{as}$ of asymmetric Bivacuum fermions. The stochastic jumps of $\mathbf{CVC}^{\pm}$ of [W] phase of particle from one anchor site to another and the ability of interference of single particle with its own *anchor site* explains two slit experiment;

14. The new general presentation of wave function, based on our wave-corpuscle duality model, takes into account not only the external *translational* dynamics of particle, but also the internal *rotational-translational* one, responsible for the rest mass and charge origination;

15. The *eigen wave functions*, as a solutions of Shrödinger equation, describe the linear superposition of *multiple anchor site,* as a possible alternatives for realization of particle's [C] phase;

16. A possible Mechanism of Quantum entanglement between remote coherent elementary particles: electrons and nuclears of atoms of Sender(S) and Receiver(R) via Virtual Guides of spin, momentum and energy ($\mathbf{VirG_{S,M,E}}$) is proposed. The single $\mathbf{VirG^{BVB^{\pm}}_{S,M,E}}$ can be assembled from Bivacuum bosons ($\mathbf{BVB}^{\pm}$)$^{i}$ by 'head-to-tail' principle. The doubled $\mathbf{VirG^{BVF^{\uparrow}\bowtie BVF^{\downarrow}}_{S,M,E}}$ from the adjacent microtubules, rotating in opposite directions, can be formed by Cooper pairs of Bivacuum fermions $[\mathbf{BVF}^{\uparrow}\bowtie \mathbf{BVF}^{\downarrow}]^{i}$, polymerized by the same principle. The spin/information transmission via Virtual Guides is accompanied by reorientation of spins of tori and antitori of Bivacuum dipoles. The momentum and energy transmission from S to R is realized by the instant pulsation of diameter of such virtual microtubule with frequency of beats, equal to difference between frequencies of $\mathbf{C} \rightleftharpoons \mathbf{W}$ pulsation of S and R. The length of $\mathbf{VirG_{S,M,E}}$, connecting fluctuating in space particles of (S) and (R), also can correspondingly vary, because of immediate self-assembly/disassembly of $\mathbf{VirG_{S,M,E}}$ from the infinitive source of Bivacuum dipoles. The Virtual Guides of both kinds represent the quasi 1D virtual Bose condensate with nonlocal properties, similar to that of 'wormholes'. The bundles of $\mathbf{VirG}_{SME}$, connecting coherent atoms of Sender (S) and Receiver (S), as well as nonlocal component of $\mathbf{VRM(r,t)}$, determined by interference pattern of Virtual Spin Waves, are responsible for nonlocal weak interaction;

17. The introduced *Bivacuum Mediated Interaction* ($\mathbf{BMI}$) is a new fundamental interaction, resulting from superposition of Virtual replicas of Sender and Receiver, because of $\mathbf{VRM(r,t)}$ mechanism, and connection of their coherent elementary particles in coherent atoms via Virtual Guides bundles:

$$\left[ \mathbf{N(t,r)} \times \sum^{n} \mathbf{VirG}_{SME} (\mathbf{S} <=> \mathbf{R}) \right]^{i}_{x,y,z}$$

Just $\mathbf{BMI}$ is responsible for remote ultraweak nonlocal interaction. The system: [S + R] should be in nonequilibrium state.